\documentclass[12pt, epsfig]{article}
\usepackage[dvips]{color}
\usepackage{epsfig}%
\usepackage[active]{srcltx}%
\usepackage{amsmath,amsfonts,amssymb,amsthm,amstext,amscd,eucal,srcltx}
\usepackage{epsfig,graphicx,bm}
\usepackage{epstopdf, epsf}
\usepackage{dcolumn}
\usepackage{hyperref}
\usepackage{subfigure}
\usepackage{textcomp}

\definecolor{darkgreen}{rgb}{0,0.3,0}
\definecolor{darkblue}{rgb}{0,0,0.3}
\definecolor{darkred}{rgb}{0.7,0,0}

\newcommand{\be}{\begin{equation}}
\newcommand{\bse}{\begin{subequations}}
\newcommand{\ese}{\end{subequations}}
\newcommand{\bea}{\begin{eqnarray}}
\newcommand{\eea}{\end{eqnarray}}
\newcommand{\ba}{\begin{array}}
\newcommand{\ea}{\end{array}}
\newcommand{\ee}{\end{equation}}
\newcommand{\nn}{\nonumber}

\def\VEV#1{\langle{ #1} \rangle}

\makeatletter \@addtoreset{equation}{section}

\makeatletter\renewcommand\section{\@startsection {section}{1}{\z@}%
                                   {-3.5ex \@plus -1ex \@minus -.2ex}
                                   {2.3ex \@plus.2ex}%
                                   {\normalfont\large\bfseries}}
\renewcommand\subsection{\@startsection{subsection}{2}{\z@}%
                                     {-3.25ex\@plus -1ex \@minus -.2ex}%
                                     {1.5ex \@plus .2ex}%
                                     {\normalfont\bfseries}}

\def\kk{\frac{\kappa g^2\phi^4}{a^4}}
\def\dk{\frac{\kappa \dot\phi^2}{a^2}}
\def\dre_g{\delta\rho_g}
\def\dpe_g{\delta P_g}
\def\dqe_g{\delta q_g}
\def\dre{\delta\rho}
\def\dpe{\delta P}
\def\dqe{\delta q}
\def\tM{\tilde M}
\def\mH{\mathcal{H}}

\def\YM1{\frac{\dot\phi^2}{a^2}}
\def\YM2{\frac{g^2\phi^4}{a^4}}
\newcommand{\ie}{{\em i.e. }}

\newcommand{\mpl}{M_{\rm pl}}
\def\tchi{\tilde\chi}
\def\cL{\mathcal{L}}
\def\Fmn{F_{\mu\nu}}

\def\gmn{g_{\mu\nu}}

\begin{document}
\begin{titlepage}
\begin{flushright}\vspace{-3cm}
{\small
IPM/P-2012/049\\
KUNS-2430\\
\today }\end{flushright}

\begin{center}

{\LARGE{\bf Gauge Fields and Inflation}}
\vskip 1cm

{\Large{\bf{{A. Maleknejad$^{\dag,}$\footnote{azade@ipm.ir}, M.M. Sheikh-Jabbari$^{\dag,}$\footnote{jabbari@theory.ipm.ac.ir}, J. Soda$^{\ddag,}$\footnote{jiro@tap.scphys.kyoto-u.ac.jp}}}}}

\vskip 1cm
$^\dag$\textit{School of Physics, Institute for research in fundamental sciences
(IPM), P.O.Box 19395-5531, Tehran, Iran }
\vskip 5mm

$^\ddag$\textit{Department of Physics, Kyoto University, Kyoto, 606-8502, Japan}

\date{\today}

\end{center}

\vspace{8mm}
\setcounter{footnote}{0}


\begin{abstract}
\noindent The isotropy and homogeneity of the cosmic microwave background (CMB) favors ``scalar driven'' early Universe inflationary models. However, gauge fields and other non-scalar fields are far more common at all energy scales, in particular at high energies seemingly relevant to inflation models. Hence, in this review we consider the role and consequences, theoretical and observational, that gauge fields can have during inflationary era. Gauge fields may be turned on in the background during inflation, or may become relevant at the level of cosmic perturbations. There have been two main class of models with gauge fields in the background, models which show violation of cosmic no-hair theorem and those which lead to isotropic FLRW cosmology, respecting the cosmic no-hair theorem. Models in which  gauge fields are only turned on at the cosmic perturbation level, may source primordial magnetic fields. We also review specific observational features of these models on the CMB and/or the primordial cosmic magnetic fields. Our discussions will be mainly focused on the inflation period, with only a brief discussion on the post inflationary (p)reheating era.

\end{abstract}


\vspace{0.5in}

\end{titlepage}

\tableofcontents

\section{Introduction}

The prospects of research in cosmology has been in an ever accelerating expansion since the discovery of Hubble expansion in late 1920's. The cosmological and astrophysical observations since then has been in favor of the Big Bang model  which has been  used to explain the presence of Cosmic Microwave Background (CMB), cosmic abundances of light nuclei and to reconstruct the thermal history of the Universe especially after the Big Bang Nucleosynthesis (BBN) when the temperature of the Universe  was around $1$ MeV.
Gaining a detailed knowledge and picture of the history of Universe before that is limited since the main messengers from the early University are the electromagnetic waves reaching to us, and that they can only (almost) freely propagate after the recombination time, when the CMB released.

Despite of the successes,  some basic questions about the early Universe remain unanswered within the (hot) Big Bang model. These questions which are horizon problem, flatness problem, relic problem, the source for CMB anisotropy (CMB temperature fluctuations) and seeds for the large structures, can find a suitable and natural answer once the hot Big Bang model is augmented with an inflationary era, a period of accelerated expansion \cite{Guth:1980zm, Inflation-Books}. This constitutes the Standard Model of cosmology. In this review we will focus on the inflationary period.

Many different models of inflation have been proposed and studied in the literature. In these models inflation is generically driven by the coupling of one or more scalar fields to gravity, and the dynamics during inflation is such that generically the potential energy of the fields dominate over their kinetic term and the potential is flat enough
to ensure the so-called slow-roll inflation (to be defined in section \ref{Inflation-review-section}). The current observations indicate that the Hubble parameter during inflation $H$ has an upper bound, $H\lesssim 10^{-5}\mpl$, corresponding to inflaton energy density of order $\rho_{inf}\lesssim 10^{-10}\mpl^4\sim  (10^{16} \mathrm{GeV})^4$. ($\mpl$ is the reduced Planck mass. Our conventions and notations are summarized in Appendix \ref{Convention-appendix}.) The inflaton energy density is at least two orders of magnitude smaller than $\mpl$ and of order of the energy scale of the Grand Unified Theories (GUT's) of particle physics. One would hence naturally expect that inflation should be formulated within the proposed particle physics models working in the same energy range.

Models of particle physics  are (chiral) gauge field theories the matter content of which generically include gauge fields, (chiral) fermions and scalar (Higgs-type) fields, which are in various representations of the gauge group. The scalar field(s) of these models are  used for spontaneous symmetry breaking (SSB) and hence giving masses to chiral fermions through Yukawa couplings. Therefore, their potential is generically tuned for the symmetry breaking purposes. This latter condition is, however, at odds with the slow-roll inflation requirement demanding a lower scalar masses.
In other words, generically, the same potential used for the Higgs-type fields of (beyond) particle physics standard model is not fit for inflation.\footnote{There are, however, ways to get around this general argument in specific models by relaxing the assumptions this result is based on. For example the above argument was based on the implicit assumption that gravity at the inflation scale is described by the Einstein GR and that Higgs-type fields are minimally coupled to gravity \cite{Higgs-inflation}. Another way to get around this argument is to explore the supersymmetric models, which have a huge wealth of scalar fields as superpartners of standard model fermions, and more options for fulfilling the flatness of the potential, e.g. see \cite{inflection-point}. In this review our discussions will mainly be within Einstein GR with minimally coupled fields.}

The above general features would hence prompt the idea of using fields other  than scalars in particle physics models for inflationary model building. Turning on non-scalar fields during inflation appears at first to be incompatible with isometry and rotational symmetry of the observed Universe; anisotropy at cosmological scales, if exists, should remain small enough to be compatible with the observations. This article is partly devoted to reviewing different ways proposed in the literature so far, in which gauge fields can be turned on in the inflationary background such that the observational anisotropy constraints are also satisfied. We will also review the observational signatures and features of these models.

Effects of gauge fields may arise at the level of perturbations, without being turned on at the background level, or gauge fields may be turned on at the background level, as well as arising through perturbations.  The former group, is what we first analyze in section \ref{prim-magnet-NG-section} and the latter is what we study in the rest of the review. In the latter group of models with vector \emph{gauge} fields in the background, one can distinguish two general classes.\footnote{In this review we will not consider the vector inflation models \cite{vector-inflation}. Having a vector field with standard kinetic term, \emph{without the gauge symmetry}, i.e. when the vector field is not a gauge field, we expect to have ghost instability \cite{vector-inflation-ghost} (unless the gauge symmetry is broken via SSB mechanism) and hence not generically a theoretically viable framework.}
The first class includes models where contribution of the gauge field to the energy budget during inflation is small, i.e. although gauge fields are turned on in the inflationary background,  inflation is mainly driven by other scalar fields. These models can have a controlled/controlable, but yet non-exponentially suppressed, anisotropy at both background and perturbation levels.
The other class mainly consists of gauge-flation and chromo-natural inflation models where the contribution of the gauge field sector has the dominant contribution to the energy budget of the Universe during inflation, while the background geometry is isotropic.

This review is organized as follows. Section \ref{Inflation-review-section} is a very quick review of basics of inflationary cosmology. In this section we also fix our conventions and notations, as well as summarizing the current observational data which is used to constrain inflationary models. We also mention CMB anomaly which motivated people to study the statistical anisotropy or the effect of gauge fields on inflation. In section \ref{prim-magnet-NG-section}, we analyze models with gauge fields at perturbation level, which can be used to provide seeds for primordial magnetic fields and may also source non-Gaussianity (NG) on the CMB power spectrum and primordial gravitational waves. In section \ref{Anisotropic-inflation-section}, we discuss anisotropic inflationary models, with gauge fields in the background. In these models contribution of the gauge field to the inflationary energy budget  remains small. Consequently, the anisotropy in the expansion is also small. Nonetheless, observational effects of gauge fields  may not be negligible. We explain how anisotropic interaction induces interesting phenomenology. We also discuss the possibility of testing anisotropic inflation. In section \ref{Gauge-flation-section}, we review the gauge-flation model, a model
in which inflation is driven by a non-Abelian gauge theory minimally coupled to Einstein GR. We review and study inflationary background and study the stability of the inflationary trajectories in this model. We also discuss cosmic perturbation theory in gauge-flation and confront the model with the observational data. In section \ref{Chromo-natural-section}, we analyze an extension of the gauge-flation model, where besides the non-Abelian gauge fields we have an axion field,  ``chromo-natural inflation''. We present both background and perturbation theory of the chromo-natural model. The last section is devoted to discussion and outlook. To make the review self-contained in some appendices  we have gathered  discussions and analysis which are related to the topic of this article.  Appendix \ref{Convention-appendix} contains our conventions. In Appendix \ref{deltaN-appendix} we present a quick review of $\delta N$ formulation for cosmic perturbation theory analysis. Appendix \ref{Bianchi-appendix} contains a review of Bianchi cosmologies relevant to the anisotropic models.  In Appendix \ref{No-hair-extension-appendix},  we present Wald's cosmic no-hair theorem \cite{Wald:1983ky} and the way it should be extended in inflationary cosmology \cite{Wald-extended-theorem}.

\section{Preliminaries of inflationary cosmology}\label{Inflation-review-section}

Inflationary paradigm has secured its status as the leading candidate with the currently available cosmic data.
Moreover, inflation has the appealing feature that it can be formulated within the
standard existing Einstein GR and quantum field theory frameworks. In this section we will briefly discuss basics of  FLRW cosmology  and review some simple models of inflation, through this we also fix the notations we will be using throughout the article. To this end, we first review background (classical) slow-roll inflationary  trajectories, and then review cosmic perturbation theory. We also present a summary of the current observation data, coming from combined CMB and related astrophysical data.

Our discussions in this section will be very brief. For more detailed discussion the reader is encouraged to consult the books and reviews on inflation, an incomplete list include \cite{Inflation-Books}.
Throughout this note we use natural units where the reduced Planck mass $\mpl$ is 1.

\subsection{Inflationary backgrounds}\label{Inf-background-review}

The standard cosmology is based on the Einstein general relativity which states that the dynamical evolution of our universe is described by a 4-dimensional geometry, $g_{\mu\nu}$, governed by the Einstein equations, as
\be%
G_{\mu\nu}=R_{\mu\nu}-\frac12g_{{\mu\nu}}R=T_{\mu\nu}.%
\ee%
Here $R_{\mu\nu}$ is the Ricci tensor, $R=R_{\mu\nu}g^{\mu\nu}$ is the Ricci scalar and $T_{\mu\nu}$ is the energy-momentum tensor of the matter in the Universe.
Furthermore, modern cosmology starts with the assumption of cosmological principle \cite{Inflation-Books}:
Universe is homogeneous and isotropic on cosmological large
distances, 100 Mpc and larger. Thus, the line element of the space-time is given by the Friedman-Lema\v{i}tre-Robertson-Walker (FLRW) metric
\be\label{FLRW}%
ds^2=-dt^2+a^2(t)\biggl(\frac{dr^2}{1-Kr^2}+r^2(d\theta^2+\sin^2\theta d\phi^2)\biggr),%
\ee%
where $t$ is the cosmic time, $a(t)$ is the scale factor and $K$ is the curvature constant and takes $+1,\ 0$ and $-1$ values respectively corresponding to close, flat and open Universes. Due to the spatial isotropy and homogeneity of the metric \eqref{FLRW}, the most general form of $T_{\mu\nu}$ is a perfect fluid
 \be\label{Tmunu}%
 T_{\mu\nu}=(\rho+P) u_\mu u_\nu+Pg_{\mu\nu},
 \ee%
where $\rho$ and $P$ are the energy and pressure densities.  The Einstein equations for the isotropic homogeneous FLRW background reduces to the Friedman and Raychaudhuri equations, respectively
\bse
\begin{align}\label{H}
& H^2=\frac13\rho-\frac{K}{a^2},\\
\label{dda}
 &\frac{\ddot{a}}{a}=-\frac{1}{6}(\rho+3P),
\end{align}\ese
where a dot represents derivative with respect to cosmic time and $H\equiv\frac{\dot a}{a}$ is the Hubble rate of the expansion.
Ordinary matter (e.g. dust, radiation) have positive energy density and non-negative pressure and satisfy strong energy condition\footnote{Strong energy condition states that for all time-like $t^\mu$, we have $(T_{\mu\nu}-\frac12g_{\mu\nu}T)t^\mu t^\nu\geq0$.} ($\rho+3P>0$) which makes the RHS of equation \eqref{dda} positive. Thus, in conventional Big Bang theory which assumes $\rho+3P>0$, the Universe always exhibits a decelerated expansion ($\ddot a < 0$) \cite{Inflation-Books}.

Standard Big Bang theory, however, is not capable of explaining the current state of the Universe as we observe it today.  Some of the most important problems with  observations are
the \textit{flatness problem}
(why our Universe is so nearly
spatially flat),
the \textit{horizon problem} (why the temperature of the CMB on the whole
sky is so accurately the same),
the \textit{monopole or heavy relic problem}. Inflation, an accelerated expansion, $\ddot a>0$, phase was proposed to overcome these problems \cite{Guth:1980zm}. For instance, inflation offers an explanation of flatness problem without any fine-tuning, since due to inflation the spatial curvature
term, $\frac{K}{a^2}$ in \eqref{H}, will be negligible. Furthermore, the fact that our entire
observable Universe might have arisen from a single early causal patch can solve the
horizon problem. The inflationary paradigm not only solves the above cosmological puzzles, but more importantly, also offers an explanation for the origin of large scale structures as well as CMB temperature anisotropy. See \cite{Inflation-Books} for more detailed discussions. The simplest and most extensively studied inflation scenario is a \emph{quasi de Sitter} inflation, almost exponential expansion,  where equation of state of cosmic fluid is $P\simeq-\rho$, $H$ is almost a constant, and the Friedman equations has the solution
\be
a(t)\simeq a(0)\exp(Ht)\,.
\ee

Based on the matter which drives accelerated expansion, inflation may have many realizations. The simplest inflationary model is a single scalar field, $\varphi$, called inflaton field, minimally coupled to Einstein gravity
\be\label{ssmodel}
S=\int d^4x\sqrt{-g}(\frac{R}{2}-\frac12(\partial_\mu\varphi)^2-V(\varphi)).
\ee
Neglecting spatial gradients, the energy and pressure density are given by
\be\label{ssT}
\rho=\frac12\dot\varphi^2+V,\quad
P=\frac12\dot\varphi^2-V.
\ee
Also, the Friedman equation and the equation of motion of the scalar field are respectively
\bea
&~&3H^2=\frac12\dot{\varphi}^2+V(\varphi),\\
&~&\ddot\varphi+3H\dot\varphi+V_{\varphi}=0,
\eea
where $V_{\varphi}\equiv\frac{\partial V}{\partial\varphi}$.
Note that since the spatial curvature term, $\frac{K}{a^2}$ in \eqref{H}, damps exponentially during the inflation, we dropped it.
As we can see from \eqref{ssT},  in the limit that $\dot\varphi^2\ll V$, the
the equation of state satisfies $\rho\simeq-P$, and an inflationary phase is possible with
\be
H^2 \simeq\frac13V(\varphi).
\ee
In fact, in this simple model inflation can only
occur if the scalar field satisfies the slow-roll conditions. \textit{Slow-roll inflation} is
quantified  in terms of the  slow-roll parameters, $\epsilon$ and $\eta$
\be\label{epsilon-eta-def}
\epsilon\equiv-\frac{\dot H}{H^2},\quad \eta\equiv-\frac{\ddot H}{2H\dot H},
\ee
and  requires $\epsilon,\vert\eta\vert\ll1$. Smallness of $\epsilon$ and $\eta$ is usually needed to make sure that inflation lasted long enough to solve the horizon and flatness problems.
A useful quantity to measure the amount of inflation is the number of e-folds $N_e$
\be\label{Number-e}
N_e\equiv\ln(\frac{a_{end}}{a_0})=\int^{t_f}_{t_0} H(t)dt\,,
\ee
where $t_0$ and $t_f$ denote the initial and the final time of inflation respectively, and $a_0$ is the value of the scale factor at the beginning of the inflation.
As an often-stated rule of thumb, in order to solve the horizon and flatness problems, we need about 60
e-folds \cite{Inflation-Books}. Thus,
$N_e=60$ is taken as a standard minimum number
of e-folds for inflationary models. However, the precise value of the required $N_e$ depends on the energy scale of the inflation and the details of the reheating era \cite{Inflation-Books}. Furthermore,  perturbations (temperature anisotropy) observed in the CMB have left the horizon $N_{CMB}\simeq40-60$ e-folds before the end of inflation.

Slow-roll parameters defined in \eqref{epsilon-eta-def} are based on the time derivative of the Hubble parameter $H$ during inflation. For specific models of inflation, however, it is often more useful to define the slow-roll parameters directly in terms of the inflaton field Lagrangian. For the simple  single scalar field model given in \eqref{ssmodel}, one can easily show that slow-roll parameters \eqref{epsilon-eta-def} can also be written as
\be\label{epsilon-eta-single-field}
\epsilon=\frac12\frac{\dot\varphi^2}{H^2},\quad\eta=-\frac{\ddot\varphi}{H\dot\varphi}.
\ee
It is also common to express the slow-roll parameters in terms of the potential $V(\varphi)$ as
\footnote{Note that $\epsilon$ and $\eta$ are called the Hubble slow-roll parameters while $\epsilon_V$ and $\eta_V$ are called potential slow-roll parameters. During the slow-roll these parameters are related as $\epsilon\simeq\epsilon_V$ and $\eta_V=\epsilon+\eta$.}
\be\label{epsilon-eta-Potential}
\epsilon_V=\frac12(\frac{V_{\varphi}}{V})^2,\quad \eta_V=\frac{V_{\varphi\varphi}}{V}.
\ee
Finally, inflation ends when $\epsilon(\varphi_{end})=1$ and number of e-folds $N_e$ is given as
\be\label{N}
N_e\simeq\int^{\varphi_0}  _{\varphi_{end}}\frac{V}{V_{\varphi}}d\varphi\simeq\int^{\varphi_0}  _{\varphi_{end}}\frac{1}{\sqrt{2\epsilon}}d\varphi.
\ee

Depending on the form of the potential, and also possibly the kinetic term which can be non-canonical in general, we may have different single-field models. A useful, but not exhaustive, classification of  \textit{single-field inflationary models} is as follows.
\begin{itemize}
\item{\textit{Large field models:} The initial value of the inflaton field is large, generically super-Planckian, and it
rolls slowly down toward the potential minimum at smaller $\varphi$ values. For instance, chaotic inflation is one of the representative models of this class.
The typical potential of large-field models has a monomial form as
\be\label{chaotic-potential}
V (\varphi) = V_0\varphi^n .
\ee
A simple analysis using the dynamical equations reveals that for  number of e-folds $N_e$ larger than
$60$, we require super-Planckian initial field values\footnote{In the presence of another natural cutoff $\Lambda$ in the model, smallness or largeness of the inflaton field should be compared to $\Lambda$; $\Lambda$ could be sub-Planckian and in general $\Lambda\lesssim\mpl$. For a discussion on this see \cite{cutoff-species,Nflation-eternal-infltion}.}, $\varphi_0>3 M_{pl}$. For these models typically $\epsilon\sim \eta \sim N_e^{-1}$. }

\item{\textit{Small field models:} Inflaton field is initially small and slowly evolves toward the potential
minimum at larger $\varphi$ values.  The small field models are characterized by the following
potential
\be\label{small-field}
V (\varphi) = V_0(1-(\frac{\varphi}{\mu})^n),
\ee
which corresponds to a Taylor expansion about
the origin, but more realistic small field models also have a
potential minimum at $\varphi\neq0$ which the system falls in at the end of inflation.
A typical property of small field models is that a sufficient number of e-folds, requires a sub-Planckian inflaton initial value. For this reason they are called small field models. Natural inflation is an example of this type \cite{Freese:1990rb}.}
\item{\textit{Hybrid inflation models:}  These models involve more than one scalar
field while inflation is mainly driven by a single inflaton field $\phi$. Inflaton starts from a large value rolling down until it reaches a bifurcation point, $\phi=\phi_e$, after which
the field becomes unstable and undergoes a waterfall
transition towards its global minimum. Its prime example is the Linde's hybrid inflation model with the following potential \cite{Linde:1993cn}
\be\label{Hybrid}
V(\phi,\chi)=\frac{\lambda}{4}(\chi^2-\frac{M^4}{\lambda})^2+\frac12g^2\phi^2\chi^2+\frac12m^2\phi^2.
\ee
During
the initial inflationary phase the potential of the hybrid
inflation is effectively described by a single field $\phi$ while
inflation ends by a phase transition triggered
by the presence of the second scalar field, the waterfall field $\chi$.
In other words, when the effective mass squared of a waterfall field becomes negative,
the tachyonic instability makes waterfall field roll down toward the true vacuum state and
the inflation suddenly ends.

Number of e-folds $N_e$ is given as
\be
N_e\simeq\frac{M^4}{4\lambda m^2}\ln(\frac{\phi_0}{\phi_e}),
\ee
where $\phi_e=\frac{M}{g}$ is the critical
value of the inflaton below which, due to tachyonic instability, $\chi=0$ becomes unstable
and $m^2_\chi$ gets negative.}

\item{\textit{K-inflation:} This is the prime example of models with non-canonical Kinetic term we discuss here. They are described by the action \cite{ArmendarizPicon:1999rj}
\be\label{K-inflation}
S=\int d^4x\sqrt{-g}\big(\frac{R}{2}+P(\varphi,X)\big),
\ee
where $\varphi$ is a scalar field and $X:=-\frac12(\partial_\mu\varphi)^2$. Here, $P$ plays the rule of the effective pressure, while the energy density is given by
\be
\rho=2X P_{,X}-P.
\ee
Thus, the slow-roll parameter is given as $$\epsilon=\frac{3XP_{,X}}{2XP_{,X}-P}.$$ The characteristic feature of these models is that in general they have a non-trivial sound speed $c_s^2$ for the propagation of perturbations (\emph{cf.} our discussion in section \ref{cosmic-perturbation-review-section})
\be\label{K-inflation-speed-sound}
c_s^2\equiv\frac{P_{,X}}{P_{,X}+2XP_{,XX}}\,.
\ee
Finding K-inflation actions $P(\varphi, X)$ which are well-motivated and consistently embedded in high-energy theories is the main challenge of this class of models \cite{Baumann:2009ds}.} Nonetheless,
{\textit{DBI inflation} is a special kind of K-inflation, which is  well-motivated from  string theory with the action \cite{DBI-original}
\be
S=\int d^4x\sqrt{-g}\bigg[\frac{R}{2}-\frac{1}{f(\varphi)}\left((\sqrt{\mathcal{D}}-1)+V(\varphi^I)\right)\bigg],
\ee
where $\mathcal{D}=1-2f(\varphi)X$.}
\end{itemize}

\subsection{Cosmic perturbation theory}\label{cosmic-perturbation-review-section}

The FLRW Universe is just an approximation to the Universe we see, as it ignores all the structure and other observed
anisotropies e.g. in the CMB temperature. One of the great achievements of inflation is having a naturally embedded mechanism
to account for these anisotropies. The idea is based on the fact that inflaton(s) and metric are indeed quantum fields and hence
have quantum fluctuations. These quantum fluctuations are, however, ``virtual'' in the sense that they do not carry energy.
Nonetheless, the cosmic expansion  and  existence of (cosmological) horizon makes it possible for some of these quantum
fluctuations become ``real'' and classical.

Quantum fluctuations appear in all wavelengths and \emph{comoving}
momentum numbers $k$. For the \emph{subhorizon} modes $k\gg aH$ these fluctuations essentially appear as they are in flat spacetime, while
for \emph{superhorizon} modes $k\lesssim aH$ fluctuations do not show oscillatory behavior, they freeze and behave as
classical perturbations on the FLRW background. Given that $k$ is constant and scale factor $a(t)$ expands (almost exponentially),
if we wait long enough any mode becomes classical. Nevertheless, inflation ends and not all the modes have had time
to cross the horizon during inflation.\footnote{
The above intuition, that modes become essentially classical at superhorizon scales, has been the basis for developing a powerful technique, the $\delta N$ formulation \cite{delta-N}, for dealing with the perturbations. We have reviewed this in Appendix \ref{deltaN-appendix}.}
Moreover,
termination of inflation also opens the crucial possibility that some of the modes which have become superhorizon modes \emph{during inflation}
to become \emph{subhorizon} again in a later time after inflation ended and \emph{reenter} the horizon. In particular, some of these modes, which are the modes
left the horizon during $40-60$ e-folds before the end of inflation, have reentered our cosmological horizon after the surface
of last scattering and have been imprinted on the CMB. These classical modes are responsible for both structure formation and
the CMB anisotropy. The precision measurements  on the CMB anisotropies can then be used to restrict inflationary models.
In this section, we review cosmic perturbation theory for inflationary models and extract data from the models which can be
compared with the CMB observations to be reviewed in the next subsection.

The standard cosmic perturbation theory starts with the assumption that during
most of the history of the Universe deviation from homogeneity and
isotropy at cosmological scales have been small, such that they can be treated as first-order perturbations
\cite{Inflation-Books}. Furthermore, the distribution of these perturbations (in the first order) is assumed to be \emph{statistically}
homogeneous and isotropic. Since the observable Universe is nearly homogeneous, and its
spatial curvature either vanishes or is negligible until very near the present epoch,
we will take the unperturbed metric to be flat FLRW \eqref{FLRW} with $K=0$.

The most general perturbed FLRW metric can be written as%
\be\label{metric-pert}%
ds^2=-(1+2A)dt^2+2a(\partial_iB+V_i)dx^idt
+a^2\left((1-2C)\delta_{ij}+2\partial_{ij}E+2\partial_{(i}W_{j)}+h_{ij}\right)dx^idx^j\,,
\ee
where $\partial_i$ denotes a derivative with respect to $x^i$ where lower case Latin indices run over the three spatial coordinates. Due to the spatial isotropy and homogeneity of the unperturbed FLRW metric and energy-momentum tensor it is convenient to decompose the perturbations
into scalars, divergence-free vectors, and divergence-free traceless
symmetric tensors, which are not coupled to each other by the field equations
or conservation equations, up to the linear order.
In the perturbed metric, $A, B,C$ and $E$ are four scalars, while $V_i$ and $W_i$ are two divergence-free vectors and $h_{ij}$ is a divergence-free traceless symmetric tensor.

The most general first order perturbed energy-momentum tensor around the perfect fluid \eqref{Tmunu} can be described as \cite{Inflation-Books}
\begin{subequations}\label{pert-Tmunu}
\begin{align}
\delta T_{ij}=&\bar P\delta g_{ij}+a^2\left(\delta_{ij}(\delta P-\frac13\nabla^2\pi^S)+\partial_{ij}\pi^S+\partial_i\pi^V_j+\partial_j\pi^V_i+\pi^T_{ij}\right)\,,\\
\delta T_{i0}=&\bar P\delta g_{i0}-(\partial_i\delta q+\delta q_i^V)\,,\\
\delta T_{00}=&-\bar\rho\delta g_{00}+\delta \rho\,.
\end{align}
\end{subequations}
where $\bar\rho$ and $\bar P$ are the unperturbed energy density, pressure, respectively while $\delta\rho$ and $\delta P$ are their corresponding perturbations. Furthermore, $\pi^S$, $\pi^V_{~i}$ and $\delta\pi^T_{~ij}$ represent scalar, divergenceless vector and divergence-free, traceless tensor parts of dissipative corrections to the inertia tensor, respectively.
In addition, $\delta q$ is the scalar part of the perturbed 3-momentum, while $\delta q^V_i$ represents its divergence-free vector part. Note in particular that, the conditions $\pi^S=\pi^V_{~i}=\pi^T_{~ij}=0$ describe a
{perfect fluid} and $\delta q^V_{~i}=0$ represents an \textit{ irrotational flow}.
Being a perfect fluid or having irrotational flows are physical properties,
thus their corresponding conditions are gauge-invariant. In other words, $\pi^S, \pi^V_{~i}, \pi^T_{~ij}$
and $\delta u^V_{~i}$ are all invariant under infinitesimal space-time coordinate transformations.

 We have already used the rotational symmetry of the system for decomposing perturbations into scalar, vector,
and tensor modes. The equations have also  a symmetry under spatial translations which make it possible to work with the Fourier components of perturbations. Once we treat perturbations as infinitesimal, Fourier components of different wave number are decoupled from each other \cite{Inflation-Books}.

One important issue we need to care about is gauge degrees of freedom.
In fact, when we define perturbed quantity, we have to make the following difference
\begin{eqnarray}
  \delta Q (x^i ,t) = \tilde{Q}(x^i ,t ) - Q(t) \ ,
\end{eqnarray}
where $\tilde{Q} (x^i ,t)$ is a real physical variable and $Q(t)$ is a fiducial background variable. These two quantities reside in different spacetimes. Hence, we need to identify two points in different spacetimes to take the difference. Apparently, this is ambiguous,
which is the source of the gauge degrees of freedom.
Because of the presence of the gauge transformations which are generated by spacetime diffeomorphisms, not all metric and energy-momentum tensor perturbations are physically meaningful \cite{Inflation-Books}.
The gauge degrees of freedom may be removed by gauge-fixing (working in a specific gauge) or working with gauge-invariant combinations of the perturbations. In what follows we work out the gauge-invariant combinations of these modes.

\subsubsection*{$\bullet$ Scalar modes}

Let us first focus on the scalar perturbations. This sector is the most involved and interesting one, involving  eight scalars  $A$, $B$, $C$, $E$, $\delta\rho$, $\delta P$, $\delta q$ and $a^2\pi^S$. Of course, not all of
these quantities are gauge invariant.  The gauge transformation in the scalar sector
can be generated by the  infinitesimal ``scalar'' coordinate transformations
\be\label{coor}
\begin{split}
t&\rightarrow \tilde{t}=t+\delta t\,,\\
x^i&\rightarrow \tilde x^i=x^i+\delta^{ij}\partial_j \delta x\,,
\end{split}
\ee
where $\delta t$ determines the time slicing and $\delta x$ the spatial threading. In order to remove these gauge degrees of freedom, we can construct two gauge invariant combinations from the metric perturbations, the Bardeen potentials,
\bse\label{Bardeen-potentials}
\begin{align}
\Psi=&C+a^2H(\dot{E}-\frac{B}{a})\,,\\
\Phi=&A-\frac{d}{dt}\left(a^2(\dot{E}-\frac{B}{a})\right)\,,
\end{align}
\ese
and the four gauge invariant scalar parts of $\delta T_{\mu\nu}$
\bse\label{diff-inv-Tmunu}
\begin{align}
\delta\rho_g=&\delta\rho-\dot{\bar \rho}a^2(\dot{E}-\frac{B}{a})\,,\\
\delta P_g=&\delta P-\dot{\bar P}a^2(\dot{E}-\frac{B}{a})\,,\\
\delta q_g=&\delta q+(\bar\rho+\bar P)a^2(\dot{E}-\frac{B}{a})\,,
\end{align}\ese
and the anisotropic stress\footnote{Since the background energy-momentum tensor has the form of a perfect fluid, $a^2\pi^S$ is a  gauge invariant quantity \cite{Inflation-Books}.}  $a^2\pi^s$  \cite{Inflation-Books}, where $\delta q=(\bar\rho+\bar P)\delta u$.

 Out of the ten perturbed Einstein equations, there are four scalars, two (divergence-free) vectors, and one massless tensor mode (gravitons). Among the four scalar perturbed Einstein equations, one  is dynamical and three are constraints
\bea
\label{piS-sec2}&~&a^2\pi^S=\Psi-\Phi,\\
\label{dq-sec2} &~&\delta q_g+2(\dot{\Psi}+H\Phi)=0\,,\\
\label{drho-sec2} &~&\delta\rho_g-3H\delta q_g+2\frac{k^2}{a^2}\Psi=0\,,\\
\label{dP-sec2} &~&\delta P_g+\delta\dot{q}_g+3H\delta q_g+(\bar\rho+\bar P)\Phi=0\,.
\eea
 Note that the above equations do not form a complete set, unless the pressure $P$ and anisotropic
inertia $a^2\pi^s$ are given as independent equations \cite{Inflation-Books}. For a general hydrodynamical fluid with pressure $P(\rho, S)$, the pressure perturbation $\delta P$ can be decomposed as
\be\label{P-decomposition}
\delta P=c_s^2\delta\rho+\delta S,
\ee
where $\delta S$ is the entropy perturbation and the sound speed is defined as $c_s^2\equiv(\frac{\delta P}{\delta \rho})_S$. For the adiabatic case $\delta S=0$, $\delta P$ is given directly by $\delta\rho$.

It is common to construct two further gauge invariant combinations in terms of metric and energy-momentum perturbations. The comoving curvature perturbation $\mathcal{R}_c$  and curvature perturbation on the uniform density hypersurfaces $\zeta$:
\be\label{R}
\mathcal{R}_c
\equiv\Psi-\frac{H}{\bar\rho+\bar{P}}\delta q_g,\quad
\zeta \equiv-\Psi-\frac{H}{\dot{\bar{\rho}}}\delta\rho_g\,.
\ee
Since $$\mathcal{R}_c =-\zeta-\frac{2H}{3(\bar\rho+\bar{P})}\frac{k^2}{a^2}\Psi,$$
$\mathcal{R}_c$ and $\zeta$ become identical in the super-horizon scales $\frac{k}{a}\ll1$.
The crucial property of $\mathcal{R}_c$ and $\zeta$ is that in case of \textit{adiabaticity} of the primordial perturbations, they are conserved outside the horizon. If cosmological fluctuations are described by such
a solution during inflation, then as long as the perturbation
is outside the horizon, $\mathcal{R}_c$ and $\zeta$ will remain equal and constant \cite{Inflation-Books}.
Later in this section we discuss more about the adiabatic and isocurvature fluctuations.


\subsubsection*{$\bullet$ Vector modes}
We now consider the four divergence-free vector modes $V_i$, $W_i$, $\delta q^V_i$ and $\delta\pi^V_i$.
Again, not all of the above quantities are gauge invariant, but transform under infinitesimal vector gauge transformations induced by
\be  \label{veccoor}
x^i\rightarrow \tilde x^i=x^i+\delta x_V^i\,,
\ee
where $\partial_i\delta x^V_i=0$.
The three gauge invariant divergence-free vector perturbations may be identified as%
\be
\mathcal{Z}_i=a\dot{W}_i-V_i\,,
\ee
$\delta q^V_i$ and $\pi^V_i$.\footnote{Since the background fluid is a perfect and irrotational fluid, $\delta q^V_i$ and $\pi^V_i$ are gauge invariant quantities \cite{Inflation-Books}.}
The perturbed Einstein equations have two independent vector equations, one constraint and one dynamical equation. These equations are%
\bse \label{firstV-2}
\begin{align}
&~\partial_i\left(2a^2\pi^V_j-\frac{1}{a}(a^2\mathcal{Z}_j\dot{)}\right)=0\,,\\
&~2\delta q_i^V+\frac{k^2}{a^2}a\mathcal{Z}_i=0\,.
\end{align}
\ese
Although not independent of the Einstein equations, it is useful to also present the vector part of the momentum conservation equation
\be\label{vec-momentum-conserv}
\delta\dot{q}^{V}_{i}+3H\delta q^{V}_{i}+\nabla\pi^{V}_i=0,
\ee
which implies that, regardless of the matter content and the specific form of the anisotropic stress $\pi^V_i$, $\delta q^V_i$ always damps like $a^{-3}$ at large (superhorizon)  scales. Thus, after horizon crossing, the flow always gets \textit{irrotational} in inflationary systems.
 However, from \eqref{firstV-2} we see that in order to determine the dynamics of $\mathcal{Z}_i$ we need to know $\pi^V_i$. This term which can be non-zero in models of inflation involving gauge (vector) fields in general may change the usual picture.

For the specific case of a perfect fluid in which $\pi^V_i=0$, equation (\ref{firstV-2}a) implies that $\mathcal{Z}_j$ as the only physical (observable) combination of metric components for vector perturbations, decays
as $\frac{1}{a^2}$. Consequently, $\delta q_i^V$ decays like $\frac{1}{a^3}$.
In other words, vector perturbations are washed away by Hubble expansion, unless they are driven by a vector anisotropic stress $\pi^V_i$. This source term is identically zero in scalar driven inflationary models.
Due to their (exponential) decay, vector modes have not played a large role in these models.

\subsubsection*{$\bullet$ Tensor modes}\label{review-section-Tensor modes}
Here we have two traceless divergence-free symmetric tensors $h_{ij}$
and $\pi^T_{~ij}$ and it is straightforward to show that both are gauge invariant.
The only field equation we have in this sector is the wave equation for $h_{ij}$
(gravitational radiation), which is sourced by the contribution of $\pi^T_{~ij}$
\be \label{T}
\ddot{h}_{ij}+3H\dot{h}_{ij}+\frac{k^2}{a^2}h_{ij}=2\pi^T_{ij}\,.
\ee
Being traceless and divergence-free, $h_{ij}$ has 2 degrees of freedom which are usually decomposed
into plus and cross ($+$ and $\times$) polarization states  with the polarization tensors $e_{ij}^{+,\times}$.
Since we assumed to have no parity-violating interaction terms in the action, the equations for both of these polarization have the same time
 evolution\footnote{Cases with parity-violating contributions to tensor perturbations, which can lead to ``cosmological birefringence'' have been considered in the literature \cite{Lue:1998mq,APS-leptogenesis}. We will return to this point in section \ref{prim-magnet-NG-section}.} and one may then introduce $h$ variable instead%
\be
h_{ij}^{+,\times}=\frac{h}{a} e_{ij}^{+,\times}\,.
\ee
Then, in the computation of the power spectrum we consider this variable, treating it effectively as a scalar
but multiply the power spectrum by a factor of 2 to account for the two polarizations.

In the case that $\pi^T_{ij}$ is zero, $h$ becomes constant after horizon crossing ($\frac{k}{a}<H$), then as long as the perturbation
is outside the horizon, $h$ will remain a constant \cite{Inflation-Books}.

\subsubsection{Characterizing the primordial statistical fluctuations}

\paragraph{$\bullet$ Power spectrum of scalar models:}
The power spectrum of primordial curvature perturbation $\mathcal{R}_c$ \eqref{R} is one of the most practical and crucial statistical observables which may be used to distinguish  models of inflation
\be
\VEV{\mathcal{R}_c(\bold{k}) \mathcal{R}_c ({\bold{k'}})}=(2\pi)^3\delta(\bold{k}+\bold{k'})P_{\mathcal{R}_c}(k),\quad  \Delta^2_{s}\equiv\frac{k^3}{2\pi^2}P_{\mathcal{R}_c}(k),
\ee
where $\VEV{~\cdot~}$ denotes the ensemble average of the fluctuations. Furthermore, the scale-dependence of the power spectrum is described in terms of the scalar spectral index
\be
n_s-1\equiv\frac{d\ln\Delta^2_{~s}}{d\ln k},
\ee
and the running of the spectral index by
\be
\alpha_s\equiv\frac{d n_s}{d\ln k}.
\ee
The special case with $n_s=1$ and $\alpha_s=0$ corresponds to a scale-invariant spectrum.

The power spectrum, $\Delta^2_{s}(k)$, is often approximated by the following power-law form \cite{Kosowsky:1995aa}
\be\label{Ps}
\Delta^2_s(k)=A_s(k_*)(\frac{k}{k_*})^{n_s(k_*)-1+\frac12\alpha_s(k_*)\ln(\frac{k}{k_*})},
\ee
where $k_*$ is a pivot scale.

\paragraph{$\bullet$ Adiabaticity of the power spectrum:} Another quantity which can potentially offer an important test for models of inflation is the adiabaticity of the primordial perturbations. Roughly speaking, the \emph{adiabaticity} may be defined as the following relation between the perturbations of Cold Dark Matter (CDM) or its various components and photons \cite{Inflation-Books}
\be\frac{\delta\rho_m}{\rho_m}-\frac34\frac{\delta\rho_r}{\rho_r}=0\,,\qquad (\delta(\frac{n_m}{n_r})=0).
\ee
Then, any deviation from the above is defined as the \emph{isocurvature}, or entropic, perturbation
\be
\mathcal{S}_m\equiv\frac{\delta\rho_m}{\rho_m}-\frac34\frac{\delta\rho_r}{\rho_r}.
\ee
Since adiabatic and isocurvature perturbations lead to different peak structure in the CMB, they are distinguishable.
CMB observations show that, if exists at all, isocurvature perturbations have to be subdominant \cite{Komatsu:2010fb}.

Whatever the constituents of the
Universe, there is always an adiabatic solution\footnote{Note that for
these scalar modes,  all individual constituents $\alpha$ of the Universe we have equal $\frac{\delta\rho_\alpha}{\dot{\bar{\rho}}_\alpha}$, whether or not energy is separately conserved for these
constituents. For this reason, such perturbations are called adiabatic.
Then, any other solutions are called entropic.} of the field equations for which $\mathcal{R}_c$ and $\zeta$
are conserved outside the horizon. If cosmological fluctuations are described by such
a solution during inflation, then as long as the perturbation
is outside the horizon, $\mathcal{R}_c$ and $\zeta$ will remain a constant \cite{Inflation-Books}.

\paragraph{$\bullet$ Non-Gaussianity:}
If the distribution of $\mathcal{R}_c$ is Gaussian with random phases, then the power spectrum contains all the statistical information. The level of deviation from a Gaussian distribution is called \emph{non-Gaussianity} and is encoded in higher order correlations functions of  $\mathcal{R}_c$.
A basic diagnostic of non-Gaussian statistics is the \emph{bispectrum}, the Fourier transform of the the three-point function of $\mathcal{R}_c$:
\be
\VEV{\mathcal{R}_c(\bold{k})\mathcal{R}_c ({\bold{k'}})\mathcal{R}_c ({\bold{k''}})}=(2\pi)^3\delta(\bold{k}+\bold{k'}+\bold{k''})B_{\mathcal{R}}(\bold{k},\bold{k'},\bold{k''}).
\ee
which its momentum dependence is a powerful probe of the inflationary action.
Note that here the $\delta$-function is a consequence of translation invariance of the background and indicates that these three Fourier modes form a triangle. Different inflationary models predict different maximal signal for different triangle configurations and hence in principle the shape of non-Gaussianity can be a powerful probe of the physics of inflation  \cite{Babich:2004gb}.

One way to parameterize non-Gaussianity phenomenologically is through $f_{NL}$ \cite{Komatsu:2001rj}
\footnote{The factor $\frac35$ in \eqref{LCNG} is conventional. Non-Gaussianity may be defined in terms of the
Bardeen potential $\Phi$, $\Phi(\bold{x})=\Phi_g(\bold{x})+f^{local}_{NL}(\Phi^2_g(\bold{x})-\VEV{\Phi^2_g(\bold{x})})$, which during the matter era and at the linear order is $\Phi=\frac35\mathcal{R}$ \cite{Kodama:1985bj}.}
\be\label{LCNG}
\mathcal{R}_c (\bold{x})=\mathcal{R}_g(\bold{x})+\frac35f^{local}_{NL}\left[ \mathcal{R}^2_g(\bold{x})-\VEV{\mathcal{R}^2_g(\bold{x})}\right] \ .
\ee
where $\mathcal{R}_g(\bold{x})$ denotes the Gaussian curvature perturbation. Being local in real space, $f_{NL}^{local}$ is a measure for
the so-called local non-Gaussianity.

In addition to $f^{local}_{NL}$, the other commonly discussed parameter is the equilateral non-linear coupling parameter, $f^{equil}_{NL}$ which is defined through the power-spectrum normalized  bispectrum for the equilateral configuration ($k\sim k'\sim k''$) e.g. see \cite{{Baumann:2009ds},Babich:2004gb}.

Large non-Gaussianity can only arise if we have significant inflaton interactions during inflation. Therefore, in single-field slow-roll inflation, non-Gaussianity is predicted to be unobservably small \cite{Acquaviva:2002ud, Maldacena:2002vr}, while it
can be significant in models with multiple fields, higher-derivative interactions or nonstandard
initial states \cite{Babich:2004gb,{Paolo-NG},Bartolo:2004if}.

\paragraph{$\bullet$ Tensor modes:}
Power spectrum of the two tensor polarization modes $h\equiv h_{+}=h_{\times}$, is defined as
\be
\VEV{h(\bold{k})h({\bold{k'}})}=(2\pi)^3\delta(\bold{k}+\bold{k'})P_{h}(k),\quad  \Delta^2_{h}=\frac{k^3}{2\pi^2}P_{h}(k),
\ee
and the primordial gravitational waves spectrum quantify as
\be
\Delta^2_{T}=2\Delta^2_{h}.
\ee
The tensor scale-dependence is parameterized in terms of $n_T$ as
\be
n_T\equiv\frac{d\ln\Delta_T^2}{d\ln k},\\ \label{Pt},
\ee
and the power spectrum itself is approximated as the following power-law form
\bea
\Delta^2_{T}=A_t(k_*)(\frac{k}{k_*})^{n_T(k_*)}.
\eea

Finally, from  the combination of \eqref{Ps} and \eqref{Pt}, we define another practical quantity, the tensor-to-scalar ratio $r$, as
\be
r\equiv\frac{\Delta^2_{T}}{\Delta^2_{s}}.
\ee
From the practical point of view, a tensor signal will be observable in CMB polarization in the upcoming observations
if this ratio is bigger than $0.01$ \cite{Baumann:2008aq}.

\subsubsection{Computation of observable perturbations in inflation models}

Up to now we studied the cosmic perturbations without considering a specific model of inflation.
At this point, considering the simple model of single-field inflation, we study the spectra of scalar
and tensor perturbations generated in \eqref{ssmodel}. In terms of the gauge invariant \emph{Sasaki-Mukhanov variable} $v$
\cite{Mukhanov:1988jd, Sasaki:1986hm}
\be\label{SM-variable-def}
v=a(\delta\varphi+\frac{\dot\varphi}{H}C)\,,
\ee
perturbed scalar field equation of motion for
a single scalar field takes the simple form
\be
v''+(k^2-\frac{z''}{z})v=0,
\ee
where $z=\frac{a\dot\varphi}{H}$ and $'$ means derivative with respect to the conformal time $\tau$ ($dt=ad\tau$). Thus, the effective mass term $\frac{z''}{z}$ is approximately
\be
\frac{z''}{z}\simeq(aH)^2(2+3\epsilon-2\eta)\simeq (aH)^2 (2+5\epsilon_V-3\eta_V)\,.
\ee

The general solution of this equation is then expressed as a linear combination of Hankel functions
\be\label{v-solution}
v\simeq\frac{\sqrt{\pi(k\vert\tau\vert)}}{2}e^{i(1+2\nu_{_R})\pi/4}\left(c_1H^{(1)}_{\nu_{_R}}+c_2H^{(2)}_{\nu_{_R}}\right),\,
\ee
where $\nu_{_R}\simeq\frac32+2\epsilon-\eta=\frac32+3\epsilon_V-\eta_V$. Imposing the usual Minkowski vacuum state,
\be
v\rightarrow\frac{e^{-ik\tau}}{\sqrt{2k}},
\ee
in the asymptotic past $(k\tau\rightarrow-\infty)$, we obtain $c_1=1$ and $c_2=0$ in \eqref{v-solution}. From the definition of the comoving curvature perturbation \eqref{R} we find
\be
\mathcal{R}_c=\frac{H}{a\dot\varphi}v.
\ee
During the inflationary phase,  fluctuations in $v$ generate scalar perturbations in the
curvature $\mathcal{R}_c$. Then, the curvature perturbation $\mathcal{R}_c$ becomes constant on the superhorizon
scales and the power spectrum  of the scalar fluctuations at the end of inflation is
\be\label{Power-single-scalar}
\Delta_s^2=\frac{1}{2\epsilon}\frac{H^2}{(2\pi)^2}\bigg|_{k=aH}\,.
\ee
The spectral index of the scalar perturbations, to the leading orders in the slow-roll parameters, is
\be
n_{s}-1 =-4\epsilon+2\eta=-6\epsilon_V+2\eta_V\,.
\ee

In the single-field slow-roll models non-Gaussianity is predicted to be  small with $f_{NL}$ of order slow-roll parameters $\epsilon$ and $\eta$ \cite{Maldacena:2002vr}. Hence,  we do
not discuss  them here.

Gravitational waves are also generated during inflation which introduce a tensor
contribution to the fluctuations. For the simple scalar models  the wave equation for tensor perturbations \eqref{T} reduces to
a source-free graviton equation of motion, which is
\be\label{eq-u-single-scalar}
u''+(k^2-\frac{a''}{a})u=0\,,
\ee
where $u\equiv ah$ and
\be
\frac{a''}{a}=(aH)(2-\epsilon)\,.
\ee
Using the slow-roll approximation and the Minkowski vacuum normalization, we have the same solution as \eqref{v-solution} with $c_1=1$ and $c_2=0$, with $\nu_T\simeq\frac32+\epsilon$.
Finally, we get the power spectrum of the fluctuations in tensor components at the end of inflation
\bea\label{power-tensor-single-scalar}
\Delta^2_{T}\simeq 8\frac{H^2}{(2\pi)^2}\bigg|_{k=aH}\,.
\eea
One can determine the tensor spectral index $n_{t}$ which is
\be\label{tilt-tensor-single-scalar}
n_T=-2\epsilon\,.
\ee
Consequently, the ratio of tensor to scalar, $r$, is given as
\be\label{consistency-relation-single-scalar}
r\simeq16\epsilon=-8 n_T\,.
\ee
Recalling \eqref{N}, we can relate it to the total inflaton field excursion during the inflation $\Delta\varphi$,
\be\label{Lythbound}
\frac{\Delta\varphi}{M_{pl}}\simeq\mathcal{O}(1)\times(\frac{r}{0.01})\,.
\ee
As mentioned before, a tensor signal is observable (in CMB polarization) only if $r>0.01$.
First showed by Lyth \cite{Lyth:1996im}, equation \eqref{Lythbound} implies that this level of tensor modes corresponds to super-Planckian $\Delta\varphi$ values. Using the effective theory of inflation, in \cite{Baumann:2011ws}, the Lyth-bound is generalized to all single-field models with two-derivative kinetic terms. They showed that the bound is always stronger than the above bound for slow-roll models.

\subsection{Observational constraints on CMB power spectrum }

We showed that the simplest (single-field)  inflation models predict nearly Gaussian, scale-invariant and adiabatic
scalar fluctuations.
In this part, we briefly review the latest quantitative constraints from the seven-year Wilkinson Microwave Anisotropy Probe (WMAP) observations \cite{Komatsu:2010fb}.
Current observations provide values for power spectrum of curvature
perturbations $\Delta_{s}^2$, its spectral index $n_{s}$, and impose an upper bound on the power spectrum of tensor modes $\Delta_T^2$, or equivalently an upper bound on tensor-to-scalar ratio $r$. Moreover, it puts quantitative limits on physical motivated primordial non-Gaussianity parameters $f^{local}_{NL}$ and $f^{equil}_{NL}$.
These values vary (mildly) depending on the details of how the data analysis has been carried out. Here we
use the best fit values of Komatsu et al. \cite{Komatsu:2010fb} which is based on WMAP seven-years data combined
with other cosmological data, within standard $\Lambda$CDM framework. Amplitude of curvature perturbations
is obtained to be
\be\label{PR-WMAP7}
\Delta^2_s\simeq 2.5\times 10^{-9},
\ee
and the data indicates a red tilted spectrum ($n_s<1$)
\be\label{ns-WMAP7}
n_{s}= 0.968\pm0.012\,,
\ee
as well as a small tensor-to-scalar ratio
\be\label{r-bound-WMAP7}
r<0.24\,,
\ee
with no evidence of the running index, $\frac{dn_s}{d\ln k}$.
Furthermore, from the WMAP temperature fluctuation data we have the following constraints on the primordial
non-Gaussianity  parameters
\be
-9<f^{local}_{NL}<111,\quad -151<f^{equil}_{NL}<253\,.
\ee
The (non-)Gaussianity tests show that the primordial fluctuations are Gaussian to the $0.1\%$ level \cite{Komatsu:2010fb},
a strong evidence for the \emph{perturbative} quantum origin of these perturbations.

In \cite{Komatsu:2010fb}, they explored the possibility of any deviations from the simplest picture: \emph{Gaussian}, \emph{adiabatic}, \emph{power-law power spectrum} $\Lambda$CDM. However, they have not detected any convincing deviations from that.

\begin{figure}[h]
\begin{center}
\includegraphics[angle=0, width=90mm, height=65mm]{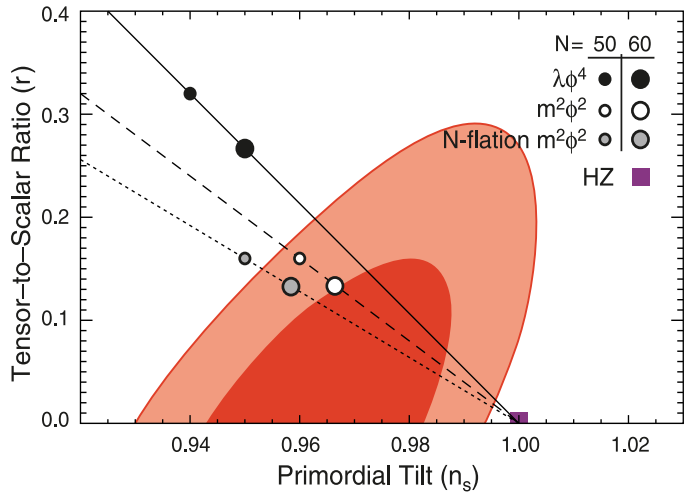}
\caption{In this figure, we have the $1\sigma$ and $2\sigma$ observational contour bounds from the combined data of 7-year WMAP+BAO+H0. The points here represent the theoretical predictions of inflation models with monomial potentials, $V (\varphi)=V_0\varphi^n$ . The solid line represents the model with $n$ = 4, the dashed line has $n$= 2 and the dotted line denotes the multi-axion field models with $n$= 2 and $\beta$= 1/2 in \cite{Easther:2005zr}, with e-folds $N_e=50$ and $60$ (from top to bottom). This figure is taken from \cite{Komatsu:2010fb}.}\label{ns-r}
\end{center}
\end{figure}

Since the release of WMAP 3-year, the accessible parameter space for inflationary models has significantly shrunk. In other words, the CMB data constrains inflationary models, and even disfavored or totally ruled out some of them. For instance, in Fig.\ref{ns-r} which is taken from \cite{Komatsu:2010fb}, the 7-year WMAP+BAO+H0 observational contours on the spectral tilt, $n_s$, and the tensor-to-scalar ratio, $r$, have been compared with the theoretical predictions of large field models with the monomial potential \eqref{chaotic-potential}. The predicted point of the quadratic potential ($n=2$) with $N_e=60$ is on the boundary of the $1\sigma$ contour, thus observationally preferable. However, predicting a large $r$, the quartic potential ($n=4$) with $N_e=60$ is totally out side of the $2\sigma$ contour, unless $N_e$ is sufficiently large.
In fact, models with rather large tensor modes, such as $\lambda\phi^4$ chaotic inflation are disfavored now. Moreover, models with blue-tilted spectrum ($n_s>1$), such as hybrid inflation models are ruled out now.
Besides that, the possible parameter space for the models which can provide significant non-Gaussianity, such as K-inflation models with small sound speeds $c_s^2\ll 1$ \cite{Chen:2006nt} are mainly shrunk by the constraints on the non-Gaussianity parameters.

\subsection{CMB anomaly and the eta problem}

As we mentioned, WMAP data strongly supports the inflationary scenario. It is believed
that the primordial fluctuations are statistically homogeneous, isotropic, and Gaussian.
However, there seems to be some anomalies in the data prompting considering and analyzing various inflationary models. These ``anomalies'' include non-Gaussianity which we already discussed,
a low power in the quadrapole moment~\cite{de OliveiraCosta:2003pu,Copi:2006tu},
the alignment of the lower multipoles~\cite{Land:2005ad},
a five-degree scale cold spot with suppressed power~\cite{Vielva:2003et},
an asymmetry in power between the Northern and Southern ecliptic hemisphere~\cite{Hansen:2004mj},
and broken rotational invariance~\cite{Groeneboom:2008fz}. These observational signature, although their statistical
significance is still controversial~\cite{Bennett:2010jb}, stimulated activity in the research of statistical
anisotropy~\cite{Hajian:2003qq,ArmendarizPicon:2005jh,Gumrukcuoglu:2006xj,Ackerman:2007nb,Pullen:2007tu,
ArmendarizPicon:2008yr,Campanelli:2009tk,Hanson:2009gu,arXiv:1107.4304,Shiraishi:2011ph}.
If the statistical anisotropy exists,  a primary candidate of source of the anomaly should be gauge fields.
Thus, from the observational point of view, it is well motivated to consider gauge fields in inflationary stage.
Actually, as we will see in sections \ref{prim-magnet-NG-section} and \ref{Anisotropic-inflation-section}, there are many phenomena induced by gauge fields.

On the other hand, from the theoretical point of view, there is a reason to seriously explore a role of gauge fields during inflation. That is the eta problem, namely, scalar fields easily get radiative corrections which generically spoil slow roll conditions \cite{eta-N}. Hence, it is natural to explore a possibility to incorporate fields other than scalar fields to realize slow roll inflation. If we move beyond scalars, gauge fields are prime candidates. Indeed, as we will see below, there are at least two different mechanisms by which we can achieve slow roll inflation through the use of gauge fields.

\section{Quantum gauge fields in inflationary background}\label{prim-magnet-NG-section}

\def\bk{{\bf k}}

In this section, we explore possible effects of gauge fields on inflation at the level of fluctuations.
First, we review how to quantize gauge fields coupled to the inflaton in inflationary background.
We then discuss the old but important topic whether primordial magnetic fields can be generated during inflation~\cite{Turner:1987bw,Ratra:1991bn}.
Recently, it turned out that there are other effects induced by gauge fields.
First of all, the presence of gauge fields induces the statistical anisotropy in curvature perturbations~\cite{Yokoyama:2008xw}.
Indeed, the statistical anisotropy appears both in the power spectrum and the bispectrum.
Interestingly, gauge fields can induce detectable non-gaussianity in contrast to the single inflaton model
 with the conventional kinetic term~\cite{Maldacena:2002vr}.
We review several mechanisms producing the non-gaussianity in primordial curvature perturbations from gauge fields.
Moreover, gauge fields can generate primordial gravitational waves even in the low energy inflationary scenarios.

\subsection{Quantum gauge fields in cosmic background}

It is well known that gauge fields in 4-dimensions are conformally invariant at classical level.
Due to the conformal invariance,  gauge field fluctuations, which are damped by the Hubble expansion as the vector perturbations discussed in the previous section, do not survive in the minimal set up.
However,   the non-minimal coupling described by the action
\begin{eqnarray}
  S = - \frac{1}{4} \int d^4 x \sqrt{-g} f^2 (\phi ) F_{\mu\nu} F^{\mu\nu} \ ,
\label{action-1}
\end{eqnarray}
may produce cosmological gauge field fluctuations. Here, $\phi $ is the inflaton field and
$F_{\mu\nu} = \partial_\mu A_\nu - \partial_\nu A_\mu$
is a field strength of an Abelian gauge field $A_\mu$.
Let us consider gauge fields in isotropic and homogeneous background spacetime
\begin{eqnarray}
ds^2 = a^2 (\tau ) \left[ -d\tau^2 + \delta_{ij} dx^i dx^j \right] \ ,
\end{eqnarray}
where $\tau$ is the conformal time and $a$ is the scale factor. We also assume the inflaton is homogeneous, $\phi=\phi(\tau)$ in the background.
Therefore, $f(\phi )$ can be regarded as a time dependent coupling.
Since the system is invariant under the gauge transformation $ A_\mu \rightarrow A_\mu + \partial_\mu \chi $ with an  arbitrary function
$\chi$, we can work in the temporal gauge $A_0 =0$ and a gauge field can be expanded as
\begin{eqnarray}
    A_i =  \int \frac{d^3 k}{(2\pi)^{3/2}}  A_i (\tau, {\bf k}) e^{i{\bf k}\cdot {\bf x}}
        = \sum_\sigma \int \frac{d^3 k}{(2\pi)^{3/2}}  A_{\bk}^\sigma (\tau ) \epsilon_i^\sigma ({\bf k}) e^{i{\bf k}\cdot {\bf x}} \ ,
\label{expansion-1}
\end{eqnarray}
where polarization vectors  $\epsilon_i^\sigma ({\bf k})$ satisfy the transverse and normalization conditions
\begin{eqnarray}
   k_i \epsilon_i^\sigma ({\bf k}) =0 \ , \qquad
    \epsilon_i^\sigma (-{\bf k})\epsilon_i^{\sigma'} ({\bf k})  = \delta_{\sigma \sigma'} \ .
\end{eqnarray}
Here, $\sigma$ represents two independent polarizations. The wavenumber vector and polarization vector form a complete basis
\begin{eqnarray}
  \sum_\sigma  \epsilon_i^\sigma (-{\bf k})\epsilon_j^{\sigma} ({\bf k})
    = \delta_{ij}  - \frac{k_i k_j}{k^2} \ .
\end{eqnarray}
Substituting the expansion (\ref{expansion-1}) into the action (\ref{action-1}),
we obtain
\begin{eqnarray}
  S = \frac{1}{2} \sum_\sigma \int d^3 k f^2 (\phi) \left[ A_k^{\sigma\prime} (\tau ) A_{-k}^{\sigma\prime} (\tau )
 - k^2 A_k^\sigma (\tau ) A_{-k}^\sigma (\tau ) \right] \ ,
\end{eqnarray}
where a prime represents a derivative with respect to $\tau$. The coefficient $A_k^{\sigma}$ satisfies the equation
\begin{eqnarray}
  A_k^{\sigma\prime\prime} + 2 \frac{f'}{f} A_k^{\sigma\prime} + k^2 A_k^{\sigma} = 0  \ .
\end{eqnarray}
The canonical conjugate momentum is defined by
\begin{eqnarray}
  \pi^\sigma_k = \frac{\delta S}{\delta A_k^{\sigma\prime}} = f^2 A_{-k}^{\sigma\prime} \ .
\end{eqnarray}
Now, we can quantize the system by promoting the fields to operators and imposing  canonical commutation relations
\begin{eqnarray}
      \left[  A_{k_1}^{\sigma_1}  \ ,  \pi^{\sigma_2}_{k_2} \right] = i \delta^{\sigma_1 \sigma_2} \delta ({\bf k}_1 -{\bf k}_2)
\ .
\label{can-com}
\end{eqnarray}
In terms of creation and annihilation operators, we can expand the operator as
\begin{eqnarray}
   A_{k_1}^{\sigma}  = u_k a_{k}^{\sigma}  + u_k^* a_{-k}^{\sigma\dagger} \ ,
\end{eqnarray}
where $[a_{k_1}^{\sigma_1} , a_{k_2}^{\sigma_2\dagger} ] =\delta^{\sigma_1 \sigma_2 } \delta  ({\bf k}_1 -{\bf k}_2)$.
The mode function obeys the equation
\begin{eqnarray}
  u''_k +2 \frac{f'}{f} u_k' + k^2 u_k = 0  \ .
\end{eqnarray}
In order to satisfy the canonical commutation relation (\ref{can-com}), the mode function have to be normalized as
\begin{eqnarray}
  u_k \frac{\partial}{\partial \tau} u_k^* - u_k^* \frac{\partial}{\partial \tau} u_k = \frac{i}{f^2} \ .
\end{eqnarray}
Once the mode function is determined, the vacuum can be defined by $a_{k}^{\sigma} |0\rangle =0$.
Then, it is easy to calculate two point function
\begin{eqnarray}
    \langle 0| A_i (x) A^i (0) |0\rangle = \frac{2}{a^2} \int \frac{dk}{2\pi^2} k^2 |u_k (\tau )|^2 e^{i{\bf k}\cdot {\bf x}}
       \equiv \int \frac{dk}{k} P_A (k,\tau ) e^{i{\bf k}\cdot {\bf x}} \ ,
\end{eqnarray}
where we have defined the power spectrum
\begin{eqnarray}
    P_A (k,\tau )  = \frac{k^3}{\pi^2 a^2} |u_k (\tau )|^2  \ .
\end{eqnarray}
Similarly, we can define the power spectrum of the magnetic field $B_i = \epsilon_{ijk} F^{jk}/2$ by
\begin{eqnarray}
  P_B (k,\tau ) = \frac{k^5}{\pi^2 a^4} |u_k (\tau )|^2
\label{B-spectrum}
\end{eqnarray}
and that of the electric field $E_i = - A_i^\prime/a$ by
\begin{eqnarray}
  P_E (k,\tau ) = \frac{k^3}{\pi^2 a^4} |u_k^\prime (\tau )|^2  \ .
\label{E-spectrum}
\end{eqnarray}
The energy density of a gauge field can be expressed as
\begin{eqnarray}
  \rho_{em} (\tau ) = \frac{f^2}{2}\int\frac{dk}{k}  \left[ P_E (k ,\tau ) + P_B (k , \tau )  \right] \ .
\label{em-energy}
\end{eqnarray}

Now, we need to determine mode function $u_k$. In the subhorizon limit $k\tau \rightarrow -\infty $, we have a general solution
\begin{eqnarray}
   u_k \sim c_1 \frac{1}{f\sqrt{2k}} e^{-ik\tau} + c_2 \frac{1}{f\sqrt{2k}} e^{ik\tau}  \ .
\end{eqnarray}
As usual we choose the standard Bunch-Davis vacuum, i.e. to take the positive frequency mode functions as
\begin{eqnarray}
   u_k \sim \frac{1}{f\sqrt{2k}} e^{-ik\tau} \ .
\end{eqnarray}
In the superhorizon limit $k\tau \rightarrow 0$, we have a general solution
\begin{eqnarray}
    u_k \sim d_1 + d_2 \int \frac{d\tau}{f^2} \ .
\end{eqnarray}
The constants $d_1 , d_2$ have to be determined by matching conditions. To perform this matching, we need to specify
the time dependence of the coupling $f(\phi(\tau))$. Here, we parametrize it as
\begin{eqnarray}
     f= \left( \frac{a}{a_f} \right)^{-2c} \ ,
\label{gauge-kinetic}
\end{eqnarray}
where $c$ is a constant and $a_f$ is the scale factor at the end of inflation.
We assume that the background spacetime is de Sitter and the scale factor is given by $a=1/(-H\tau)$.
Then, with new constants, we obtain
\begin{eqnarray}
   u_k = d_1 + \tilde{d}_2 a^{4c-1} \ .
\label{general-u}
\end{eqnarray}
Apparently, the two  $c>1/4$ and $c<1/4$ cases should be discussed separately.

For $c>1/4$, the second term is a growing mode. Hence, matching at the horizon crossing $a_k H=k$
gives
\begin{eqnarray}
   \tilde{d}_2 = \frac{1}{f_k \sqrt{2k}} \left( \frac{1}{a_k}  \right)^{4c-1} \ .
\end{eqnarray}
Here, the suffix $k$ represents the horizon crossing time of the mode $k$. Hence, $f_k$ should be
\begin{eqnarray}
    f_k = \left(  \frac{a_k}{a_f} \right)^{-2c} \ .
\end{eqnarray}
Thus, the power spectrum of the gauge field reads
\begin{eqnarray}
   P_A (k ,\tau_f ) = \frac{H^2}{2\pi^2} \left( \frac{a_f H}{k}\right)^{4c-4} \ .
\end{eqnarray}
Therefore, the power spectrum of the electric field is given by
\begin{eqnarray}
   P_E (k ,\tau_f ) = (4c-1)^2\frac{H^4}{2\pi^2} \left( \frac{a_f H}{k}\right)^{4c-4} \ ,
\end{eqnarray}
and that of the magnetic field can be written as
\begin{eqnarray}
   P_B (k ,\tau_f ) = \frac{H^4}{2\pi^2} \left( \frac{a_f H}{k}\right)^{4c-6} \ .
\label{pri-B}
\end{eqnarray}
From these results, we see the gauge field and electric field survive during inflation for the parameter region $c\geq 1$.
On the other hand, the magnetic field has no power on large scales unless $c\geq 3/2$.

Next, we consider cases with $c<1/4$. For these cases, the matching condition gives
\begin{eqnarray}
   u_k = \frac{1}{f_k \sqrt{2k}}  \ .
\end{eqnarray}
Thus, the power spectrum of the gauge field reads
\begin{eqnarray}
   P_A (k ,\tau_f ) = \frac{H^2}{2\pi^2} \left( \frac{a_f H}{k}\right)^{-2c-1} \ .
\end{eqnarray}
The power spectrum of the electric field is given by
\begin{eqnarray}
   P_E (k ,\tau_f ) = 0 \ ,
\end{eqnarray}
and that of the magnetic field can be written as
\begin{eqnarray}
   P_B (k ,\tau_f ) = \frac{H^4}{2\pi^2} \left( \frac{a_f H}{k}\right)^{-2c-2} \ .
\end{eqnarray}
Thus, for the parameter region $c \leq -1/2$, gauge fields survive. In order for magnetic fields to survive, the parameter should be $c\leq -1$.

\subsection{Primordial magnetic fields}\label{prim-mag-field-subsection}

Based on the results in the previous subsection, let us first examine if primordial magnetic fields can be generated during inflation. Noting that the magnetic field has mass dimension two, namely $1 {\rm Gauss} = 10^{-20} {\rm GeV}^2$,
and assuming that Hubble during inflation is close to its current observational bounds, e.g. $H\sim 10^{-6} \mpl$,  just on dimensional grounds  we can expect
\begin{eqnarray}
   B \sim H^2 \sim 10^{-12} \mpl^2 \sim 10^{26} {\rm GeV}^2 \sim 10^{46} {\rm Gauss}  \ .
\end{eqnarray}
After reheating, the energy density of magnetic fields evolves as $\rho_B \sim B^2 \propto 1/a^4$.
Assuming the instantaneous reheating with maximal efficiency, i.e. all the energy of the inflaton has turned into the thermal energy of a relativistic gas of particles \cite{Inflation-Books},  we have $T_0^2 = \mpl^2 H^2 (a_f /a_0)^4$ or
\begin{eqnarray}
  \frac{a_0}{a_f}  = \frac{(\mpl H)^{1/2}}{T_0}  \sim 10^{29} \ ,
\end{eqnarray}
where $a_f$ and $a_0$ are the scale factor at the end of inflation and the present and
 $T_0$ is the observed CMB temperature, $T_0 = 2.4 \times 10^{-13} {\rm GeV}$.
Thus, the expected magnetic fields at present should be
\begin{eqnarray}
    B_0 = B \times 10^{-58} = 10^{-12} {\rm Gauss}.
\end{eqnarray}
To make the above estimation we have assumed the maximal efficiency for reheating. In reality, we should have $a_0 /a_f \ll 10^{-29}$.
Hence, on dimensional grounds, we can obtain $B_0 \sim 10^{-9} $ Gauss.
Thus, it is a natural idea to generate primordial magnetic fields during inflation~\cite{Turner:1987bw,Ratra:1991bn}.
There are many works on primordial magnetic fields from inflation~\cite{Yokoyama:2008xw,Bamba:2003av,Martin:2007ue,Dimopoulos:2008em,Emami:2009vd,Byrnes:2011aa,
{Demozzi:2009fu},Suyama:2012wh,Kandus:2010nw},
although there exists no convincing model so far.

In the previous subsection, we have discussed gauge fields in a fixed background.
Here, we should note that previous analysis can be readily  used in inflationary models.
The most generic action for a single field inflation in this setup reads
\begin{eqnarray}
  S =    \int d^4 x \sqrt{-g} \left[ \frac{\mpl^2}{2}R -\frac{1}{2} \partial_\mu \phi \partial^\mu \phi  - V(\phi )
- \frac{1}{4} f^2 (\phi ) F_{\mu\nu} F^{\mu\nu} \right]  \ ,
\end{eqnarray}
In the slow roll regime, we need to solve
\begin{eqnarray}
   3 \mpl^2 H^2 = V (\phi )   \ , \qquad  3 H \dot{\phi} = - V_\phi (\phi )   \ .
\end{eqnarray}
Combining both equations, we get
\begin{eqnarray}
    \frac{d\phi}{d\log a}  = - \mpl^2 \frac{V_\phi}{V } \ ,
\end{eqnarray}
which yields
\begin{eqnarray}
     a \propto \exp \left[ - \frac{1}{\mpl^2}  \int \frac{V}{V_\phi} d\phi \right]  \ .
\label{a-phi}
\end{eqnarray}
Recalling (\ref{gauge-kinetic}) an appropriate choice for the coupling $f(\phi)$ is
\begin{eqnarray}
  f(\phi) = \exp \left[  \frac{2c}{\mpl^2}  \int \frac{V}{V_\phi} d\phi \right]  \ ,
\label{f-phi}
\end{eqnarray}
or more explicitly  $f (\phi) \propto a^{-2c}$. For example, for a polynomial function $V=\phi^n$, we obtain
\begin{eqnarray}\label{f(phi)-mag-field}
   f (\phi ) = \exp \left[ \frac{c}{n}  \frac{\phi^2}{\mpl^2} \right] \ .
\end{eqnarray}
Thus, with this choice for $f$ we can directly use analysis of the previous subsection and inflation.

Since $f$ can be interpreted as an inverse of the coupling constant, large $f$ means weak coupling.
Hence, for positive $c$, the coupling starts from weak and becomes strong. We can normalize the coupling at the
end of inflation. Then, the system stays in the weak coupling regime during inflation and the perturbative analysis is reliable.

In the $c>0$ weak coupling case the electric fields dominates energy density of electromagnetic fields.
On the other hand a scale invariant primordial magnetic fields requires $c=3/2$. In this case, however, the energy density of electromagnetic fields grows as $\rho_{em} \sim (a_f /a_i )^2$ and we cannot neglect backreaction of the electric field produced through quantum fluctuations during inflation on the background inflationary trajectory.
To avoid the backreaction we should set $c=1$ and hence magnetic field cannot be scale invariant. In this case the primordial magnetic field has a blue spectrum
\begin{eqnarray}
   B \sim \sqrt{P_B} \sim H^2 \left( \frac{k}{a_f H}\right)^2\,.
\end{eqnarray}
In the comoving scale 1 Mpc corresponds to $k/(aH)\sim 10^{-25}$ and hence
\begin{eqnarray}
    B \sim 10^{21} {\rm Gauss}
\end{eqnarray}
at the end of inflation and
\begin{eqnarray}\label{B-field-pos-c}
    B \sim 10^{-37} {\rm Gauss}
\end{eqnarray}
at present. Apparently, this is too small.

For negative $c$ the effective coupling is a decreasing function of time. If we identify the coupling with the measured value at the end of inflation,
the coupling during inflation is too strong and one may not rely on perturbative analysis. Despite the strong coupling issue, let us proceed with the order of magnitude estimate for the magnetic field.
In the strong coupling cases $c<0$ the electric field can be negligible and to circumvent the backreaction problem
we take $c=-1$. Then, we obtain the primordial magnetic fields
\begin{eqnarray}
    B \sim 10^{46} {\rm Gauss}
\end{eqnarray}
at the end of inflation and
\begin{eqnarray}
    B \sim 10^{-12} {\rm Gauss}
\end{eqnarray}
at present. Although the value of the primordial magnetic field in this case is much bigger than the $c>0$ case \eqref{B-field-pos-c} and close to the desired value, we should again stress that this result is not  reliable due to the strong coupling problem.

In conclusion, it is difficult to generate primordial magnetic fields if we want to avoid the strong coupling and backreaction problems.
Later, we will discuss what happens in the analysis of primordial magnetic fields when we take into account the backreaction. Apart from a possible primordial magnetic field generation mechanism, one may explore the evolution of the primordial magnetic fields  after inflation and the effects they may have on other observables. Such  analysis has been carried out in~\cite{Kahniashvili:2012vt,Urban:2012ib,Caldwell:2011ra}.

\subsection{Statistical anisotropy in power spectrum and bispectrum from gauge fields}

Besides the primordial magnetic fields, one may wonder if presence of gauge fields can in principle have observable effects on the CMB data. As discussed, for generic models of inflation vector fluctuations are generically suppressed by the cosmic expansion and to explore such a possibility we need to consider specific models of inflation with appropriate gauge field-inflaton couplings. Analysis of the previous subsection already provides a suggestion. To increase such effects one may consider models which allow a sharp change in the inflaton field, something similar to what happens in models of hybrid inflation, at the end of inflation (\emph{cf.} discussions of section \ref{Inf-background-review}).

Let us consider a generic hybrid model with an inflaton $\phi$, a waterfall field $\chi$, and a vector field
$A_{\mu} (\mu=0,1,2,3)$ which couples to the waterfall field~\cite{Yokoyama:2008xw}.
The action can be written as
\begin{eqnarray}
S &\! = \!& {1 \over 2}\int d^4x \sqrt{-g} R
- \int d^4 x \sqrt{-g} \left[ {1 \over 2}g^{\mu\nu}
\left(\partial_\mu \phi \partial_\nu \phi + \partial_\mu \chi \partial_\nu \chi\right) + V(\phi,\chi,A_\mu)\right] \nonumber\\
&&
\qquad\qquad\qquad
-{1 \over 4}\int d^4 x \sqrt{-g}g^{\mu\nu}g^{\rho\sigma}f^2(\phi) F_{\mu\rho}F_{\nu\sigma}~,
\end{eqnarray}
where $F_{\mu\nu} \equiv \partial_\mu A_\nu - \partial_\nu A_\mu$, $V(\phi,\chi,A_\mu)$ is the potential of fields
and an arbitrary function $f(\phi)$ represents gauge coupling.
Here, we do not restrict ourselves to the potential \eqref{Hybrid}. The value of the inflaton at the end of inflation $\phi_e$ is determined by parameters of this potential; in any case $\phi_e$ and hence the coupling of the gauge field at the end of inflation $f(\phi_e)$ are fixed and do not fluctuate. As we discuss below, coupling other fields with  the waterfall field $\chi$ can, however, change the situation.
To avoid standard problems in dealing with gauge field potentials, we assume that the vector field is massless
 and have a small expectation value compared to the inflaton.
We hence neglect the terms coming from the coupling with the vector field
in the background homogeneous equations of motion for the scalar fields
and we treat the gauge field perturbatively.

The curvature perturbation on superhorizon scales is given by the perturbation of e-folding number $\delta N$.
In the standard single scalar inflation or hybrid inflation, in each Universe (causally connected Hubble patch),
inflation ends when the inflaton $\phi$ reaches its critical value $\phi_e$ which is determined by the inflationary potential.
On the other hand, in the multi-component inflation, the critical value
$\phi_e$ may be different in each Hubble patch due to a light field other than inflaton $\phi$.
Hence, in such situation there is a possibility of generating the curvature perturbations
because of the fluctuation of $\phi_e$.
This mechanism is first proposed in ~\cite{Lyth:2005qk}.
We generalize their work and introduce a massless vector field $A^\mu$ as another light field.

From the above discussion it becomes apparent that to analyze the effects of fluctuations of $\phi_e$ at superhorizon scales it is convenient to use the $\delta N$ formulation \cite{delta-N}. Within the $\delta N$ formalism
the curvature perturbation generated at the end of inflation ($t=t_{e}$) can be expressed as
\begin{eqnarray}
\zeta_{end} &=& {\partial N \over \partial \phi_e} \delta \phi_e
+ {1 \over 2}{\partial^2 N \over \partial \phi^2_e} \delta \phi_e^2 \nonumber\\
&=& {\partial N \over \partial \phi_e} {d \phi_e(A) \over d A^i}\delta A^i + {1 \over 2}
\left[{\partial N \over \partial \phi_e} {d^2 \phi_e(A) \over d A^i d A^j} + {\partial^2 N \over \partial \phi^2_e} {d \phi_e(A) \over d A^i}{d \phi_e(A) \over d A^j}\right]\delta A^i \delta A^j~,
\end{eqnarray}
where we set $A^0=0$.
That is, $\phi_e$ can fluctuate due to the perturbation of the vector field $A^i$.
Let us take the hypersurface at the end of inflation $t=t_e$ to be a uniform energy density one.
Then,
the total curvature perturbation at the end of inflation $t=t_{e}$ is given by
\begin{eqnarray}
\zeta (t_{e}) = \zeta_{inf} + \zeta_{end}~,
\end{eqnarray}
where
\begin{eqnarray}
&&\zeta_{inf} = {\partial N \over \partial \phi_*}\delta \phi_* + {1 \over 2}{\partial^2 N \over \partial \phi^2_*}\delta \phi_*^2~
\end{eqnarray}
is the conventional  curvature perturbation generated by quantum fluctuations of the inflaton.
Here, $\phi_*$ represents a value at the initial hypersurface.

The power spectrum of curvature perturbation At the leading order is given by
\begin{eqnarray}
\langle \zeta_{{\bf k}_1} \zeta_{{\bf k}_2}\rangle
&=& P_\zeta({\bf k}_1)\delta^{(3)}\left({\bf k}_1 + {\bf k}_2\right)  \nonumber\\
&=& N_*^2 P_\phi(k_1)
\delta^{(3)}\left({\bf k}_1 + {\bf k}_2\right)
+ N_e^2 {d \phi_e(A) \over d A^i}{d \phi_e(A) \over d A^j}
 \langle \delta A^i_{*} ({\bf k}_1) \delta A^j_{*}({\bf k}_2)\rangle~,
 \label{curvps}
\end{eqnarray}
where we defined  $P_\phi(k) = {H_*^2}/2k^3$, $N_* = \partial N /\partial \phi_*$ and $N_e = \partial N / \partial \phi_e$.
Here, we assumed that the scalar and the vector fields are statistically independent and Gaussian.
 The scale invariant power spectrum of the vector field $\langle \delta A^i ({\bf k}_1) \delta A^j ({\bf k}_2) \rangle$
is obtained by setting  $c=1$ in section 3.1 and  given by
\begin{eqnarray}
\langle \delta A^i_{*} ({\bf k}) \delta A^j_{*} ({\bf k}') \rangle &\!=\!& {H^2_* \over 2k^3f^2_*}\left(\delta^{ij} - {k^ik^j \over k^2}\right)\delta^{(3)}({\bf k} + {\bf k}') \nonumber\\
&\!=\!& P_\phi(k)f^{-2}_*
\left(\delta^{ij} - {k^ik^j \over k^2}\right)\delta^{(3)}({\bf k} + {\bf k}')~.
\label{finalpsvec}
\end{eqnarray}
We note that to compute the above we have used  canonically normalized field $\delta A_i = A_i /a$ where $A_i$ is the coordinate basis field used in subsection 3.1.
Using (\ref{finalpsvec}),
we obtain the expression for the power spectrum of curvature perturbations
in the modified hybrid inflation model (\ref{curvps}) as
\be\label{Anisotropic-power-spectrum}
P_\zeta({\bf k}_1) =
\left[ N_*^2+ \left({N_e \over f_*}\right)^2 {q_i}{q_j}
\left(\delta^{ij} - {k^i_1k^j_1 \over k^2_1}\right)
\right]
 P_\phi(k_1)    \ ,
\ee
where we have defined $q_i \equiv d\phi_e/dA^i$, $q_{ij} \equiv d^2\phi_e/dA^idA^j$.
The second term in the bracket above leads to statistical anisotropy in the power spectrum. It is usual to parameterize  this anisotropy as \cite{Ackerman:2007nb},
\begin{eqnarray}
 P_\zeta({\bf k}) = P^{iso}_{\zeta}(k)\left[ 1 + g_\beta (\hat{\bf q}\cdot \hat{\bf k})^2 \right]~,
\end{eqnarray}
where we used the notation $\hat{\bf q} = {\bf q}/\left|{\bf q}\right|$, $\hat{\bf k} = {\bf k}/\left|{\bf k}\right|$ and the isotropic part
\begin{eqnarray}
P^{iso}_{\zeta}(k) = {P_\phi(k) \over 2\epsilon_*}(1+\beta)
\label{anisp}
\end{eqnarray}
is separated. The coefficient of the anisotropic part reads
\begin{eqnarray}
~g_\beta = - {\beta \over 1+\beta}~, \quad
\beta \equiv \left({N_e \over N_* f_*}\right)^2\left|{\bf q}\right|^2~,
\end{eqnarray}
where we defined $N_* = 1/\sqrt{2\epsilon_*}$.

The bispectrum  to the leading order is  given by
\begin{eqnarray}
\langle \zeta_{{\bf k}_1} \zeta_{{\bf k}_2} \zeta_{{\bf k}_3}\rangle
&\! \equiv &\! (2\pi)^{-3/2}B_\zeta({\bf k}_1, {\bf k}_2, {\bf k}_3) \delta^{(3)}\left({\bf k}_1 + {\bf k}_2 + {\bf k}_3 \right) \nonumber\\
&\!=&\! (2\pi)^{-3/2}\Biggl\{N_*^2 N_{**} \left[P_\phi(k_1)P_\phi(k_2) + 2~ {\rm perms} \right] \delta^{(3)}\left({\bf k}_1 + {\bf k}_2 + {\bf k}_3 \right)\nonumber\\
&& \qquad\qquad
+  N_e^4 {d \phi_e(A) \over d A^i}{d \phi_e(A) \over d A^j}
\left({1 \over N_e}{d^2 \phi_e(A) \over d A^{\ell_1}dA^{\ell_2}}+{N_{ee} \over N_e^2}
 {d \phi_e(A) \over d A^{\ell_1}}{d \phi_e(A) \over d A^{\ell_2}}
\right) \nonumber\\
&&\qquad\qquad \ \ \times
\left[\langle  \delta A^i_{*} ({\bf k}_1) \delta A^j_{*}({\bf k}_2)
\left(\delta A^{\ell_1} \star \delta A^{\ell_2}\right)_{*}({\bf k}_3) \rangle  + 2~ {\rm perms}\right] \Biggr\}~,
\label{curvbi}
\end{eqnarray}
where $\star$ in the last line denotes the convolution.
From these expressions,  we see the curvature perturbation has the direction-dependence due to the vector field.
Substituting the expression (\ref{finalpsvec})  into (\ref{curvbi}), we  can deduce the bispectrum  (\ref{curvbi}) as
\begin{eqnarray}
\langle \zeta_{{\bf k}_1} \zeta_{{\bf k}_2} \zeta_{{\bf k}_3}\rangle
&\!=&\! (2\pi)^{-3/2} B_\zeta({\bf k}_1, {\bf k}_2, {\bf k}_3) \delta^{(3)}\left({\bf k}_1 + {\bf k}_2 + {\bf k}_3 \right) \nonumber\\
B_\zeta({\bf k}_1, {\bf k}_2, {\bf k}_3)
&\!=&\! \Biggl[
N_*^2N_{**}+ \left({N_e \over f_*}\right)^4{q_i}{q_j}
\left({1 \over N_e}{q_{\ell_1\ell_2}}
+ {N_{ee} \over N_e^2}
{q_{\ell_1}}{q_{\ell_2}}
\right)
\left(\delta^{i\ell_1} - {k^i_1k^{\ell_1}_1 \over k^2_1}\right)
\left(\delta^{j\ell_2} - {k^j_2k^{\ell_2}_2 \over k^2_2}\right)
 \Biggr] \nonumber\\
&& \qquad\qquad\qquad\qquad\qquad\qquad\qquad\qquad
\times P_\phi(k_1)P_\phi(k_2)
+ 2~{\rm perms}~,
\end{eqnarray}
where  we assumed that $\delta A_i$ is  Gaussian.

Recalling the anisotropy in the power spectrum \eqref{Anisotropic-power-spectrum} it is convenient to define
the non-linear parameter $f_{NL}$ as the bispectrum normalized by the isotropic part of power spectrum
$P_\zeta^{iso}(k)$
\be
{6 \over 5}f_{NL}({\bf k}_1,{\bf k}_2,{\bf k}_3)
\!\equiv \! {B_\zeta({\bf k}_1,{\bf k}_2,{\bf k}_3) \over P_\zeta^{iso}(k_1)P_\zeta^{iso}(k_2) + 2~{\rm perms}}
=
\left[
{F({\bf k}_1,{\bf k}_2)k^3_3 \over \sum_{i}{ k_i^3}}
+ 2~{\rm perms}\right]~,
\ee
where we assumed the scale-invariant power spectrum and defined
\begin{eqnarray}
F({\bf k}_1,{\bf k}_2) &\!\equiv\!& (1+\beta)^{-2}{N_{**} \over N_*^2} + g_\beta^2{N_{ee} \over N_e^2}
\left[ 1 - (\hat{\bf q} \cdot \hat{\bf k}_1)^2 \right]\left[ 1 - (\hat{\bf q} \cdot \hat{\bf k}_2)^2 \right] \nonumber\\
&& \qquad\qquad
+{g_\beta^2 \over N_e}{q_{\ell_1\ell_2} \over  \left|{\bf q}\right|^2}\left[ \hat{q}^{\ell_1} - (\hat{\bf q} \cdot \hat{\bf k}_1)\hat{k}^{\ell_1}_1 \right]\left[\hat{q}^{\ell_2} - (\hat{\bf q} \cdot \hat{\bf k}_2)\hat{k}^{\ell_2}_2 \right]~.
\end{eqnarray}
Strictly speaking, one should define the non-Gaussianity parameter $f_{NL}$
normalizing the bispectrum with the full power spectrum instead of the isotropic one.
However, since we expect the anisotropic part in the power spectrum to be small,
the above definition provides a good approximation. We will discuss this point further in the end of this subsection.

Under the slow-roll approximation, we obtain $N_{**}/N_*^2 \sim N_{ee}/N_e^2 = O(\epsilon)$.
Hence, neglecting the first and second terms in the right hand side of the above equation,
we have more simplified expression as
\begin{eqnarray}
F({\bf k}_1, {\bf k}_2) &\simeq& {g_\beta^2 \over N_e}{\hat{q}_{\ell_1\ell_2}}\left[ \hat{q}^{\ell_1} - (\hat{\bf q} \cdot \hat{\bf k}_1)\hat{k}^{\ell_1}_1 \right]\left[\hat{q}^{\ell_2} - (\hat{\bf q} \cdot \hat{\bf k}_2)\hat{k}^{\ell_2}_2 \right]~,
\end{eqnarray}
where $\hat{q}_{\ell_1 \ell_2} = {q}_{\ell_1 \ell_2}/\left|{\bf q}\right|^2$.
We can decompose the non-linear parameter into the isotropic part and the anisotropic part as
\begin{eqnarray}
{6 \over 5}f_{NL}({\bf k}_1, {\bf k}_2, {\bf k}_3)
&\! = \!&
{6 \over 5}f_{NL}^{iso}\left( 1 + f^{ani}( {\bf k}_1, {\bf k}_2, {\bf k}_3) \right)~,
\label{anisofnl}
\end{eqnarray}
where the isotropic part reads
\begin{eqnarray}
{6 \over 5}f_{NL}^{iso} \equiv {\sqrt{2 \epsilon_e} g_\beta^2}
\hat{q}_{ij}\hat{q}^i\hat{q}^j~, \label{fnliso}
\end{eqnarray}
and the anisotropic part is deduced as
\be
f^{ani}({\bf k}_1,{\bf k}_2, {\bf k}_3) = {\hat{q}_{ij} \over \hat{q}_{kl}\hat{q}^k\hat{q}^l}
\Bigl\{ \left[
- \hat{q}^i\left(\hat{k}_1^j (\hat{\bf q}\cdot \hat{\bf k}_1)
+ \hat{k}_2^j (\hat{\bf q}\cdot \hat{\bf k}_2)
\right)
+
\hat{k}^i_1\hat{k}^j_2 (\hat{\bf q}\cdot \hat{\bf k}_1)
(\hat{\bf q}\cdot \hat{\bf k}_2) \right]
{k_3^3
 \over \sum_{i}{ k_i^3}}  + 2~{\rm perms}\Bigr\}  \ .
\label{fnlani}
\ee
Here, we used the relation $N_e = 1/\sqrt{2\epsilon_e}$.
Taking a look at the above formula, we notice that the statistical anisotropy
gives a non-trivial shape to the bispectrum even for the local model.

Let us now consider a specific hybrid model \eqref{Hybrid} with a charged waterfall field $\chi$ in the unitary gauge. The potential for this model is given as
\begin{eqnarray}
V(\phi,\chi,A^i) = {\lambda \over 4}\left(\chi^2 - v^2\right)^2 + {1 \over 2}g^2 \phi^2 \chi^2 + {1 \over 2}m^2\phi^2
+ {1 \over 2}h^2 A^iA_i \chi^2,
\end{eqnarray}
where $A_i = \delta_{ij}A^j$ is the gauge field.
The coupling constants are denoted by $\lambda, g, h$, the inflaton mass is given by $m$,
and the vacuum expectation value for $\chi$ is represented by $v$.
For this model the effective mass squared of the waterfall field $\chi$ is given by
\begin{eqnarray}
m^2_\chi \equiv - \lambda v^2 + g^2 \phi^2 + h^2 A^iA_i~.
\end{eqnarray}
Inflation ends when
\begin{eqnarray}
\lambda v^2 = g^2 \phi^2_e + h^2 A^i A_{i}.
\end{eqnarray}
As discussed the critical value $\phi_e$ depends on $A^i$. We hence have
\begin{eqnarray}
\hat{q}^i = - { A^i \over \left|{\bf A}\right|}~,
~\hat{q}^{ij} = -{1 \over \phi_e}\left({g^2 \phi_e^2 \over h^2 \left|{\bf A}\right|^2} \delta^{ij} + \hat{q}^i \hat{q}^j \right)
\label{form1}
\end{eqnarray}
and
\begin{eqnarray}
\beta \simeq {1 \over f_*^2}\left({h^2\left|{\bf A}\right| \over g^2 \phi_e}\right)^2~,
\label{form2}
\end{eqnarray}
where we used the approximation $N_* \simeq N_e$.
We can now write down the power spectrum
\begin{eqnarray}
 P_\zeta({\bf k}) = P^{iso}_{\zeta}(k)\left[ 1
 - \frac{{1 \over f_*^2}\left({h^2\left|{\bf A}\right| \over g^2 \phi_e}\right)^2}
 {1+{1 \over f_*^2}\left({h^2\left|{\bf A}\right| \over g^2 \phi_e}\right)^2}
 (\hat{\bf q}\cdot \hat{\bf k})^2 \right]~.
\end{eqnarray}
Notice that despite being direction-dependent, the power spectrum is scale invariant. The magnitude of statistical anisotropy is determined
 by the parameter $g_\beta$.

Let us now study the bispectrum which is more interesting.
Substituting the expressions (\ref{form1}) and (\ref{form2}) into (\ref{fnliso}),
we have the isotropic part of the $f_{NL}$ parameter
\begin{eqnarray}
{6 \over 5}f_{NL}^{iso} \simeq
 -\eta_e g_\beta^2\left(1 + {g^2 \phi_e^2 \over h^2 \left|{\bf A}\right|^2}\right) \ .
\end{eqnarray}
where $\eta = V_{\phi\phi}/V$. Similarly, the anisotropic part (\ref{fnlani}) reads
\begin{eqnarray}
f^{ani}({\bf k}_1,{\bf k}_2,{\bf k}_3) &\!=\!& -\left\{ \left[\left(\hat{\bf q}\cdot \hat{\bf k}_1\right)^2
 + \left(\hat{\bf q}\cdot \hat{\bf k}_2\right)^2\right]{k_3^3
 \over \sum_{i}{ k_i^3}}
 + 2~{\rm perms} \right\}\nonumber\\
&&\hspace{-3cm}
+ \left(1 + {g^2 \phi_e^2 \over h^2 \left|{\bf A}\right|^2}\right)^{-1}\Biggl\{
\left[\left(\hat{\bf q}\cdot \hat{\bf k}_1\right)^2\left(\hat{\bf q}\cdot \hat{\bf k}_2\right)^2 + {g^2 \phi_e^2 \over h^2 \left|{\bf A}\right|^2}\left(\hat{\bf k}_1\cdot \hat{\bf k}_2\right)\left(\hat{\bf q}\cdot \hat{\bf k}_1\right)\left(\hat{\bf q}\cdot \hat{\bf k}_2\right) \right]{k_3^3
 \over \sum_{i}{ k_i^3}}  \nonumber\\
&&   \hspace{4cm}  + 2~{\rm perms}\Biggr\}~,
\label{fani}
\end{eqnarray}
where as usual $\hat{\bf k}_1+\hat{\bf k}_2+\hat{\bf k}_3 = 0$.
From the above expression we see that
the amplitude of the non-Gaussianity $f_{NL}^{iso}$ depends also on
the magnitude of statistical anisotropy $g_\beta$.
For a large $\beta$ ($\beta \gg 1$), $g_\beta \simeq 1$ and hence
the statistical anisotropy appearing in the primordial power spectrum (\ref{anisp}) is large.
On the other hand, for a small $\beta$ ($\beta \ll 1$),
the statistical anisotropy is small and the non-linear parameter $f_{NL}$ is also small.

Eq.(\ref{fnliso}) provides the possibility of generating large
non-Gaussianity even for the cases with the small statistical anisotropy in the power spectrum
if we choose $f_*$ to be small and ${g^2 \phi_e^2 \over h^2 \left|{\bf A}\right|^2}$ to be large while $g_\beta \ll 1$.
From (\ref{anisofnl}) and (\ref{fani}) we see that for this case the anisotropic and anisotropic parts of the bispectrum are of the same order,
in contrast to the power spectrum.
Hence, it may be possible to detect the statistical anisotropy
in the bispectrum with the future experiments and
it will give us information about
a new physics in the early universe associated with the violation
of the rotational invariance. The  mechanism explained in this subsection
has been expanded and extended in many different ways which may be found in \cite{Dimopoulos:2009vu,Bartolo:2009pa,arXiv:1107.2779,
arXiv:1107.3186,Emami:2011yi,Dimopoulos:2012av,Lyth:2012br,Dey:2011mj,Jain:2012vm}.

\subsection{Non-Gaussiantiy induced from gauge fields}

It is well known that a single field inflation with standard kinetic term cannot generate the sizable non-Gaussianity~\cite{Maldacena:2002vr}.
Presence of other fields, including gauge fields, may however change this result and generate detectable non-Gaussianity. In this subsection we consider
two such setups in which gauge fields can enhance the bispectrum of the inflaton fluctuations. One such possibility is
when  gauge fields have an axionic coupling. The other one is when the gauge kinetic term  has a non-trivial inflaton dependence. As we will discuss these two cases lead to two different shape non-Gaussianities, the former to an equilateral shape and the latter to a local shape.

\subsubsection{Non-Gausssianity from gauge fields with an axion coupling}

In the presence of axion coupling~\cite{Lue:1998mq}, there exists tachyonic instability in the positive helicity polarization mode of gauge fields.
The amplitude of gauge fields can hence be large, inducing  large non-Gaussianity in curvature perturbations~\cite{Barnaby:2010vf,Barnaby:2011vw, Sorbo-Anber}.
To see explicitly how this happens, let us consider the Abelian U(1) gauge theory coupled to an axion $\phi$:
\begin{eqnarray}
  S =    \int d^4 x \sqrt{-g} \left[ \frac{\mpl^2}{2}R -\frac{1}{2} \partial_\mu \phi \partial^\mu \phi  - V(\phi )
- \frac{1}{4}  F_{\mu\nu} F^{\mu\nu} - \frac{\alpha}{8f_a} \phi \epsilon^{\mu\nu\alpha\beta} F_{\mu\nu} F_{\alpha\beta} \right]  \ ,
\label{axion-action}
\end{eqnarray}
where $f_a$ is an axion decay constant and $\alpha$ is a constant.
In the inflationary background, the gauge field in the Coulomb gauge obeys the following equations
\begin{eqnarray}\label{gauge-field-axion-e.o.m}
   A''_i - \nabla^2 A_i - \frac{\alpha}{f_a}\phi' \epsilon_{ijk} \nabla_j A_k   = 0 \ .
\end{eqnarray}
To diagonalize equations we use circular polarization vectors $\epsilon^i_\lambda$ satisfying
\be\begin{split}
k^i \epsilon^i_\lambda =0 \ &,\qquad  \epsilon_{ijk} k_j \epsilon^k_{\pm} = \mp ik\epsilon^i_{\pm} \ , \\
\epsilon^i_{\pm} (-{\bf k})=\epsilon^i_{\pm} ({\bf k})^* \ &,\qquad \epsilon^i_{\lambda} ({\bf k})^* \epsilon^i_{\lambda'} ({\bf k}) = \delta_{\lambda \lambda'}\,.
\end{split}
\ee
In terms of the polarization vectors, we can expand the gauge field as
\begin{eqnarray}
    A_i = \sum_{\lambda =\pm} \int \frac{d^3 k}{(2\pi)^{3/2}}
\left[ a_\lambda ({\bf k}) A_\lambda (\tau , k) \epsilon^i_\lambda ({\bf k}) e^{i{\bf k}\cdot {\bf x}} + h.c. \right] \ ,
\label{expansion-2}
\end{eqnarray}
where annihilation and creation operators $a_\lambda ({\bf k}), a_\lambda^\dagger ({\bf k})$ satisfy the standard canonical commutation relations.
The equations of motion \eqref{gauge-field-axion-e.o.m} boils down to
\begin{eqnarray}
   \left[ \frac{\partial^2}{\partial \tau^2} + k^2 \mp \frac{2k\xi}{\tau} \right] A_{\pm} (\tau , k) = 0 \ ,
   \qquad \xi \equiv \frac{\alpha \dot{\phi}}{2f_a H}  \ .
\end{eqnarray}
Here, we neglected the deviation from de Sitter expansion and also
assumed $\xi$ is almost constant during slow-roll inflation.

In the interval $(8\xi)^{-1} \lesssim k/aH \lesssim 2\xi$, there exists tachyonic instability in the positive helicity polarization
modes where we can approximate the mode function as
\begin{eqnarray}
  A_+ (\tau ,k ) = \frac{1}{\sqrt{2k}} \left( \frac{k}{2\xi a H}\right)^{1/4} e^{\pi \xi - 2\sqrt{\frac{2\xi k}{ aH}}} \ .
\end{eqnarray}
Since negative helicity polarization modes do not suffer from the tachyonic instability, hereafter we will neglect them.
The exponential growth caused by the tachyonic instability  is physically the same phenomenon as the usual mode-freezing at the horizon crossing and thus
we have production of gauge fields due to axion coupling.

The produced gauge fields in their own turn can source  scalar fluctuations. To see this, let us decompose the inflaton fluctuations as
\begin{eqnarray}
    \delta \phi = \int \frac{d^3 k}{(2\pi)^{3/2}} \frac{Q_k (\tau )}{a(\tau )} e^{i{\bf k}\cdot{\bf x}} \ .
\end{eqnarray}
For our analysis below it is enough to consider zeroth order in slow-roll parameter where $a(\tau)=-\frac{1}{H\tau}$, and hence the modes $Q_k(\tau)$ satisfy
\begin{eqnarray}\label{Q-J-section-3.2}
   \left[ \frac{\partial^2}{\partial \tau^2}  + k^2 - \frac{2}{\tau^2} \right] Q_k (\tau ) = J_k (\tau )  \ ,
\end{eqnarray}
where the source term $J_k$, which is induced from the axion coupling, is
\begin{eqnarray}
  J_k (\tau ) = a^3 (\tau ) \frac{\alpha}{f_a} \int \frac{d^3 k}{(2\pi)^{3/2}} e^{-i{\bf k}\cdot {\bf x}} E^i B^i \ .
\end{eqnarray}
Because of the tachyonic instability, this source term can induce significant effect on the statistics of curvature perturbations.
The solution consist of homogeneous $Q_k^{\rm vac}$ and inhomogeneous $Q_k^{\rm J}$ ones
\begin{eqnarray}
  Q_k (\tau ) =   Q_k^{\rm vac} (\tau ) +  Q_k^{\rm J} (\tau ) \ .
\end{eqnarray}
The homogeneous solution can be expressed by the positive frequency mode function
\begin{eqnarray}
   \varphi_k (\tau) = i \frac{\sqrt{\pi}}{2} \sqrt{-\tau} H_{3/2}^{(1)} (-k\tau ) \ .
\end{eqnarray}
In order to obtain inhomogeneous solution, we use the retarded Green's function
\begin{eqnarray}
G_k (\tau , \tau' ) = i \theta (\tau - \tau' ) \left[ \varphi_k (\tau ) \varphi_k^* (\tau' )
                        -  \varphi_k^* (\tau ) \varphi_k (\tau' )  \right]
\end{eqnarray}
satisfying
\begin{eqnarray}
   \left[ \frac{\partial^2}{\partial \tau^2}  + k^2 - \frac{2}{\tau^2} \right] G_k (\tau , \tau') = \delta (\tau -\tau')  \ .
\end{eqnarray}
Using the Green function, we obtain
\begin{eqnarray}
     Q_k^{\rm J} ( \tau ) = \int^0_{-\infty} d\tau' G_k (\tau ,\tau' ) J_k (\tau'  )  \ .
\end{eqnarray}
Recalling the relation
\begin{eqnarray}
  {\cal R}_c = - \frac{H}{\dot{\phi}} \frac{Q_k}{a}  \ ,
\end{eqnarray}
we can calculate the power spectrum
\begin{eqnarray}
 \langle {\cal R}_c ({\bf k}_1) {\cal R}_c ({\bf k}_2) \rangle = \frac{H^2}{a^2 \dot{\phi}^2}
  \left[ \VEV{Q_{k_1}^{\rm vac} Q_{k_2}^{\rm vac}} + \VEV{Q_{k_1}^{\rm J} Q_{k_2}^{\rm J}} \right] \ .
\end{eqnarray}
Here, we used the fact that $Q_{k}^{\rm vac}$ and $Q_{k}^{\rm J}$ have no statistical correlation.
Apparently, the first term is the conventional one
\begin{eqnarray}
\VEV{{\cal R}_c^{\rm vac} ({\bf k}_1) {\cal R}_c^{\rm vac} ({\bf k}_2) }
=  \frac{H^4}{2 \dot{\phi}^2}  \frac{1}{k_1^3} \delta^{3} ({\bf k}_1+{\bf k}_2) \ .
\end{eqnarray}
The contribution from the inhomogeneous terms is given by
\begin{eqnarray}
\VEV{{\cal R}_c^{\rm J} ({\bf k}_1) {\cal R}_c^{\rm J} ({\bf k}_2) }
=  \frac{H^2}{\dot{\phi}^2} \int d\tau' d\tau'' \frac{G_k (\tau , \tau' )}{a(\tau)} \frac{G_k (\tau , \tau'' )}{a(\tau)}
        \VEV{ J_{k_1} (\tau_1 ) J_{k_2} (\tau_2 ) }    \ .
\end{eqnarray}
For large $\xi$, we can deduce
\begin{eqnarray}
   P_{{\cal R}_c} = P \left[ 1+  P f_2 (\xi) e^{4\pi \xi }  \right] \ ,
\end{eqnarray}
where $f_2 (\xi )$ can be numerically calculated. For $2<\xi <3$, the best fit becomes
\begin{eqnarray}
   f_2 (\xi ) = \frac{3\times 10^{-5}}{\xi^{5.4}} \ .
\end{eqnarray}

Similarly, one may compute the correction to the bispectrum
\begin{eqnarray}
\VEV{ {\cal R}_c^{\rm J} ({\bf k}_1) {\cal R}_c^{\rm J} ({\bf k}_2) {\cal R}_c^{\rm J} ({\bf k}_3) }
&=&  -\frac{H^3}{\dot{\phi}^3} \int d\tau_1 d\tau_2 d\tau_3 \frac{G_k (\tau , \tau_1 )}{a(\tau)}
               \frac{G_k (\tau , \tau_2 )}{a(\tau)}  \frac{G_k (\tau , \tau_3 )}{a(\tau)}  \nonumber\\
   &&  \hspace{2cm}   \times     \VEV{ J_{k_1} (\tau_1 ) J_{k_2} (\tau_2 )J_{k_3} (\tau_3 ) }    \ .
\end{eqnarray}
Defining new variables
\begin{eqnarray}
   |{\bf k}_1| = k \ ,  |{\bf k}_2| = x_2 k \ ,  |{\bf k}_3| = x_3 k \ ,
\end{eqnarray}
we obtain
\begin{eqnarray}
 \VEV{ {\cal R}_c^{\rm J} ({\bf k}_1) {\cal R}_c^{\rm J} ({\bf k}_2) {\cal R}_c^{\rm J} ({\bf k}_3)}
=  -\frac{3}{10} (2\pi)^{5/2} P^3 e^{6\pi \xi}
     \frac{1+x_2^3 + x_3^3}{x_2^3 x_3^3}
                f_3 (\xi , x_2 , x_3 )\ \frac{\delta ({\bf k}_1 + {\bf k}_2 + {\bf k}_3)}{k^6}     ,
\end{eqnarray}
where $f_3 (\xi , x_2 , x_3 )$ is a complicated function. Since the bispectrum should have equilateral configuration, we only need
\begin{eqnarray}
   f_3 (\xi ,1,1) = \frac{7.4\times 10^{-8}}{\xi^{8.1}} \ .
\end{eqnarray}
From the bispectrum, we can read off
\begin{eqnarray}
  f_{\rm NL}^{\rm eff} = \frac{f_3 (\xi , 1 , 1 ) P^3 e^{6\pi \xi}}{P_{{\cal R}_c} (k)^2} \ .
\end{eqnarray}
Taking into account the current observational bound $-214 < f_{\rm NL}^{\rm equil } < 266 $ obtained from the WMAP data,
we get the constraint $\xi \lesssim 2.65$~\cite{Barnaby:2011vw}.

\subsubsection{Non-Gaussianity from the gauge kinetic function}

Another possibility for generating sizable non-Gaussianity from gauge fields discussed in the literature is through the gauge kinetic term~\cite{Barnaby:2012tk,Namba:2012gg}.
Let us consider the following model
\begin{eqnarray}
  S =    \int d^4 x \sqrt{-g} \left[ \frac{\mpl^2}{2}R -\frac{1}{2} \partial_\mu \phi \partial^\mu \phi  - V(\phi )
- \frac{1}{4} f^2 (\phi ) F_{\mu\nu} F^{\mu\nu} \right]  \ ,
\end{eqnarray}
where instead of tachyonic instability, the gauge kinetic function generates gauge field quanta. Let us again consider the slow roll limit of inflation. Thus,
the calculation is mathematically similar to the previous case and the difference is the source term. That is, the perturbations $Q_k(\tau)$ are governed by the equation \eqref{Q-J-section-3.2} but now the source term is
\begin{eqnarray}
  J_k (\tau ) =  \frac{a^3}{2} \frac{f_\phi^2}{f}\int \frac{d^3 k}{(2\pi)^{3/2}} e^{-i{\bf k}\cdot {\bf x}} \left( E^2- B^2 \right) \ .
\end{eqnarray}
In the case of axion coupling, the exponential growth caused by tachyonic instability enhances the amplitude of the gauge field, which works at the horizon crossing. This is why the shape of the bispectrum is equilateral. Here, instead, the non-minimal coupling makes the gauge fields survive, which works on superhorizon scales and persists during inflation.  Thus, the shape of the bispectrum should be local.

The procedure for the calculation is exactly the same as the axion coupling cases.
Defining the power spectrum
\begin{eqnarray}
\VEV{ {\cal R}_c ({\bf k}_1) {\cal R}_c ({\bf k}_2) } = \frac{2\pi^2}{k_1^3} P_{{\cal R}_c}(k_1) \delta ({\bf k}_1-{\bf k}_2) \ ,
\end{eqnarray}
we obtain the result
\begin{eqnarray}
   P_{{\cal R}_c} = P \left[ 1+ 192 P N_{\rm CMB} \left( N_{\rm tot} - N_{\rm CMB} \right) \right] \ ,
\end{eqnarray}
where $P$ is the conventional power spectrum, $N_{\rm CMB}$ is the number of e-folds from the CMB scale to the end of inflation,
and $N_{\rm tot}$ is the total number of e-folds of inflation.
In order for the perturbative analysis to be valid, the correction must be subdominant
\begin{eqnarray}
  192 P N_{\rm CMB} \left( N_{\rm tot} - N_{\rm CMB} \right) <1 \ .
\label{P-constraint}
\end{eqnarray}
Since $P\sim 10^{-9}$, this can be easily satisfied.

Next, we can calculate the bispectrum
\begin{eqnarray}
   \VEV{ {\cal R}_c ({\bf k}_1) {\cal R}_c ({\bf k}_2) {\cal R}_c ({\bf k}_3) }
  = B_{{\cal R}_c} (k_i ) \delta  ({\bf k}_1 + {\bf k}_2 + {\bf k}_3)
\end{eqnarray}
and the local non-Gaussianity  can be characterized by $f_{\rm NL}$ defined in \eqref{LCNG}.
The result is intriguing
\begin{eqnarray}
  f_{\rm NL} (k_i ) \simeq f_{\rm NL}^{\rm equiv. local} \times \frac{3}{4} \left[ 1 +
              \frac{k_1^3 \cos^2 (k_2 ,k_3) + k_2^3 \cos^2 (k_3 ,k_1) + k_3^3 \cos^2 (k_1 ,k_2) }{k_1^3 + k_2^3 + k_3^3}  \right]  \ ,
\end{eqnarray}
where
\begin{eqnarray}
f_{\rm NL}^{\rm equiv. local}  \simeq
 1280 P N_{\rm CMB}^3 \left( N_{\rm tot} - N_{\rm CMB} \right)
\simeq 0.7 \left( \frac{N_{\rm CMB}}{60} \right)^3 \left( N_{\rm tot} - N_{\rm CMB} \right)
\end{eqnarray}
In spite of the constraint (\ref{P-constraint}), we can obtain sizable non-Gaussianity in the range
\begin{eqnarray}
    10 \leq f_{\rm NL} \leq 100 \ .
\end{eqnarray}
Moreover, for $k_3 \ll k_1 \sim k_2 $, the $k$-dependence of the bispectrum becomes
\begin{eqnarray}
   k_1^3 k_3^3  \VEV{ {\cal R}_c ({\bf k}_1) {\cal R}_c ({\bf k}_2) {\cal R}_c ({\bf k}_3) }
 &\propto& 1+ \cos^2 (k_1 ,k_3 )   \nonumber \\
 &\propto&  \cos \epsilon Y^0_0 + \sin \epsilon Y^0_2 \ , \qquad \epsilon \equiv \tan^{-1} \frac{1}{2\sqrt{5}} \simeq 0.22 \ .
\end{eqnarray}
That is, there is anisotropy in the bispectrum.

\subsection{Primordial gravitational waves from gauge fields}

Having a non-zero contribution from the gauge fields to the energy momentum tensor, they will appear as the source term for (primordial) gravity waves.
This could be more pronounced when the gauge fields are produced due to the tachyonic instability  induced by a time dependent axion, as discussed in section 3.4.1. So, let us consider the action (\ref{axion-action}). It is straightforward to show that the gravity waves are governed by
\begin{eqnarray}
     \frac{1}{2a^2}\left[ \frac{\partial^2}{\partial \tau^2} + 2 \frac{a'}{a} \frac{\partial}{\partial \tau} + k^2 \right]h_{ij}
= -\frac{1}{\mpl^2}  \left[ E_i E_j + B_i B_j \right] \ ,
\end{eqnarray}
where the source term should be calculated taking into account the tachyonic instability. As we see the two graviton polarizations appear with different source terms. This will cause a birefringent (chiral) gravity wave production which has its own interesting observational prospects~\cite{Barnaby:2010vf,Sorbo-Anber,Sorbo-parity-violation,Barnaby-monodromy}. Although there  are several mechanisms to generate birefringent gravitational waves~\cite{Alexander:2004wk,Satoh:2007gn}, the tachyonic instability provides a natural and simple setup. In particular, it gives rise to primordial gravitational waves even if the energy scale of inflation is low. Hence, it is interesting to see the signature in the CMB~\cite{Saito:2007kt}.

Here we only review the effects of the source on the total gravity wave power spectrum, which is calculated to be
\begin{eqnarray}
  P_{GW} = \frac{2H^2}{\pi^2 \mpl^2} \left(\frac{k}{k_0}\right)^{n_T}  \left[ 1+ \frac{H^2}{\mpl^2} f_L (\xi) e^{4\pi \xi} \right]
\end{eqnarray}
where the approximate fitting function $f_L$ for the relevant range $2< \xi <3$ reads
\begin{eqnarray}
    f_L (\xi) \simeq \frac{2.6\times 10^{-7}}{\xi^{5.7}}  \ .
\end{eqnarray}
Here, the first factor is the standard amplitude for source-free gravity waves \eqref{power-tensor-single-scalar} while the extra contribution is due to the gauge fields. This extra contribution could be large even in the low energy inflation.
Besides the parity violation in gravity, one may also analyze non-Gaussianity induced by such gravity wave. Such an analysis has been carried out in~\cite{Soda:2011am,Shiraishi:2011st}.

\section{Anisotropic inflation with gauge fields}\label{Anisotropic-inflation-section}

To be consistent with the current observations the degree of anisotropy at cosmological scales and on the CMB data should be quite small,  of the order of $10^{-5}$ or less. Hence,  the accelerated expansion during inflation should be practically isotropic. This point has been put as the cosmic no-hair conjecture \cite{no-hair-conjecture, Kitada}.
The classic attempt in providing a proof for the conjecture was made by R. Wald \cite{Wald:1983ky} which goes under the name of cosmic no-hair theorem. This theorem, as discussed in \cite{Wald-extended-theorem} and reviewed in Appendix \ref{No-hair-extension-appendix}, need not be obeyed by the inflationary models. So, it is interesting to explore possibility of having inflationary models with anisotropic hair.

In general, since turning on vector gauge fields in the background gives rise to a preferred direction, one would naturally explore building anisotropic inflation models involving gauge fields.
In the previous section, we discussed that gauge fields may be used to describe a possible statistical anisotropy in the CMB data. In our discussions, however, the gauge fields were turned on at the fluctuations level and their  existence led to the anisotropic couplings between gauge fields and curvature perturbations.
It is then natural to explore  presence of gauge fields  at classical background level and study how they can affect inflationary trajectory and other observables within anisotropic inflationary models.

\subsection{Historical remarks}

The first attempt to use vector fields as inflatons was proposed by Ford~\cite{Ford:1989me}. He used  a potential of vector fields
to realize exponential expansion of the universe.
After that, there appeared other attempts to construct anisotropic inflation.
However,  because of the prejudice due to the cosmic no-hair conjecture, and the assumptions of Wald's theorem in view \cite{Wald:1983ky}, people tried to make
contrived models which generically break energy conditions. For example, higher curvature terms are used to
violate the cosmic no-hair conjecture~\cite{evading-no-hair-old-papers}.
Recently, a non-minimal coupling between vector fields and curvature
is adopted to make the vector fields slow roll~\cite{vector-inflation, Kanno:2008gn}.
Unfortunately, these models suffer from (ghost) instability, or a fine tuning problem~\cite{vector-inflation-ghost}.
After all, it turns out that violation of energy conditions, more than what is already required in inflationary models, is not necessary to realize anisotropically accelerating cosmologies \cite{Wald-extended-theorem} (see Appendix \ref{No-hair-extension-appendix}).  The gauge kinetic term in supergravity is a necessary ingredient to realize anisotropically accelerating expansion.
Indeed, anisotropic inflation could be ubiquitous in supergravity~\cite{Watanabe:2009ct,Soda:2012zm}.



As  shown in the previous section, gauge fields can play important roles
in inflation. In these calculations, typically, the backreaction is not negligible.
Indeed, the energy density of gauge fields is kept almost constant during ``anisotropic'' inflation models and anisotropic inflation provides a self-consistent framework for calculating various observables.
A a matter of terminology, the inflationary setup in which the gauge kinetic term is $f(\phi)^2 F^2$ is named anisotropic inflation or ``hairy inflation'' \cite{Watanabe:2009ct}.

\subsection{Inflation with anisotropic hair}\label{Hairy-inflation-section}

In this subsection, we study the bosonic part of the action in supergravity and explain how anisotropic inflation appears in the models with the action
\begin{eqnarray}
S&=&\int d^4x\sqrt{-g}\left[~\frac{1}{2}R
-\frac{1}{2}\left(\partial_\mu\phi\right)\left(\partial^{\mu}\phi\right)
-V(\phi)-\frac{1}{4} f(\phi )^2 F_{\mu\nu}F^{\mu\nu}
~\right] \ ,
\label{eq:action}
\end{eqnarray}
where we set $\mpl =1$ and $V(\phi)$ and $f(\phi)$ represent a potential of the inflaton and the gauge kinetic function, respectively.
Note that the gauge kinetic function can contribute to the potential energy as is found in a different
context~\cite{hep-th/0604192,Soda:2006wr}.
The picture we envisage is as follows. As we have seen in the previous section, gauge fields can be produced at superhorizon scales and their spectrum could be red. Therefore, a coherent, space independent gauge field can be consistently turned on in the background and affect background evolution.
We now explore the consequence of the coherent vector field.

Without loosing the generality, one can take
$x$-axis in the direction of the vector field. Using the gauge invariance,
we can express the vector field as
\begin{eqnarray}
 A_{\mu}dx^{\mu} = v(t) dx  \ .
\end{eqnarray}
Thus, there exists the rotational symmetry in the $y$-$z$ plane.
 Given this configuration, it is convenient to parameterize the metric as follows:
\begin{equation}
\hspace{-2cm}
ds^2 = -{\cal N}(t)^2dt^2 + e^{2\alpha(t)} \left[ e^{-4\sigma (t)}dx^2
+e^{2\sigma (t)}(e^{2\sqrt{3}\sigma_{-}(t)}dy^2
+e^{-2\sqrt{3}\sigma_{-}(t)}dz^2) \right] \ ,
\end{equation}
where $e^\alpha$, $\sigma$ and $\sigma_{-}$ are an isotropic scale factor
 and spatial shears, respectively.
 Here, the lapse function ${\cal N}$ is introduced to obtain the Hamiltonian constraint.
With the above ansatz, the action becomes
\begin{equation}
\hspace{-2cm} S=\int d^4x \frac{1}{\cal N}e^{3\alpha} \left[ 3 (-\dot{\alpha}^2+\dot{\sigma}^2
+\dot{\sigma}_{-}^2)+\frac{1}{2}\dot{\phi}^{2}-{\cal N}^2 V(\phi)+\frac{1}{2}f(\phi )^2\dot{v}^2 e^{-2\alpha (t) +4\sigma(t) } \right],
\end{equation}
where an overdot denotes  derivative with respect to the comoving time $t$.
First, its variation with respect to $\sigma_{-}$ yields
\begin{equation}
\ddot{\sigma}_{-}=-3\dot{\alpha}\dot{\sigma}_{-} \ .
\end{equation}
This gives $\dot{\sigma}_{-}\propto e^{-3\alpha}$, hence,
 the anisotropy in the $y$-$z$ plane rapidly decays as the Universe expands.
 Hereafter, for simplicity, we assume $\sigma_{-}=0$ and set the metric to be
\begin{eqnarray}\label{axi-metric}
 ds^2 = -dt^2 + e^{2\alpha(t)} \left[ e^{-4\sigma (t)}dx^2
 +e^{2\sigma (t)}(dy^2+dz^2) \right] \ ,
\end{eqnarray}
which is a Bianchi type-I model (see Appendix \ref{Bianchi-appendix} for more discussions).

The equation of motion for gauge field $v$ is easily solved as
\begin{equation}
\dot{v} =f(\phi)^{-2}  e^{-\alpha -4\sigma}  p_A  \ ,
\label{eq:Ax}
\end{equation}
where $p_A$ is a constant of integration. Taking the variation of the action with respect to
${\cal N}, \alpha, \sigma$ and $\phi$ and substituting the solution (\ref{eq:Ax}) into them,
we obtain the following basic equations
\begin{eqnarray}
\dot{\alpha}^2 &=& \dot{\sigma}^2+\frac{1}{3} \left[ \frac{1}{2}\dot{\phi}^2+V(\phi )+\frac{p_A^2}{2}f(\phi )^{-2}e^{-4\alpha -4\sigma} \right] \ , \label{eq:hamiltonian}\\
\ddot{\alpha} &=& -3\dot{\alpha}^2+ V(\phi ) +\frac{ p_A^2}{6}f(\phi )^{-2}e^{-4\alpha-4\sigma} \ , \label{eq:alpha}\\
\ddot{\sigma} &=&  -3\dot{\alpha}\dot{\sigma}+\frac{ p_A^2}{3}f(\phi)^{-2}e^{-4\alpha -4\sigma} \ , \label{eq:sigma}\\
\ddot{\phi} &=& -3\dot{\alpha}\dot{\phi}-V_{\phi}+p_A^2f(\phi)^{-3} f_{\phi}e^{-4\alpha -4\sigma} \label{eq:inflaton} \ ,
\end{eqnarray}
where as before the subscript in $V_\phi$ denotes  derivative with respect to its argument $\phi$.

Let us now check whether inflation occurs in this model. Using (\ref{eq:hamiltonian}) and (\ref{eq:alpha}), the equation for acceleration of the cosmic expansion is given by
\begin{equation}
\frac{(e^{\alpha})^{\cdot \cdot}}{e^{\alpha}} = \ddot{\alpha}+\dot{\alpha}^2 = -2\dot{\sigma}^2-\frac{1}{3} \dot{\phi}^2 + \frac{1}{3} \left[ V - \frac{p_A^2}{2}f^{-2}e^{-4\alpha -4\sigma} \right] \ .
\end{equation}
We see that to have (slow-roll) inflation the potential energy of the inflaton $\phi$ should  dominate over  the energy density of the vector field
\begin{eqnarray}\label{rho-v-section-4}
\rho _v \equiv \frac{1}{2} p_A^2 f(\phi)^{-2}e^{-4\alpha -4\sigma}\,,
\end{eqnarray}
and the shear $\Sigma \equiv \dot{\sigma}$.

First, we need to look at the shear to the expansion rate ratio $\Sigma/H$, which characterizes the anisotropy of the inflationary universe. Notice that (\ref{eq:sigma}) reads
\begin{eqnarray}
 \dot{\Sigma} = -3H\Sigma+\frac{2}{3}\rho _v \ .
\end{eqnarray}
If the anisotropy converges to a value, i.e. $\dot{\Sigma}$ becomes negligible, the terminal value should be given by
\begin{equation}\label{R-def}
\frac{\Sigma}{H} = \frac{2}{3}\frac{\rho _v }{ V(\phi)} \ ,
\end{equation}
where  we used (\ref{eq:hamiltonian}) in the slow-roll approximation, i.e.
\begin{eqnarray}
\dot\alpha^2=H^2 = \frac{1}{3} V(\phi)   \ .
\label{slow-roll:hamiltonian}
\end{eqnarray}

In order to realize the above situation, $\rho_v$ must be almost constant.
Assuming the standard slow-roll approximation and
that the vector field is subdominant in the evolution equation of
the inflaton field (\ref{eq:inflaton}), one can show the coupling function $f(\phi)$ should be
proportional to $e^{-2\alpha}$ to keep $\rho_v$ almost constant.
 In the slow roll phase, the e-fold number $\alpha$ is related to the inflaton
 field $\phi$ as $d\alpha = -  V(\phi) d\phi / V_\phi$, as usual.
Then,  the functional form of $f(\phi)$ is determined as
\begin{equation}
f(\phi) = e^{-2\alpha} = e^{2 \int \frac{V}{V_\phi} d\phi} \ .
\label{eq:function}
\end{equation}
For the polynomial potential $V\propto \phi ^n$, for example, we have $f=e^{\frac{ \phi ^2}{n}}$. The above case is, in a sense, a critical one. What we want to consider is super-critical cases.
For simplicity, we parameterize $f(\phi)$ by
\begin{equation}
f(\phi) = e ^{2c \int \frac{V}{V_\phi}d\phi},
\label{formula:f}
\end{equation}
where $c$ is a constant parameter. (The above may be compared with the analysis in subsection \ref{prim-mag-field-subsection} and in particular to \eqref{f(phi)-mag-field}.)

Let us consider $c>1$ case.
Note that (\ref{formula:f}) can be written as
\begin{eqnarray}
  \frac{f_\phi}{f} = 2c  \frac{V}{V_\phi}
\label{formula:f1}\ .
\end{eqnarray}
Then, the condition $c>1$ can be promoted to the condition
\begin{eqnarray}
\frac{1}{2}\frac{f_\phi V_\phi}{f V} >1  \ .
\label{general-condition}
\end{eqnarray}
Thus, any functional pairs $f$ and $V$ which satisfies (\ref{general-condition})
in some range could produce the vector-hair during inflation.
The equation for the inflaton  becomes
\begin{eqnarray}
\ddot{\phi} = -3\dot{\alpha}\dot{\phi}-V_\phi \left[ 1-\frac{2c}{\epsilon_V}
\frac{\rho_v}{V(\phi)}\right] \ , \label{eq:inflaton2}
\end{eqnarray}
where we used (\ref{formula:f1}) and  the slow-roll parameter $\epsilon_V$ is defined as \eqref{epsilon-eta-Potential}.
In this case, if the vector field is initially small ${\rho_v}/{V(\phi)} \ll \epsilon_V /2c $, then the conventional single field slow-roll inflation is realized. During this stage $f\propto e^{-2c\alpha}$ and the vector field grows as $\rho_v \propto e^{4(c-1)\alpha}$. Therefore, the vector field eventually becomes relevant to the inflaton dynamics (\ref{eq:inflaton2}). Nevertheless, the accelerating expansion of the Universe  continues because  ${\rho_v}/{V(\phi)}$ does not exceed $\epsilon_{V}/2c$. In fact, if ${\rho_v}/{V(\phi)}$ exceeds $\epsilon_{V}/2c$, the inflaton field $\phi$
does not roll down, which makes $\rho _v = p_A^2 f(\phi)^{-2}e^{-4\alpha-4\sigma} /2$
 decrease. Hence,  $\rho_v \ll V(\phi)$ always holds.
 In this way, there appears an attractor where
 the inflation continues even when the vector field affects the inflaton dynamics.
We stress that even if new infrared perturbations become relevant, those may be renormalized to the background
vector field and the anisotropy stays at the attractor value.


 Let us make the above statement more precise.
Note that the inflaton dynamics is determined by solving the slow-roll equation:
\begin{equation}
 -3\dot{\alpha}\dot{\phi}-V_{\phi}+p_A^2f^{-3}f_{\phi}e^{-4\alpha -4\sigma} =0 \ .
 \label{eq:inflaton3}
 \end{equation}
Here, we see gauge kinetic function changes the effective potential.
Using the slow-roll equation (\ref{slow-roll:hamiltonian}), this yields
\begin{equation}
\frac{d \phi}{d \alpha}=\frac{\dot{\phi}}{\dot{\alpha}}=-\frac{V_\phi}{ V} + 2c \frac{p_A^2}{V_\phi}e^{-4\alpha -4\sigma -4c  \int\frac{V}{V_\phi}d\phi} \ .
\label{eq:inflaton4}
\end{equation}
This can be integrated by neglecting evolutions of $V,V_\phi$ and $\sigma$
\begin{eqnarray}
  e^{4\alpha +4\sigma +4c  \int\frac{V}{V_\phi}d\phi}
  = \frac{2c^2 p_A^2}{c-1} \frac{ V}{V_{\phi}^2}
  \left[ 1+ \Omega e^{-4(c-1)\alpha +4\sigma} \right]  \ ,
\label{exponential}
\end{eqnarray}
where $\Omega$ is a constant of integration.
Substituting this back into the slow-roll equation (\ref{eq:inflaton4}), we obtain
\begin{eqnarray}
\frac{d\phi}{d\alpha} &=& - \frac{V_{\phi}}{ V}+\frac{c-1}{c}\frac{V_{\phi}}{V} \left[ 1+ \Omega e^{-4(c-1)\alpha+4\sigma}   \right] ^{-1} \ .
\end{eqnarray}
Initially $\alpha \rightarrow -\infty $, the second term can be neglected. While, in the future $\alpha \rightarrow \infty$,
the term containing $\Omega$ disappears.
This clearly shows a transition from the conventional single field slow-roll inflationary phase, where
\begin{eqnarray}
\frac{d\phi }{d\alpha} = -  \frac{V_{\phi}}{ V}\,,
\label{first-stage}
\end{eqnarray}
to what we refer to as the second inflationary phase, where the vector field is relevant to the inflaton dynamics and the inflaton gets $1/c$ times slower as
\begin{eqnarray}
\frac{d\phi }{ d\alpha} = - \frac{1}{c}  \frac{V_{\phi} }{ V} \ .
\label{second-stage}
\end{eqnarray}
 In the second inflationary phase, we can use (\ref{exponential}) discarding $\Omega$ term
and rewrite the energy density of the vector field as
\begin{equation}
\rho _v =\frac{p_A^2}{2} e^{-4\alpha-4\sigma-4c  \int\frac{V}{V_\phi}d\phi}
 = \frac{1}{2} \frac{c-1}{c^2} \epsilon_{V} V(\phi) \ ,
\end{equation}
which yields the anisotropy
\begin{eqnarray}
\frac{\Sigma}{H} =  \frac{1}{3} \frac{c-1}{c^2} \epsilon_{V} \ .
\end{eqnarray}
Moreover, from (\ref{eq:hamiltonian}) and (\ref{eq:alpha}) the scale-factor slow-roll parameter \eqref{epsilon-eta-def} is
\begin{equation}
\epsilon\equiv -\frac{\ddot{\alpha}}{\dot{\alpha}^2}=-\frac{1}{\dot{\alpha}^2} \left(-\frac{1}{2}  \dot{\phi}^2-\frac{2}{3} \rho_v \right) = \frac{1}{c} \epsilon_{V} \ ,
\end{equation}
where we neglected the anisotropy and used  (\ref{slow-roll:hamiltonian}) and (\ref{second-stage}).
Thus we have a remarkable result~\cite{Watanabe:2009ct}
\begin{equation}\label{Sigma-over-H-generic}
\frac{\Sigma}{H} = \frac{1}{3}\frac{c-1}{c}\epsilon.
\end{equation}
Therefore, for a broad class of potential and gauge kinetic functions, there exist
anisotropic inflationary solutions, and the anisotropy is of order the slow-roll parameter $\epsilon$.

\subsubsection{Example: chaotic inflation}

In order to make the statement more precise, we consider chaotic inflation with
the potential
\begin{eqnarray}
 V(\phi ) = \frac{1}{2} m^2 \phi^2  \ ,
\end{eqnarray}
where $m$ is mass of the inflaton.
For this potential, the coupling function \eqref{formula:f} becomes
\begin{eqnarray}
f(\phi)=e^{c \phi^2 /2}  \ .
\end{eqnarray}
It is instructive to see what happens by solving (\ref{eq:hamiltonian})-(\ref{eq:inflaton}) numerically~\cite{Watanabe:2009ct}.
\begin{figure}[h]
\begin{center}
\includegraphics[width=9cm]{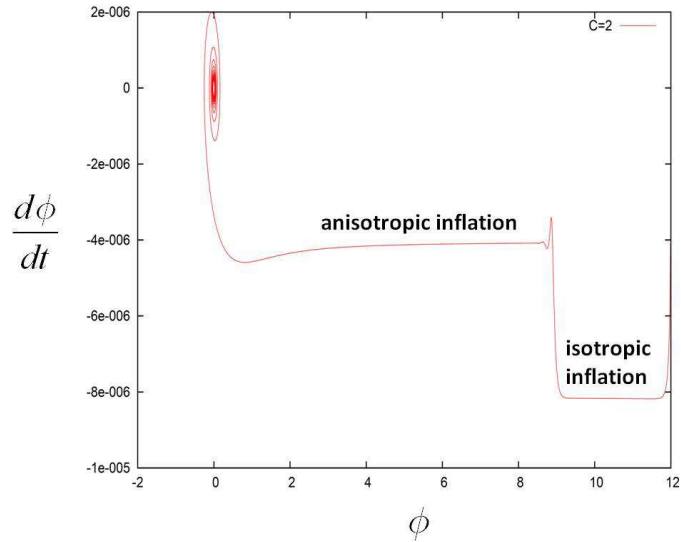}
\caption{Trajectory of the inflaton field $\phi$ in phase space.
This trajectory is drawn for parameters $c=2$ and
  $ m=10^{-5} $ and with  initial conditions
 $\phi_i=12$ and $\dot{\phi}_i=0$.
 There are two different slow-roll phases, namely isotropic and anisotropic inflation.
 The transition occurs around $\phi= 9$.}
\label{fg:phase}
\end{center}
\end{figure}
\begin{figure}
\begin{center}
\includegraphics[width=9cm]{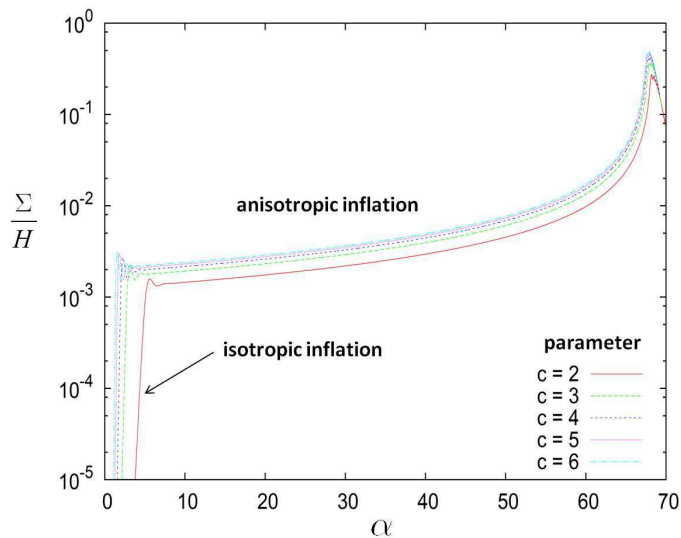}
\caption{
The time evolutions of the anisotropy $\Sigma/H$ for various $c$
with respect to the number of $e$-folds. The anisotropic inflation  phase can be sufficiently long.
 }
\label{fg:ce-ratio}
\end{center}
\end{figure}
In Fig.\ref{fg:phase}, we have shown a trajectory in $\phi-\dot{\phi}$ space
where, as our earlier analytical discussion also showed, we see two slow-roll phases. The first one is the
  conventional isotropic inflationary phase  and the second one is the anisotropic inflationary phase.
As usual, inflation ends with oscillation around the bottom of the potential.
This tells us that isotropic inflation corresponds to a saddle point
and ansiotropic inflation is an attractor in the slow roll phase.
 In Fig.\ref{fg:ce-ratio},
 we have calculated the evolution of the degree of anisotropy
 $\Sigma/H \equiv \dot{\sigma}/\dot{\alpha}$ for various parameters $c$
 under the initial conditions $\sqrt{c}\phi_i=17$.
As expected, all of solutions show a rapid growth of anisotropy
in the first slow-roll phase which corresponds to the conventional inflation.
However, the growth of the anisotropy eventually stops at the order of the slow roll parameter.
Notice that this attractor like behavior is not so sensitive to the parameter $c$.
As one can see there is a sufficient amount of number of e-folds during an anisotropic inflation.
We should emphasize that the anisotropy of the expansion disappears after the inflation.
Hence, the anisotropy affects only the generation process of primordial fluctuations and the evolution of fluctuations during inflation.

As we have proved for general cases, the anisotropy satisfies the inequality
\begin{eqnarray}
   \frac{\Sigma}{H} \leq  \epsilon \ .
\end{eqnarray}
This inequality holds universally for any potential functions.
This result exemplifies the extension of Wald's cosmic no-hair theorem in inflationary setup discussed in \cite{Wald-extended-theorem} and reviewed in Appendix \ref{No-hair-extension-appendix}. It is noteworthy that deviations from a de Sitter space during a slow-roll inflation, the isotropic and anisotropic deviations alike, are labeled and  characterized by the slow-roll parameter $\epsilon$.


\subsubsection{Anisotropic inflation in a variety of models}

There are many models which realize anisotropic inflation.
We can generalize single field inflation to multi-field inflation models~\cite{Emami:2010rm} or readily show that the small field or hybrid inflation models also admit anisotropic extension along the line we discussed here. A wider range of anisotropic inflationary models are discussed in ~\cite{Dimopoulos-anisotropic}.
We can extend the standard kinetic term to the Born-Infeld type~\cite{More-anisotropic-models-Born-Infeld}
which may be useful in finding stringy realization of anisotropic inflation.
We can introduce a mass term to the vector field, that is, a vector curvaton~\cite{Dimopoulos-vector-curvaton}.
It is interesting to study cosmological consequences of the vector curvaton scenario in detail.
It is also possible to extend the model to non-Abelian gauge fields~\cite{Bartolo:2009pa,arXiv:1103.6164}.
In this case, we have  more complicated dynamics which would lead to interesting phenomenology.
In a straightforward way the Bianchi type-I model we discussed here can be extended to  other Bianchi type models~\cite{Aniso-Inf-Other-Bianchi}. (For a discussion on Bianchi cosmologies see \ref{Bianchi-appendix}.)

Instead of a gauge field, which is a one-form, one may realize anisotropic inflation through higher form fields
~\cite{AnIso-form-fields}. For example, we can consider the action for a 2-form field $B_{\mu\nu}$
\begin{eqnarray}
     S = \int d^4x\sqrt{-g}\left[~\frac{1}{2}R
-\frac{1}{2}\left(\partial_\mu\phi\right)\left(\partial^{\mu}\phi\right)
-V(\phi)-\frac{1}{12} f^2 (\phi ) H_{\mu\nu\lambda}H^{\mu\nu\lambda}
~\right] \ ,
\end{eqnarray}
where $H_{\mu\nu\lambda} =\partial_\mu B_{\nu\lambda} + \partial_\nu B_{\lambda\mu}+\partial_\lambda B_{\mu\nu}$
 is the field strength of the 2-form.
Moreover, the parity violating term
\begin{eqnarray}
     S = \int d^4x\sqrt{-g}\left[~\frac{1}{2}R
-\frac{1}{2}\left(\partial_\mu\phi\right)\left(\partial^{\mu}\phi\right)
-V(\phi)-\frac{1}{4} F_{\mu\nu}F^{\mu\nu}-\frac{1}{8}  j (\phi ) \epsilon^{\mu\nu\lambda\rho} F_{\mu\nu} F_{\lambda\rho}
~\right]
\end{eqnarray}
can also induce anisotropy in the expansion. Here, the axionic coupling function $j(\phi)$ should be chosen appropriately.

\subsection{Exact power-law anisotropic inflation}

In the previous subsection, we have seen anisotropic inflation appears in many different forms and scenarios.
To gain more profound understanding of the mechanism of anisotropy,
it would be useful to have exact solutions which exhibit anisotropic expansion~\cite{Kanno:2010nr}.
We know there exists exact isotropic power law inflation for the exponential potential
\begin{eqnarray}\label{V-power-law}
  V(\phi ) = e^{\lambda  \phi } \ .
\end{eqnarray}
Hence, it is natural to consider the exponential kinetic function
\begin{eqnarray}\label{f-power-law}
   f(\phi ) = e^{\rho  \phi}
\end{eqnarray}
in order to obtain exact anisotropic power law inflation.

\subsubsection{Exact solutions}\label{section-4.3.1}
We start by recalling that isotropic power-law solutions can be obtained through the ansatz
\begin{eqnarray}
  \alpha = \zeta \log t \ , \qquad
   \phi = \xi \log t + \phi_0 \ ,
\end{eqnarray}
where
\begin{eqnarray}
  \zeta = \frac{2}{\lambda^2} \ , \hspace{1cm}
  \xi = - \frac{2}{\lambda} \ , \hspace{1cm}
   V_0 e^{\lambda  \phi_0} = \frac{2(6-\lambda^2)}{\lambda^4} \ ,
  \label{iso-power}
\end{eqnarray}
corresponding to FLRW geometry
\begin{eqnarray}
 ds^2 = -dt^2 + t^{4/\lambda^2} \left( dx^2 +dy^2 + dz^2 \right) \ .
 \label{isotropic}
\end{eqnarray}
Thus, for $\lambda \ll 1$, we have power-law slow-roll inflation.

We now seek exact anisotropic solutions. We start with the power-law ansatz
\begin{eqnarray}
  \alpha = \zeta \log t \ , \hspace{1cm}
  \sigma = \eta \log t \ , \hspace{1cm}
   \phi = \xi \log t + \phi_0 \ .
\label{ask}
\end{eqnarray}
From the hamiltonian constraint (\ref{eq:hamiltonian}),
we get two relations
\begin{eqnarray}
  \lambda \xi = -2  \ , \hspace{1cm} \rho \xi +2 \zeta + 2\eta =1\,,
  \label{A}
\end{eqnarray}
to have the same time dependence for each term.
The latter relation is necessary
only in the non-trivial vector case, $p_A \neq 0$.
Then, for the amplitudes to be balanced, we need
\begin{eqnarray}
  -\zeta^2 +\eta^2 +\frac{1}{6} \xi^2 + \frac{1}{3} u +\frac{1}{6} w =0 \ ,
  \label{B}
\end{eqnarray}
where we have defined variables
\begin{eqnarray}
   u =  V_0 e^{\lambda \phi_0} \  ,  \qquad
   w =  p_A^2 f_0^{-2} e^{-2\rho \phi_0}\,.
\end{eqnarray}
The equations for the scale factor (\ref{eq:alpha}) and the anisotropy (\ref{eq:sigma}),
under (\ref{A}) yields
\begin{eqnarray}
 -\zeta + 3\zeta^2 -u - \frac{1}{6} w &=&0 \ ,  \label{C}\\
 -\eta + 3 \zeta\eta - \frac{1}{3} w &=& 0 \ .
  \label{D}
\end{eqnarray}
Finally, from the equation for the scalar (\ref{eq:inflaton})
we obtain
\begin{eqnarray}
  -\xi + 3\zeta \xi + \lambda u -\rho w = 0  \ .
  \label{E}
\end{eqnarray}
Using (\ref{A}),~(\ref{C}) and (\ref{D}), we can solve $u$ and $w$ as
\begin{eqnarray}
u &=& \frac{9}{2} \zeta^2 - \frac{9}{4} \zeta - \frac{3\rho}{2\lambda} \zeta
                 + \frac{1}{4} + \frac{\rho}{2\lambda}  \ , \label{u}
\\
w &=& -9 \zeta^2 + \frac{15}{2} \zeta + \frac{9\rho}{\lambda} \zeta
                 -\frac{3}{2} -\frac{3\rho}{\lambda} \,.
                 \label{w}
\end{eqnarray}
Substituting these results into (\ref{E}), we obtain
\begin{eqnarray}
  \left( 3\zeta-1 \right) \left[ 6 \lambda \left( \lambda + 2\rho \right)\zeta
  - \left( \lambda^2 + 8\rho \lambda + 12 \rho^2 + 8 \right) \right] =0 \ .
\end{eqnarray}

If $\zeta=1/3$, we have $u=w=0$ which is not our desired anisotropic
solution. Thus, we have to choose
\begin{eqnarray}
 \zeta = \frac{\lambda^2 + 8 \rho \lambda + 12 \rho^2 +8}{6\lambda (\lambda + 2\rho)} \ .
\label{expansion-rate}
\end{eqnarray}
Substituting this result into (\ref{C}) and noting \eqref{A}, we obtain
\begin{eqnarray}
 \eta &=& \frac{\lambda^2 + 2\rho \lambda -4 }{3\lambda (\lambda + 2\rho)}\,,\\
 \xi &=& - \frac{2}{\lambda}\,,
\end{eqnarray}
Finally, (\ref{u}) and (\ref{w}) reduce to
\begin{eqnarray}
 u = \frac{(\rho \lambda + 2\rho^2 +2)(-\lambda^2 + 4\rho \lambda +12 \rho^2 +8)}
      {2\lambda^2 (\lambda +2\rho )^2 }\,,\quad
 w = \frac{(\lambda^2 + 2\rho \lambda -4)(-\lambda^2 + 4\rho \lambda +12 \rho^2 +8)}
      {2\lambda^2 (\lambda +2\rho )^2 } \ .
\end{eqnarray}
Note that (\ref{B}) is automatically satisfied.
Thus, we have obtained exact anisotropic power-law solutions.

 Recalling the definition (\ref{ask}), we see $\zeta \gg 1$ or $|\lambda|\ll 1$ is necessary for slow-roll inflation.
From the solution (\ref{expansion-rate}), it turns out that this requirement can be achieved
if $\lambda \ll \rho$.
For these cases, $u$ is always positive and since $w$ is by definition also positive,
\begin{eqnarray}
\lambda^2 + 2\rho \lambda > 4 \ .
\end{eqnarray}
Hence, $\rho $ must be much larger than one.

The spacetime metric is
\begin{eqnarray}
  ds^2 = -dt^2 + t^{2\zeta-4\eta} dx^2 + t^{2\zeta +2\eta}
\left( dy^2 + dz^2 \right) \ .
  \label{anisotropic}
\end{eqnarray}
The average expansion rate is determined by $\zeta$ and the average slow roll parameter \eqref{epsilon-eta-def}
is given by
\begin{eqnarray}
\epsilon \equiv -\frac{\ddot{\alpha}}{\dot\alpha^2}
= \frac{6\lambda (\lambda + 2\rho)}{\lambda^2 + 8 \rho \lambda + 12 \rho^2 +8}
\ ,
\label{epsilon-section4}
\end{eqnarray}
In the slow-roll limit  $\lambda \ll 1$ and $\rho \gg 1$,
this reduces to $\epsilon = \lambda /\rho$.

The anisotropy is characterized by
\begin{eqnarray}
  \frac{\Sigma}{H} \equiv \frac{\dot{\sigma}}{\dot{\alpha}}
  = \frac{2(\lambda^2 + 2\rho \lambda -4) }
  {\lambda^2 + 8 \rho \lambda + 12 \rho^2 +8} \ ,
\label{soverh}
\end{eqnarray}
where $H=\dot\alpha$ is the isotropic expansion rate.
One may rewrite the above in the form \eqref{Sigma-over-H-generic}
\begin{eqnarray}
\frac{\Sigma}{H} = \frac{c-1}{3c}\epsilon \,,\qquad
c=\frac{\lambda^2 + 2\rho\lambda}{4}>1 \ .
\end{eqnarray}
We see the anisotropy is positive and proportional to the slow roll parameter $\epsilon$, with a proportionality constant which is always less than $1/3$.

To summarize, we have constructed a model of anisotropic power-law inflation which can be exactly solved. Although the anisotropy is always small, it persists during inflation.
Clearly these exact solutions give rise to counter examples to the cosmic no-hair
conjecture~\cite{no-hair-conjecture, Kitada}.


\subsubsection{Stability of the anisotropic inflation}

We showed that a model with \eqref{V-power-law} and \eqref{f-power-law} admits both isotropic \eqref{iso-power} and anisotropic power-law solutions. Here, we investigate the phase space structure and stability of these solutions. It is convenient to choose  e-fold number as the time coordinate $d\alpha = \dot{\alpha} dt$ and
use the dimensionless variables
\begin{eqnarray}
  X = \frac{\dot{\sigma}}{\dot{\alpha}} \ , \hspace{1cm}
  Y =  \frac{\dot{\phi}}{\dot{\alpha}} \ , \hspace{1cm}
  Z =  f(\phi) e^{-\alpha +2\sigma} \frac{\dot{v}}{\dot{\alpha}} \ .
\end{eqnarray}
With these definitions, we can write the hamiltonian constraint equation
as
\begin{eqnarray}
  -  \frac{V}{\dot{\alpha}^2}
  = 3(X^2 -1) + \frac{1}{2} Y^2 + \frac{1}{2} Z^2 \ .
  \label{hamconst}
\end{eqnarray}
Since we are considering a positive potential, we have the inequality
\begin{eqnarray}
3(X^2 -1) + \frac{1}{2} Y^2 + \frac{1}{2} Z^2 < 0 \ .
\label{XYZ-constraint}
\end{eqnarray}
Using the hamiltonian constraint (\ref{hamconst}) we can eliminate $\phi$
from the equations of motion
and write them  in the autonomous form
\begin{eqnarray}
\frac{dX}{d\alpha} &=& \frac{1}{3} Z^2 (X+1)
+ X\left\{ 3(X^2 -1) + \frac{1}{2} Y^2 \right\}
\label{eq:X} \,,\\
\frac{dY}{d\alpha} &=& (Y+\lambda) \left\{ 3(X^2 -1) + \frac{1}{2} Y^2 \right\}
+ \frac{1}{3} YZ^2 + \left( \rho + \frac{\lambda}{2} \right)Z^2
\label{eq:Y} \,,\\
\frac{dZ}{d\alpha} &=& Z \left[ 3(X^2 -1) + \frac{1}{2} Y^2
-\rho Y +1 -2X + \frac{1}{3} Z^2 \right]
\label{eq:Z} \ .
\end{eqnarray}
Therefore, we have a 3-dimensional space with a constraint (\ref{XYZ-constraint}).
 A fixed point in this phase space is defined by $dX/d\alpha=dY/d\alpha=dZ/d\alpha=0$.

We first analyze  the isotropic fixed point $X=0$.
From (\ref{eq:X}) we learn $Z=0$.
The remaining equation (\ref{eq:Y}) yields $Y=-\lambda$ or $Y^2=6$.
The latter solution does not satisfy the constraint (\ref{XYZ-constraint}).
Thus, the isotropic fixed point is
\begin{eqnarray}
  (X , Y, Z) = ( 0, -\lambda , 0)
\label{fixedpoint1}\ .
\end{eqnarray}
This fixed point corresponds to the isotropic power-law solution (\ref{isotropic}).
Indeed, one can check that the solution (\ref{iso-power}) leads to the above fixed point.\footnote{Note that the fixed curve
$Z=0$ and $6 X^2 + Y^2 =6$ does not satisfy the constraint (\ref{XYZ-constraint}).}

Next, let us discuss an anisotropic fixed point.
From (\ref{eq:X}) and (\ref{eq:Y}), we have
\begin{eqnarray}
  Y &=& \left( 3\rho +\frac{\lambda}{2} \right) X - \lambda \ ,
  \label{Y0}\\
Z^2 &=& - \frac{3X}{X+1} \left[ 3(X^2 -1) + \frac{1}{2} Y^2 \right] \ .
  \label{Z2}
\end{eqnarray}
Using the above results in (\ref{eq:Z}) we obtain
\begin{eqnarray}
  \left( X-2 \right) \left[ \left( \lambda^2 + 8 \rho \lambda + 12 \rho^2 +8 \right)X
              - 2 \left( \lambda^2 +2\rho \lambda -4 \right) \right] =0 \ .
\end{eqnarray}
The solution $X=2$ does not make sense because it implies $Z^2 = -18-36 \rho^2 <0$ by (\ref{Y0}) and (\ref{Z2}).
Thus, an anisotropic fixed point is expressed by
\begin{eqnarray}
X &=& \frac{2 \left( \lambda^2 +2\rho \lambda -4 \right)}
            {\lambda^2 + 8 \rho \lambda + 12 \rho^2 +8} \ ,\\
Y&=& - \frac{12 \left( \lambda +2\rho  \right)}
            {\lambda^2 + 8 \rho \lambda + 12 \rho^2 +8} \ ,\\
Z^2 &=&  \frac{ 18 \left( \lambda^2 +2\rho \lambda -4 \right)
             \left(-\lambda^2 + 4\rho \lambda +12 \rho^2 +8\right) }
            {\left( \lambda^2 + 8 \rho \lambda + 12 \rho^2 +8 \right)^2}  \ .
\end{eqnarray}
Note that from the last equation, we find that $\lambda^2 +2\rho \lambda > 4$ is
required for this fixed point to exist under the condition $\lambda \ll \rho$.
This nicely dovetails with the discussions of previous subsection \ref{section-4.3.1} and
this fixed point corresponds to the anisotropic power-law solution (\ref{anisotropic}).

Now, we examine the linear stability of the fixed points.
Eqs.~(\ref{eq:X}), (\ref{eq:Y}), (\ref{eq:Z}) in the linearized form are given by
\begin{eqnarray}
\frac{d\delta X}{d\alpha} &=&
  \left( \frac{1}{3}Z^2 + 9 X^2 + \frac{1}{2} Y^2 -3 \right) \delta X
  + XY \delta Y + \frac{2}{3} \left( X+1 \right)Z \delta Z \,,\\
\frac{d\delta Y}{d\alpha} &=& 6X\left( Y + \lambda \right) \delta X
 + \left\{ 3\left( X^2 -1 \right)
    + \frac{1}{2}Y^2 +Y\left( Y+\lambda \right)
                + \frac{1}{3} Z^2 \right\} \delta Y \nonumber\\
    &&  \qquad          + \left( \frac{2}{3}Y +2\rho + \lambda \right) Z\delta Z
\,,\\
\frac{d\delta Z}{d\alpha} &=&
2(3X -1)Z\delta X +\left( Y-\rho \right)Z \delta Y \nonumber\\
&&   \quad +\left(3X^2 + \frac{1}{2} Y^2 + Z^2
-2X -\rho Y -2 \right)\delta Z\,.
\end{eqnarray}
First, we consider the stability of isotropic fixed point (\ref{fixedpoint1}). In this case
the above equations reduce to
\begin{eqnarray}
 \frac{d\delta X}{d\alpha} &=&
  \left( \frac{1}{2} \lambda^2 -3 \right) \delta X \,,\\
\frac{d\delta Y}{d\alpha} &=&
  \left(  \frac{1}{2} \lambda^2 -3 \right) \delta Y  \,,\\
\frac{d\delta Z}{d\alpha} &=&
  \left[ \frac{1}{2}\lambda^2 -2 + \rho \lambda \right]\delta Z\,.
\end{eqnarray}
We see that the coefficient in the right hand side of above equations becomes
negative when $\lambda^2 +2\rho \lambda < 4$ during inflation
$\lambda \ll 1$, which means
the isotropic fixed point is an attractor under these conditions
and the isotropic fixed point becomes stable in this parameter
region.
In the opposite case, $\lambda^2 +2\rho \lambda > 4$,
the isotropic fixed point becomes a saddle point and unstable.
In the latter case, existence of a background gauge field, no matter how small,
destabilizes isotropic inflation.
\begin{figure}
\begin{center}
\includegraphics[width=11cm]{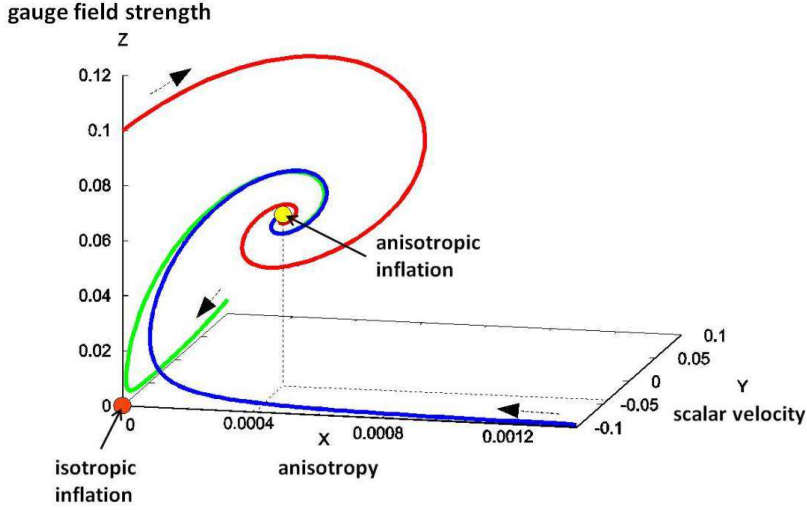}
\caption{Several trajectories in $X$-$Y$-$Z$ space are shown for $\lambda = 0.1, \rho=50$.
 The trajectories converge to the attractor corresponding to anisotropic inflation.}
   \label{fg:flow}
\end{center}
\end{figure}
Next, let us discuss the stability
around the anisotropic fixed point.
Since we are considering the inflationary universe
$\lambda \ll 1$, the condition $\lambda^2 +2\rho \lambda > 4$
implies $\rho \gg 1$. Under these conditions the linearized equations are approximated as
\begin{eqnarray}
 \frac{d\delta X}{d\alpha} &=&
  -3 \delta X \,,\\
\frac{d\delta Y}{d\alpha} &=&
  -3 \delta Y +  \sqrt{6(\lambda ^2 + 2\rho\lambda- 4)} \delta Z \,,\\
\frac{d\delta Z}{d\alpha} &=&
  - \frac{1}{2}\sqrt{6(\lambda ^2 +2\rho\lambda - 4) } \delta Y \ .
\end{eqnarray}
The stability can be analyzed by setting
\begin{eqnarray}
  \delta X = e^{\omega \alpha} \delta \tilde{ X} \ , \hspace{1cm}
   \delta Y = e^{\omega \alpha} \delta \tilde{ Y} \ , \hspace{1cm}
    \delta Z = e^{\omega \alpha} \delta \tilde{ Z} \ .
\end{eqnarray}
Then we find the eigenvalues $\omega$ are given by
\begin{eqnarray}
\omega = -3 \ ,
-\frac{3}{2} \pm i \sqrt{3(\lambda ^2 +2\rho\lambda -4)-\frac{9}{4}} \ .
\end{eqnarray}
As the eigenvalues have negative real part,
the anisotropic fixed point is stable.
Thus, the end point of trajectories around the unstable isotropic power-law inflation
are anisotropic power-law inflation.

In Fig.\ref{fg:flow}, we depicted the phase flow in $X$-$Y$-$Z$ space for $\lambda = 0.1,\ \rho=50$.
We see that the trajectories converge to the anisotropic fixed point indicated by yellow circle.
The isotropic fixed point indicated by orange circle is a saddle point
 which is an attractor only on $Z=0$ plane.
Thus, anisotropic power-law  inflation is an attractor
solution for parameters satisfying $\lambda^2 +2\rho \lambda > 4$~\cite{Kanno:2010nr}.

\subsection{Inflation with multi-vector hair}
\label{sc:basic}

In the previous subsection, we have considered a single Abelian vector gauge field. It turns out that there are
qualitative differences in the dynamics when the number of Abelian vector fields becomes more than two~\cite{Yamamoto:2012tq}.
For example, \emph{isotropic} inflation with vector field hair can be realized dynamically (as an attractor solution) in the multi-vector field system with a uniform coupling between an inflaton and gauge fields~\cite{Yamamoto:2012tq,Yamamoto:2012sq}.

We consider  the exponential potential with parameters $V_0$ and $\lambda$ and
$N$ copies of Abelian gauge field $A_\mu^{(m)}$ with field strength
$F_{\mu\nu}^{(m)} = \partial_\mu A_\nu^{(m)} -\partial_\nu A_\mu^{(m)} $ described by the action
\begin{equation}
 S = \int d^4 x \sqrt{-g}\left[ R-\frac{1}{2}\partial _{\mu }\phi \partial ^{\mu}\phi - V_0 e^{\lambda\phi }
-\frac{1}{4}\sum  _{m=1}^N e^{ g_m \phi } F^{(m)}_{\mu\nu}F^{(m)\mu\nu} \right]\,.
\end{equation}
The $N$ copies of Abelian gauge field $A_\mu^{(m)}$ are coupled to an inflaton $\phi$ with coupling constants $g_m$.
As before, the Greek letters denote the full space-time indices and the Latin letters are reserved for spatial parts.

For the analysis in this subsection, it is convenient to use the tetrad formalism~\cite{Ellis:1968vb}.
As used in Appendix \ref{No-hair-extension-appendix}, let us introduce the unit normal
$n^\mu$ of the homogeneous hypersurface in a generic Bianchi I spacetime.
The evolution of the spatial slice is governed by the equation
\begin{equation}
 n_{\mu;\nu} = H(g_{\mu\nu}+n_\mu n_\nu ) + \sigma _{\mu\nu} \ ,
\end{equation}
where $H$ is the averaged expansion rate and the trace-free tensor $\sigma _{\mu\nu}$ represents anisotropy of the universe. Similarly to \eqref{tildeT}, the energy-momentum tensor can be decomposed as
\begin{equation}
T_{\mu\nu}= \rho n_\mu n_\nu + 2q_{(\mu}n_{\nu)} + P (g_{\mu\nu}+n_\mu n_\nu ) + \pi _{\mu\nu} \ ,
\end{equation}
where $\rho ,p, q_\mu$ and $\pi _{\mu\nu}$ are respectively energy density, pressure, energy flux
and anisotropic pressure seen by the  observer with 4-velocity $n^\mu$.
The electric fields seen by an observer with the four-velocity $n^\mu$ are defined by
\begin{equation}
E^{(m)}_\mu = F_{\mu\nu}^{(m)}n^\nu
\end{equation}
and we assume,  for simplicity, that the magnetic parts vanish
\begin{equation}
\epsilon _{\mu\nu\alpha\beta}F^{(m)\alpha\beta}n^\nu = 0 \ ,
\end{equation}
where $\epsilon _{\mu\nu\alpha\beta}$ is the volume four-form (see Appendix \ref{Convention-appendix} for our conventions).
The Einstein equations then take the form
\begin{eqnarray}
\rho &=& \frac{1}{2}\dot{\phi }^2 +V_0 e^{\lambda \phi } + \frac{1}{2}\sum _{m=1}^N e^{g_m \phi } E_\mu^{(m)}E^{(m)\mu} \\
q_\mu &=& 0 \\
P &=& \frac{1}{2}\dot{\phi }^2 - V_0 e^{\lambda\phi } +\frac{1}{6}\sum _{m=1}^N e^{g_m \phi } E_\mu^{(m)}E^{(m)\mu} \\
\pi _{ab} &=& \sum _{m=1}^N e^{g_m \phi }\left[- E^{(m)}_\mu E^{(m)}_\nu + \frac{1}{3}E_\alpha^{(m)}E^{(m)\alpha}(g_{\mu\nu}+n_\mu n_\nu) \right] \ .
\end{eqnarray}
Here, an overdot denotes a derivative with respect to the proper time $t$ associated with an observer with proper velocity
$n^\mu$.

In an orthonormal frame, the Einstein equations take the form given in  \cite{WE}.
We assume that $E^{(1)}_{i}$ and $E^{(2)}_{i }$ are non-vanishing and not parallel to each other
and choose the spatial frame such that
\begin{equation}
E_{i }^{(1)} = ( E,0,0) , \ \ \ \ \ E_3^{(2)}= 0.
\end{equation}
This fixes the choice of directions in spatial sections and results in
\begin{equation}
\Omega _1 = \sigma _{23} , \ \ \ \ \ \Omega _{2} =-\sigma _{13} , \ \ \ \ \ \Omega _3 = \sigma _{12} ,
\end{equation}
where $\Omega _{i }$ represents rotational velocity of the frame with respect to the Fermi propagated one (for a discussion
on Fermi transport see e.g. \cite{Padi-book}).
Let us adopt the following dimensionless variables
\begin{equation}
\Sigma _{\pm } = \frac{1}{2H}(\sigma _{22}\pm \sigma _{33}), \ \ \ \ \ \Sigma _{ij } = \frac{\sigma _{ij }}{H} \ \ \ \ \ {\rm for} \ \ \ \ \ i \neq j ,
\end{equation}
and
\begin{equation}
 \Omega \equiv \frac{\dot{\phi}}{H} , \ \ \ \ \ \Pi \equiv \frac{V_0 e^{\lambda\phi }}{3H^2 } ,\ \ \ \ \ \mathcal{E} = \frac{e^{\frac{g_1}{2}\phi }E}{\sqrt{6}H}, \ \ \ \ \ \mathcal{E}_{i }^{(2)} = \frac{e^{\frac{g_2}{2}\phi }E_{i }^{(2)}}{\sqrt{6}H} \ ,
 \quad \mathcal{E}_{i }^{(A)} = \frac{e^{\frac{g_A}{2}\phi }E_{i }^{(A)}}{\sqrt{6}H}  \ ,
\end{equation}
where capital Latin superscript indices from $3$ to $N$.
Then, the evolution of geometry is governed by the following equations
\be\begin{split}\label{eq:start}
 \Sigma _{+} ^{\prime } =& (q-2)\Sigma _{+} -(\Sigma _{12}^2 + \Sigma _{13}^2 ) +2\mathcal{E}^2 + 2(\mathcal{E}_1^{(2)})^2  - (\mathcal{E}_2^{(2)})^2  \\
  +& \sum _{A = 3}^N (2(\mathcal{E}_1^{(A)} )^2 - (\mathcal{E}_2^{(A)} )^2 -(\mathcal{E}_3^{(A)})^2 ) \ ,  \\ \Sigma _{-}^{\prime } = &(q-2)\Sigma _{-} - \Sigma _{12}^2 +\Sigma _{13}^2 +2\Sigma _{23}^2 -3(\mathcal{E}_2^{(2)})^2 -3 \sum _{A = 3}^N ((\mathcal{E}_2^{(A)})^2 - (\mathcal{E}_3^{(A)})^2 ) \ ,  \\
 \Sigma _{12} ^{\prime } = &(q-2 +3\Sigma _{+}+\Sigma _{-})\Sigma _{12} +2\Sigma _{13}\Sigma _{23} -6\mathcal{E}_1^{(2)} \mathcal{E}_2^{(2)} -6 \sum _{A = 3}^N \mathcal{E}_1^{(A)} \mathcal{E}_2^{(A)} \ , \\
 \Sigma _{13} ^{\prime }  =& (q-2+3\Sigma _{+}-\Sigma _{-})\Sigma _{13}  -6 \sum _{A = 3}^N \mathcal{E}_1^{(A)} \mathcal{E}_3^{(A)} \ , \\
 \Sigma _{23} ^{\prime } =& (q-2-2\Sigma _{-})\Sigma _{23}-2 \Sigma _{12}\Sigma _{13}  -6
 \sum _{A = 3}^N \mathcal{E}_2^{(A)} \mathcal{E}_3^{(A)} \ ,
\end{split}\ee
where a prime represents a derivative with respect to the  ``e-fold number'' time coordinate $\tau $ defined by
\begin{equation}
d\tau = Hdt \ ,
\end{equation}
(note that, being in an anisotropic background, $\tau$ is not the conformal time) and
$q= -1-\dot{H}/H^2 $ is the  deceleration parameter
\begin{equation}
 q = 2\Sigma _{+}^2 + \frac{2}{3}(\Sigma _{-}^2 + \Sigma _{12}^2 + \Sigma _{13}^2 + \Sigma _{23}^2 ) + \frac{1}{3}\Omega ^2 - \Pi + \mathcal{E}^2 + (\mathcal{E}_1^{(2)})^2 + (\mathcal{E}^{(2)}_2)^2 + \sum _{i =1}^3  \sum _{A = 3}^N  ( \mathcal{E}_{i }^{(A)})^2   .
\end{equation}
One may also work out equations of motion for the scalar and gauge fields
\be\begin{split}
 \Omega ^{\prime } &= (q-2)\Omega - 3\lambda \Pi + 3g_1 \mathcal{E}^2+ 3g_2 \{ (\mathcal{E}_1^{(2)})^2 + (\mathcal{E}_2^{(2)})^2 \}
+3g_A \sum _{i =1}^3  \sum _{A = 3}^N  ( \mathcal{E}_{i }^{(A)})^2   \ ,\\
\Pi ^{\prime } &= (2q+2+ \lambda\Omega ) \Pi\,,
\end{split}\ee
\be\begin{split}
\mathcal{E}^{\prime } =& (q-1-\frac{g_1}{2}\Omega -2\Sigma _{+})\mathcal{E}, \\
(\mathcal{E}_1^{(2)})^{\prime } =& (q-1-\frac{g_2}{2}\Omega - 2\Sigma _{+})\mathcal{E}_1^{(2)} + 2\Sigma _{12}\mathcal{E}_2^{(2)} ,\\
(\mathcal{E}_2^{(2)})^{\prime } =& (q-1-\frac{g_2}{2}\Omega+\Sigma _{+}+\Sigma _{-}) \mathcal{E}_2^{(2)} \ ,
\end{split}\ee
\be\begin{split}
(\mathcal{E}_1^{(A)})^{\prime } =& (q-1-\frac{g_A}{2}\Omega - 2\Sigma _{+})\mathcal{E}_1^{(A)} + 2\Sigma _{12}\mathcal{E}_2^{(A)}+2\Sigma _{13}\mathcal{E}_3^{(A)} , \\
(\mathcal{E}_2^{(A)})^{\prime } =& (q-1-\frac{g_A}{2}\Omega +\Sigma _{+}+\Sigma _{-}) \mathcal{E}_2^{(A)}
+2\Sigma _{23} \mathcal{E}_3^{(A)}, \\
(\mathcal{E}_3^{(A)} )^{\prime } =& (q-1-\frac{g_A}{2}\Omega + \Sigma _{+}-\Sigma _{-})\mathcal{E}_3^{(A)} \ .\label{eq:end}
\end{split}\ee
Moreover, there exists a constraint among the variables
\begin{eqnarray}
   && \Sigma _{+}^2 + \frac{1}{3}( \Sigma _{-}^2 + \Sigma _{12}^2 + \Sigma _{13}^2 + \Sigma _{23}^2 ) +
\frac{1}{6}\Omega ^2 + \Pi  \nonumber\\
&& \hspace{2cm} + \mathcal{E}^2 +(\mathcal{E}_1^{(2)})^2 + (\mathcal{E}^{(2)}_2)^2  + \sum _{i =1}^3  \sum _{A = 3}^N (
 \mathcal{E}_{i }^{(A)})^2=1 \ .
\end{eqnarray}
We are now in a position to study the fate of anisotropy in inflation with multiple vector fields hair.

\subsubsection{Universe tends to be isotropic}
\label{dynamical}

Here, we consider the cases of uniform coupling constants $g\equiv g_m$
and show that isotropic inflation is an attractor
in the phase space when the number of vector fields is greater than two.

The equations (\ref{eq:start}) to (\ref{eq:end}) form a dynamical system of dimension $3(N+1)$.
To understand the dynamics, it is useful to find out equilibrium points and their linear stability.
Let us look for equilibrium points with non-vanishing $\mathcal{E}$, $\mathcal{E}_2^{(2)}$ and $\mathcal{E}_3^{(A)}$.
For their time derivatives to vanish, we need
\begin{equation}
q-1 = \frac{g}{2}\Omega, \ \ \ \ \ \Sigma _{+} = \Sigma _{-} = 0  \ .
\end{equation}
From the evolution equations for $\mathcal{E}_1^{(2)}$ and $\mathcal{E}_{1,2}^{(A)} $, we also obtain
\begin{equation}
\Sigma _{12}= \Sigma _{13}= \Sigma _{23} =0  \ .
\end{equation}
Defining the total energy density parameter of the electric fields
\begin{equation}
\bar{\mathcal{E}}^2 = \mathcal{E}^2 + (\mathcal{E}_1^{(2)})^2 + (\mathcal{E}_2^{(2)})^2 + \sum _A (\mathcal{E}_{i }^{(A)})^2
\end{equation}
and using the equilibrium conditions for $\Omega $ and $\Pi $,  we can deduce relations
\begin{equation}
q= \frac{\lambda-g}{\lambda+g} , \ \ \ \ \ \Omega = -\frac{4}{\lambda+g}, \ \ \ \ \ \bar{\mathcal{E}}^2 = \frac{\lambda(\lambda+g)-4}{(\lambda+g)^2 } \ .
\end{equation}
Apparently, we need the condition $\lambda(\lambda+g) \geq 4$.
In order to have an accelerating universe, we also require $q<0$ or, equivalently $\lambda^2 -g^2 <0$.
What remains is to determine the relative strengths and angles among the vectors through $\Sigma _{\alpha \beta } ^{\prime } =0$. There are six equations for the remaining $3(N-1)$ variables
\be\begin{split}
&2 \mathcal{E}^2 + 2 ( \mathcal{E}_1^{(2)})^2 + 2 \sum _A (\mathcal{E}_1^{(A)})^2  = (\mathcal{E}_2^{(2)})^2 + \sum _A (\mathcal{E}_2^{(A)})^2 + \sum _A (\mathcal{E}_3^{(A)})^2, \\
&\mathcal{E}_1^{(2)} \mathcal{E}_2^{(2)} + \sum _A \mathcal{E}_1^{(A)} \mathcal{E}_2^{(A)} = 0 ,\quad
\sum _A \mathcal{E}_1^{(A)} \mathcal{E}_3^{(A)}  = 0 ,\quad
\sum _A \mathcal{E}_2^{(A)} \mathcal{E}_3^{(A)} = 0 \\
&(\mathcal{E}_2^{(2)})^2 + \sum _A (\mathcal{E}_2^{(A)})^2 = \sum _A (\mathcal{E}_3^{(A)})^2 ,\quad
\mathcal{E}^2 + (\mathcal{E}_1^{(2)})^2 + (\mathcal{E}_2^{(2)})^2
+ \sum _{A, i } (\mathcal{E}_{i }^{(A)})^2  = \frac{\lambda(\lambda+g)-4}{(\lambda+g)^2 }
\end{split}\ee
For $N=3$, they lead to an orthogonal solution
\begin{equation}
\mathcal{E}^2 = (\mathcal{E}_2^{(2)})^2 = (\mathcal{E}_3^{(3)})^2 = \frac{\lambda(\lambda+g)-4}{3(\lambda+g)^2 },\ \ \ \ \ \mathcal{E}_1^{(2)} = \mathcal{E}_1^{(3)} = \mathcal{E}_2^{(3)} = 0 \ .
\end{equation}
For more general cases $N>3$, it is convenient to introduce three $N-2$ dimensional vectors
\begin{equation}
\vec{\mathcal{E}}_{i } = \left(   \begin{array}{c}
    \mathcal{E}_{i }^{(3)} \\
     \mathcal{E}_{i }^{(4)} \\
     \vdots \\
    \mathcal{E}_{i }^{(N)}  \\
  \end{array} \right) \ .
\end{equation}
Then, the magnitude of $\vec{\mathcal{E}}_3 $ is given by
\begin{equation}
|\vec{\mathcal{E}}_3|^2 =  \frac{\lambda(\lambda+g)-4}{3(\lambda+g)^2 } \ ,
\end{equation}
and, the magnitudes of $\vec{\mathcal{E}}_2$ and $\vec{\mathcal{E}}_1$ are given through relations
\be\begin{split}
(\mathcal{E}_2^{(2)})^2 + |\vec{\mathcal{E}}_2|^2 &=  \frac{\lambda(\lambda+g)-4}{3(\lambda+g)^2 } \ ,\\
\mathcal{E}^2 + (\mathcal{E}_1^{(2)})^2 + | \vec{\mathcal{E}}_1|^2&=  \frac{\lambda(\lambda+g)-4}{3(\lambda+g)^2 } .
\end{split}\end{equation}
They introduce three arbitrary constant parameters. We also know from
\begin{equation}
\vec{\mathcal{E}}_1 \cdot \vec{\mathcal{E}}_3 = \vec{\mathcal{E}}_2 \cdot \vec{\mathcal{E}}_3 = 0
\end{equation}
that $\vec{\mathcal{E}}_3$ is perpendicular to $\vec{\mathcal{E}}_1$ and $\vec{\mathcal{E}}_2$. The angle between $\vec{\mathcal{E}}_1$ and $\vec{\mathcal{E}}_2$ is fixed once we choose the three parameters controling their magnitudes by using
\begin{equation}
\mathcal{E}_1^{(2)} \mathcal{E}_2^{(2)} + \vec{\mathcal{E}}_1 \cdot \vec{\mathcal{E}}_2 = 0.
\end{equation}
Apart from these geometrical conditions, we can take arbitrary combinations of the componenets for these three $N-2$ dimensional vectors.
Due to the $O(N-2)$ invariance of the equilibrium values of $\mathcal{E}_{\alpha }^{(A)}$, they are a family of equilibrium points that spans a $3(N-3)$ dimensional submanifold in the state space, for which the expansion rate becomes isotropic even though the vector fields have non-zero background values.

To examine local stability of an equilibrium point, we need to examine linearized equations around it. It is not so difficult to
see that the problem can be reduced to the stability analysis of the orthogonal equilibrium point with three electric fields.
Hence, we shall do the stability analysis for $N=3$. Since $\Pi $ can be eliminated by using the Hamiltonian constraint,
there remain the twelve independent varibales which are grouped into six pairs: $\Sigma _{12}$,$\mathcal{E}_1^{(2)}$; $\Sigma _{13}$,$\mathcal{E}_1^{(3)}$; $\Sigma _{23}$,$\mathcal{E}_2^{(3)}$; $\Sigma _{-}$,$\mathcal{E}_2^{(2)} - \mathcal{E}_3^{(3)}$; $\Sigma _{+}$,
$2\mathcal{E}-\mathcal{E}_2^{(2)}-\mathcal{E}_3^{(3)}$ and, $\Omega$, $\mathcal{E}+\mathcal{E}_2^{(2)}+ \mathcal{E}_3^{(3)}$.
The stability analysis gives the following results.
The first five pairs, which represent anisotropic perturbations,
 share common eigenvalues given by
\begin{equation}
\omega_{\Sigma , \mathcal{E}}= \frac{q-2 \pm \sqrt{(q-2)^2 - 16\bar{\mathcal{E}}^2}}{2} \ .
\end{equation}
Their real parts are negative since $-1 \leq q \leq 2$.
Furthermore, the isotropic mode has
\begin{equation}
\omega_{\Omega ,\mathcal{E}} = \frac{q-2 \pm\sqrt{(q-2)^2 - 4 \bar{\mathcal{E}}^2 (4+3g(\lambda+g))}}{2} .
\end{equation}
If we require $q = \frac{\lambda-g}{\lambda+g}<0$, it follows $4+3g(\lambda+g)>0$, which means that both of the eigenvalues have  again a negative
real part. In other words, these isotropic equilibrium points are local sinks for a range of the parameters for which the universe undergoes accelerated expansion.

One may confirm the above result beyond linearized stability level by  solving the dynamical equations (\ref{eq:start}) - (\ref{eq:end}) numerically,
for the case of three vector fields.
We solved the basic equations with the parameters $\lambda= 2$, $g=5$,
and the initial conditions  $\Sigma_\pm = \Sigma_{12} = \Sigma_{13} = \Sigma_{23} =0$,
$\Omega =0.6 , \mathcal{E} = 0.1, \mathcal{E}^{(2)}_1 = 0.2 \ ,  \mathcal{E}^{(2)}_2  = 0.001 \ ,
\mathcal{E}^{(3)}_1  = 0.1 \ , \mathcal{E}^{(2)}_2  = 0.01 \ , \mathcal{E}^{(2)}_3  = 0.0001 $.
The initial condition for  $\Pi$ is determined by the constraint equation.
The parameters are chosen so that the universe is accelerating, namely $q <0$.
\begin{figure}
\begin{center}
\includegraphics[width=9cm]{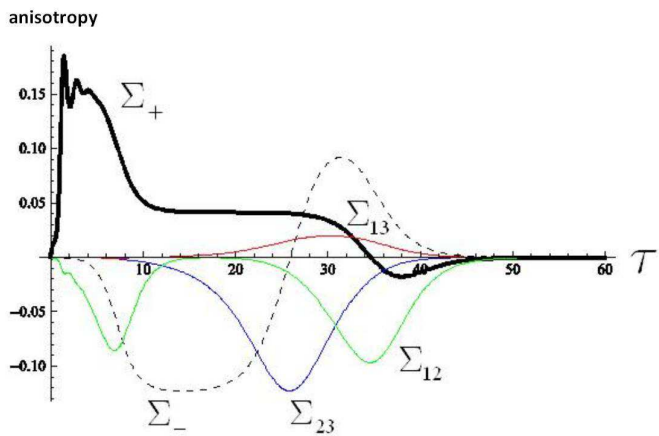}
\caption{For a uniform coupling $g=5$, we plotted time evolutions of anisotropic expansion rate normalized by the Hubble parameter
with respect to the e-fold number time $\tau$.
The components $\Sigma_+$,  $\Sigma_-$,  $\Sigma_{23}$,
$\Sigma_{13}$, and  $\Sigma_{12}$ correspond to thick, dashed, blue, red, and green lines, respectively.}
\label{fg:sigma-a}
\end{center}
\end{figure}
From Fig.\ref{fg:sigma-a}, we see the anisotropy disappears after transient anisotropic inflationary phases.
This confirms that the isotropic inflation is an attractor. It should be noted that
the vector fields possess a nontrivial orthogonal configuration.
We have solved the basic equations for other sets of parameters and initial conditions.
Although the transient behavior depends on the parameters and the initial conditions,
the system always  asymptotically approaches the isotropic inflationary final state.
The transient anisotropic phases correspond to the saddle points.
However, since duration of anisotropic inflation is sufficiently long, the anisotropy at each of the saddle points
would be relevant to CMB observations in the realistic cases where inflation ends with a finite duration.
It is worth mentioning that non-Abelian gauge fields also exhibit a similar isotropization~\cite{Maeda:2012eg}.

\subsubsection{The cosmic minimum-hair conjecture}

Based on the results and experiences of anisotropic models presented so far, we  put forward the ``cosmic minimum-hair''
conjecture stating that the hair in inflation tend to be minimized by the dynamics of the system in a quasi-de Sitter expanding Universe.
We provide supportive evidence for this conjecture through studying more general anisotropic trajectories.
A closely related model independent analysis is also presented in Appendix \ref{No-hair-extension-appendix}.

It is natural to ask whether isotropic inflation with non-vanishing vector fields is a unique attractor of the system. It is difficult to extract a clear conclusion for a system of this complexity. It is also a parameter dependent problem. However, we can argue that multiple vector fields are expected to repel each other and try to become isotropic in an expanding Universe.
Let us first look at the stability of axisymmetric inflating solutions discovered in \cite{Kanno:2010nr}. In the present model, they are located on lower dimensional boundaries of the full state space. For example, we have an equilibrium point
\begin{equation}\begin{split}
 q = \frac{5\lambda^2+2\lambda g -3g^2 -8}{(\lambda+3g)(\lambda+g)+8}, \quad
\Sigma _{+} = \frac{2(\lambda^2 +\lambda g -4)}{(\lambda +3g)(\lambda +g)+8}&, \quad
\Omega = - \frac{12(\lambda +g)}{(\lambda +3g)(\lambda +g)+8},\\
 \Sigma _{-} = \Sigma _{12} = \Sigma _{13} = \Sigma _{23} = \mathcal{E}_{1,2}^{(2)} &= \mathcal{E}_{\alpha }^{(A)} =  0 \\
 \Pi = - \frac{3((\lambda-3g)(\lambda +g)-8)(g(\lambda +g)+ 4)}{[(\lambda +3g)(\lambda +g)+8]^2 }, \quad
\mathcal{E}^2  = -&\frac{3(\lambda (\lambda +g)-4)((\lambda -3g)(\lambda +g)-8)}{[(\lambda +3g)(\lambda +g)+8]^2 }\ .\nonumber
\end{split}\end{equation}
There are many others which represent physically the same spacetime, but lie on different boundaries. This was a stable attractor solution for the single-vector-field model. For this equilibrium state, we notice that $\Sigma _{+} \geq 0$ by looking at the evolution equation for $\Sigma _{+}$ and requiring $\mathcal{E}^2 \geq 0$.  More generally, any electric field in 1-direction tends to support positive $\Sigma _{+}$ by generating a tension along that direction. It is also easy to see that for any electric perturbation in 2- or 3-direction, the eigenvalue is given by $3\Sigma _{+} \geq 0$. This axisymmetric solution is hence a saddle point in the general multi-field state space. These points together with the mathematical structure of the Maxwell's equations imply that  the positive $\Sigma _{+}$ created by the 1-component of a field comes with negative sign in the evolution equation in 1-direction while it has plus signs in the other directions. Thus, if there are more than one vector fields and one of them is dominant, it creates such an anisotropy that destabilizes orthogonal components of the other fields. This instability does not show up for single field models since it would merely cause a rotation of that vector. In summary, we expect  multiple vector fields to rearrange their orientations  to minimize anisotropy of the space which would cause an instability in some of their components.

In the present model the vectors cannot decouple from cosmic dynamics because they are kept excited through coupling to the
dominant scalar inflaton field. However, they still redistribute themselves to achieve as much isotropy as possible within the given circumstances.  As an example, if $N=2$, the attractor solution contains two orthogonal electric fields with the same amplitude:
\begin{equation}
 q = \frac{2\lambda^2 + 2\lambda g -3g^2 -2}{(\lambda + 3g)(\lambda +g)+2}, \ \ \ \ \
\Sigma _{+}= \frac{\lambda (\lambda +g)-4}{2(\lambda + 3g)(\lambda +g)+4}, \ \ \ \ \
\Sigma _{-} = - \frac{3\lambda (\lambda +g)-12}{2(\lambda +3g)(\lambda +g)+4} ,\nn
\end{equation}
\begin{equation}
\Omega = - \frac{6(\lambda +2g)}{(\lambda +3g)(\lambda +g)-2}, \ \ \ \ \
\Pi =  \frac{3(g(\lambda +g)+2)(g(2\lambda +3g)+2)}{((\lambda + 3g)(\lambda + g)+2)^2},\nn
\end{equation}
\begin{equation}
\mathcal{E}^2 = (\mathcal{E}_2^{(2)})^2 = \frac{3(\lambda (\lambda +g)-4)(g(2\lambda +3g)+2)}{2((\lambda +3g)(\lambda +g)+2)^2} ,\nn
\end{equation}
\begin{equation}
\Sigma _{12}= \Sigma _{13}=\Sigma _{23} = \mathcal{E}_1^{(2)} = 0 .\nn
\end{equation}
Shall we included a third gauge field, it becomes unstable because of the eigenvalue for $\mathcal{E}_3^{(3)}$
\begin{equation}
\omega_{\mathcal{E}_3} = 6\Sigma _{+} >0 .
\end{equation}

From the above argument, we can understand the global structure of the phase space for cases of uniform coupling constants. There are a bunch of saddle points consisting of those axisymmetric signle-field and orthogonal two-field equilibrium solutions on the boundaries, which are attractors when restricted in properly chosen invariant subsystems. Since these invariant sets have dimensions smaller than $3(N+1)$, the initial condition from which the orbits are attracted to any of the anisotropic attractors is of measure zero in the entire
state space. Whenever an orbit starts from a point not included in these subsystems, it is attracted
to the isotropic final state after a sufficiently long time. For example, in order for an orbit to be attracted
to the two field attractor, it must satisfy $\mathcal{E}_{i }^{(A)}=0$ for all $A=3,\cdots ,N$. Any small deviation from this subsystem
would lead it to isotropy due to the linear instability demonstrated above. A typical orbit is first attracted towards a nearby saddle point. Then the instability explained above kicks in and another vector component rises. It might come across another instability and go to another saddle point. An orbit continues this routine until it finally settles down to one of the isotropic attractors.

\begin{figure}
\begin{center}
\includegraphics[width=9cm]{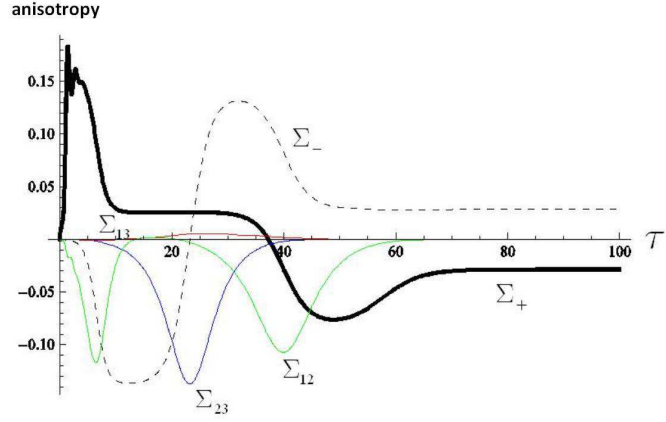}
\caption{Time evolutions of anisotropic expansion rate normalized
by the Hubble parameter versus the e-fold number time $\tau$. The plots are for the couplings
$g_1=4.8, \  g_2=5.0, \  g_3=5.2$.
The components $\Sigma_+$,  $\Sigma_-$,  $\Sigma_{23}$,
$\Sigma_{13}$, and  $\Sigma_{12}$ correspond to thick, dashed, blue, red, and green lines, respectively.}
\label{fg:sigma}
\end{center}
\end{figure}

\begin{figure}
\begin{center}
\includegraphics[width=9cm]{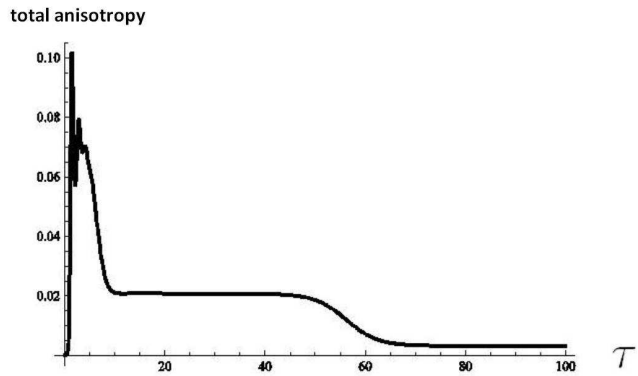}
\caption{Time evolution of the total anisotropy $\Sigma^{\mu\nu} \Sigma_{\mu\nu} /2$  with respect to
the e-fold number time $\tau$.}
\label{fg:total}
\end{center}
\end{figure}

In order to see what happens for general couplings, we took couplings
$g_1= 4.8, \ g_2= 5.0, \  g_3 = 5.2 $ for the three gauge fields.
As one can see in Fig.\ref{fg:sigma}, the attractor is an anisotropic inflation.
In this general case, the behavior is even more interesting. Since there are several saddle points,
we  have different anisotropic inflationary stages. During inflation the preferred direction in the Universe is changing, which would lead to interesting observational consequences. For example, the anisotropy
can be scale dependent.  Even if we change initial conditions, the attractor is the same.
Hence, the predictive power of the theory is maintained.
In Fig.\ref{fg:total}, we plotted the square of the magnitude of anisotropy $\Sigma^{\mu\nu} \Sigma_{\mu\nu} /2$.
While the system is not close to any of the fixed points, the orbit follows a complicated dynamical trajectory.
However, once it starts to be attracted towards one of the inflating solutions, the evolution
is fairly regular.
The result shows monotonic decrease of the total anisotropy after the orbit goes into this regular
tracking regime despite the fact that each component of the anisotropy may exhibit a complicated behavior.
It is inferred that when an orbit goes from a fixed point to another, the latter is always less anisotropic
than the former.

The above evidence  supports the cosmic minimum-hair conjecture:  the Universe organizes itself
so that any anisotropic feature in the spacetime during inflation becomes minimum.

\subsection{Primordial magnetic fields revisited}

Noting that space-independent background gauge field configurations we
discussed in this section and the backreaction of superhorizon gauge field configurations analyzed in
section \ref{prim-magnet-NG-section} have the same mathematical
appearance in the equations of motion,
we can return to the question of primordial magnetic fields of section \ref{prim-magnet-NG-section} and address the  backreaction issue
discussed there
using the elaborate computational tools of previous subsections.

Whenever the produced primordial magnetic field has a scale invariant spectrum $c=3/2$,
(\ref{pri-B}) implies that the electric fields have a red spectrum. Hence, the  dominant contribution to the energy
density comes from large scales.
We can assume the coherent electric fields with a definite direction dominate the energy density of the electromagnetic fields. Contrary to the simple expectation that the energy of the electromagnetic field will eventually overcome
that of the inflaton and terminate inflation~\cite{Demozzi:2009fu}, as we discussed, growth of the electric field
initiates anisotropic inflation~\cite{Watanabe:2009ct,Kanno:2009ei} and one should revise and improve this simple picture
by considering its backreaction effects.
Explicitly, before the backreaction becomes important, we have the relation
\begin{eqnarray}
     f \propto \left(\frac{a}{a_f} \right)^{-2c}  \ .
     \label{before}
\end{eqnarray}
However, backreaction effects changes the dynamics of the inflaton such that (\emph{cf.}(\ref{exponential}))
we find the attractor behavior
\begin{eqnarray}
     f \propto \left(\frac{a}{a_f} \right)^{-2} \ .
     \label{after}
\end{eqnarray}
That is, the effective $c_{\rm eff} $ changes from
$c_{\rm eff} =c$ to the critical value $c_{\rm eff}=1$ due to backreaction.
In other words, the scale invariant spectrum of a gauge field is an attractor~\cite{Dimopoulos-anisotropic,Kanno:2009ei}.
Using (\ref{E-spectrum}), (\ref{em-energy}) and (\ref{general-u}),
before the backreaction becomes important we have
\begin{eqnarray}
  \rho_{em}  \sim
   H^4 \left( \frac{a_b}{a_i} \right)^{4c-4} \ ,
\end{eqnarray}
where $H$ is the Hubble during inflation which is almost constant, and hence the transition point $a_b$ occurs at
$\rho_{em} \sim \rho_A \sim 10^{-2} \rho_\phi
\sim 10^{-2} \mpl^2 H^2 $, i.e.
\begin{eqnarray}
    \left( \frac{a_b}{a_i} \right)^{4c-4}
   \sim 10^{-2} \frac{\mpl^2}{H^2} \ .
\end{eqnarray}

We can now calculate the power spectrum of magnetic fields.
First, we consider the modes which exit the horizon before $a_b$.
The superhorizon evolution of the mode function before $a_b$ is given by
\begin{eqnarray}
   u_{\bf k} (\tau)
  = \frac{1}{f_k \sqrt{2k}}  \left( \frac{a}{a_k} \right)^{4c-1} \ ,
  \label{u:ev}
\end{eqnarray}
where we should note $f_k$ is defined by $f_k = (a_b /a_f)^{2c-2}(a_k /a_f)^{-2c}$
from the continuity at $a_b$.
Since the evolution after $a_b$ becomes $u_{\bf k} \propto a^3$,
we obtain the mode function after $a_b$ as
\begin{eqnarray}
  u_{\bf k}
  = \frac{1}{f_k \sqrt{2k}}  \left( \frac{a_b}{a_k} \right)^{4c-1}
               \left( \frac{a}{a_b} \right)^3 \ .
\end{eqnarray}
From (\ref{B-spectrum}) and that $B^2=H^4P_B$, we obtain magnetic fields
\begin{eqnarray}
  B (\lambda_p , \tau_f )
  = \frac{H^2}{\sqrt{2} \pi} \left( {\lambda_p}{H}\right)^{2c -3}
            \left( \frac{a_b}{a_f} \right)^{2c-2}  \ ,
            \label{main}
\end{eqnarray}
where $\tau_f$ denotes end of inflation and $\lambda_p=a_f/k$ is the physical wavelength of the mode at the end of inflation.

Compared to the cases with no backreaction (\ref{pri-B}), the amplitude is
reduced by the factor
\begin{eqnarray}
   \left( \frac{a_b}{a_f} \right)^{2c-2}
  \sim 10^{-1}\frac{\mpl}{H} \left( \frac{a_i}{a_f} \right)^{2c-2} \ .
\end{eqnarray}
For the flat spectrum $c=3/2$, we can deduce magnetic fields at the end of
inflation as
\begin{eqnarray}
   B (\lambda_p , \tau_f )
  =  10^{-1}\frac{\mpl}{H} \left( \frac{a_i}{a_f} \right)
    \frac{H^2}{\sqrt{2} \pi}         \ .
\end{eqnarray}
Without backreaction, we anticipated $10^{-12}$Gauss for the scale invariant case $c=3/2$.
However, by taking into account the backreaction, we
have a suppression factor $a_b /a_f$ in (\ref{main}) which is  about $10^{-24}$.
Hence, we can expect at most $10^{-36}$Gauss on Mpc scales at present.

For modes which exit the horizon after the transition time $a_b$,
by setting $c=1$ in (\ref{u:ev}), we obtain
\begin{eqnarray}
u_{\bf k} (\tau)
  = \frac{1}{f_k \sqrt{2k}}  \left( \frac{a}{a_k} \right)^3 \ ,
\end{eqnarray}
where $f_k$ is now defined by $f_k = (a_k /a_f)^{-2}$.
Thus, we can calculate magnetic fields at the end of inflation as
\begin{eqnarray}
 B (\lambda_p , \tau_f )
  =   \frac{H^2}{ \sqrt{2}\pi} \frac{1}{H\lambda_p}
             \ .
\end{eqnarray}
\begin{figure}[ht]
\begin{center}
\includegraphics[height=7cm, width=9.5cm]{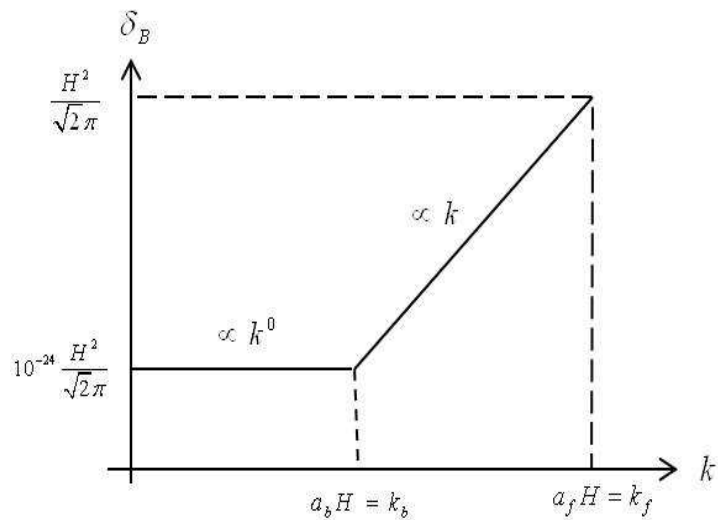}
\caption{The magnitude of magnetic fields is schematically
depicted as a function of wavenumber $k$ for the case $c=3/2$.
There exists a break at $k_b = a_b H $ in the spectrum due to the backreaction.
As can be seen, the amplitude of magnetic fields on Mpc scales gets
 a suppression by $10^{-24}$ due to this break. }
\label{fig:cond}
\end{center}
\end{figure}
The resultant spectrum for $c=3/2$ is schematically depicted in Fig.\ref{fig:cond}.
As expected, we have a flat spectrum on large scales before the backreaction
becomes relevant. However, once the backreaction becomes important,
the spectrum becomes blue.
Thus, after taking into account the backreaction, we realized that
 primordial magnetic fields on large scales from inflation cannot be expected~\cite{Kanno:2009ei}.

\subsection{Phenomenology in anisotropic inflation}

So far, we have shown the possibility of  anisotropic hair in a variety of inflationary models.
To explore observational consequences of this anisotropy we need to work out anisotropic cosmic perturbation theory.
Since the background is anisotropically expanding, we cannot use
the standard cosmological perturbation theory reviewed in section \ref{cosmic-perturbation-review-section}. This analysis has been carried out in litereature  ~\cite{Anisotropc-cosmic-pert-refs}.
Note that, even in a fixed anisotropic background, the behavior of fields is non-trivial~\cite{Kim-minami}.
In this subsection, in a direct extension of standard analysis review in section \ref{cosmic-perturbation-review-section} and motivated by the Bianchi type-I background isometries, we classify perturbations under the
2-dimensional rotational symmetry and obtain the quadratic actions
for 2-dimensional scalar and vector sectors.

\subsubsection{Classification of fluctuations}
In order to gain some intuition let us start with the spatially homogeneous and isotropic FLRW Universe.
For simplicity, we consider $K=0$ flat space which in the conformal time coordinate $\tau$ its metric is given as
$$
ds^2 = a^2 (\tau) \left[ -d\tau^2 + \delta_{ij} dx^i dx^j \right] \ .
$$
In the $C=E=0, W_i=0$  gauge (\emph{cf.} \eqref{metric-pert})\footnote{If we ignore vector and tensor perturbations $V_i , h_{ij}$,
this gauge is called \emph{flat slicing gauge}.} the metric perturbations are parameterized as
\begin{eqnarray}
ds^2 = a^2 \left[ -(1+2A)d\tau^2 +2 (B_{,i}+V_i ) d\tau dx^i
+  (\delta_{ij} + h_{ij} ) dx^i dx^j \right] \ ,
\end{eqnarray}
where we imposed $V_{i,i}=0$ and $ h_{ij,j}=h_{ii}=0$. As usual, dealing with linear perturbations it is convenient to work in the Fourier space. Since there exists 3-dimensional
rotation symmetry, we can take  the wave vector to be
${\bf k} = (k,0,0)$. Then, the perturbed metric has the following form
\begin{equation}
\delta g_{\mu \nu} =\left(
\begin{array}{ccccc}
& -2 a^2 A~ &  ~a^2 B_{,x}~ & ~a^2 V_2~ & ~a^2 V_3~
\vspace{1mm}\\\vspace{1mm}
& \ast  & 0 &    0            &   0
\\\vspace{1mm}
& \ast & \ast & a^2 h_{+}      &  -a^2 h_{\times}
\\
& \ast     & \ast & \ast &  -a^2 h_{+}
\end{array}
\right) \ . \hspace{5mm} * {\rm ~is ~symmetric ~part}.
\end{equation}
Here, we utilized the special choice ${\bf k} = (k,0,0)$ to solve
the constraints $V_{i,i}=0$ and $ h_{ij,j}=h_{ii}=0$. With the same token only $B_{,x}$ remains, and $h_{yz}=-h_{\times},
h_{yy}=-h_{zz}=h_{+}$.

To prepare the setup for studying the cases with only 2-dimensional rotation symmetry in $y-z$-plane we consider ${\bf k}=(k_x, k_y ,0)$. This case can be obtained from the previous case upon the rotation
\begin{eqnarray}
  \left(
\begin{array}{c}
k_x \\
k_y \\
0
\end{array}
\right)
= \frac{1}{k}\left(
\begin{array}{cccc}
& k_x~ & ~-k_y~ & ~0~ \\\vspace{1mm}
& k_y & k_x & 0 \\
& 0 & 0 & 0
\end{array}
\right)
 \left(
\begin{array}{c}
 k \\
 0 \\
 0
\end{array}
\right)
\ ,
\end{eqnarray}
where  $k^2 = k_x^2 + k_y^2$.
Under this rotation, the perturbed metric becomes
\begin{equation}
\hspace{-.5cm} \delta g_{\mu \nu} =\left(
\begin{array}{ccccc}
& -2 a^2 A~ & ~\frac{k_x}{k} a^2 B_{,x} -\frac{k_y}{k} a^2 V_2~
& ~\frac{k_y}{k} a^2 B_{,x} +\frac{k_x}{k} a^2 V_2~
& a^2 V_3~
\vspace{2mm}\\\vspace{2mm}
& \ast &  a^2 \frac{k_y^2}{k^2}h_{+}
& -a^2 \frac{k_x k_y}{k^2}  h_{+} & a^2 \frac{k_y}{k} h_{\times}
\\\vspace{2mm}
& \ast & \ast
& a^2 \frac{k_x^2}{k^2} h_{+} &  -a^2 \frac{k_x}{k} h_{\times}
\\
& \ast & \ast & \ast &  -a^2 h_{+}
\end{array}
\right).
\end{equation}
To simplify the perturbations, we use the remaining part of gauge transformations induced by spatial dependent diffeomorphisms
$$\delta g_{\mu\nu} \rightarrow
\delta g_{\mu\nu} + \xi_{\mu ; \nu} +\xi_{\nu;\mu} \ ,
$$
where
$$
\xi^0 =0 \ , \quad \xi^x = \frac{k_x}{2i k^2 }h_{+} \ , \quad
\xi^y = \frac{k_y}{2i k^2} h_{+} \ , \quad
\xi^z = \frac{k_x}{ik_y k } h_{\times} \ ,
$$
we obtain
\begin{equation}
\hspace{-2cm} \delta g_{\mu \nu} =\left(
\begin{array}{ccccc}
& -2 a^2 A~
& ~\frac{k_x}{k}a^2B_{,x}+\cdots ~
& ~\frac{k_y}{k}a^2 B_{,x}+\cdots ~
& ~a^2 V_3+\cdots ~
\vspace{2mm}\\\vspace{2mm}
& \ast &  a^2 h_{+}
&  0  &   a^2 \frac{k}{k_y} h_{\times}
\\\vspace{2mm}
& \ast &  \ast
&  a^2  h_{+}   &  0
\\
& \ast & \ast
& \ast &  -a^2 h_{+}
\end{array}
\right) \ ,
\label{gauge-A}
\end{equation}
where we have omitted some unimportant parts.
It should be noted that the above gauge transformation does not change flat slicing gauge for which
the 3-dimensional scalar curvature vanishes.

In the anisotropic models with the background metric
\begin{equation}
ds^2_b =a(\tau) ^2(-d\tau ^2+dx^2)+b(\tau) ^2(dy^2+dz^2),
\end{equation}
that is, $ a=e^{\alpha -2\sigma}, b=e^{\alpha +\sigma}, d\tau=dt/a$, \footnote{Notice that the conformal time in anisotropic inflation is the conformal
time in 2-dimensional part $(\tau , x)$.} the isometries are smaller than the previous case.
Nonetheless one can show that, as we have done in (\ref{gauge-A}), in appropriate gauge the most general metric perturbation can be brought to the form
\begin{equation}
\delta g_{\mu \nu} =\left(
\begin{array}{ccccc}
& \delta g_{00}~ & ~\delta g_{0x}~ & ~\delta g_{0y}~ & ~\delta g_{0z}~
\vspace{1mm}\\\vspace{1mm}
& *  &  \delta g_{xx}  &  0 &  \delta g_{xz}
\\\vspace{1mm}
&  * &  * &  \delta g_{yy}  &  0
\\\vspace{1mm}
&  *  &  *   & * &  \delta g_{zz}
\end{array}
\right) \ ,
\end{equation}
where we can impose further conditions so that the perturbed metric goes back
to (\ref{gauge-A}) in the isotropic limit.

One can classify the perturbed metric using the rotational symmetry in $y-z$-plane.
In 2-dimensional flat space, an arbitrary vector $m^a $ where $a=y,z$ can be
decomposed into the scalar part $m^a_{,a}\neq 0$ and the vector part $m^a_{,a} =0$.
In 2-dimensions the tensor can be constructed from 2-dimensional vector and scalar. Thus, the most general metric perturbations
can be classified into the scalar sector and the divergence-free vector sector: There are three 2-dimensional vectors and seven 2-dimensional scalars. There are, however, three scalar and one vector gauge transformations and hence we deal with two gauge invariant vectors and four gauge invariant scalar combinations.
Thanks to the symmetry in the $y-z$ plane,  without loss of generality,
 we can take the wavenumber vector to be ${\bf k} = (k_x , k_y ,0)$.
Hence, the vector sector in 2-dimensional classification can be identified as
$\delta g_{0z} , \delta g_{xz}$ in the above perturbed metric.
The remaining components $\delta g_{00} , \delta g_{0x} , \delta g_{0y} ,
\delta g_{xx}, \delta g_{yy}, \delta g_{zz} $ belong to the scalar sector. (Note that out of these six scalars, only four are independent.)

In order to calculate the statistical properties
of primordial fluctuations from anisotropic inflation
~\cite{Aniso-pert-stat-analysis,Gumrukcuoglu:2010yc,{Watanabe:2010fh}},
we need to reduce the action to the one for physical variables.
Then, we can quantize the system and specify the vacuum state. We note that besides metric perturbations, one has to consider gauge field(s) and inflaton scalar field perturbations too. As in the 3-dimensional case and due to rotation symmetry, the scalar and vector parts do not mix at linear level and we can hence analyze the vector and scalar sectors, separately.

In what follows we start with the anisotropic inflationary model introduced and discussed in section \ref{Hairy-inflation-section}.
\paragraph{$2d$ vector perturbations.} The metric perturbations that belong to 2d vector sector
 can be written as
\begin{equation}
\delta g_{\mu \nu}^{\rm vector} =\left(
\begin{array}{ccccc}
& 0~ & ~0~ & ~0~ & ~b^2 \beta_3~ \vspace{1mm}\\\vspace{1mm}
& *  &  0  &  0 & b^2 \Gamma \\\vspace{1mm}
&  * &  * &  0  &  0 \\
&  *  &  *   & * &  0
\end{array}
\right) \ ,
\end{equation}
where we have incorporated the anisotropy while keeping the spatial
scalar curvature to be zero, and the 2d vector part of gauge field perturbations can be taken as
\begin{eqnarray}
\delta A_{\mu}^{\rm vector} = \left(0 \ , 0 \ , 0 \ ,  D \right) \ .
\end{eqnarray}
Note that we have no residual gauge transformation and, in particular,
 $D$ is a gauge invariant under Abelian gauge transformations.
 And, as we have seen in (\ref{gauge-A}), $\Gamma$ corresponds to
  the cross-mode polarization of gravitational waves in the isotropic limit $a=b$.
Using this gauge, we can calculate the quadratic action as
%
\be\begin{split}
 S^{{\rm vector }}
 =&
\int d\tau d^3 x \left[~
\frac{b^4}{4a^2} \beta^2_{3,x} + \frac{b^2}{4} \beta^2_{3,y}
- \frac{b^4}{2a^2} \Gamma' \beta_{3,x}
+ \frac{f^2 v' b^2}{a^2} \beta_3 D_{,x}     \right. \\
\hspace{-1.5cm}&\left. \qquad\qquad
-\frac{b^2}{4} \Gamma^2_{,y} + \frac{b^4}{4a^2} \Gamma^{\prime 2}
-\frac{f^2a^2}{2b^2} D_{,y}^2
-\frac{1}{2}f^2 D_{,x}^{2}+\frac{f^2}{2} D^{\prime 2}
-\frac{f^2 v' b^2}{a^2} D' \Gamma ~\right] \ .
\end{split}
\ee
Since the perturbed shift function $\beta_3$ does not have a time derivative,
  it is not dynamical and its equation of motion is a constraint.
There are two propagating dynamical degrees of freedom $\Gamma$ and $D$ in this
2-dimensional  vector sector.
Eliminating the non-dynamical variable $\beta_3$
and defining canonically normalized variables
\begin{eqnarray}
\bar{\Gamma}  \equiv  \frac{b |k_y|}{\sqrt{2}k}\Gamma, \qquad
\bar{D}  \equiv  fD   \ ,
\end{eqnarray}
we obtain the reduced action for physical variables
\be\begin{split}
S^{\rm vector} =& \int d\tau d^3 k \left[
\frac{1}{2} |\bar{\Gamma}^{'}|^2
+\frac{1}{2}\left( \frac{(b/k)^{''}}{(b/k)}-k^2 \right) |\bar{\Gamma}|^2
\right. \\
\hspace{-2cm} &\qquad \qquad \left.
 +\frac{1}{2}|\bar{D}^{'}|^2
 +\frac{1}{2}\left( \frac{f^{''}}{f}-k^2
 -2\frac{f^2v^{'2}}{a^2}\frac{k_x^2}{k^2}
                         \right) |\bar{D}|^2  \right. \\
\hspace{-2cm} & \qquad \qquad \left.
 + \frac{1}{\sqrt{2}} \frac{fv^{'}}{a}\frac{a}{b}\frac{k_y}{k}
 \left\{ \bar{\Gamma}^{'}\bar{D}^{*} +  \bar{\Gamma}^{*'}\bar{D}
 +\frac{(k/b)^{'}}{(k/b)}\left( \bar{\Gamma}\bar{D}^{*} + \bar{\Gamma}^{*} \bar{D} \right)
                              \right\} \right]  \ ,
\label{vec-action}
\end{split}\ee
where $k$ is time dependent and given by
\begin{eqnarray}
k(\tau) \equiv \sqrt{k_x^2+ \frac{a^2 (\tau)}{b^2 (\tau )} k_y^2} \ ,
\end{eqnarray}
and becomes constant in the isotropic limit $a=b$.
In the isotropic limit $a=b$,
 $\bar{\Gamma} $ and $\bar{D}$ represent the cross-mode of gravitational waves
and vector waves, respectively. The third line in the action (\ref{vec-action})
describes how these modes  interact with each other.

Next, we use the slow roll approximation to simplify the action. To this end we start with \eqref{Sigma-over-H-generic}
\begin{eqnarray}\label{Sigma-I}
-\frac{\dot{H}}{H^2} = \epsilon, \qquad
\frac{\Sigma}{H}  = \frac{1}{3} I \epsilon \ ,
\end{eqnarray}
where
\be\label{I-aniso-def}
I=\frac{c-1}{c}\,,
\ee
and integrate the above assuming the slow-roll approximation $\epsilon^{'}/\epsilon \ll a^{'} /a$. The resultant expressions are
\begin{eqnarray}
a = (-\tau )^{-1-\epsilon  }, \qquad
b = (-\tau )^{-1-\epsilon - I \epsilon } \ .
\end{eqnarray}
Note that the range $(1,\infty )$ for $c$ corresponds to $(0,1)$ for $I$.
Recalling \eqref{rho-v-section-4} and \eqref{R-def} we have
\begin{eqnarray}
\frac{f^2 v^{'2}}{a^2} &=& 3(-\tau )^{-2} I \epsilon \ .
\end{eqnarray}
From (\ref{eq:Ax}), the background equation for the vector can be found as
\begin{eqnarray}
\left[ \frac{f^2v^{'}b^2}{a^2} \right] ^{'} = 0  \ ,
\end{eqnarray}
and hence
\begin{eqnarray}
\frac{f^{'}}{f} = (-\tau )^{-1}\left[ -2 -3\epsilon + \eta -2 I \epsilon  \right] \ ,
\end{eqnarray}
where $\eta$ is a slow-roll parameter \eqref{epsilon-eta-def}
\begin{equation}
\frac{\epsilon  ^{'}}{\epsilon }
= 2 \frac{(e^{\alpha})^{'}}{e^{\alpha}} \left( 2 \epsilon - \eta \right)
= 2 (2 \epsilon  -\eta) (-\tau )^{-1} \ .
\end{equation}
Furthermore, we obtain
\begin{eqnarray}
\frac{f^{''}}{f} &=&
(-\tau )^{-2} \left[ 2+9\epsilon  -3 \eta + 6 I\epsilon  \right] \ .
\end{eqnarray}
Note that all the above equalities are in the first order in slow-roll approximation.

Substituting these results into the action, we obtain the action
in the slow roll approximation~\cite{Watanabe:2010fh}
\begin{eqnarray}
\hspace{-1.5cm}S^{\rm vector}
&=& \int d\tau d^3 k \left[
 \frac{1}{2} | \bar{\Gamma}^{'} | ^2
+\frac{1}{2} \left[ -k^2+ (-\tau )^{-2} \left\{ 2+3\epsilon +3I \epsilon
+ 3 I\epsilon  \sin^2 \theta \right\} \right] |\bar{\Gamma}| ^2
\right. \nonumber \\
 \hspace{-2.5cm}&& \qquad \qquad + \frac{1}{2}|\bar{D}^{'}|^2
 +\frac{1}{2} \left[ -k^2+ (-\tau)^{-2} \left\{ 2+9\epsilon  -3\eta
  +6 I\epsilon  \sin^2 \theta \right\} \right] |\bar{D}|^2
\nonumber \\
\hspace{-2.5cm}&& \!\!\!\! \left. + \frac{\sqrt{6I\epsilon}}{2}(-\tau)^{-1}
\sin \theta (\bar{\Gamma}^{'}\bar{D}^{*}+\bar{\Gamma}^{*'}\bar{D})
-\frac{\sqrt{6I\epsilon}}{2}(-\tau)^{-2}
\sin \theta (\bar{\Gamma}\bar{D}^{*}+\bar{\Gamma}^{*}\bar{D}) \right]  , \quad
\label{eq:lagbegin}
\end{eqnarray}
where we have defined
\begin{eqnarray}
\sin \theta \equiv \frac{k_y a}{ k b} \ .
\end{eqnarray}
This $\theta$ represents the direction dependence.
In the isotropic limit $I=0$,
the Lagrangian for $\bar{\Gamma}$ becomes the familiar one for gravitational
waves in an FLRW Universe,  \emph{cf.} discussion of section \ref{cosmic-perturbation-review-section}.

\paragraph{$2d$ scalar perturbations.}
In a similar way, we can derive the quadratic action for physical variables
in the 2-dimensional scalar sector for the model discussed in section \ref{Hairy-inflation-section}.
The four gauge invariant  metric perturbations may be parameterized as
\begin{equation}
\delta g_{\mu \nu}^{\rm scalar} =\left(
\begin{array}{ccccc}
& -2 a^2 \Phi~ & ~a \beta_1~  & ~a \beta_2~ & ~0~ \vspace{2mm}\\\vspace{1mm}
& *  &  2 a^2 G  &  0 & 0 \\\vspace{1mm}
&  * &  * &  2 b^2 G  &  0 \\
&  *  &  *   & * & -2 b^2 G
\end{array}
\right) \ ,
\end{equation}
where we have kept the spatial scalar curvature vanishing.
The perturbations of the scalar (inflaton field $\phi$) will be represented by $\delta \phi$.
The variable $G$ and $\delta \phi$ are the gauge invariant variables
that correspond to the plus mode of gravitational waves
and the scalar perturbations, respectively, in the isotropic  limit $a=b$.
The gauge field $A_\mu$ will also have two 2-dimensional scalar perturbations
\begin{eqnarray}
\delta A^{\rm scalar}_{\mu} = \left( \delta A_0 \ , 0 \ ,   J \ , 0 \right)  \ ,
\end{eqnarray}
where we have fixed the Abelian gauge by putting the longitudinal component
to be zero.

From these ansatz, we can calculate the quadratic action as
\begin{eqnarray}
\hspace{-0.5cm} S^{{\rm scalar}}
  &=& \int d^3 x d\tau
 \left[~
 \frac{b^2}{2a^2} f^2 \delta A_{0,x}^2 + \frac{f^2}{2} \delta A_{0,y}^2
 +\frac{b^2}{a^2} f^2 v' \left(G+\Phi \right) \delta A_{0,x} -f^2 J' \delta A_{0,y}
   \nonumber \right. \\
\hspace{-2cm}&& -2 \frac{b^2}{a^2} ff_\phi v'\delta \phi \delta A_{0,x}
+\frac{1}{4} \beta_{1,y}^2
 -\frac{1}{2} \beta_{2,x} \beta_{1,y} + 2\frac{bb'}{a} \Phi_{,x} \beta_1
 - \frac{b^2}{a} \phi' \delta\phi_{,x} \beta_1 + \frac{1}{4} \beta_{2,x}^2
   \nonumber \\
\hspace{-2cm}&& + a \left( \frac{a'}{a}+\frac{b'}{b} \right) \beta_2 \Phi_{,y}
-a \left(\frac{a'}{a} -\frac{b'}{b} \right) \beta_2 G_{,y}
 +  \frac{f^2}{a} v' \beta_2 J_{,x} -a \phi' \beta_2 \delta\phi_{,y}
  + \frac{1}{2} f^2 J^{\prime 2}
   \nonumber\\
\hspace{-2cm}&&  -\frac{1}{2} f^2 J_{,x}^2
   +b^2 G^{\prime 2} -a^2 G_{,y}^2 -b^2 G_{,x}^2
+\frac{1}{2}b^2 \delta\phi^{\prime 2}
 -\frac{a^2}{2} \delta\phi_{,y}^2 - \frac{b^2}{2} \delta\phi_{,x}^2
 -\frac{1}{2} a^2 b^2 V_{\phi\phi} \delta\phi^2
 \nonumber \\
\hspace{-2cm}&&
  + \frac{b^2}{2a^2} \left( f_\phi^2 +ff_{\phi\phi} \right)
 v^{\prime 2} \delta\phi^2
 - a^2 b^2 V  \Phi^2 +  \frac{b^2}{2 a^2} f^2 v^{\prime 2}  G^2
 - 2a^2 b^2 V \Phi G    -2bb' \Phi' G   \nonumber \\
\hspace{-2cm}&&  \left.  \qquad \qquad
 - \left( \frac{b^2}{a^2} ff_\phi v^{\prime 2} + a^2 b^2 V_\phi \right)
  \delta\phi \left( G+\Phi \right) +b^2 \phi' \delta\phi' \left( G-\Phi \right)
  ~\right]    \ .
\end{eqnarray}
$S^{{\rm scalar}}$ governs the dynamics of seven scalar perturbations $\Phi, \beta_1, \beta_2, G, \delta A_0 , \delta \phi$
 and $J $. Among these, $\Phi , \beta_1, \beta_2$ and $\delta A_0 $
 are non-dynamical and can be eliminated.
In the slow roll approximation, we obtain the following reduced action~\cite{Watanabe:2010fh}
\begin{eqnarray}
\hspace{-2cm}&&S^{\rm  scalar} = \int d\tau d^3k
\left[ L^{GG}+L^{JJ}+L^{\phi\phi}+L^{\phi G}
+L^{\phi J}+L^{JG} \right] \ ,
\label{sca-action}
\end{eqnarray}
where diagonal parts are given by
\begin{eqnarray}
L^{GG}   &=& \frac{1}{2}|\bar{G}^{'}|^2
+\frac{1}{2} \left[-k^2+(-\tau) ^{-2} \left\{ 2+3\epsilon +3I\epsilon
 + 3 I\epsilon\sin^{2} \theta \right\} \right] |\bar{G}|^2,
\label{GG} \\
L^{JJ} &=& \frac{1}{2}|\bar{J}^{'}|^2
+\frac{1}{2} \left[ -k^2 +(-\tau )^{-2} \left\{ 2+9\epsilon  -3\eta
          - 6 I\epsilon  \sin^2 \theta \right\} \right] |\bar{J}|^2,
\label{JJ} \\
 L^{\phi \phi} &=& \frac{1}{2} | \delta\bar{ \phi}^{'}|^2    +\frac{1}{2} \Bigl[ -k^2   +(-\tau) ^{-2} \Bigl\{ 2 + 9\epsilon
 -\frac{3\eta }{1-I}-\frac{12I}{1-I} \nonumber\\
&&  \hspace{5.5cm} +\left( 12I \epsilon  +\frac{24I}{1-I} \right)
  \sin^2 \theta \Bigr\} \Bigr] | \delta\bar{\phi}|^2,
\label{dphi}
\end{eqnarray}
and the non-diagonal ``interaction'' parts reads
\begin{eqnarray}
\hspace{-2cm}&&L^{\phi G} = -3 I \sqrt{\frac{\epsilon }{1-I}}
(-\tau )^{-2} \sin^2\theta
\left(\bar{G} \delta\bar{\phi}^* +\bar{G}^* \delta\bar{\phi} \right) \ ,
\label{phi:G} \\
\hspace{-2cm}&& L^{\phi J} = \sqrt{ \frac{6I}{1-I}} \sin\theta
\left[(-\tau )^{-1} \left(\delta\bar{ \phi}^{*'}\bar{J}+\delta\bar{\phi}^{'}\bar{J}^{*}\right)
-(-\tau )^{-2}
\left(\delta\bar{ \phi}^{*} \bar{J}+\delta\bar{\phi}\bar{J}^*\right)\right],
\label{phi:J} \\
\hspace{-2cm}&& L ^{JG} = -\frac{\sqrt{6I \epsilon }}{2} \sin\theta\left[ (-\tau ) ^{-1}
\left(\bar{G}^{*'}\bar{J} + \bar{G}^{'} \bar{J}^{*}\right)
- (-\tau )^{-2}\left(\bar{G}^{*}\bar{J}+\bar{G}\bar{J}^*\right)\right] \ .
\label{J:G}
\end{eqnarray}
Here, we defined canonical variables
\begin{equation}
\bar{G}\equiv\sqrt{2}bG \ , \quad
\bar{J}\equiv \frac{f |k_x| }{k}J \ , \quad
\delta\bar{ \phi} \equiv b \delta \phi      \ .
\end{equation}
Note that $\bar{G} \ , \bar{J}$ and $\delta\bar{\phi}$ represent
the gravitational waves, the vector waves, and the scalar perturbations, respectively.
The above action shows there exist the interaction among these variables.
We notice the scalar part (\ref{dphi})
contains anisotropic factor $I$ \eqref{Sigma-I} without suppression by a slow-roll parameter $\epsilon$.
Therefore, to obtain the quasi-scale invariant spectrum of curvature perturbation, $I$ itself has to be small. Note also that the interaction terms all vanish in the isotropic $I=0$ limit.

\subsubsection{Anisotropic coupling vs. anisotropic expansion}

We are now in a position to calculate corrections to power spectrum of various variables due to the
 anisotropy~\cite{Aniso-pert-stat-analysis,Gumrukcuoglu:2010yc,Watanabe:2010fh}.
From the actions (\ref{eq:lagbegin}) and (\ref{sca-action}),
we see there are two sources of statistical anisotropy for fluctuations.
The first comes from the anisotropic
expansion itself and is encoded in (\ref{GG}), (\ref{JJ}), and (\ref{dphi}) while the second comes from the couplings
(\ref{phi:G}), (\ref{phi:J}) and (\ref{J:G}) due to the background vector field. As we discuss below, the latter dominate over the former source.

The first source of statistical anisotropy in fluctuations can be understood intuitively, recalling that the amplitude of canonical perturbations
at horizon crossing is equal to the effective Hawking temperature $H_{\rm eff}/2\pi$, where $H_{\rm eff}$ is the effective expansion rate. In the anisotropic case this $H_{\rm eff}$ depends on direction and hence we expect to see a direction dependent power spectrum.
The essential structure of  the statistical anisotropy of fluctuations due to the couplings can also be understood without complicated calculations, recalling the gauge field-inflaton coupling term
$$
  \sqrt{-g} g^{\mu\alpha} g^{\nu\beta} f^2(\phi) F_{\mu\nu} F_{\alpha\beta} \ ,
$$
and the order of magnitude of background quantities
$$
 \frac{f^2 v^{\prime 2}}{a^2} \sim I \epsilon \ , \quad
 \frac{f_\phi}{f} \sim \frac{ V}{V_\phi}
 \sim \frac{1}{\sqrt{\epsilon}} \ .
$$
For example, to obtain the $J$-$G$ coupling, one of the $F_{\mu\nu}$'s has to be replaced
by the background quantity $v'$. Hence, the coefficients in the $J$-$G$ coupling
is proportional to $f v'$ which is of the order of $\sqrt{I\epsilon}$.
This explains the strength of the coupling in (\ref{J:G}).
Similarly, $J$-$\delta\phi$ coupling is proportional to $f_\phi v'$
 because we have to take the variation with respect to $\phi$.
 Hence, we can estimate its magnitude to be $\sqrt{I}$.
 This explains the interaction term (\ref{phi:J}). Finally, the coupling
$G$-$\delta\phi$ has a magnitude of the order of $f_\phi v^{\prime 2}$
which is proportional to $I \sqrt{\epsilon}$.
This shows a good agreement with the coupling (\ref{phi:G}).
Thus, we can understand why there is a hierarchy among the couplings of
the gravitational waves, the vector waves and the scalar field.

To proceed further, we  should quantize this system
by promoting canonical variables to operators which satisfy the canonical commutation relations, with appropriately chosen vacuum state.
In the deep subhorizon limit  $k\tau \to -\infty$, the actions (\ref{eq:lagbegin})
and (\ref{sca-action}) reduce to those of independent harmonic oscillators where
we choose the Bunch-Davis vacuum state $|0 \rangle$
by imposing the conditions $a_{a,\bf k}|0 \rangle  =0$ at an initial time $\tau_i$.
Here,  $a_{a,{\bf k}}$ is an annihilation operator whose commutation relations are given by
\begin{eqnarray}
\left[ a_{a,\bf k} , a^\dagger _{b,\bf k^\prime } \right]
 = \delta_{ab}\delta ^{(3)}({\bf k-k^\prime}), \qquad
\left[ a_{a, \bf k} , a_{b,\bf k^\prime } \right]
= 0 \ .
\end{eqnarray}
We are interested in the power spectrum of
 the scalar perturbations
\begin{eqnarray}
\langle 0 \big|\delta \bar{\phi}_{\bf k}(\tau)
               \delta \bar{\phi}_{\bf p}(\tau) \big| 0 \rangle
\equiv P_{\delta\phi} ({\bf k}) \delta({\bf k} + {\bf p})
    \ ,
\end{eqnarray}
and the power spectrum of  the cross and  plus mode of
gravitational waves
\begin{eqnarray}
&&  \langle 0 \big| \bar{\Gamma}_{\bf k}(\tau) \bar{\Gamma}_{\bf p}(\tau) \big| 0 \rangle
 \equiv P_{\Gamma} ({\bf k}) \delta({\bf k} + {\bf p}) \ , \\
&&  \langle 0 \big| \bar{G}_{\bf k}(\tau) \bar{G}_{\bf p}(\tau) \big| 0 \rangle
 \equiv P_{G} ({\bf k}) \delta({\bf k} + {\bf p})     \ .
\end{eqnarray}
We can also calculate the cross correlation between the plus mode
of gravitational waves and the scalar perturbations
\begin{eqnarray}
\langle 0 \big| \delta \bar{\phi}_{\bf k}(\tau) \bar{G}_{\bf p}  (\tau) \big| 0 \rangle
\equiv P_{\delta\phi G} ({\bf k}) \delta({\bf k} + {\bf p})  \ .
\end{eqnarray}

As is mentioned in the previous subsection,  $I$ has to be small.
And, the anisotropy in the expansion rate is much more small.
Hence, we can treat the anisotropy perturbatively and estimate its magnitude by
using perturbation in the interaction picture.
In the interaction picture,
the expectation value for a physical quantity ${\cal O} (\tau)$ is given by
\begin{equation}
\langle in \left| {\cal O} (\tau) \right |in \rangle
= \left< 0 \left|
\left[ \bar{T}\exp \left( i \int ^{\tau}_{\tau_i} H_I(\tau ^{'}) d\tau ^{'}
\right) \right] {\cal O} (\tau )
\left[ T \exp \left( -i\int ^{\tau}_{\tau_i} H_I(\tau ^{'}) d\tau ^{'} \right) \right]
\right| 0 \right> \ ,
\end{equation}
where $|in \rangle$ is an in vacuum in the interaction picture,
 $T$ and $\bar{T}$ denote a time-ordered and an anti-time-ordered product
 and $H_I$ denotes  the interaction Hamiltonian.
 This is equivalent to the following
\begin{eqnarray}
\hspace{-2cm}\langle in \left| {\cal O} (\tau) \right| in \rangle
&=& \sum _{N=0}^{\infty} i^N \int _{\tau_i}^{\tau} d\tau _N\int _{\tau_i}^{\tau_N}
d\tau _{N-1} \cdots \int _{\tau_i}^{\tau_2} d\tau _{1} \nonumber \\
&&  \qquad \times \left<0 \left|
\left[ H_I(\tau_1),\left[H_I(\tau_2), \cdots \left[H_I(\tau _N),{\cal O}(\tau) \right]
\cdots \right] \right]
\right| 0 \right>.
\end{eqnarray}
In our analysis, we assume the noninteracting part of Hamiltonian to be that of free fields in de Sitter spacetime, which for each mode ${\bf k}$ it is
\begin{equation}
L_0 = \sum _{n} \left[ \frac{1}{2} |Q^{'}_{n,{\bf k}}|^2-\frac{1}{2}
                    \left( k^2 - 2(-\tau )^{-2} \right) |Q_{n,{\bf k}}|^2 \right] \ ,
\end{equation}
where
\begin{eqnarray}
Q_{n,{\bf k}}(\tau) &=& u(\tau)  a_{n,{\bf k}}
+ u(\tau)^{*} a_{n,{\bf -k}}^{\dagger},
\label{eq:nonint}\\
u(\tau) &\equiv &\sqrt{\frac{1}{2k}}e^{-ik\tau}\left(1-\frac{i}{k\tau} \right) \ ,
\label{eq:modefunc}
\end{eqnarray}
where $Q_n$ represent the physical variables $\bar{D}, \bar{\Gamma}, \bar{G}, \bar{J}, \delta\bar{\phi}$.
The rest of the Lagrangian (\ref{eq:lagbegin})-(\ref{J:G}) may then be regarded as the interaction part $-H_I = L^{(2)}-L_0$.
To see the leading effect on the anisotropy in the scalar perturbation,
which is of the order of $I$, we evaluate the correction due to the interaction
 given by
\begin{eqnarray}
H_I^{\phi J} = \int d^3k \ \sqrt{ \frac{6I}{1-I}} \sin\theta \left[-(-\tau )^{-1}
\left(\bar{\delta \phi}^{\dagger'}\bar{J} +\bar{\delta\phi}^{'}\bar{J}^{\dagger}\right)
+ (-\tau )^{-2}
\left(\bar{\delta \phi}^{\dagger} \bar{J} +\bar{\delta\phi} \bar{J}^{\dagger}\right)
\right]
\ .
\end{eqnarray}
From  \eqref{dphi} and in the analogy with the slow-roll parameter in the ordinary slow-roll inflation, we find that the mass term for $\delta\bar\phi$ is proportional to
\begin{eqnarray}
I\sin^2 \theta \delta\bar{\phi}
 \delta\bar{\phi}^{\dagger}
\end{eqnarray}
and is expected to give the anisotropy
\begin{eqnarray}
\frac{\delta \langle in \left| \delta \bar{\phi}_{\bf k}
                         \delta \bar{\phi}_{\bf p} \right| in \rangle }{
 \langle 0 \left| \delta \bar{\phi}_{\bf k} \delta \bar{\phi}_{\bf p} \right| 0 \rangle }
\sim \sin ^2 \theta I N(k) \ ,
\end{eqnarray}
where $N(k)$ is the number of  e-folds from the horizon exit of fluctuations with wavenumber $k$ to the end of the inflation.
We will show below that the corrections coming from the interaction term $H^{\phi J}_{I}$ is proportional to $N(k)^2$
and hence dominate over the correction due to the mass term.
Thus, the leading correction is given by
\begin{eqnarray}
\hspace{-1cm} &&\delta \langle in \left| \delta \bar{\phi}_{\bf k} (\tau)
                   \delta \bar{\phi}_{\bf p} (\tau) \right| in \rangle  \nonumber \\
\hspace{-1cm}&& \qquad = i^2 \int_{\tau _i}^{\tau}d\tau_2\int_{\tau _i}^{\tau _2} d\tau _1
\left< 0 \left|
 \left[ H^{\phi J}_I(\tau_1),\left[H^{\phi J}_I(\tau_2),\delta \bar{\phi}_{\bf k} (\tau)
 \delta \bar{\phi}_{\bf p} (\tau) \right] \right]
 \right| 0 \right> \ .  \
\end{eqnarray}
Using (\ref{eq:nonint}) and commutation relations for the creation and annihilation operators, we obtain the anisotropy expressed as follows
\begin{eqnarray}
\hspace{-2cm}&&\frac{\delta \langle in \left| \delta \bar{\phi}_{\bf k}
                             \delta \bar{\phi}_{\bf p} \right| in \rangle}
{\langle 0 \left| \delta \bar{\phi}_{\bf k}
                 \delta \bar{\phi}_{\bf p} \right| 0 \rangle}  \nonumber\\
\hspace{-2cm}&=& \frac{24 I}{1-I} \sin ^2 \theta
 \int ^{\tau} _{\tau_i} d\tau _2 \int ^{\tau_2}_{\tau_i} d\tau_1
  \frac{8}{|u(\tau)|^2} {\rm Im}
\left[  -(-\tau_2)^{-1}  u^{'}(\tau_2) u^{*}(\tau)
   + (-\tau _2 )^{-2} u(\tau_2 )u^{*}(\tau )  \right] \nonumber\\
\hspace{-2cm}&\ &\, \times {\rm Im}
\left[ u(\tau_1 )u^{*}(\tau _2)\left\{ -(-\tau_ 1)^{-1} u^{'}(\tau _1)u^{*}(\tau )
 + (-\tau_1)^{-2} u(\tau_1 )u^{*}(\tau) \right\}  \right] \ ,
\end{eqnarray}
where ${\rm Im}$ denotes the imaginary part.
Substituting the function form of $u$ (\ref{eq:modefunc})
and introducing time variables $\chi \equiv k\tau$, $\chi_1 \equiv k\tau_1$ and
$\chi_2 \equiv k\tau_2$, we have
\begin{eqnarray}
&&\frac{\delta \langle in \left| \delta \bar{\phi}_{\bf k}
                             \delta \bar{\phi}_{\bf p} \right| in \rangle}
{\langle 0 \left| \delta \bar{\phi}_{\bf k}
                             \delta \bar{\phi}_{\bf p} \right| 0 \rangle}=  \nonumber\\
&=& \frac{6I}{1-I} \sin ^2 \theta \int ^{\chi}_{\chi_i}d\chi_2 \int ^{\chi_2}_{\chi_i} d\chi_1
 \frac{8}{1+\frac{1}{(-\chi)^2}}\frac{1}{-\chi_1}\frac{1}{-\chi_2}\left[ \cos(-\chi_2+\chi)-\sin(-\chi_2+\chi) \frac{1}{\chi} \right] \nonumber\\
&\ &\, \times
\bigg[ \cos(-2\chi_1+\chi+\chi_2) \left( 1+\frac{1}{\chi\chi_1}-\frac{1}{\chi\chi_2}+\frac{1}{\chi_1\chi_2} \right) \nonumber\\
&&\ \ \qquad + \sin(-2\chi_1+\chi+\chi_2)\left( -\frac{1}{\chi\chi_1\chi_2}+\frac{1}{\chi_1}-\frac{1}{\chi}-\frac{1}{\chi_2} \right) \bigg].\label{eq:integrand}
\end{eqnarray}
The contribution to the integral
from the subhorizon $- \chi_1 \gg 1$ is negligible.
 In the limit of superhorizon $-\chi_1 \ll 1$, we also have
 $-\chi_2 \ll 1 , -\chi \ll 1$. Hence,  the integrand in (\ref{eq:integrand}) approximately becomes $8/\chi_1 \chi_2 $.
  Thus, the anisotropy can be evaluated as~\cite{Watanabe:2010fh}
\begin{eqnarray}
\frac{\delta \langle in \left| \delta \bar{\phi}_{\bf k}
                             \delta \bar{\phi}_{\bf p} \right| in \rangle}
{\langle 0 \left| \delta \bar{\phi}_{\bf k}
                             \delta \bar{\phi}_{\bf p} \right| 0 \rangle} (\chi)
&=& \frac{6I}{1-I} \sin ^2 \theta \int^{\chi}_{-1} d\chi_2 \int^{\chi_2}_{-1} d\chi_1 \frac{8}{\chi_1 \chi_2} \nonumber\\
&=& \frac{24I}{1-I} \sin ^2 \theta \ N^2(k),
\end{eqnarray}
where $N(k) \equiv -\ln (-k\tau)$ is the number of e-folds after the horizon exist (to the end of inflation).

Similarly one can compute  anisotropy in both polarizations of gravitational waves~\cite{Watanabe:2010fh}
\begin{equation}
\frac{\delta \langle in \left|  \bar{\Gamma}_{\bf k}
                                \bar{\Gamma}_{\bf p} \right| in \rangle}
{\langle 0 \left|  \bar{\Gamma}_{\bf k}
                                \bar{\Gamma}_{\bf p} \right| 0 \rangle}
 = \frac{\delta \langle in \left| \bar{G}_{\bf k} \bar{G}_{\bf p} \right| in \rangle}
 {\langle 0 \left| \bar{G}_{\bf k} \bar{G}_{\bf p} \right| 0 \rangle}
 = 6 I \epsilon \sin ^2 \theta \ N^2(k)  \ ,
\end{equation}
where we used the interaction term in the action (\ref{eq:lagbegin})
for $\bar{\Gamma}$ and that in (\ref{J:G}) for $\bar{G}$.
It is interesting to calculate the cross correlation.
The leading contribution comes from $H_I^{JG}$ and $H_I^{\phi J}$.
The result is as follows~\cite{Watanabe:2010fh}:
\begin{eqnarray}
\frac{\langle in \big| \delta\bar{\phi}_{\bf k} \bar{G}_{\bf p} \big| in \rangle}
{  \langle 0\big| \delta\bar{\phi}_{\bf k} \delta\bar{\phi}_{\bf p} \big| 0 \rangle}
\simeq - 24 I \sqrt{\frac{\epsilon}{1-I}}  N^2(k)  \ .
\label{cross}
\end{eqnarray}
As we will soon show, this might be detectable.
We remind the reader that in these results  we have ignored the anisotropy of the background expansion
while considered the anisotropic coupling due to the existence of the background gauge field.

\subsubsection{Statistical anisotropy from anisotropic inflation}

We found anisotropic inflation is an attractor
in supergravity motivated inflaton-gauge field coupling given in \eqref{eq:action} with a wide range of gauge kinetic functions. The metric during inflation approximately reads
\begin{eqnarray}
   ds^2 = -dt^2 + e^{2Ht} \left[ e^{-4\Sigma t} dx^2 + e^{2\Sigma t} \left( dy^2 + dz^2 \right) \right] \ ,
\end{eqnarray}
where $H$ and $\Sigma$ describe the average expansion rate and the anisotropic expansion rate, respectively and their values are specified by \eqref{Sigma-I}.
Remarkably, the degree of anisotropy is
at most of the order of the slow-roll parameter $\epsilon $. The point is the existence of anisotropic coupling
during inflation.

We can now discuss cosmological implication
of an anisotropic inflationary scenario.
There are many interesting phenomenology in anisotropic inflation.
We start with the anisotropy in the power spectrum we is parameterized as
\begin{eqnarray}
  P({\bf k} ) = P(k) \left[ 1 + g_* \sin^2 \theta \right] \ ,
\label{anisotropy-parametrization}
\end{eqnarray}
where $g_*$ will be replaced by $g_s$ for (scalar) curvature perturbations and $g_t$ for tensor perturbations.
Then, we can predict the following
\begin{itemize}
\item There exists statistical anisotropy in curvature perturbations
of the order
\begin{eqnarray}
 g_s = 24 I N^2(k)  \ .
\end{eqnarray}
In \cite{Gumrukcuoglu:2010yc}, it is pointed out that the sign of
$g_s$ predicted by our models is different from the observed one~\cite{Groeneboom:2008fz}. Note that our parametrization
in (\ref{anisotropy-parametrization}) is different from that used in ~\cite{Ackerman:2007nb}.
The positive sign in our definition means the anisotropy is oblate, on the other hand, the negative sign
means the anisotropy is prolate. In \cite{Groeneboom:2008fz}, they found prolate anisotropy.
 However, it is possible to modify the model so that the sign of $g_s$ is flipped.
As reviewed in subsection \ref{dynamical} the dynamics of vector fields tends to minimize the anisotropy in
 the expansion of the universe and leading to the orthogonal dyad~\cite{Yamamoto:2012tq}.
 Then, the orthogonal direction to the plane determined by two vectors
 becomes a preferred direction. In this case, the sign of $g_{s}$ becomes opposite. 
The reason is that the vector field tend to make the expansion slow along the direction of 
the vector. Hence, for a single vector, we had oblate anisotropy. While, the same mechanism
gives rise to prolate anisotropy for the dyad.
We can also utilize anti-symmetric tensor fields \cite{AnIso-form-fields}
to achieve the same goal.
\item There exists statistical anisotropy in gravitational waves
of the order
\begin{eqnarray}
g_t = 6I \epsilon N^2(k)  \ .
\end{eqnarray}
\item These exists the cross correlation
between scalar perturbations and gravitational waves
of the order of
$-24 I \sqrt{\epsilon} N^2(k) $.
Using the definition of curvature perturbations
${\cal R}_c = \delta\bar{\phi}/\sqrt{2 \epsilon}$, one can translate
the cross correlation (\ref{cross}) between the scalar perturbations and
gravitational waves to that between the curvature perturbations and
gravitational waves normalized by the power spectrum of curvature perturbations:
\begin{eqnarray}
r_c = \frac{\langle in \big| {\cal R}_c ({\bf k}) \bar{G}_{\bf p} \big| in \rangle}
{  \langle 0\big| {\cal R}_c ({\bf k}) {\cal R}_c ({\bf p}) \big| 0 \rangle}
= - 24 \sqrt{2} I N^2(k) \epsilon  \ .
\end{eqnarray}
\end{itemize}
Here, we should notice that there appears an enhancement factor $N^2(k)$ in the above formula.
This is because the interaction on superhorizon scales persists after horizon crossing during inflation.
Because of this enhancement, even when the anisotropy of the spacetime
is quite small, say $\Sigma/H \sim 10^{-7}$ in our example,
the statistical anisotropy imprinted in primordial fluctuations
can not be negligible in  precision cosmology. In other words, we have to assume $I\ll 1$.
In fact, from the observational upper bound  $g_s < 0.3$, there is
 a cosmological constraint
\begin{eqnarray}
 I < \frac{0.3}{24 N^2 (k)} \ ,
\end{eqnarray}
where we used the result in \cite{Pullen:2007tu}.
 Since $I$ is derived from the gauge kinetic function \eqref{I-aniso-def} and the $e$-fold number $N(k)$ can be
determined once reheating process is clarified, the constraint on $g_s$ implies
the constraint on the gauge kinetic function.

After taking into account the observational constraint, the anisotropy in the background metric of {\sl anisotropic inflation} is negligibly small. Nonetheless, as we discussed the anisotropic effects may arise through perturbations which can have observational prospects.

It is useful to notice that
there exist consistency relations between observables
\begin{eqnarray}
   4 g_t = \epsilon \  g_s   \ , \quad r_c = -\sqrt{2} \epsilon\ g_s \ , \quad r=16 \epsilon  \ .
\end{eqnarray}
The consistency relations allows us to test anisotropic inflation in a model independent way.
Let us explain how to use it.
It is known that the current observational limit of the statistical anisotropy
for the curvature perturbations is given by
$g_s < 0.3$~\cite{Pullen:2007tu}. Note that, according to \cite{Pullen:2007tu}, a signal as small as 2\% can be detected with the PLANCK.
Now, suppose that we detected $g_s =0.3$. We also assumed the tensor-to-scalar ratio to be $r = 0.3$.
Note that this particular number is not important, as is explained later, as long as we can detect gravitational waves.
Then, the consistency relations would give us predictions.
Namely, anisotropic inflation implies the anisotropy in the gravitational waves
\begin{eqnarray}
g_t \simeq 10^{-3}
\end{eqnarray}
and the cross correlation
\begin{eqnarray}
r_c = -  \sqrt{2} g_s \epsilon \sim -  4 \times 10^{-3} \ ,
\end{eqnarray}
where we used $g_{s} \sim 0.3 $ and $\epsilon \sim 10^{-2}$.
If these predictions are confirmed by the CMB observations, that must be a strong evidence of anisotropic inflation.

\begin{figure}
\begin{center}
\includegraphics[width=9cm]{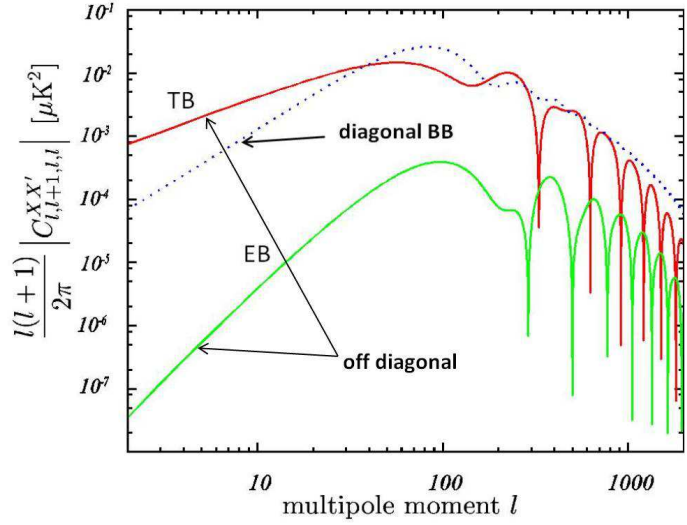}
 \caption{The off diagonal TB and EB spectra $C^{TB}_{\ell , \ell +1}$, $C^{EB}_{\ell , \ell +1}$ induced by the cross correlation. As a reference, the conventional diagonal BB spectrum induced by isotropic part of the tensor perturbations is plotted with a dotted line. The parameters $g_s =0.3,\ r=0.3$ are taken.}
 \label{fg:tbeb}
\end{center}
\end{figure}
We showed in the above how the consistency relations are used to make predictions.
So, the next task is to clarify how we see these features in the CMB.
The answer is that
the anisotropy induces off-diagonal $TB,EB$ spectra $C_{\ell , \ell +1}$.
Here, we define the angular power spectrum
$C_{\ell, \ell'} = \langle a_{\ell,m} a_{\ell' , m'} \rangle$ with coefficients $a_{\ell m}$ of the spherical harmonic expansion.
In Fig. \ref{fg:tbeb}, we have depicted examples of $TB$ and $EB$ correlations $C^{TB}_{\ell , \ell +1}$, $C^{EB}_{\ell , \ell +1}$~\cite{Watanabe:2010bu} with the parameters $r=0.3$ and $g_s =0.3$.
As a reference, the conventional $BB$ diagonal spectrum induced by the isotropic part of tensor perturbations is also plotted with a dotted line.
Note that unlike parity violating cases for which odd parity correlations $C_{ll}^{TB}, C_{ll}^{EB}$ exist~\cite{Lue:1998mq,Alexander:2004wk,Satoh:2007gn,Saito:2007kt},
 our model predicts even parity correlations such as $C_{l,l+1}^{TB}$ as the result of parity symmetry of the system.
The ratio of $TB$ correlation induced by cross correlation to the isotropic $BB$ correlation is not dependent on $\epsilon$ (or $r$) for a fixed value of $g_s$ in our anisotropic inflation model. For the optimistic value of $g_s \sim 0.3$,
 both amplitudes become comparable. This simple order estimation implies
 that the $TB$ signal could be comparable to that
of $B$ mode correlation induced by primordial gravitational waves.
Hence, anisotropic inflation can be a potential source of off-diagonal $TB$ correlation.
Since the current constraints on the $TB/TE$ ratio is of the order of $10^{-2}$~\cite{Komatsu:2010fb}, we need to improve
the accuracy by one more order, which might be achieved by the PLANCK.

It should be stressed that the statistical anisotropy is a natural consequence of the fact that gauge fields are
ubiquitous~\footnote{There are other mechanism to produce the statistical anisotropy~\cite{Libanov:2010nk}.
However, the consistency relations can be used to discriminate
anisotropic inflation from others.} and there remains many other phenomenological aspects  inflation models with gauge field hair to be explored.

\subsubsection{Non-Gaussianity in anisotropic inflation}

As we explained in section \ref{prim-magnet-NG-section}, we know the vector field can induce a non-trivial non-Gaussianity in the presence of a gauge kinetic function.
Since the anisotropy in the expansion is quite small in realistic cases and the dominant effect is a coupling between scalar and vector,
one can calculate non-Gaussianity in the isotropic universe by neglecting the metric backreaction.
Concrete calculations in this regards has been carried out in~\cite{Bartolo:2012sd} with the statistical anisotropy in the power spectrum
\begin{eqnarray}
  |g_s|_{\rm CMB} \gtrsim 0.1 \frac{N_{\rm tot}-N_{\rm CMB}}{37} \ .
\end{eqnarray}
Moreover, they have evaluated the non-Gaussianity as
\begin{eqnarray}
  f_{\rm NL} \simeq 26 \frac{|g_s|_{\rm CMB}}{0.1}  \ .
\end{eqnarray}
This is another consistency relation which anisotropic inflation predicts.
Namely, there exists a large non-Gaussianity if the statistical anisotropy is also large $|g_s|\sim 0.1$.

The authors in~\cite{Bartolo:2012sd} pointed out quantum fluctuations of the electro-magnetic field are comparable to the classical background when $c-1$ is severely constrained by the observations. For these cases, quantum diffusion process would be important and those affect the classical background in the long wavelength limit. Nevertheless, they succeeded in making a statistical prediction for the non-gaussianity.
Namely, the consistency relations are robust and the models are testable.

\section{Gauge-flation: Inflation driven by  non-Abelian gauge fields}\label{Gauge-flation-section}

As mentioned in the previous sections  the energy density which drives the slow-roll inflation dynamics is generically provided by one or more scalar fields and that scalar fields are the natural choices if we are interested in homogeneous and isotropic FLRW models. Nonetheless, as we will review in this section, it is possible to drive isotropic inflation by non-scalar fields and in particular vector gauge fields. This model/scenario will hence be called gauge-flation \cite{arXiv:1102.1513,arXiv:1102.1932}. In this section we spell out how it is possible to get quasi-de Sitter expansion from a gauge field theory minimally coupled to Einstein gravity and study the stability of the inflationary trajectories against classical isotropic and anisotropic perturbations \cite{Maleknejad:2011jr}. We then study gauge-flation cosmic perturbation theory and compute the information which could be extracted from our second order (quadratic) perturbation analysis and finally compare these results with the cosmological observation data.

\subsection{The theoretical setup, how to get an isotropic model}\label{gauge-flation-general-setup}

Our starting point in gauge-flation  is a non-Abelian gauge theory which is minimally coupled to Einstein gravity in four dimensions. The gauge group can be any non-Abelian compact group $G$ which could be simple or a generic product of simple groups. Generators of algebra of $G$ will be denoted by $T^A,\ A=1,2,\cdots dimG$. Lagrangian of the models of our interest is of the generic form
\be\label{gauge-flation-generic-action}
\cL=\cL(g_{\mu\nu}, \Fmn)=\frac12 R+\cL_G(\Fmn)\,,
\ee
$\Fmn$ is the field strength of the gauge fields $A_\mu^A$:
\be\label{Fmn-def}
\Fmn^A=\partial_\mu A_\nu^A-\partial_\nu A_\mu^A-g f^{ABC} A_\mu^BA_\nu^C\,,
\ee
where $g$ is the gauge coupling and $f^{ABC}$ are the structure constants of the gauge group $G$, $[T^A,T^B]=if^{ABC}T^C$. We choose the standard normalization $Tr(T^AT^B)=\frac12\delta^{AB}$.
Under a generic local gauge transformation $U\in G$
\be\label{gauge-transformation}
\begin{split}
A_\mu\ &\longrightarrow\ A_\mu'=-i U^{-1}\partial_\mu U+ U^{-1} A_\mu U\,,\\
\Fmn\ &\longrightarrow\ \Fmn'=U^{-1} \Fmn U\,,
\end{split}
\ee
and  $\cL_G(\Fmn)$ is a  generic gauge invariant action made out of powers of $\Fmn$ whose Lorentz indices are properly summed over using \emph{only} metric $g_{\mu\nu}$ and/or the Levi-Civita tensor $\epsilon^{\alpha\beta\mu\nu}$ and the gauge indices are summed over by taking trace. That is, $\cL_G$ is a minimally coupled diff+gauge invariant action. Moreover, with our choice ($\cL_G$ being only a function of $\Fmn$) gauge field equations of motion will be at most of second order in time derivatives.
In particular, we stress that $\cL_G$ may include Yang-Mills term $-\frac12 Tr\Fmn^2$ but is not limited to that.

We would like to get an inflationary background driven by gauge fields as inflaton. If we ignore the gauge indices for the moment, i.e. considering an Abelian gauge field, turning on a vector gauge field in the background will break rotational symmetry of the flat FLRW geometry, unless only the temporal component is turned on and homogeneity implies that this component should be only time dependent. This gauge field will hence have a vanishing field strength, it is a pure gauge. Now, let us consider non-Abelian gauge fields. Considering gauge indices will change the situation: Gauge fields are defined up to gauge transformations  and two gauge fields which are related by gauge transformations \eqref{gauge-transformation} are physically equivalent. To be precise, let us fix the background metric coordinates to the standard spatially flat ($K=0$) FLRW one \eqref{FLRW}. In this coordinate system we may turn on a generic vector gauge field with spatial components $A_i$ and homogeneity implies $A_i=A_i(t)$. Note that $A_0$ may always be set to zero in a particular gauge, the temporal gauge; i.e. one can always find $U=U(t)$ for which $A'_\mu=0$. This choice, therefore, fixes the gauge transformations up to \emph{time independent gauge transformations}.\footnote{We may set $dim G$ number of relations between gauge fields and their first derivatives through gauge choice, and in the temporal gauge $A_0^A=0$ we have used all $dim G$ choices. Note also that one may consider a generic space-dependent gauge transformations $U=U(x_i)$, will not change $A_0=0$, however, will introduce $x^i$ dependence into the spatial components $A_i$ and hence we will not consider these gauge transformations. Therefore, going to temporal gauge completely fixes the \emph{local} gauge freedom.} We have hence fixed the gauge freedom up to  space-time independent \emph{global} gauge transformations. Therefore, up to global gauge transformations, our background gauge field is (note that for the ease of notation we have suppressed the explicit gauge indices)
\be\label{background-gauge-field}
A_0=0\,,\quad A_i=A_i(t)\,.
\ee

As discussed, turning on a vector gauge field $A_i$ will break rotational invariance. We use the remaining global gauge transformations to remedy this: Two fields related by gauge transformations are physically equivalent, i.e.
\be\label{gauge-equiv}
(A_i)_G=U^{-1} A_i U \equiv_G A_i,\quad \mathrm{with\ constant\ } U\in G\,.
\ee
The above gauge transformation acts upon the gauge indices which are suppressed in the above equation.
On the other hand, upon global spatial rotations
\be\label{gauge-rotation}
A_i\to (A_i)_R=R_{ij}A_j \,.
\ee
If the background configuration $A_i$ is chosen appropriately such that $A_R=A_G$ then our physical gauge configuration will preserve rotational symmetry. In other words, rotational non-invariance caused by turning on vector fields in the background may be compensated by the global gauge transformation.  To see this, we first note that $A_i$ is in vector (triplet) representation of the rotation group $SO(3)_R$. On the other hand, any non-Abelian gauge group has an $SU(2)$ subgroup. This $SU(2)$ may be \emph{identified} with $SO(3)_R$ for some particular gauge field configurations. As far as our current discussion is concerned, without loss of generality, we can choose the gauge group $G$ to be $SU(2)$ or $SO(3)$ and choose the $T^A$'s to be $SU(2)$ generators in the triplet (adjoint) representation $\frac12 \sigma^a,$ $a=1,2,3$ where $\sigma^a$ are Pauli matrices. Let us choose the background to be
$$
A_i^a=a(t)\psi(t) \delta_i^a\,,
$$
where $a(t)$ is the scale factor and $\psi(t)$ is a scalar under rotations. This latter may be seen immediately, noting that one can rewrite the above as the following form
\be\label{gauge-ansatz-triad}
A_i^a=\psi(t) e^a_{~i}\,,
\ee
where $e^a_{~i}$ are the triads of the spatial hyper-surface.
It is readily seen that for a gauge field of the form \eqref{gauge-ansatz-triad}, the effects of any generic rotation
by angle $\vec{\theta}$, $R=e^{\frac{i}{2}\vec{\sigma}\cdot\vec{\theta}}$, can be viewed as gauge transformation
$U=e^{\frac{i}{2}\vec{\sigma}\cdot\vec{\lambda}}$ with  $\vec{\lambda}=-\vec{\theta}$.
To summarize, we have shown that the ansatz
\be\label{gauge-ansatz-summary}
A^a_\mu=\left\{\begin{array}{cc} 0&\quad \mu=0\\
a(t)\psi(t)\delta^a_i&\quad \mu=i
\end{array}\right.
\ee
leads to a rotationally invariant, homogeneous background.

The ``homogeneous-isotropic'' non-Abelian gauge field configurations in the context of FLRW cosmology (and not necessarily inflationary models) has been discussed in the literature before, e.g. see \cite{homo-iso-gauge-config-1,homo-iso-gauge-config-2}.\footnote{For the closed Universe case with $K=1$ \eqref{FLRW}, there are other choices for the homogeneous isotropic gauge field configuration \cite{homo-iso-gauge-config-2}.}
Here we discussed the non-Abelian gauge fields which as we will see are suitable for building inflationary models. From our discussions above, however, one can readily see that it is possible to maintain homogeneity and isotropy with non-Abelian global symmetry (i.e. having several Abelian $U(1)$ gauge fields rotating among each other by a global symmetry $G$, which has an $SO(3)$ subgroup). Such Abelian gauge field configurations in the context of cosmology has been called ``cosmic triads'' \cite{Triads,vector-inflation,Ford:1989me}.

In the above, and hereafter, we will only consider $SU(2)$ gauge theory, however, our arguments can be directly generalized to an $SU(2)$ subgroup of a generic non-Abelian gauge group $G$. With the above choice \eqref{gauge-ansatz-summary} all gauge freedom, local and global, has been employed. Note also that, although one can always set $A_0=0$ by a gauge choice, taking $A_i^a$ as we have, is our choice, ansatz, for the gauge field. That is, we have chosen to work with a very particular configuration of the gauge fields. We hence need to study two issues: 1) The choice \eqref{gauge-ansatz-summary} is compatible with the equations of motion of the action \eqref{gauge-flation-generic-action} with an FLRW ansatz for the metric \cite{arXiv:1102.1513,arXiv:1102.1932} and, 2) this choice is stable under both classical \cite{Maleknejad:2011jr} and/or quantum fluctuations or perturbations. In what follows we will study these two questions. The first question will be dealt with in \ref{consistency-of-reduction-section} and the second in \ref{stability-section} and in \ref{gauge-flation-cosmic-perturbation-theory-section}.

\subsubsection*{Consistency of reduction}\label{consistency-of-reduction-section}

Here we show that if we start with a gauge field of the form \eqref{gauge-ansatz-summary} at, say $t=t_i$, the dynamics of the system does not take us out of this ``homogeneous and isotropic sector'' (where the gauge field is of the form \eqref{gauge-ansatz-summary} and metric is FLRW). To this end we should analyze the equations of motion of the theory:
\bse
\begin{align}
G_{\mu\nu}=T_{\mu\nu}&=-\frac{2}{\sqrt{-g}}\frac{\delta (\sqrt{-g}\cL_G)}{\delta g^{\mu\nu}}\label{e.o.m-metric}\\
D_\mu\frac{\delta \cL_G}{\delta F_{\mu\nu}}&=0 \label{e.o.m-gauge-field}\,,
\end{align}
\ese
where $D_\mu$ is the covariant derivative and is defined as
$$
D_\mu X^A=\nabla_\mu X^A-gf^{ABC} A^B X^C\,,
$$
for a generic tensor $X^A$ in the adjoint representation of the gauge group $G$ and $\nabla_\mu$ is the diffeomorphism-covariant derivative computed for the FRLW metric \eqref{FLRW}. Restricting to the $G=SU(2)$ case,
\be\label{gauge-field-e.o.m-SU(2)}
(D_\mu \frac{\delta \cL_G}{\delta F_{\mu\nu}})^a=\nabla_\mu \frac{\delta \cL_G}{\delta F^a_{\mu\nu}}-g\epsilon^{abc} A^b \frac{\delta \cL_G}{\delta F^c_{\mu\nu}}\,.
\ee

We are now ready to insert the ansatz \eqref{gauge-ansatz-summary} into the above equations. We first note that
\be\label{F-components}
F_{0i}^a=a(t)(\dot\psi+H\psi)\delta_i^a\,,\qquad F_{ij}^a=-ga(t)^2 \psi^2\epsilon^a_{\ ij}\,,
\ee
where dot is derivative w.r.t. comoving time, and that
\be\label{T-mumnu-generic}
T_{\mu\nu}=2\frac{\delta\mathcal{L}_G}{\delta
F_{~\sigma}^{a~\mu}}F^a_{~\sigma\nu}+g_{\mu\nu}\mathcal{L}_G\,,
\ee
in the above $\cL_G=\cL_G(\Fmn, \gmn)$ (the minimal coupling condition) has been used.\footnote{For $\cL_G$ which is a polynomial in $\Fmn$ one can show that the first term in \eqref{T-mumnu-generic} is symmetric in $\mu$,$\nu$ indices.}
Inserting \eqref{F-components} into \eqref{T-mumnu-generic} one finds
\be
T^{\mu}_{\,\,\nu}=diag(-\rho,P,P,P)\,,
\ee
where
\be\label{rho-P}
\rho=\frac{\partial\mathcal{L}_{red.}}{\partial\dot{\phi}}\dot{\phi}-\mathcal{L}_{red.}\,,\qquad
P = \frac{\partial(a^3 {\cal L}_{red.})}{\partial a^3}\,,
\ee
and
\be\label{phi-psi}%
\phi\equiv a(t)\psi(t)\,.
\ee
(Note that $\phi$, unlike $\psi$, is not a scalar.) ${\cal L}_{red.}$ is the \emph{reduced Lagrangian} density, which is obtained from calculating ${\cal L}(F_{\mu\nu}^a; g_{\mu\nu})$ for field strengths $F_{\mu\nu}^a$ given in \eqref{F-components} and FLRW metric \eqref{FLRW} \cite{arXiv:1102.1932}. Since the energy momentum tensor is of the form of a homogeneous-isotropic perfect fluid, the metric ansatz will remain of the FLRW form when we evolve in time using
\eqref{e.o.m-metric}.

To show that the gauge field ansatz \eqref{gauge-ansatz-summary} maintains its form in time we should consider \eqref{e.o.m-gauge-field}. Plugging the ansatz into the equation of motion \eqref{gauge-field-e.o.m-SU(2)} one can readily see that the $\nu=0$ components of the equation of motion are trivially satisfied and the $\nu=i$ components
become proportional to $\delta_i^a$, leading to the single equation
\be\label{red-e.o.m}%
\frac{d}{a^3 dt}(a^3\frac{\partial {\cal L}_{red.}}{\partial
\dot\phi})-\frac{\partial {\cal L}_{red.}}{\partial \phi}=0 \,,
\ee%
where $\mathcal{L}_{red.}(\dot{\phi},\phi;a(t))$ is the reduced Lagrangian introduced above.

Technically, we have shown that there exists a consistent truncation/reduction of the gauge field theory to the  homogeneous-isotropic sector specified by the scalar field $\psi$ (or $\phi$). Moreover, the Lagrangian governing the dynamics in this sector is simply obtained from computing the complete Lagrangian over the reduction ansatz.

\subsection{Choosing the gauge theory action, the gauge-flation model}

As discussed regardless of the form of the action $\cL_G$ one can always reduce the theory to a isotropic-homogeneous sector. The question we would like to tackle now is to find the appropriate action(s) which can lead to successful slow-roll inflation, and for this purpose one may only focus on the reduced Lagrangian $\cL_{red}$.
To gain some intuition about the form of reduced Lagrangian it is useful to work out explicit form of some terms appearing in the Lagrangian. The space-time indices may be summed over using metric, e.g.
\be\label{F2-metric}
\Fmn^aF^{{\mu\nu}\ b}=2\delta^{ab}[-(\dot\psi+H\psi)^2+g^2\psi^4]\,,
\ee
or using the Levi-Civita tensor
\be\label{F2-epsilon}
\epsilon^{\alpha\beta\mu\nu} \Fmn^aF_{\alpha\beta}^b=-8g\delta^{ab}(\dot\psi+H\psi)\psi^2\,.
\ee
As we see Lorentz invariance (summing over all free space-time indices) implies that  $\dot\psi$ is always accompanied by a factor of $H\psi$ and, if we use metric for summing over the indices $\dot\psi+H\psi$ term appears together with the same power of $\psi^2$. Therefore, the reduced Lagrangian is neither of form of a kinetic term plus a potential, it is nor of a simple K-inflation type where the action is only a function of $X=(\partial_\mu\psi)^2$ or powers of $\psi$; our gauge-flation model, even in the homogeneous-isotropic sector, is hence not a special K-inflation model.
Due to this particular features, one should work out the slow-roll conditions from the first principles, e.g.
to have accelerated expansion $\rho+3P<0$ and
\be\label{epsilon-eta-rho-P}
\epsilon\equiv -\frac{\dot H}{H^2}=\frac32\frac{\rho+P}{\rho}\,,\qquad \eta=\epsilon-\frac{\dot\epsilon}{2H\epsilon}\,.
\ee

As a warmup example, let us consider the Yang-Mills action, for which
\be\label{YM-reduced-rho-P}
\begin{split}
\cL_{red-YM}=-\frac12 Tr(\Fmn F^{\mu\nu})&=\frac32[(\dot\psi+H\psi)^2-g^2\psi^4]\,\\
\rho_{YM}=\frac32[(\dot\psi+H\psi)^2+g^2\psi^4]\,&,\quad P_{YM}=\frac13\rho_{YM}=\frac12[(\dot\psi+H\psi)^2+g^2\psi^4]\,.
\end{split}
\ee
As expected for the Yang-Mills theory $T_{\mu\nu}$ is traceless, yielding $\rho=3P$. For this case, $\rho+3P=2\rho>0$
and the theory does not lead to accelerated expansion. Noting the features discussed above, it turns out that satisfying
$\rho+3P<0$ and $\rho>0$ conditions is not trivially obtained for actions which only functions of $\Fmn^aF^{{\mu\nu}\ b}$
(such terms in the context of cosmology has been considered e.g. in
\cite{homo-iso-gauge-config-1,homo-iso-gauge-config-2,Galtsov-inflation}) and one may consider terms involving $\epsilon^{\alpha\beta\mu\nu}$. The simplest choice which is gauge invariant and is not a total derivative is the term\footnote{Note that $F\wedge F$ is a total
derivative and a topological term, i.e. it does not contribute to equations of motion and the energy-momentum tensor.}
$$ (F\wedge F)^2\equiv \frac14(\epsilon^{\alpha\beta\mu\nu} \Fmn^aF_{\alpha\beta}^a)^2.$$
The dependence of this metric on metric $\gmn$ is only through $\det\ g$ and is of the form $1/\sqrt{-\det\ g}$. Therefore, the contribution to the energy momentum tensor from the above $F^4$ term will have $P=-\rho$, making it particularly interesting for inflationary model building.
We therefore choose the gauge-flation action to be of the form \cite{arXiv:1102.1513,{arXiv:1102.1932}}
\be\label{The-model}%
S=\int
d^4x\sqrt{-{g}}\left[\frac{R}{2}-\frac{1}{4}F^a_{~\mu\nu}F_a^{~\mu\nu}+\frac{\kappa}{384
}(\epsilon^{\mu\nu\lambda\sigma}F^a_{~\mu\nu}F^a_{~\lambda\sigma})^2\right]\,,
\ee
with the reduced gauge field Lagrangian
\be\label{L-reduced-gauge-flation}
\cL_{red}=\frac32[(\dot\psi+H\psi)^2-g^2\psi^4+\kappa g^2(\dot\psi+H\psi)^2\psi^4]\,,
\ee
and
\be\label{rho-P-YM-kappa}
\rho=\rho_{YM}+\rho_\kappa\,,\qquad P=\frac13\rho_{YM}-\rho_\kappa\,,
\ee
where $\rho_{YM}$ is given in \eqref{YM-reduced-rho-P} and
\be\label{rho-kappa}
\rho_\kappa=\frac32\kappa g^2 (\dot\psi+H\psi)^2\psi^4\,.
\ee
In the above action $\kappa$ is a dimensionful parameter, of mass dimension $-4$. The Lagrangian \eqref{The-model} has then two parameters, a dimensionless Yang-Mills coupling $g$ and a dimensionful coupling $\kappa$. We also note that $\kappa$ always appears in combination $\kappa g^2$. Demanding  for any field configuration the energy density is positive definite (weak energy condition) we take $\kappa>0$.

Before entering into the details of inflationary trajectory analysis we would like to make some comments about the above choice of the gauge theory action. One may ask if it is possible to write the reduced Lagrangian in as a diff-invariant Lagrangian for a scalar field $\psi(t,x_i)$, such that when we restrict to $\psi=\psi(t)$ we obtain $\cL_{red}$. Let us consider the Yang-Mills term first:
\be
S_{red-YM}=\frac32\int dt a^3 [(\dot\psi+H\psi)^2-g^2\psi^4]=\frac32\int dt a^3[\dot\psi^2-g^2\psi^4-(2H^2+\dot H)\psi^2]+\frac32\int{d}(a^3H\psi^2)\,.
\ee
Recalling that the Ricci scalar $R$ for the FLRW geometry is $R=6(2H^2+\dot H)$, dropping the total derivative term
the above action may be written in a diff-invariant form
\be
S^{cov}_{YM-red}=\frac32\int d^4x [-(\partial_\mu\psi)^2-\frac16R H^2-g^2\psi^4]\,,
\ee
which is a $\lambda\phi^4$ theory with coupling $g^2$ and a conformal mass term. This is expected, because the Yang-Mills theory is at classical tree level a scale invariant theory. Finding the diff-invariant form of $\kappa$-term is more involved:
\be
S_{red-\kappa}=\frac32\kappa g^2 \int dt a^3(\dot\psi+H\psi)^2\psi^4=\frac32\kappa g^2\int dt\bigl( a^3[\dot\psi^2\psi^4+\psi^6-\frac16\psi^6\frac{{(Ha^3)}^{\cdot}}{a^3}]+tot.der.\bigr)
\ee
The first two terms have an obvious diff-invariant extension of the form $(\partial_\mu\psi)^2\psi^4$ and $\psi^6$ form, the last term which is proportional to $3H^2+\dot H$, does not have a simple form in terms of curvature invariants.\footnote{For example, one may rewrite $3H^2+\dot H=\frac16 R+\frac12 R^0_0$. To contract the $R^0_0$ indices we need to have square of time-derivative of $\psi$, which we do not. The closest diff-invariant action that one can write will be something of the form $\frac{1}{24}\psi^6 R$ which misses terms proportional to $\dot H$.} Although it is possible to write the reduced gauge-flation action in a diff-invariant form for scalar field $\psi$, this action is not in a minimally coupled form (e.g. note the conformal mass term), and the $\kappa$-term takes a
contrived, practically useless form. We should finally point out that the equivalence of the diff-invariant reduced action and gauge-flation action \eqref{The-model} on the homogeneous-isotropic trajectories is only true at classical level; quantum mechanically and once we consider quantum fluctuations this equivalence does not persist. Taking note of this point will be crucial in studying gauge-flation cosmic perturbation theory, to which we will return in section \ref{gauge-flation-cosmic-perturbation-theory-section}.

\subsubsection{Theoretical motivation for the gauge-flation}\label{gauge-flation-theoretical-motivation}

As discussed the gauge-flation action \eqref{The-model}, and in particular the $\kappa$-term, was primarily chosen in search for a model with quasi-de Sitter expansion behavior, i.e. $P\simeq -\rho$. However, one may question the presence and naturalness of $\kappa$-term from the gauge theory viewpoint. This term is a specific $F^4$ term. $F^4$ and higher power of $F$ generically appear in the gauge theory (Wilsonian) effective action in one or higher loops and their effects in the cosmological contexts have been discussed in \cite{homo-iso-gauge-config-2,Galtsov-inflation,Odintsov}.\footnote{In our setup we only work with Einstein gravity. Extensions of our discussions to cases of modified gravity, e.g. $f(R)$ gravity, or when there are terms in the action of the form $R\ TrF^2$ has been discussed in \cite{Odintsov}.} These terms  typically appear in the gauge theory effective action below the charged fermion mass scale, once we integrated out massive fermions. Such terms for QED has been extensively discussed in the literature, in one and two loops levels. For example, for  QED below the electron mass scale, at one loop level the effective action is
\be\label{QED-one-loop-action}
\cL_{QED-one\ loop}=-\frac14 F^2+\alpha (F^2)^2+\beta F^4\,.
\ee
(At $F^4$ level there are only two ways to contract Lorentz indices of photon field strength field $\Fmn$ using only metric, $(F^2)^2$ and $F^4$. Therefore, by gauge and Lorentz invariance only these two terms are expected.) The coefficients $\alpha$ and $\beta$ may be computed using standard field theory techniques and their values for QED may be found in \cite{One-loop-QED}. What is important in our discussion here, is however, the parametric dependence of $\alpha, \beta$; $\alpha, \beta$ terms are one loop effects and hence should be proportional to $\alpha_{QED}^2=(\frac{e^2}{4\pi})^2$ where $e$ is the QED coupling. Moreover, they are of mass dimension $-4$ and hence they should both be proportional to $m_e^{-4}$, where $m_e$ is the electron mass scale. That is,
\be
\alpha\sim \beta\sim \frac{e^4}{(4\pi)^2 m_e^4}\,.
\ee
One would expect to be able to repeat the same argument for higher loops, to obtain $F^{2(l+1)}$-type terms at loop $l$ level with coefficient $\alpha_l$ where $\alpha_l\sim \alpha^l$. The above effective loop expansion is of course valid for small couplings and for the low energy photons ($E_{photon}\lesssim m_e$).

After the above brief review of gauge theory loop effects, let us return to the $\kappa$-term. Given its Lorentz index structure, this term cannot arise from fermion loops. Nonetheless, the axion-type coupling ${\chi} F\wedge F$ involves the right type of indices. Explicitly, let us consider a Yang-Mills theory coupled to massive axions \cite{Weinberg-QFT-II}:
\be\label{massive-axion-YM}
\cL_{axion-YM}=-\frac12 TrF^2-\frac12(\partial_\mu\chi)^2-\frac12 M^2\chi^2+\lambda \frac{\chi}{8f}Tr(\epsilon^{\alpha\beta\mu\nu}F_{\alpha\beta}\Fmn)
\ee
where $\chi$ is a massive pseudoscalar with mass $M$, $f$ is the axion scale\footnote{The axion field $\chi$ takes values in the  $[0,\pi f]$ range.}, taken to be much larger than $M$ and $\lambda$ is the dimensionless axion coupling. For energies below axion mass one may integrate out the axion, using standard field theory techniques (for a detailed discussion see \cite{gauge-flationVs.chormo-natural,chromo-natural-long}). This leads to and action of the form \eqref{The-model} with
\be\label{kappa-axion}
\kappa=\frac{3\lambda^2}{\mu^4}\,,\qquad \mu^4=f^2M^2\,,
\ee
where we have presented the above in terms of mass scale $\mu$ which is the cutoff of the axion theory.

Such ``massive axions'' are commonplace in beyond standard model and string theory motivated particle physics models and hence our gauge-flation action is well motivated with particle physics models. Of course there remains two important questions 1) whether one can ignore all the other loop effects while having a significant contribution from certain loops and, 2) whether the range or value of parameters or scales natural to these particle physics models allow for a successful model of inflation. Here we will discuss the former and will return to the latter in section \ref{Chromo-natural-section}. To make sure that at one loop level the $\kappa$-term dominates over the fermionic loop effects, $\alpha,\ \beta$ terms, it is enough to check or demand that $\kappa\gg \alpha, \beta$, i.e.
\be\label{loop-exapsion-validity-condition}
\left(\frac{\mu}{M_f}\right)^2\ll {\frac{\lambda}{g^2}}\,,
\ee
where $M_f$ is the typical fermionic (or generic charged matter) masses in the system.\footnote{The above equations are obtained for a standard quantum field theory loop analysis on flat background. For inflationary background,
one should note that the Compton wavelength of fermions $1/M_f$ which contribute to the loops should be larger that the Hubble horizon size $H^{-1}$.} This condition can be met for small enough gauge couplings. Moreover, small gauge coupling will also be needed to suppress the higher fermionic loop effects.

The condition \eqref{loop-exapsion-validity-condition} guarantees that the $\kappa$-term dominates over all the other
dimension eight or higher contributions coming from gauge field or fermionic loops. However, (slow-roll) inflation, as was implicit in our earlier discussions, can take place if the $\kappa$-term can dominate over the Yang-Mills term. The above discussion clarifies this issue too: The $\kappa$-term, unlike all the other gauge theory loop corrections, comes from integrating out an axion field, which is nothing but eliminating  the massive axion field by evaluating the action on its classical trajectory \cite{gauge-flationVs.chormo-natural}. Therefore, the $\kappa$-term is in fact representing the potential of the axion field which can naturally dominate over Yang-Mills term and this all can consistently happen within perturbation theory. More discussion on this will be presented in section \ref{chromo-natural-vs-gauge-flation-section}.

{For completeness of the discussions, we would like to also mention that it is possible to obtain gauge-flation action, naturally, within low energy string theory setup. Explicitly, as discussed in \cite{Martinec-1}, consider compactification of ten dimensional heterotic supergravity theory (which arises as the low energy effective theory of heterotic string theory) to four dimensions. Consistency of the theory implies presence of certain $\alpha'$ corrections, which include a term like the $\kappa$-term.}

\subsection{Gauge-flation inflationary trajectories, analytic treatment}

The equations of motion \eqref{e.o.m-metric} and \eqref{e.o.m-gauge-field}, for the flat FLRW metric and gauge field ansatz \eqref{gauge-ansatz-summary}, in terms of $\phi=a(t)\psi(t)$ field, takes the form
\bse
\begin{align}
\label{cosm1}
H^2=\frac{1}{2}(&\frac{\dot{\phi}^2}{a^2}+\frac{g^2\phi^4}{a^4}+\kappa \frac{g^2\phi^4\dot{\phi}^2}{a^6})
\,,\\ \label{cosm2}
\dot{H}=&-(\frac{\dot{\phi}^2}{a^2}+\frac{g^2\phi^4}{a^4})\,,\\
\label{phi-eom}
(1+\kappa\frac{g^2\phi^4}{a^4})\frac{\ddot{\phi}}{a}+&(1+\kappa\frac{\dot{\phi}^2}{a^2})\frac{2g^2\phi^3}{a^3}+
(1-3\kappa\frac{g^2\phi^4}{a^4})\frac{H\dot{\phi}}{a}=0\,.
\end{align}
\ese
We are interested in slow-roll dynamics specified by $\epsilon,\ \eta\ll 1$. Using the Friedmann equations \eqref{cosm1}, \eqref{cosm2} and  \eqref{epsilon-eta-rho-P} we have%
\be\label{epsilon-rho0-rho1}
\epsilon= \frac{2\rho_{_{YM}}}{\rho_{_{YM}}+\rho_\kappa}\,.
\ee%
That is, to have slow-roll the $\kappa$-term contribution
$\rho_\kappa$ should dominate over the Yang-Mills contributions
$\rho_{_{YM}}$, or $\rho_\kappa\gg \rho_{_{YM}}$ during slow-roll period. As we will see the
time evolution of the system increases $\rho_{_{YM}}$ with respect to
$\rho_\kappa$ pushing inflation to its end. The end of accelerated expansion is marked by $\epsilon=1$ which happens when $\rho_{_{YM}}=\rho_\kappa$.

For having slow-roll inflation, however, it is \emph{not} enough to make sure $\epsilon\ll 1$. For the latter, time-variations of $\epsilon$ and all the other physical dynamical variables of the problem, like $\eta$ and the $\psi$ field, must also remain small over a reasonably large period in time (to result in enough number of e-folds). As discussed in section \ref{Inflation-review-section} for simple slow-roll models, $\eta$ through equations of motion also measures rolling velocity of the inflaton \eqref{epsilon-eta-single-field}. In our case, however, this relation is modified. It is useful to define%
\be\label{delta-def}
\delta\equiv-\frac{\dot{\psi}}{H\psi}\,.
\ee%
$\delta$ is not an independent slow-roll parameter and is related to $\epsilon$ and $\eta$ through equations of motion:
\bse
\begin{align}
\label{epsil}
\epsilon=&2-\kappa g^2\psi^6(1-\delta)^2, \\ \label{tilde-eta}
\eta=&\epsilon-(2-\epsilon)\left[\frac{\dot{\delta}}{H(1-\delta)\epsilon}+\frac{3\delta}{\epsilon}\right]\,.
\end{align}
\ese
To have a ``standard slow-roll'', $\dot\epsilon\sim H\epsilon^2$ and $\eta\sim\epsilon$, we should demand that $\delta\sim \epsilon^2$. Explicitly, the equations of motion \eqref{cosm1}, \eqref{cosm2} and \eqref{phi-eom} admit the solution \footnote{Based on our numerical analysis, there exists a range of initial values and parameters for which  starting with $\dot\delta/(H\delta)\sim {\cal O}(1)$, while $\psi_i^2\sim \epsilon\ll 1$, after a very short time (less than an e-fold) $\delta$ becomes very small, of order $\epsilon^2$. That is, the dynamics of our system suppresses $\dot\psi$ very fast if $\epsilon, \eta\ll 1$. Therefore, for all the inflationary period we may confidently use $\delta\simeq \frac{\gamma}{6(\gamma+1)}\epsilon^2=(\epsilon-\eta)\epsilon/6$. See section \ref{gauge-flation-numerical-analysis-section} for a more detailed discussion.}%
\bse
\begin{align} \label{epsilon-x}
\epsilon&\simeq\psi^2(\gamma+1),\\
\label{eta-x}
\eta &\simeq\psi^2\quad \Rightarrow \quad (3+\frac{\dot\delta}{H\delta})\delta\simeq\frac{\gamma}{2(\gamma+1)}\epsilon^2\,,\\
\label{kappa-x}
\kappa g^2\psi^6&\simeq2-\epsilon\,, %
\end{align}\ese%
where $\simeq$ means equality to first order in slow-roll parameter $\epsilon$ and \footnote{Note that all the dimensionful parameters, \ie $H,\psi$ and $\kappa$, are measured in units of $\mpl$; $H,\ \psi$ have dimension of energy while $\kappa$ has dimension of one-over-energy density.}
\be\label{x-def}
\gamma= \frac{g^2\psi^2}{H^2}\,,\qquad \textrm{or
equivalently}\qquad
H^2\simeq\frac{g^2\psi^4}{\epsilon-\psi^2}=\frac{g^2\epsilon}{\gamma(\gamma+1)}\,.%
\ee%
In the above $\gamma$ is a positive parameter which is slowly varying during slow-roll inflation.

Recalling \eqref{delta-def} and that $\delta\sim \epsilon^2$, \eqref{x-def} implies that $\gamma H^2$ remains almost a constant during the slow-roll inflation and hence%
\be\label{epsilon-H}%
\frac{\epsilon}{\epsilon_i}\simeq\frac{\gamma+1}{\gamma_i+1}\,,\qquad \frac{\gamma}{\gamma_i}\simeq\frac{H_i^2}{H^2}\,,
\ee%
where $\epsilon_i,\ \gamma_i$ and $H_i$ are the values of these parameters at
the beginning of inflation.  As discussed the (slow-roll) inflation ends when $\epsilon=1$, where%
\be
\gamma_f\simeq \frac{\gamma_i+1}{\epsilon_i}\,,\qquad
\frac{H_f^2}{H_i^2}\simeq \frac{\gamma_i}{\gamma_i+1}\ \epsilon_i\,.
\ee
Using the above one can compute the number of e-folds $N_e$
\bea\label{Ne-gauge-flation}
N_{e}=\int_{t_i}^{t_f} Hdt=-\int_{H_i}^{H_f} \frac{dH}{\epsilon
H}\simeq \frac{\gamma_i+1}{2\epsilon_i}
\ln\frac{\gamma_i+1}{\gamma_i}\,.
\eea

Before moving to numerical analysis some comments are in order:
\begin{itemize}
\item $\eta\ll 1$ implies that our field values $\psi$ is sub-Planckian (as $\eta\simeq \psi^2$).
\item If $\gamma_i$ is a parameter with order 1-10 values, then \eqref{x-def} implies that $H^2\sim g^2\epsilon$.
\item Number of e-folds, as in simple single field models, is proportional to $1/\epsilon_i$.
\item Further analysis of the equations of motion, also confirmed by the numerical analysis, reveals regardless of initial values of $\delta$, for the region of parameter space which leads to long-enough slow-roll period, $\dot\delta/H\delta$ remains small and hence \eqref{eta-x} leads to
\be\label{psi-roaming}
\frac{\dot\psi}{\psi^5}\simeq \frac{g^2}{6}\frac{\dot H}{H^3}\,.
\ee
The above equation may be integrated   to obtain $\psi_i^{-4}-\psi_f^{-4}\simeq-\frac13\psi_i^{-4}$, i.e. $\psi^4_f=\frac34\psi_i^4$. That is $\psi$ field displacement during inflation is of order the field itself and is sub-Planckian, and this is independent of the initial value of $\gamma$ parameter.
\item The above conclusion (field roaming being of the same order of the field itself) may also be reached noting  \eqref{kappa-x} $\kappa g^2\psi^6(1-\delta)^2=2-\epsilon$.
During slow-roll we can drop $\delta$-term, then the ratio of $\psi_f$ (which is computed for $\epsilon_f=1$) to $\psi_i$ is obtained as $\psi_f^6\simeq \frac12\psi_i^6$. This estimation matches with the above up to percent level ($\sqrt{3/4}$ vs. $\sqrt[3]{1/2}$). This is a good evidence for the validity of slow-roll approximation for the whole period of inflation (till $\epsilon$ becomes one). In other words, the above indicates that end of inflation and exit from slow-roll happens very fast in the last couple of e-folds in the end of inflation.
This is the expectation which is confirmed by our numerical analysis of the next subsection.
\item The gauge-flation model has other trajectories which could be relevant to cosmology and are not necessarily within the slow-roll trajectories discussed here. An analytical and numerical discussion of these trajectories has been discussed in \cite{Ghalee}.
\end{itemize}

\subsection{Gauge-flation inflationary trajectories, numeric analysis}\label{gauge-flation-numerical-analysis-section}
\begin{figure}
\includegraphics[angle=0, width=80mm, height=70mm]{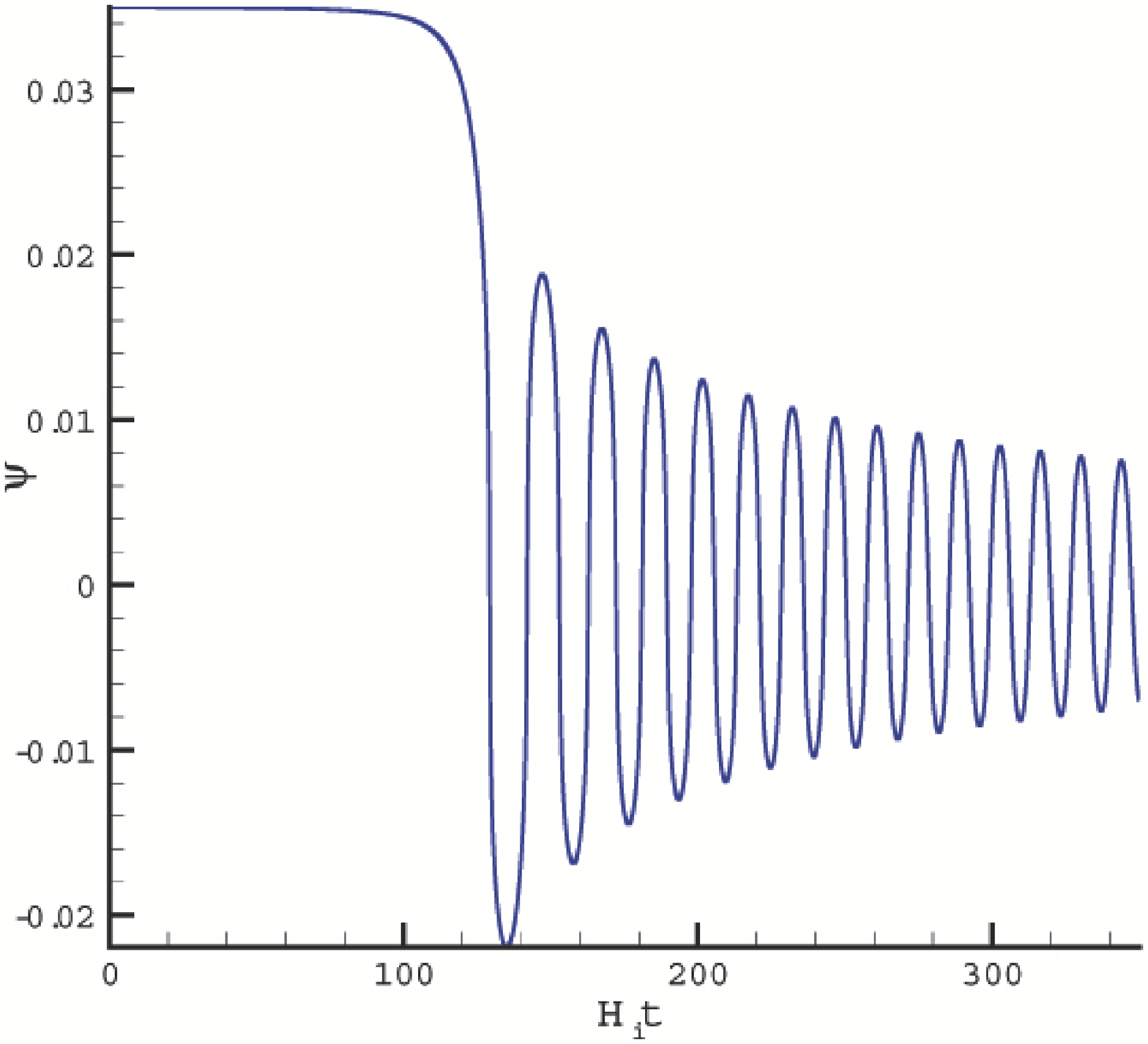}
\includegraphics[angle=0, width=80mm, height=70mm]{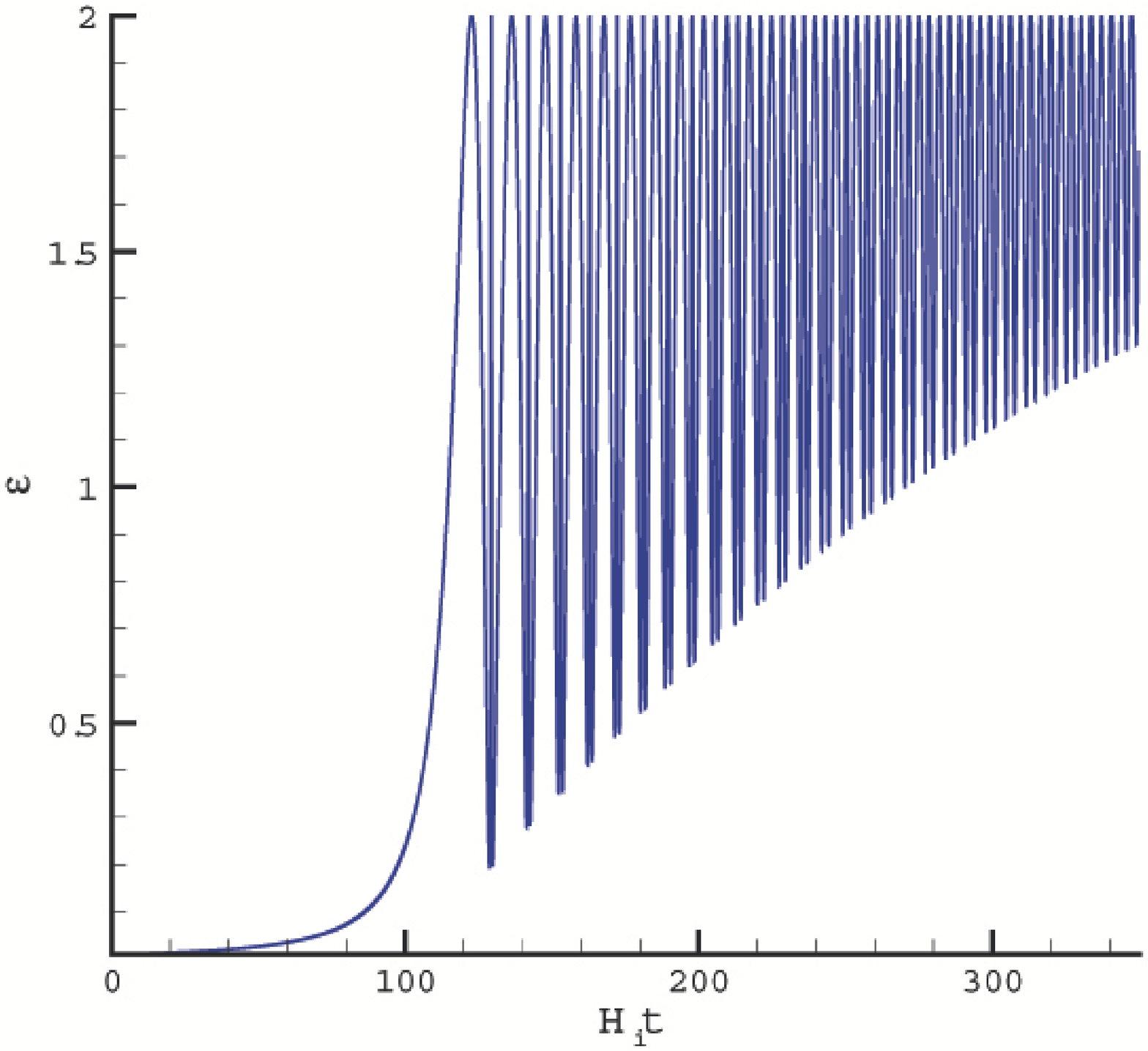}\\
\includegraphics[angle=0,width=80mm, height=70mm]{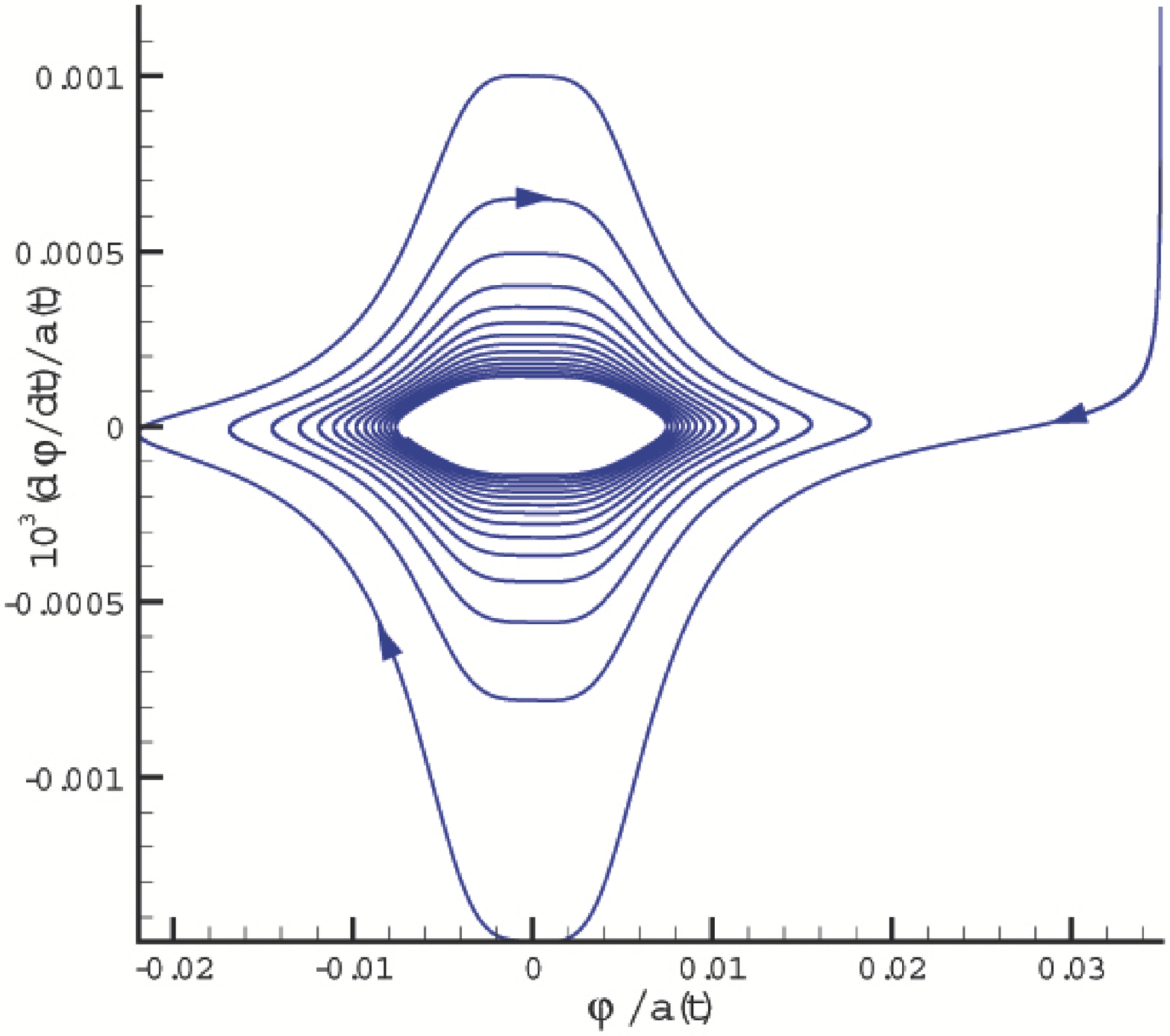}
\includegraphics[angle=0,width=80mm, height=70mm]{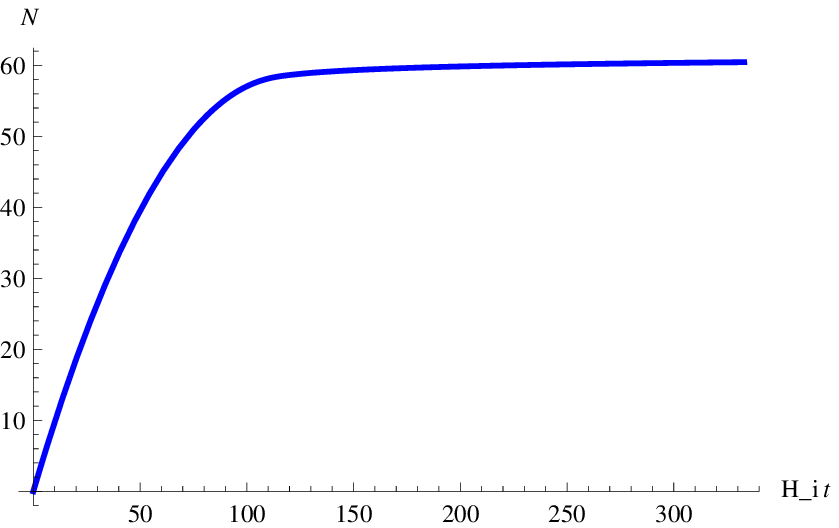}
\caption{The classical trajectory for ${\psi_i=0.035,\dot\psi_i=-10^{-10};\ g=2.5\times10^{-3}, \kappa=1.733\times 10^{14}}$. These values correspond to a slow-roll trajectory with $H_i=3.4\times 10^{-5},\ \gamma_i=6.62,\ \epsilon_i=9.3\times 10^{-3},\delta_i=8.4\times 10^{-5}$. These are the values very close to the range for which the gauge-flation is compatible with the current cosmological and CMB data (\emph{cf.} discussions of section \ref{testing-gauge-flation-section}). Note that $\kappa,\ H_i$ and $\psi_i$ are given in the units of $\mpl$.}\label{x=6.35-slow-roll-figures}
\end{figure}

In this part we make a more thorough parameter space analysis of the gauge-flation model through a numerical analysis.
Gauge-flation model has two parameters, the gauge coupling $g$ and the coefficient of the $F^4$ term $\kappa$, the former is dimensionless and the latter is of mass dimension $-4$. The degrees of freedom of the model in the homogeneous-isotropic sector are the scalar field $\psi$ and the scale factor $a(t)$. Inflationary trajectories are hence specified by four initial values of these parameters and their time derivatives. These were parameterized by $H_i,\ \psi_i$ and $\delta_i$ (or $\dot\psi_i$). (The initial value of the scale factor $a(t_i)$ is not a physical observable in flat FLRW model, as it can be absorbed into the redefinition of spatial coordinates $x_i$.) The Friedmann equations, however, provide some relations between these parameters; assuming slow-roll dynamics these relations are \eqref{epsilon-x}-\eqref{kappa-x}. As a result each inflationary trajectory may  be specified by the values of four parameters, $(\psi_i,\dot\psi_i;\ g, \kappa)$. In what follows we explore the parameter space of our model more thoroughly and present the results of the numerical analysis of the equations of motion \eqref{cosm1}, \eqref{cosm2} and \eqref{phi-eom}, for three typical sets of values for $(\psi_i,\dot\psi_i;\ g, \kappa)$.

\subsubsection{Generic features of the slow-roll trajectories}
Our numeric analysis shows that it is possible to get slow-roll trajectories with  enough  ($N_e\geq 60$) or even arbitrarily large $N_e$,  for a large region of the parameter space. The slow-roll region in the parameter space has the following features:

\begin{figure}
\includegraphics[angle=0, width=80mm, height=70mm]{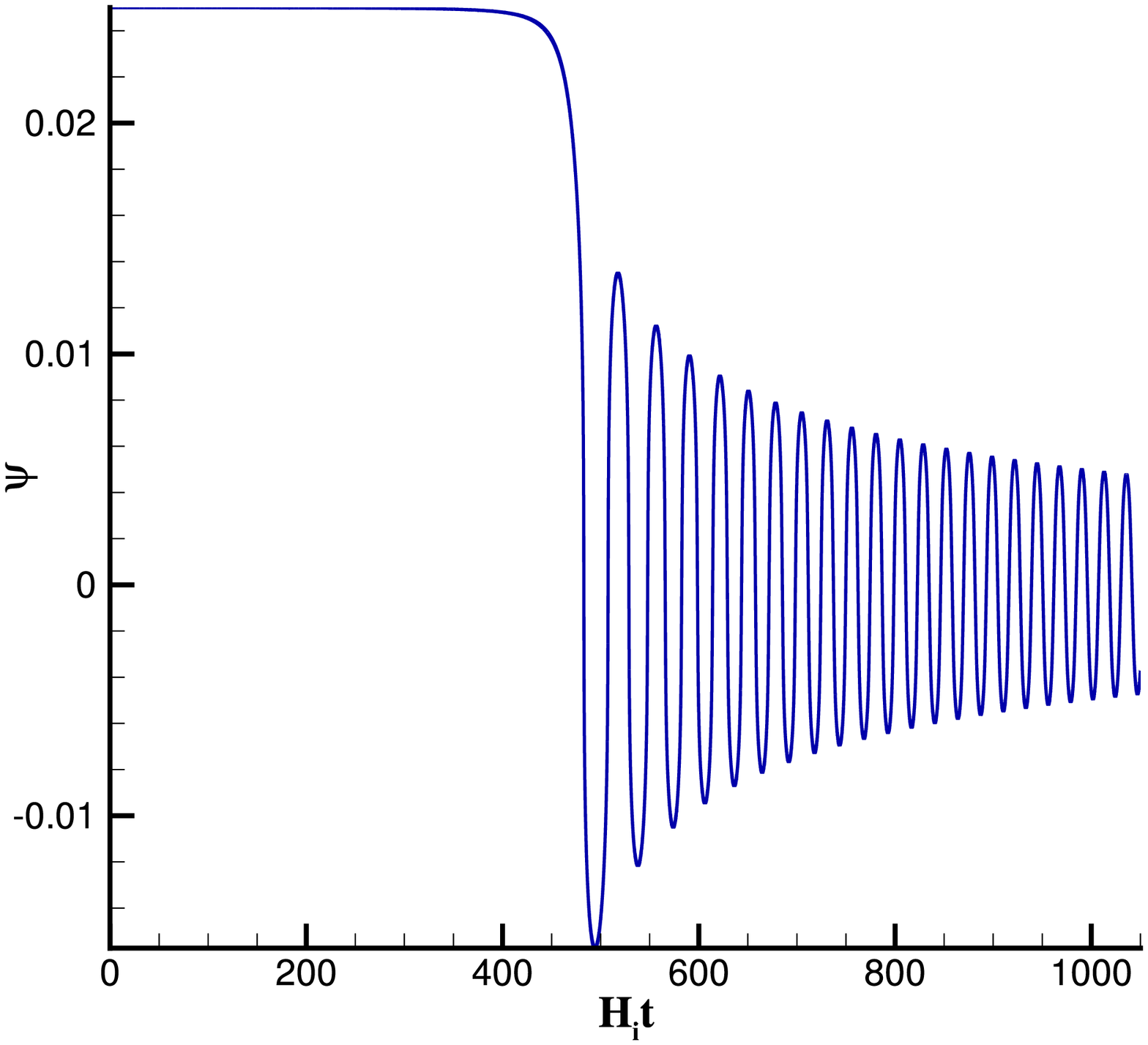}
\includegraphics[angle=0, width=80mm, height=70mm]{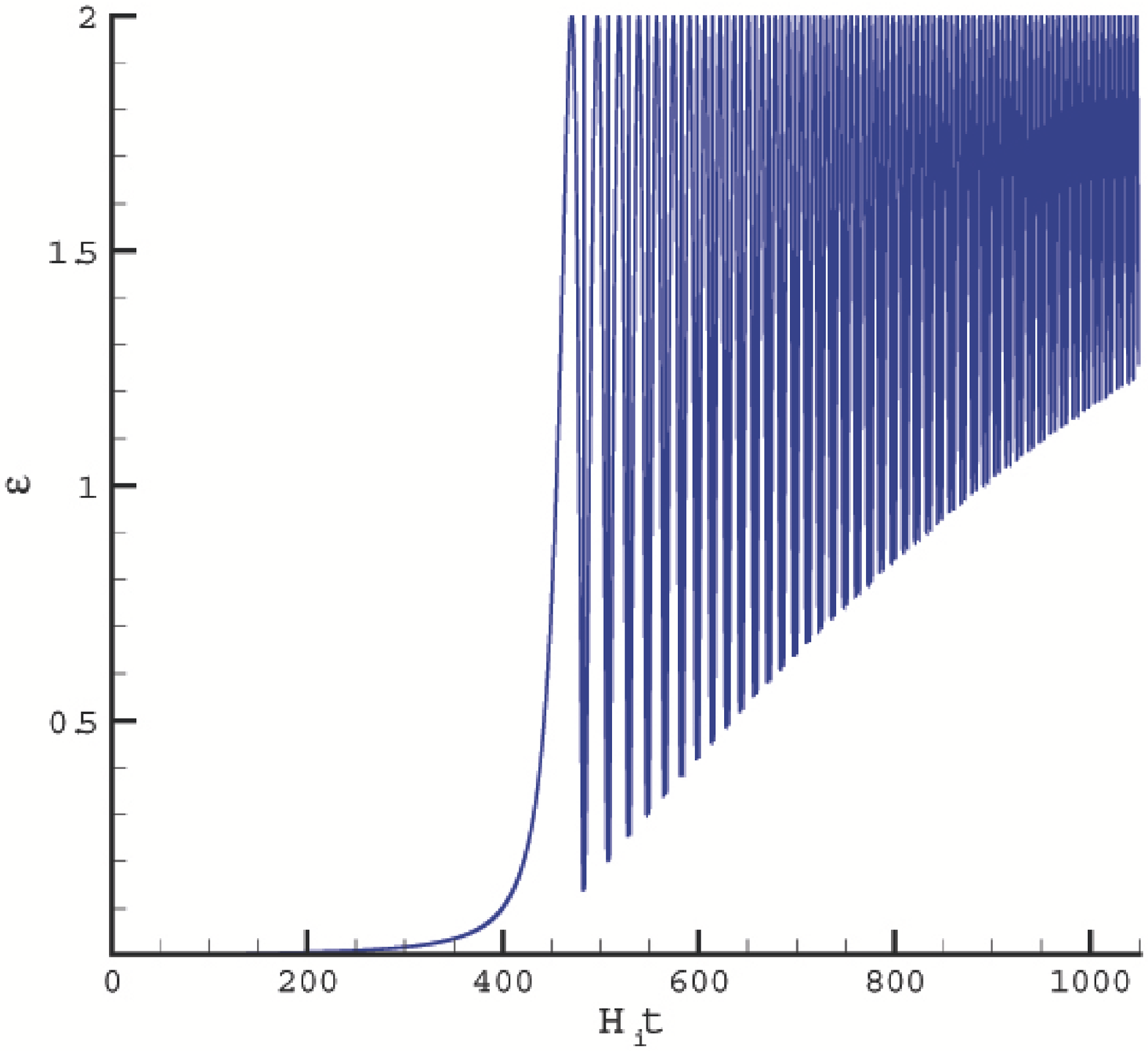}\\
\includegraphics[angle=0,width=80mm, height=70mm]{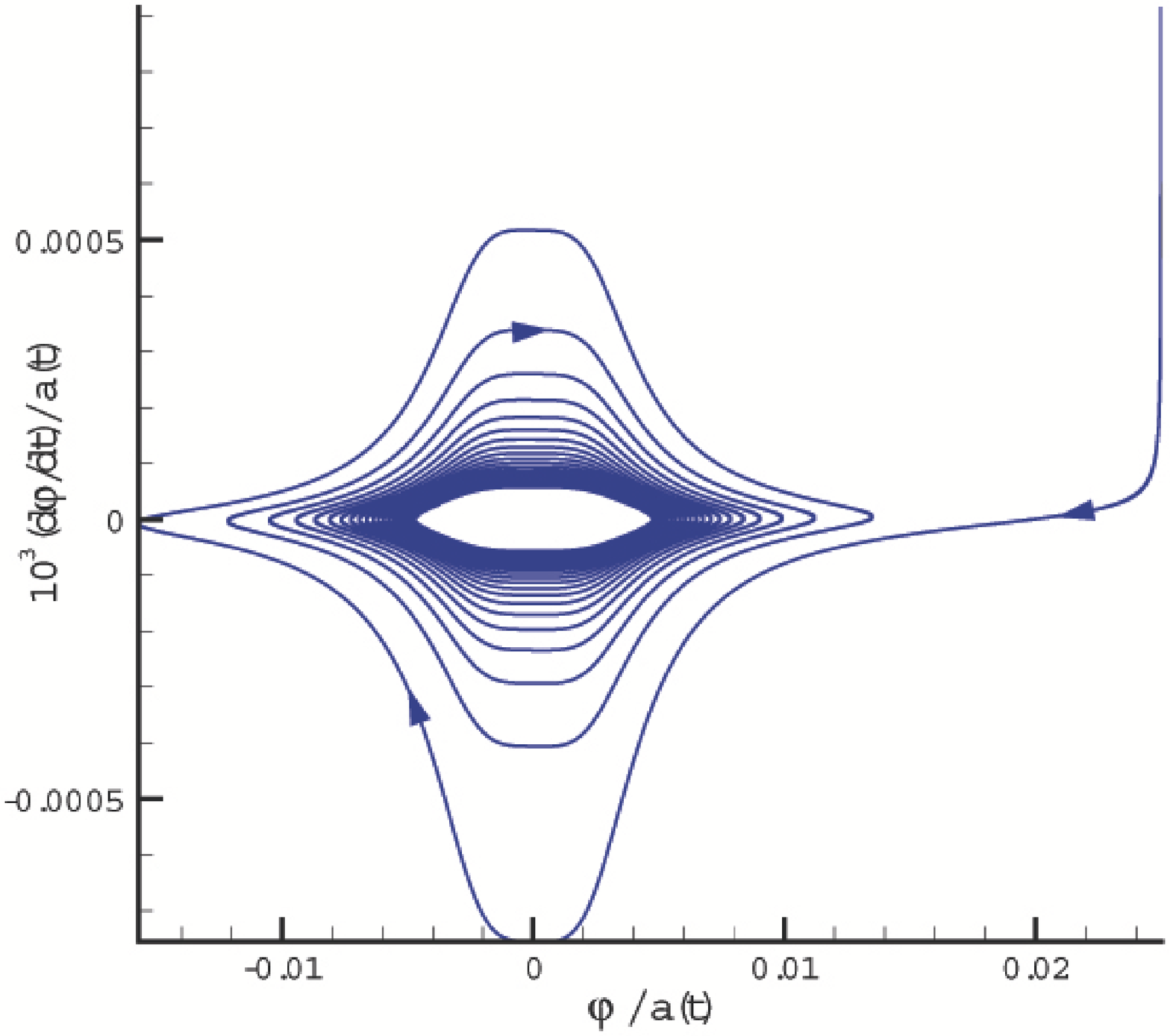}
\includegraphics[angle=0,width=80mm, height=70mm]{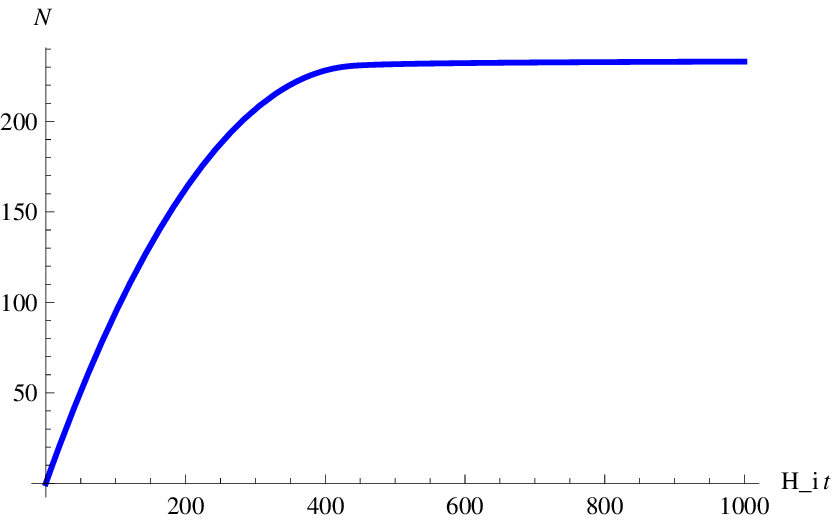}
\caption{The classical trajectory for ${\psi_i=0.025,\dot\psi_i=-10^{-10};\ g=2.507\times10^{-3}, \kappa=1.3\times 10^{15}}$. These values correspond to a \emph{slow-roll} trajectory with $H_i=3.63\times 10^{-5},\ \gamma_i=2.98,\ \epsilon_i=2.5\times 10^{-3},\ \delta_i=1.1\times 10^{-4}$.  These figures show that it is possible to get arbitrarily large numbers of e-folds within the \emph{slow-roll} phase of our gauge-flation model. 
}\label{N200-slow-roll-figures}\end{figure}

\begin{itemize}
\item Small gauge coupling $g$, $g\sim 10^{-3}-10^{-4}$. This latter is important because we are dealing with a non-Abelian gauge theory with a specific $F^4$ term, the $\kappa$-term. As was discussed in section \ref{gauge-flation-theoretical-motivation} smallness of gauge coupling $g$ is crucial for justifying the theoretical stability of gauge-flation action \eqref{The-model}.
\item Naturalness of $\kappa$. A typical value of $\kappa$ for slow-roll trajectories is $\kappa\sim 10^{13}-10^{14}$ (in Planck units). The $\kappa$ parameter may be related to the cutoff scale $\mu$ of an axion model \eqref{kappa-axion}, leading to $\mu\lesssim 10^{16}$GeV, which is a natural scale within grand unified theories.
\item Dependence on initial field values. It appears that with the above values for $g$ and $\kappa$, typical value for the field $\psi$ and its roaming is of order $10^{-2}-10^{-1}\mpl$, in accord with $\epsilon\sim \eta\sim 10^{-2}$.
\item  Slow-roll trajectories are not so sensitive to the initial value of $\dot\psi$ (or $\delta$ parameter). That is,  getting slow-roll inflation mainly depends on the values of $g,\kappa, \psi$ and to a less extent on $\delta$. This feature is also shared with the ordinary chaotic models. This latter will become more clear in the analysis of section \ref{chromo-natural-slow-roll-section}.
\end{itemize}

Moreover,  within the above range of parameter space, and regardless of their precise values, the slow-roll trajectories show the following generic features. These may also be seen from our figures and time evolution diagrams.
\begin{itemize}
\item If we start with a small $\delta$, $\delta\lesssim 10^{-2}$, the field $\psi$ remains almost constant during inflation and at the end of inflation $|\psi|$ falls off very fast and starts oscillating. However,
    if we start with a large $\delta$, $\delta\sim {\cal O}(1)$, $\psi$ has a very fast single oscillation, reducing $\delta$ to around $10^{-2}$ and then it follows a small-$\delta$ trajectory mentioned above.
\item After inflation ends the dominant term in dynamics of the system is the Yang-Mills term, the energy momentum of which behaves like a radiation gas.\footnote{Recall that for a radiation dominated Universe with $P=\rho/3$, $a(t)\sim t^{1/2},\ H=1/(2t)$ and $\epsilon=2$} Therefore, it is expected that: 1) the slow-roll $\epsilon$ parameter should  asymptote to $\epsilon=2$ (as is also indicated by \eqref{epsilon-rho0-rho1}) and, 2) the system should essentially behave as a $\lambda\phi^4$ theory. (This is expected as the Yang-Mills part of the reduced Lagrangian  behaves like a $g^2\psi^4$.) That is \cite{preheating-Linde}, $\psi$ and hence other dynamical variables, should oscillate (following a Jacobi-cosine function) with a period of the order $g\psi_f\sim H_i$ \eqref{L-reduced-gauge-flation}. Moreover, the amplitude of oscillations should decrease in time like $t^{-1/2}$.
\item The phase diagram of ``effective inflaton field $\psi$'' during slow-roll is hence like a straight line during slow-roll, while becomes very similar to that of a chaotic  $\lambda\phi^4$ model \cite{Inflation-Books}.
\end{itemize}

\subsubsection{More detailed discussion for various parameter choices}

Here we present evolution of the system for three different sets of the four parameter values
$(g, \kappa; \psi, \dot\psi)$. There are four graphs for each set. Three of them depict evolution of effective inflaton
field $\psi$, number of e-folds $N_e$ and slow-roll parameter $\epsilon$ vs. comoving time $t$ (in Hubble units
$H_i^{-1}$). The last one is the phase diagram of the effective inflaton field, for presentation purposes we have drawn
$\dot\phi/a(t)$ vs. $\phi/a(t)=\psi$. These numeric analysis match very well with our slow-roll analytical equations.

\begin{figure}
\includegraphics[angle=0, width=80mm, height=70mm]{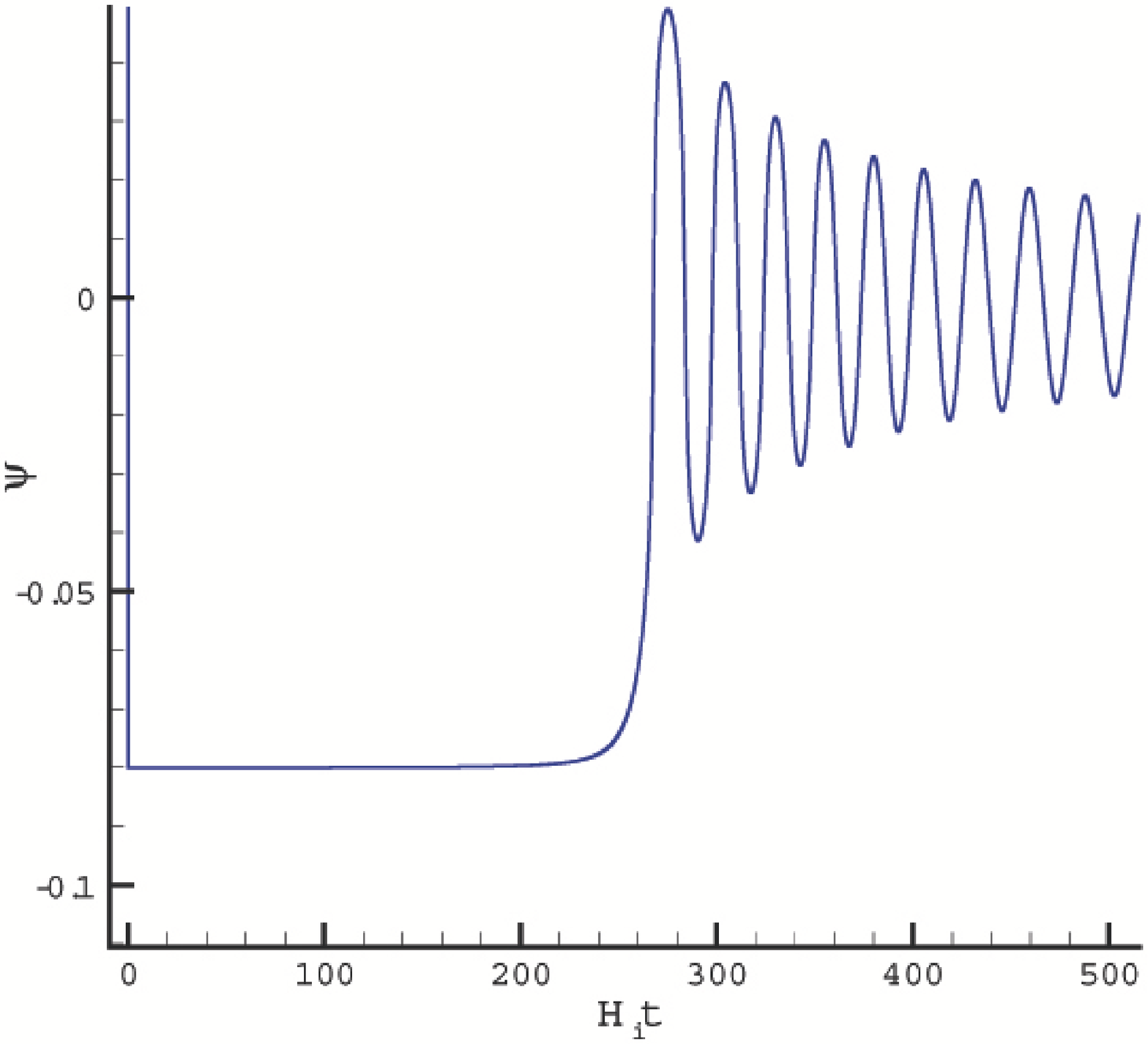}
\includegraphics[angle=0, width=80mm, height=70mm]{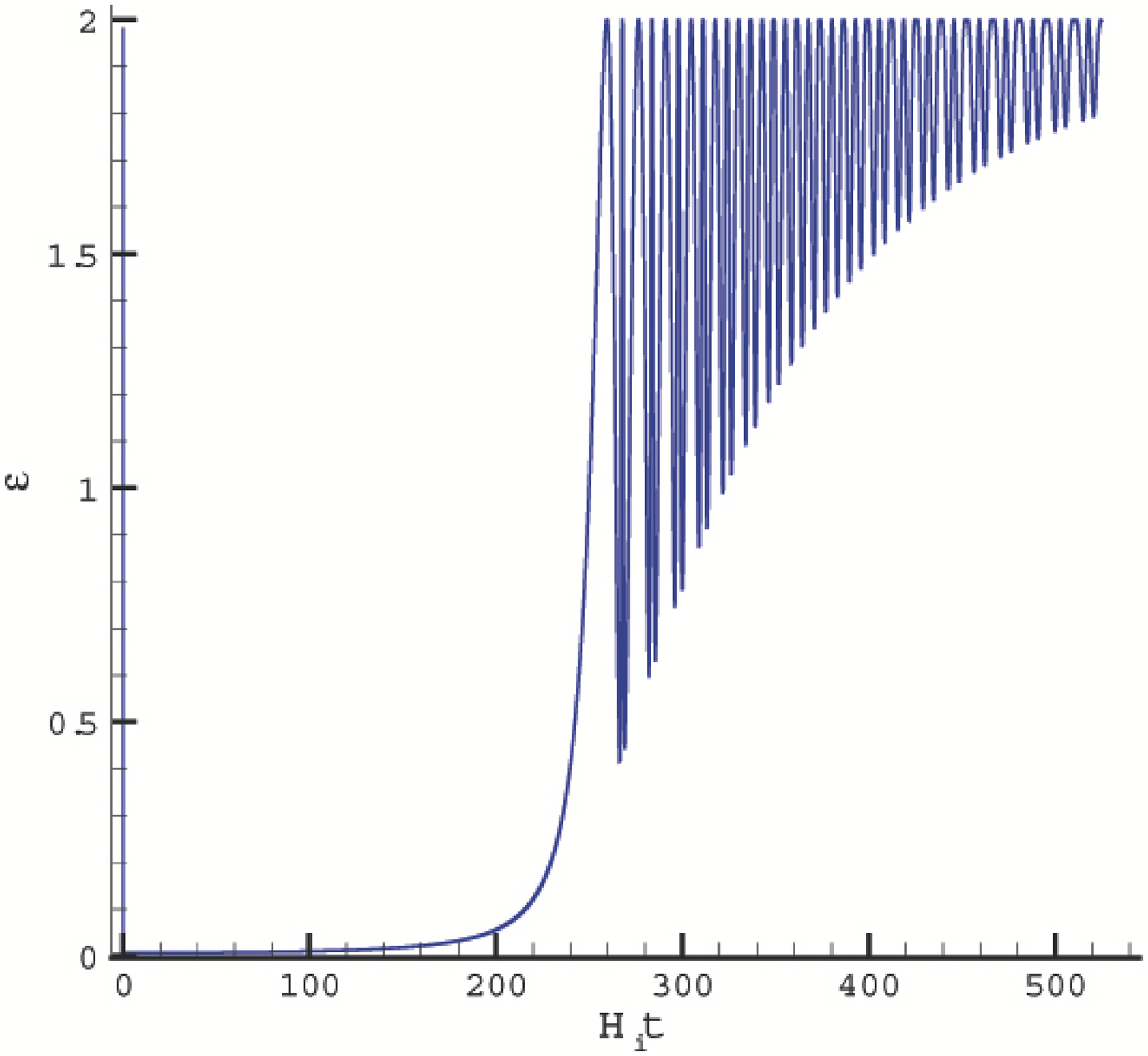}\\
\includegraphics[angle=0,width=80mm, height=70mm]{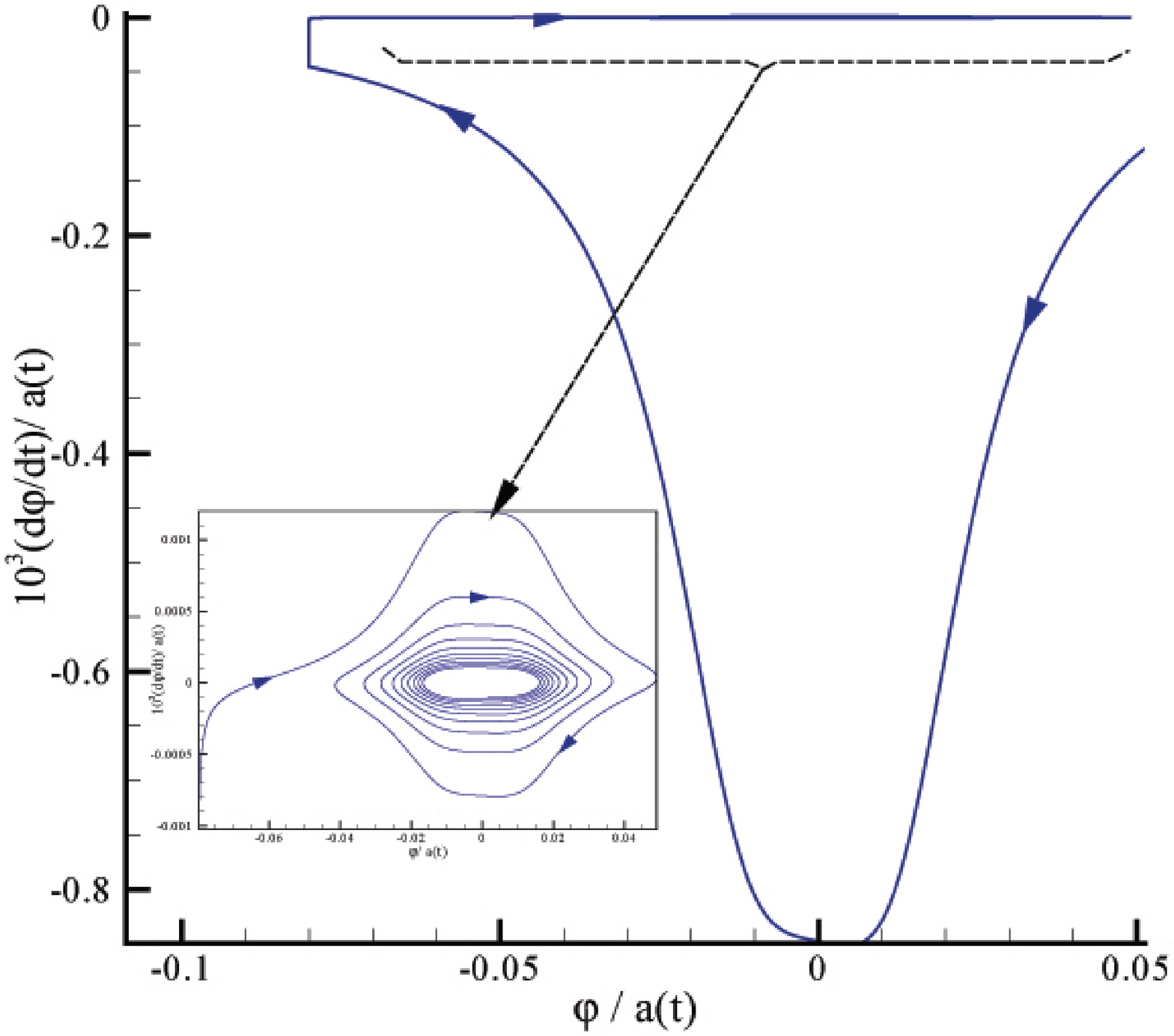}
\includegraphics[angle=0,width=80mm, height=70mm]{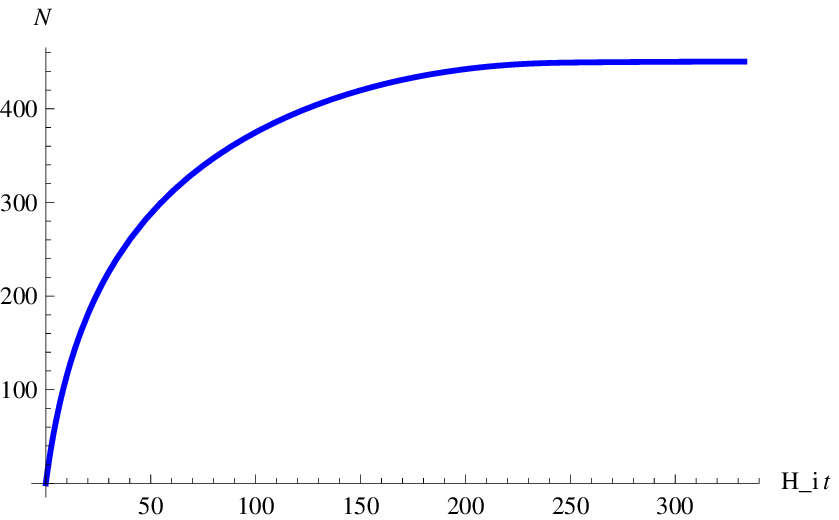}
\caption{The classical trajectory for ${\, g=4.004\times 10^{-4}\ , \kappa=4.73\times 10^{13}}\,;\
\psi_i=8.0\times 10^{-2} ,\dot\psi_i=-10^{-4}$. These values correspond to a \emph{non-slow-roll} trajectory with
$\delta\sim 2$, $H_i=6.25\times 10^{-4},\  \epsilon_i=6.4\times 10^{-3}$.
As is also seen from the phase diagram, the bottom-left figure, we start far from the slow-roll regime for which $\delta\sim \epsilon^2\ll 1$.
Despite of this, as we see from the top-left figure, after an abrupt oscillation the field $\psi$ loses its momentum and falls into the
standard slow-roll trajectory. As shown in the bottom-right figure, for this case we get a large number of e-folds which seems to be a fairly robust result not depending much on the initial value of $\delta$.
}\label{delta20-figures}\end{figure}

\subsection{Stability of the isotropic background}\label{stability-section}

We showed in  previous subsections that gauge-flation can lead to a successful isotropic and homogeneous inflationary dynamics. We demonstrated that in spite of having a non-zero gauge field value on the background, the isotropy of the background is preserved due to the time independent part of  gauge transformations (\emph{cf.} discussions of section \ref{gauge-flation-general-setup}).
However, one may ask if the isotropic gauge-flation model is a stable setup against the initial anisotropies.
In other words, due to the gauge-vector nature of inflaton field in
gauge-flation, a question that may arise naturally
is the classical stability of gauge-flation against the initial anisotropies and choice of initial conditions.
In this section, we will investigate this issue and in order to study the generality of the isotropic FLRW
gauge-flation, here, we will study gauge-flation in a homogeneous but anisotropic background. This subsection
is mainly a review of \cite{Maleknejad:2011jr}.
Here, for practical reasons we consider gauge-flation in an axially symmetric
Bianchi type-I setup but we argue that our results is expected to be valid for more general anisotropic cases.
({For a quick review on Bianchi cosmologies see \ref{Bianchi-appendix}.})
Similar question for a general inflationary setup was posed and analyzed in \cite{Wald-extended-theorem} and
reviewed in Appendix \ref{No-hair-extension-appendix}.

Consider Bianchi type-I axially symmetric metric  \eqref{axi-metric}
$$
ds^2=-dt^2+e^{2\alpha(t)}\big(e^{-4\sigma(t)}
dx^2+e^{2\sigma(t)}(dy^2+dz^2)\big),
$$
where $e^{\sigma(t)}$ represents the anisotropy and $e^{\alpha(t)}$ is the isotropic scale factor.
Due to the symmetries of the metric, as before, the temporal gauge is an appropriate choose for the gauge fields $A^a_{~0}=0$, and we consider the following modification to the isotropic ansatz \eqref{gauge-ansatz-summary}
\bea\label{axi-ansatz}
A^a_{~i}=e^a_{~i}\psi_i= diag(e^{\alpha(t)-2\sigma(t)}\psi_1,e^{\alpha(t)+\sigma(t)}\psi_2,e^{\alpha(t)+\sigma(t)}\psi_2),
\eea
where $\psi_i$ act as two scalar fields and $e^a_{~i}$ are the triads associated with spatial metric.
Without loss of generality and for some technical reasons which will be clear shortly, we parameterize $\psi_1$ and $\psi_2$ in terms of $\psi$ and $\lambda$ as
\be\label{psi-lambda}
\psi_1\equiv\frac{\psi}{\lambda^2},\qquad \psi_2=\psi_3\equiv\lambda\psi.
\ee
Furthermore, as in the isotropic case \eqref{phi-psi}, the equations take a simpler form once written in terms of $\phi$ and $a$
\bea\label{field}
\phi(t)=a(t)\psi(t),\quad a(t)\equiv e^\alpha.
\eea

Plugging the ansatz \eqref{axi-ansatz} and the metric \eqref{axi-metric} into $T_{\mu\nu}$ for the gauge fields, we obtain a diagonal homogenous tensor
\be\label{axi-Tmunu}
T^{\nu}_{~\mu}=\textmd{diag}(-\rho,P-2\tilde P,P+\tilde P,P+\tilde P).
\ee
 Following the decomposition \eqref{rho-P-YM-kappa}, we can write the energy density $\rho$ and the (isotropic) pressure density $P$ as
\be
\rho=\rho_\kappa+\rho_{_{YM}}\quad \textmd{and}\quad P=-\rho_\kappa+\frac13\rho_{_{YM}},
\ee
where as before, $\rho_\kappa$ and $\rho_{_{YM}}$ are respectively the contributions of $\kappa$ and Yang-Mills terms \bea\label{k-rho}
\rho_\kappa&=&\frac{3}{2}\frac{\kappa g^2\phi^4}{a^4}\frac{\dot{\phi}^2}{a^2},\\
\label{ym-rho}
\rho_{_{YM}}&=&\frac{3}{2}\left(\frac{1}{3\lambda^4}\big(\frac{\dot\phi}{a}-2(\dot\sigma+\frac{\dot\lambda}{\lambda})\frac{\phi}{a}\big)^2
+\frac{2\lambda^2}{3}\big(\frac{\dot\phi}{a}+(\dot\sigma-\frac{\dot\lambda}{\lambda})\frac{\phi}{a}\big)^2
+\frac{(2+\lambda^6)}{3\lambda^2}\frac{g^2\phi^4}{a^4}\right)\,,
\eea
while $\tilde P$, which  represents the anisotropic part of the pressure density, is given by
\bea
\tilde P=\frac13(1-\lambda^6)\left(\frac{1}{\lambda^4}(\frac{\dot\phi}{a}-2(\dot\sigma+\frac{\dot\lambda}{\lambda})\frac{\phi}{a})^2-\frac{1}{\lambda^2}\frac{g^2\phi^4}{a^4}\right)
-\lambda^2(\frac{\dot\lambda}{\lambda}+\dot\sigma)\left(2\frac{\dot\phi}{a}-(\dot\sigma+\frac{\dot\lambda}{\lambda})\frac{\phi}{a}\right)\frac{\phi}{a}\,.
\eea
It is readily seen that in the isotropic case  $\lambda^2=1$ ($\dot{\sigma}=0$ and $\dot\lambda=0$), $\tilde P$ vanishes and $\rho_{_{YM}}$ and $\rho_{\kappa}$ reduce to their expected values  \eqref{YM-reduced-rho-P} and \eqref{rho-kappa}.
Note that $\rho_\kappa$ is only a function of $\phi$ and not $\lambda$ which makes parametrization \eqref{psi-lambda} perfect for studying the quasi-de Sitter solutions.

The independent gravitational field equations are
 \bea
\label{lambda}
&~&\dot{\alpha}^2-\dot{\sigma}^2=\frac{\rho}{3},\\
\label{dsigma}
&~&\ddot{\sigma}+3\dot{\alpha}\dot{\sigma}=\tilde P,\\
\label{xi} &~&\ddot{\alpha}+3\dot{\sigma}^2=-\frac{\rho+P}{2}=-\frac23\rho_{_{YM}}.
\eea
From \eqref{dsigma} we learn that the evolution of $\dot\sigma$ depends on the anisotropic part of the pressure $\tilde P$, i.e. $\tilde P$ is the source for the anisotropy $\dot\sigma$. Thus in the absence of $\tilde P$, the anisotropy $\dot\sigma$ is exponentially damped, with time scale $\dot\alpha^{-1}$.
Note that the independent Einstein equations are three independent equations for four unknowns, so we need an extra equation which can be obtained from the action.

After substituting the axi-symmetric Bianchi metric and the gauge field ansatz into the total Lagrangian \eqref{The-model}, the total reduced (effective) Lagrangian is obtained as
\bea\label{axiL}
\cL_{tot}&=&\biggl(-3\dot \alpha^2+\dot\sigma^2\big(3+\frac{(2+\lambda^6)}{\lambda^4}\frac{\phi^2}{a^2}\big)
+\dot\sigma\frac{\big(\lambda^{-4}(\lambda^6-1)\phi^2\dot{\big)}}{a^2}+\frac{(1+2\lambda^6)}{2\lambda^4}\frac{\dot\phi^2}{a^2}\nonumber\\
&+&2\frac{(\lambda^6-1)}{\lambda^4}\frac{\dot\lambda}{\lambda}\frac{\dot\phi\phi}{a^2}+\frac{(2+\lambda^6)}{\lambda^4}\frac{\dot\lambda^2}{\lambda^2}\frac{\phi^2}{a^2}
-N^2\frac{(2+\lambda^6)}{2\lambda^2}\frac{g^2\phi^4}{a^4}+
\frac{3}{2}\frac{\kappa g^2\phi^4}{a^4}\frac{\dot\phi^2}{a^2}
\biggr)\,,
\eea
where  dot denotes derivative with respect to the time coordinate $t$. As we see, $\sigma$ is a cyclic variable in the above action, therefore, it's conjugate momentum is a constant of motion which gives
\be\label{const1}
\dot\sigma=-\frac{\big(\lambda^{-4}(\lambda^6-1)\phi^2\dot{\big)}}{2a^2\big(3+\lambda^{-4}(2+\lambda^6)\frac{\phi^2}{a^2}\big)}\,.
\ee
Here we used this fact that the anisotropy $\dot\sigma$ should vanish for the isotropic gauge field, i.e. when $\lambda^2=1$.

\subsubsection{Analysis in quasi-de Sitter regime}

Up to this point $T_{\mu\nu}$, the reduced action and $\dot\sigma$ have been determined without imposing the slow-roll inflation condition, i.e. the $\kappa$-term dominance. Hereafter, we simplify and analyze the equations assuming quasi-de Sitter inflation, in the sense that the parameter $\epsilon$
$$\epsilon=-\frac{\ddot{\alpha}}{\dot\alpha^2} $$
is  small during  inflation. Instead of $\dot\sigma$, it is more useful to work with its Hubble-normalized quantity $\Sigma/H$, which is defined as
 \be
\frac{\Sigma}{H}\equiv\frac{\dot\sigma}{\dot\alpha}.
\ee
In fact, one should only impose the slow-roll condition on the isotropic sector of the dynamics and, as the equations imply, the rest of variables are \emph{not} enforced to be the slow-roll.

Combining  \eqref{lambda} and \eqref{xi} we obtain $\epsilon$ in terms of $\rho_\kappa$, $\rho_{_{YM}}$ and $\dot\sigma$:
\be\label{axi-epsilon}
\epsilon=\frac{2\rho_{_{YM}}+9\dot\sigma^2}{\rho_\kappa+\rho_{_{YM}}+3\dot\sigma^2}\,,
\ee
which demanding a very small $\epsilon$, implies that $\rho_{_{YM}}$ and $\dot\sigma^2$ should be much smaller than $\rho_{\kappa}$. Moreover, noting that $\tilde P$ is a term from the contribution of Yang-Mills part, and that the energy momentum tensor of the Yang-Mills terms $T^{^{YM}}_{~\mu\nu}$ satisfies dominant energy condition (\emph{cf.} Appendix \ref{No-hair-extension-appendix}),  we learn that the absolute value of all the elements of $T^{^{YM}}_{~\mu\nu}$ are less than its energy density $\rho_{\kappa}$. Thus, from \eqref{axi-Tmunu} we always have
\be
\tilde P\leq\frac13\rho_{YM},
\ee
which combining with \eqref{dsigma} and \eqref{axi-epsilon}, gives
\be
\frac{\Sigma}{H}\lesssim\epsilon\quad and\quad \epsilon\simeq\frac{2\rho_{_{YM}}}{\rho_\kappa},
\ee
where, as mentioned before, $\simeq$ means equality to the leading order of $\epsilon$.

Using the above relation, \eqref{k-rho} and \eqref{ym-rho}, and ignoring the $\Sigma$ terms in $\rho_{_{YM}}$ we find
\bea
\epsilon\simeq\frac{\phi^2}{3a^2\lambda^4}\bigg[(2+\lambda^6)(\lambda^2\gamma+\frac{2\dot\lambda^2}{\dot\alpha^2\lambda^2})-
\frac{4\dot\lambda}{\dot\alpha\lambda}(1+\lambda^6)+(1+2\lambda^6)\bigg]\,,
\eea
where $\gamma\equiv\frac{g^2\phi^2}{a^2\dot\alpha^2}$ (\emph{cf.} \eqref{x-def}).
As we will show through analytical calculations, $\frac{\dot\lambda}{\dot\alpha\lambda}$ is at most an order one quantity. Therefore, in order to have a successful quasi-de Sitter inflation ($\epsilon\ll1$), we should have
\be\label{phi-small}
\frac{\lambda\phi}{a}\ll1 \qquad  and \qquad  \frac{\phi}{a\lambda^2}\ll1\,.
\ee
In other words, recalling \eqref{axi-ansatz}, as in the isotropic inflation case,  our field values $\psi_i$'s should have (physically
reasonable) sub-Planckian values during the quasi-de Sitter inflation.

So far we have hence shown that the quasi-de Sitter inflation still means ``$\rho_\kappa$ dominance''. Precisely, we have shown that $\epsilon\simeq\frac{2\rho_{_{YM}}}{\rho_\kappa}\,\mathrm{and}\
\Sigma^2/H^2=\frac{\dot\sigma^2}{\dot\alpha^2}\lesssim \epsilon^2\,.$
To proceed further we need to analyze dynamical field equations. Using \eqref{lambda}, \eqref{axi-epsilon} and ignoring $\dot\sigma$ terms, we obtain the following relations similar to the isotropic case
\be\label{sl-BI}
3\dot\alpha^2 \simeq\rho_\kappa \,,\qquad
\dot\phi\simeq\dot\alpha\phi.
\ee
Then, neglecting $\dot\sigma$ terms, from \eqref{axiL}, we can deduce the field equation corresponding to $\lambda$ as
\bea\label{lambdaeq}
\bigg((2+\lambda^6)(\lambda\ddot\lambda+\dot\alpha\lambda\dot\lambda+2\frac{\dot\phi}{\phi}\lambda\dot\lambda)-6\dot\lambda^2\bigg)
\frac{\phi^2}{a^2}+\lambda^2(\lambda^6-1)(\frac{\phi\ddot\phi}{a^2}+\frac{\dot\alpha\phi\dot\phi}{a^2}+\frac{\lambda^2g^2\phi^4}{a^4})=0,\quad
\eea
which, using \eqref{sl-BI} and keeping the leading orders, can be simplified to
\be\label{eqlambda1}
(2+\lambda^6)(\lambda\ddot\lambda+3\dot\alpha\lambda\dot\lambda)-6\dot\lambda^2
+\lambda^2(\lambda^6-1)(2+\lambda^2\gamma)\dot\alpha^2\simeq0.\quad
\ee
One can see that $\lambda=0$ is a singular point of the above equation and that the dynamics do not mix $\lambda>0$ and $\lambda<0$ regions. That is, if $\lambda$ is initially positive (negative), it always remains positive (negative) during inflation.

Since \eqref{eqlambda1} is a nonlinear second order differential equation which has no explicit time dependence, its solution will be of the form $\dot\lambda=\dot\lambda(\lambda)$ and hence
\be
\ddot\lambda=\frac{d\dot\lambda}{d\lambda}\dot\lambda\,.
\ee
In terms of derivatives with respect to $dN=\dot\alpha dt $, and denoting $\frac{d}{dN}$ by a prime, we obtain
\be
\frac{d\lambda'}{d\lambda}\simeq-3+\frac{1}{(2+\lambda^6)}\big(\frac{6\lambda'}{\lambda}
-\frac{\lambda(\lambda^6-1)}{\lambda'}(2+\gamma\lambda^2)\big),
\ee
which implies that $\lambda'(\lambda)$ is an odd function of $\lambda$, $\lambda'(-\lambda)=-\lambda'(\lambda)$.
We note that $\lambda\to -\lambda$ (together with $\phi\to -\phi$) is in fact a symmetry of our theory; this symmetry is nothing but the charge conjugation symmetry of the original gauge theory we start with.

\begin{figure}[ht]
\includegraphics[angle=0, width=80mm, height=70mm]{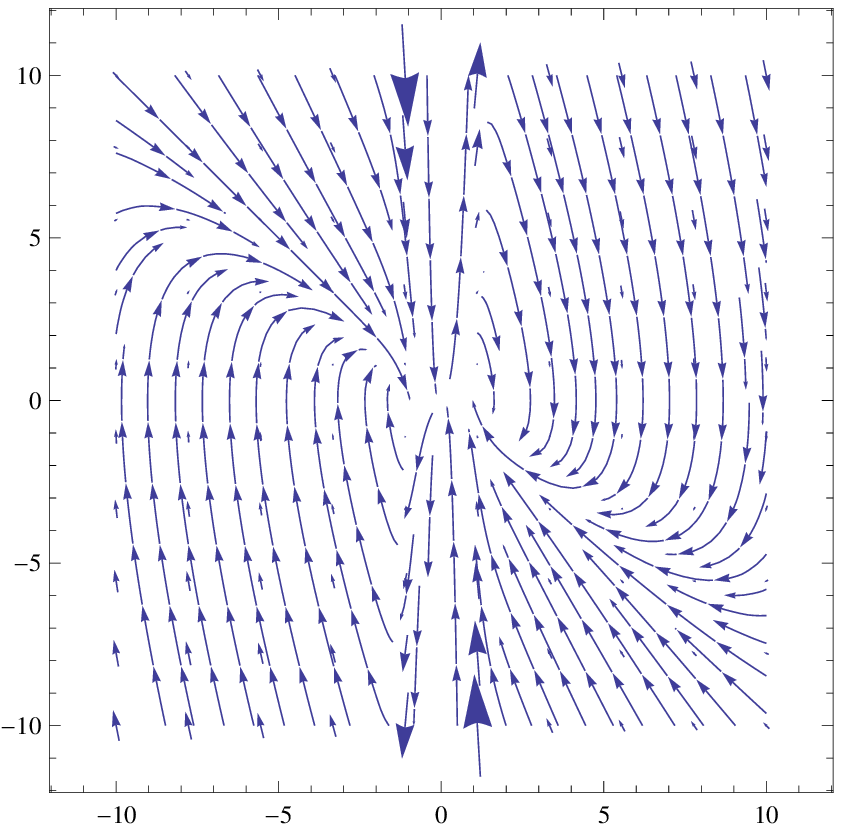}
\includegraphics[angle=0, width=80mm, height=70mm]{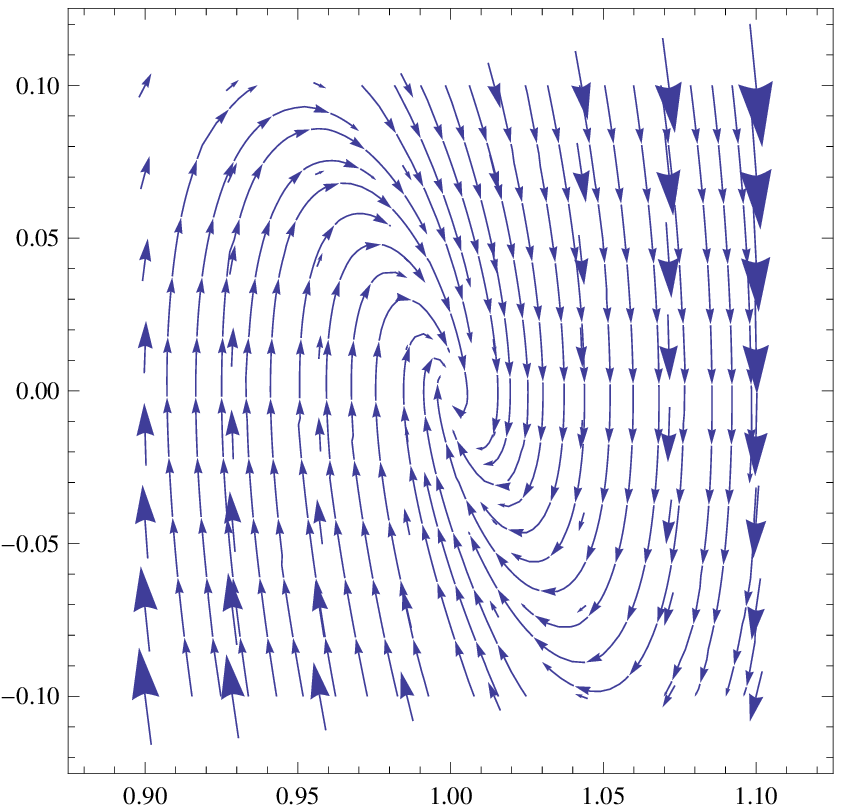}
\caption{The  phase diagram in the $\lambda'-\lambda$ plane, the vertical axis is $\lambda'$ and the horizontal is $\lambda$. Both figures show existence of attractor at $|\lambda|=1$, corresponding to the isotropic FLRW background. The left figure shows the phase diagram over a large range of values for $\lambda$, while  the right figure shows the phase diagram for $\lambda$ in the vicinity of the attractor solution $\lambda=1$. The left figure explicitly exhibits the $\lambda'(\lambda)=-\lambda'(-\lambda)$ symmetry.
}\label{phaseDiag}
\end{figure}

 Using \textsf{Mathematica}, the above equation can be studied by the phase diagram method, and in Figure \ref{phaseDiag}, we have presented the behavior of the solutions
in the $\lambda'-\lambda$ plane. Apparently, all of trajectories approach to the isotropic fixed point $\lambda^2 =1$. Next, we give the asymptotic analysis to confirm
that the isotropic inflation is an attractor in the phase space.

\subsubsection*{Asymptotic analysis}

In order to investigate the system analytically, it is convenient to rewrite  \eqref{eqlambda1} as
\be\label{eqlambda}
(2+\lambda^6)(\frac{\lambda''}{\lambda}+3\frac{\lambda'}{\lambda})-6\frac{\lambda'^2}{\lambda^2}
+(\lambda^6-1)(2+\lambda^2\gamma)\simeq0\,,
\ee
Moreover, using \eqref{sl-BI} in \eqref{const1}, we can write $\Sigma$ as
 \be\label{eqsigma}
 \frac{\Sigma}{H}\simeq-\frac13\big((\lambda^6+2)\frac{\lambda'}{\lambda}+(\lambda^6-1)\big)\frac{\phi^2}{a^2\lambda^4}\,.
 \ee
From \eqref{eqlambda}, one can distinguish three different regions for the value of $\lambda$:  $\lambda$ in the vicinity of one,  $\lambda$ close to zero and  large $\lambda$ values.
In order to have a better understanding of the system, we will solve \eqref{eqlambda} in these three  limits and can then determine $\Sigma$ using \eqref{eqsigma}.\\

I) If $\lambda$ is initially in the vicinity of one, we can approximate $\lambda$ as%
\be\label{dellambda}
\lambda=\pm1+\frac16\delta\lambda,\quad |\delta\lambda|\ll1\,,
\ee
and equation \eqref{eqlambda} as
$$
 \delta\lambda''+3\delta\lambda'+2(2+\gamma)\delta\lambda\simeq0\,.
$$
yielding
\be\label{dellamdasol}
\delta\lambda\simeq 6e^{-\frac32\dot\alpha t}\big(A_1\cos(\sqrt{\frac74+2\gamma}~\dot\alpha t)+A_2\sin(\sqrt{\frac74+2\gamma}~\dot\alpha t)\big).
\ee
As we see, the damping ratio is equal to $\frac{3}{2\sqrt{2(2+\gamma)}}$, which is less than one and hence $\delta\lambda$ is a damped oscillator, damped within one or two $e$-folds. Therefore, $\lambda=\pm 1+\delta\lambda$
exponentially approaches the attractors at $\lambda=\pm1$.

Inserting \eqref{dellambda} in \eqref{eqsigma}, we can determine $\Sigma$
\be
\frac{\Sigma}{H}\simeq-\frac{\phi^2}{3a^2}\big(\frac{1}{2}\frac{\delta\lambda'}{\delta\lambda}+1\big)\delta\lambda,
\ee
which implies that $\Sigma$ and $\delta\lambda$ have opposite signs. For $\lambda^2$  in the vicinity of one, $\dot\sigma$ has the following behavior
\be
|\frac{\dot\sigma}{\dot\sigma_0}|\leq e^{-\frac32\dot\alpha t},
\ee
here the subscript 0 denotes an initial value. Thus, $|\dot\sigma|$ is exponentially damped, with a time scale $2/(3\dot\alpha)$, as is also indicated by the cosmic no-hair theorems (see Appendix \ref{No-hair-extension-appendix}).\\

II) In the limit of very small $\lambda$ values ($|\lambda|\ll1$) and considering the leading orders, equation \eqref{eqlambda} has the following form
 \be
 \frac{\lambda''}{\lambda}+3\frac{\lambda'}{\lambda}-3\frac{\lambda'^2}{\lambda^2}-1\simeq0\,,
 \ee
which can be simplified as
\be\label{lam1}
(\frac{1}{\lambda^2}\big)''+3\big(\frac{1}{\lambda^2}\big)'+2\frac{1}{\lambda^2}\simeq0\,.
\ee
Solving the above equation, we obtain
\be\label{lambda0}
\frac{1}{\lambda^2}\simeq A_1e^{-2\dot\alpha t}+A_2e^{-\dot\alpha t}\,,
\ee
which represents an exponential increase in $|\lambda|$ value with time scale of the order $\dot\alpha^{-1}$.
Thus, in the limit of initially very small $\lambda^2$ values, $|\lambda|$ is growing very rapidly
and escaping quickly from the vicinity of zero. As a result, the above approximate solution is only applicable in first few $e$-folds where $\lambda^2$ is far from one. Despite the monotonic increase of $\lambda$, as seen from,  interestingly, this is not necessarily the case for $\dot\sigma$. To see this we evaluate  $\Sigma$ for two different initial conditions in which (i) $A_1=0$ and (ii) $A_2=0$.
\begin{itemize}
\item[(i)]{ Putting $A_1=0$ in \eqref{lambda0}, we have $\lambda=\lambda_0e^{\frac12\dot\alpha t}$, and
\eqref{eqsigma} yields
    \be
    \Sigma/H\simeq-\frac{\psi^2}{2}\lambda^2\simeq-\frac{\lambda_0^2\psi^2}{2}e^{\dot\alpha t},
    \ee
i.e. $|\Sigma|$ is exponentially increasing in time. (Recall that $\psi=\frac{\phi}{a}$ is  a constant in the leading order of $\epsilon$.) However, as mentioned above, this exponential growth of $\Sigma$ can only be sustained for the first few $e$-folds, after that $\lambda$ gets close to one and $|\Sigma|$ is exponentially damped. That is, our gauge-flation does not strictly follow cosmic no-hair theorem \cite{Wald:1983ky}, as our model does not satisfy the theorem's assumptions.
}
\item[(ii)]{Putting $A_2=0$, we obtain $\lambda=\lambda_0e^{\dot\alpha t}$, leading to
\be
\Sigma/H\simeq-\frac{\psi^2}{3\lambda^4}\simeq-\frac{\psi^2}{3\lambda_0^4}e^{-4\dot\alpha t},
\ee
which is quickly damped.
}
\end{itemize}
 Note that in this limit $\lambda^6\ll1$ and $\Sigma$ has always  a negative sign.

III) In the limit of  large $\lambda$ values ($|\lambda|\gg1$) and recalling \eqref{phi-small}, $\frac{\lambda\phi}{a}\ll1$, we have $\lambda^2\gamma\ll1$. As a result, up to the leading orders, we obtain the following approximation for \eqref{eqlambda}
\be\label{lam2}
\lambda''+3\lambda'+2\lambda\simeq0,
\ee
which is identical to \eqref{lam1} that governs the evolution of $\frac{1}{\lambda^2}$ in the limit of $|\lambda|\ll1$. Thus, the behavior of $\lambda$ in the limit of $|\lambda|\gg1$ is identical to the behavior of $\frac{1}{\lambda^2}$ in the limit of $|\lambda|\ll1$.
More detailed analysis of this case may be found in \cite{Maleknejad:2011jr}.

To summarize, assuming a system which undergoes quasi-de Sitter inflation
in the sense that $\epsilon$ is very small, we determined $\lambda$ and $\Sigma$. We see that regardless of the initial $\lambda$
values, all solutions converge to $\lambda^2=1$, within the few first e-folds.
Note that,  $\lambda^2=1$
corresponds to two values $\lambda=\pm1$, which are the isotropic solutions.
As we saw before, $\lambda$ cannot pass through zero during its evolution, so its sign does not change in time.
As a result, as we have shown analytically and will be demonstrated numerically in the next subsection, system's
trajectory eventually meets its attractor solution
$$\lambda\rightarrow1~~~\textmd{if}~~~\lambda_0>0,$$
$$\lambda\rightarrow-1~~ \textmd{if}~~~\lambda_0<0.$$
Furthermore, comparing \eqref{lam1} and \eqref{lam2} which are the approximate forms of \eqref{eqlambda} in the
limits of very small and very large $\lambda^6$ values, we find out that the behavior of $\lambda$ in the limit of
$|\lambda|\gg1$ is identical to the behavior of $\frac{1}{\lambda^2}$ in the limit of $|\lambda|\ll1$.
These results may be traced numerically too, which is depicted in Fig.\ref{phaseDiag}.

Although $\lambda^2$ evolves toward the FLRW isotropic solutions, it is shown that in some solutions, $|\Sigma|$
grows rapidly at the first few e-folds saturating our upper bound of $\Sigma\sim\epsilon$ for a short time.
However, this growth  stops fast (within a couple of e-folds) and $\Sigma$ is damped for the rest of the quasi-de Sitter
inflation.

\subsubsection{Numerical Analysis}\label{numerical-analysis-section}

The degrees of freedom in our system consist of two scalar fields $\psi$ and
$\lambda$, the isotropic expansion rate $\dot\alpha$ and the anisotropic expansion rate $\dot\sigma$. Thus, our
solutions are specified by eight initial values for these parameters
and their time derivatives. The gravitational equations, however,
provide some relations between these parameters. Altogether, each
inflationary trajectory may  be specified by the values of six
parameters, $(g, \kappa;\ \psi_0, \dot\psi_0,
\lambda_0,\dot\lambda_0)$, here $0$ subscript indicates the  initial value.

\begin{figure}[h]
\includegraphics[angle=0, width=75mm, height=65mm]{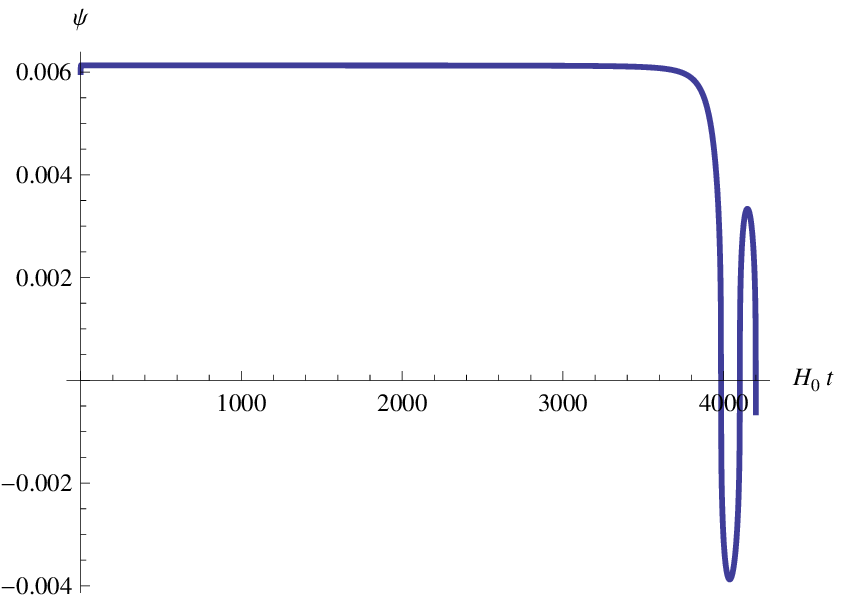}
\includegraphics[angle=0, width=75mm, height=65mm]{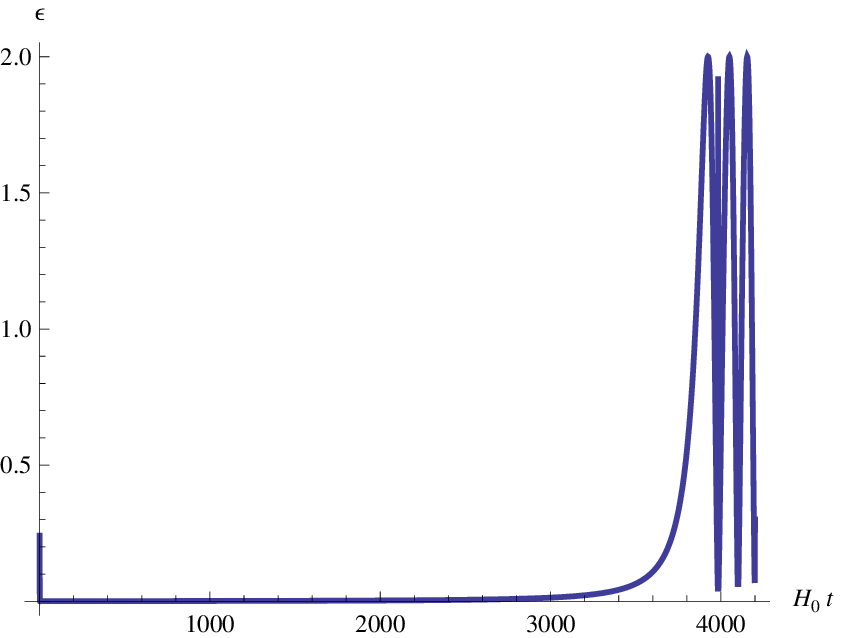}\\
\includegraphics[angle=0,width=75mm, height=65mm]{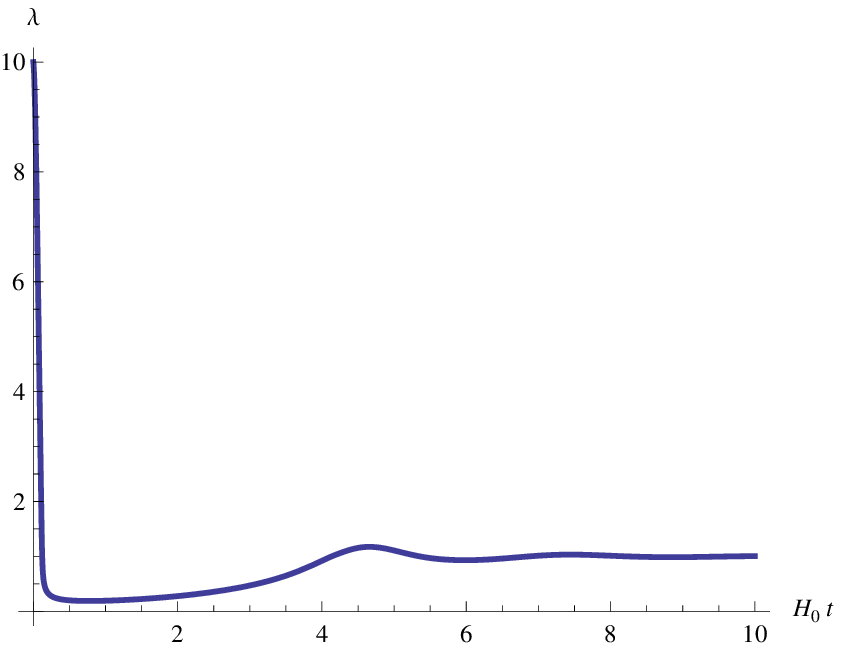}
\includegraphics[angle=0,width=75mm, height=65mm]{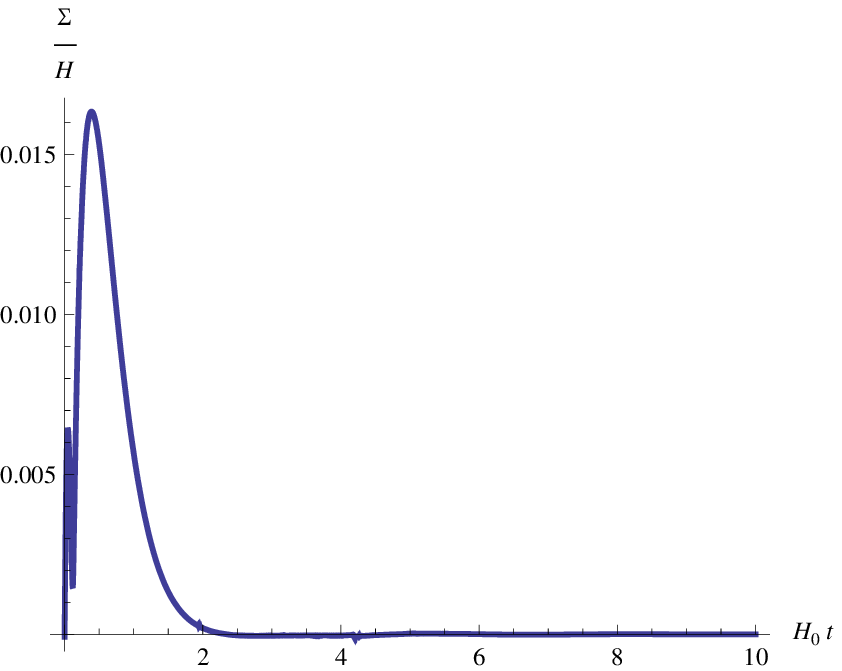}
\caption{The classical trajectory for ${\kappa = 3.77\times10^{15},\ g = 10^{-1},\ \psi_0 = 0.6\times 10^{-3},\
\dot\psi_0 = 10^{-10},}$ ${\lambda_0 = 10,\ \dot\lambda_0 = -3.6}$. Here $\dot\lambda=-\dot\alpha\lambda$ which
corresponds to  $A_1=0$ case in \eqref{lam2}. As expected from our analytical calculations there is a short period in
which $\Sigma/H$ is positive and rapidly increasing,
but this lasts very short and quickly (in a couple of e-folds) decreases to become almost zero.
These values lead to a trajectory with $\dot\alpha_0=4\times10^{-4},\ \epsilon_0=0.24$ and $\dot(\sigma/H)_0=-5\times10^{-7}$. The initial value of $\epsilon$ is rather large, but with in one number of e-folds it decreases and reaches $10^{-2}$. Note that
value of $\epsilon$ at the point of maximum  $\Sigma/H$ is equal to $0.05$ ($\Sigma/H\simeq\frac13\epsilon$), almost
saturating our upper bound for anisotropy $\Sigma/H$.
}\label{10-figures}
\end{figure}

\begin{figure}[h]
\includegraphics[angle=0, width=75mm, height=65mm]{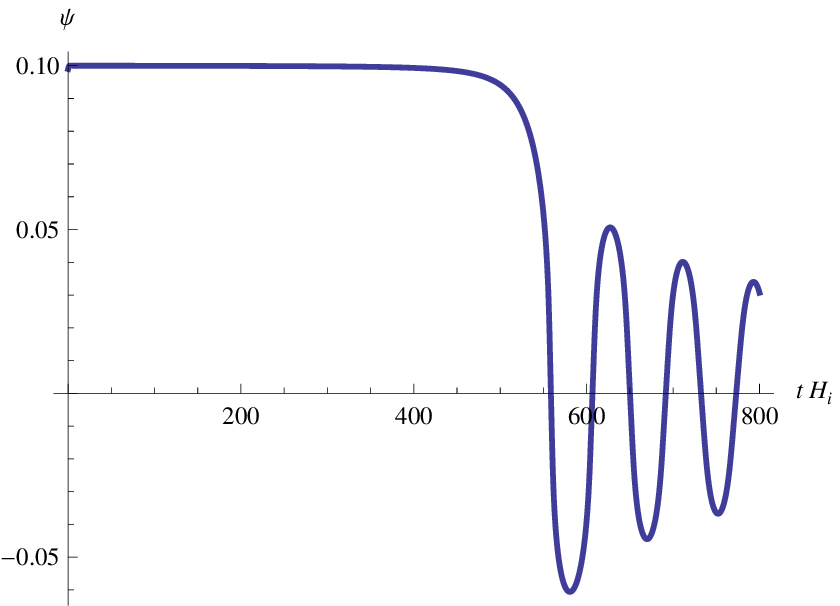}
\includegraphics[angle=0, width=75mm, height=65mm]{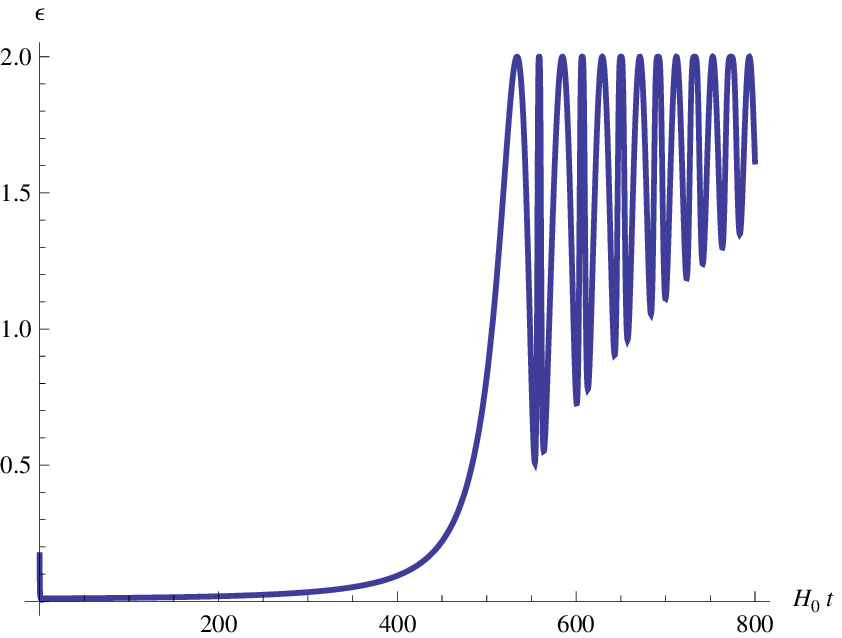}\\
\includegraphics[angle=0,width=75mm, height=65mm]{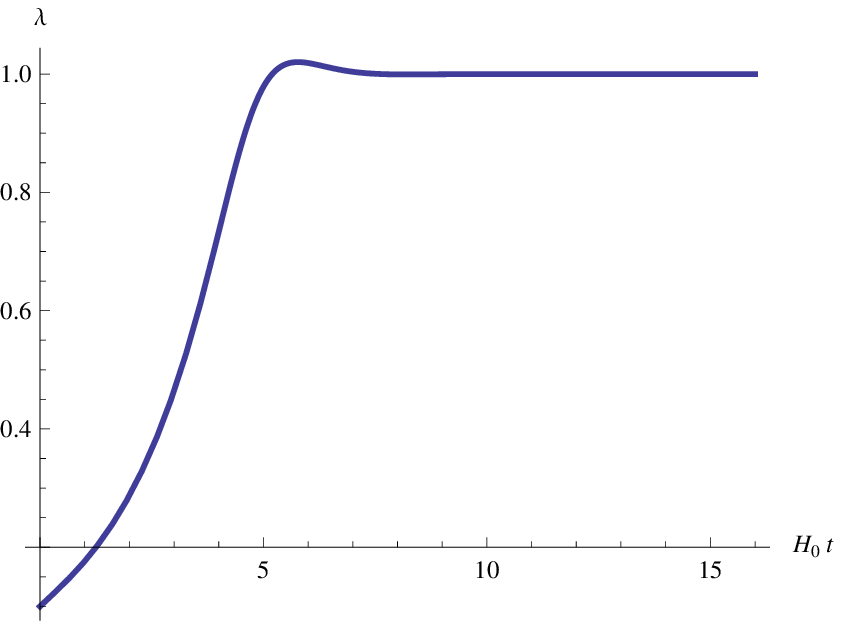}
\includegraphics[angle=0,width=75mm, height=65mm]{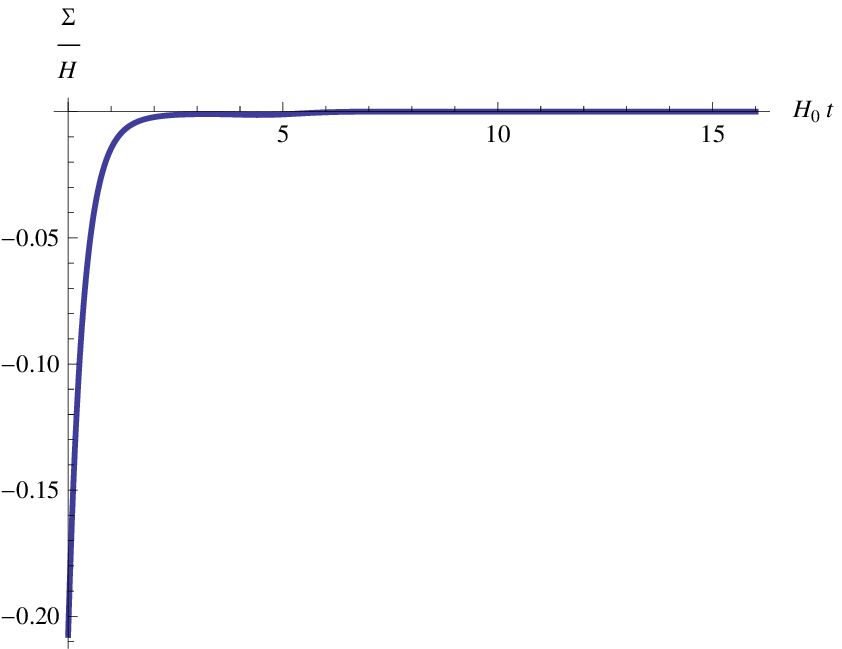}
\caption{The classical trajectory for ${\kappa = 1.99\times10^{10},\ g = 10^{-2},\ \psi_0 = 0.099,\ \dot\psi_0 = 10^{-10},}$ ${\lambda_0 = 0.1,\ \dot\lambda_0 = 1.288\times10^{-3}}$. Here $\dot\lambda=\dot\alpha\lambda$ which corresponds to the $A_2=0$ case in \eqref{lam1}. As has been predicted by the analytical calculations, $\Sigma$ is negative and monotonically, exponentially damped.
These values give a trajectory with $\dot\alpha_0=4\times 10^{-3},\  \epsilon_0=0.16$ which is a rather large initial $\epsilon$ value. As we learn from the top right figure, $\epsilon$ decreases within a few number of $e$-folds to $0.5\times10^{-2}$. For these values $\dot\sigma_0=-0.8\times10^{-4}$.}\label{0.1-figures}
\end{figure}

In what follows we present the
results of the numerical analysis of the equations of motion
\eqref{lambda}, \eqref{const1} and the $\phi$ and $\lambda$ field equations, for
two sets of parameters  corresponding to two different positive
initial $\lambda$ values ($\lambda_0=0.1$ and
$\lambda_0=10$). Similar diagrams for two more values of $\lambda_0$, $|\lambda_0-1|\sim 0.01$ has been presented
in \cite{Maleknejad:2011jr}.
The top left figures in Figures \ref{10-figures} and \ref{0.1-figures} show classical trajectories of the field
$\psi$ with respect to $\dot\alpha_0t$, while the top right figures indicate dynamics of $\epsilon$.
As we see there is a period of quasi-de Sitter inflation, where $\psi$ remains almost constant and $\epsilon$ is almost
constant and very small, which is essentially the same behavior we saw in our gauge-flation isotropic analysis of
previous subsections.
Bottom left and right figures respectively show evolutions of our two dimensionless variables $\lambda$ and $\Sigma/H$
during the first several e-folds. In the left bottom figure of Fig. \ref{10-figures}, we see that $\lambda$ which started
from $\lambda_0=10$ quickly decreases and gets close to one. The right bottom figure indicates that $\Sigma/H$, which is
initially equal to $1.2\times10^{-3}$, shows a phase of rapid growth and saturates our upper bound
$\Sigma/H \sim \epsilon$. More precisely, the peak value of $\Sigma/H$ is $\Sigma/H|_{t_{peak}}=0.016$ which is about $\frac{1}{3}\epsilon|_{t_{peak}}$. After its sharp peak, $\Sigma$ decreases quickly and within  few e-folds becomes negligible. At that point, system mimics the behavior of isotropic inflation.
The left bottom figure of Fig. \ref{0.1-figures} shows $\lambda$, initially equal to $\lambda_0=0.1$, quickly evolves
towards one. The right bottom figure indicates $\Sigma/H$, which is initially equal to $-2\times10^{-2}$, is exponentially
damped and becomes negligible. As a result, after a few e-folds,  the system undergoes an essentially isotropic
quasi-de Sitter inflation.

\subsection{Gauge-flation cosmic perturbation theory}\label{gauge-flation-cosmic-perturbation-theory-section}

We showed that the gauge-flation model can lead to a fairly standard slow-roll inflating Universe with enough number of
e-folds and examined the stability of isotropic FLRW background against initial anisotropies. However, as reviewed in section
\ref{Inflation-review-section}, we now have a wealth of
CMB and other cosmological data which can be used to test  inflationary models beyond the background dynamics and any analysis of inflationary
models is incomplete without discussing the cosmic perturbation theory. In this section we work out the gauge-flation cosmic
perturbation theory. Dealing with non-Abelian gauge fields brings many new features and complications compared to the standard
scalar-driven models reviewed in section \ref{Inflation-review-section}. Although, as discussed in previous subsections,
at the level of the classical background the gauge-flation effectively  resembles a single-scalar driven model, our consistent reduction
arguments does not extend to quantum level; in principle, all the gauge field modes
can contribute to the quantum fluctuations around the background.
Due to isotropy of the background one can still use the scalar, vector and tensor decomposition for the perturbations.
In what follows we first classify all these perturbations
and as we will see gauge invariant perturbations consist of five  scalars, three vectors and two tensor modes. We
study their dynamics, at classical and quantum levels and finally compute the power spectra for curvature perturbations and the tensor modes and the
corresponding spectral tilts.


\subsubsection{Classification of gauge invariant perturbations and field equations}\label{section-5.6.1}

The perturbed metric can be parameterized as in the standard case \eqref{metric-pert}:
\begin{equation*}
\begin{split}
ds^2&=-(1+2A)dt^2+2a(\partial_iB+V_i)dx^idt\\
&+a^2\left((1-2C)\delta_{ij}+2\partial_{ij}E+2\partial_{(i}W_{j)}+h_{ij}\right)dx^idx^j\,,
\end{split}
\end{equation*}
where $\partial_i$ denotes partial derivative respect to $x^i$ and $A,\ B,\ C$ and $E$ are scalar perturbations, $V_i,\ W_i$ parameterize vector perturbations (these are divergence-free three-vectors) and $h_{ij}$, which is symmetric, traceless and divergence-free, is the tensor mode.
The 12 components of the gauge field fluctuations may be decomposed as%
\bea \label{gauge-field-pert-1}
\delta A^a_{~0}&=&\delta^{k}_a\partial_k(Y+\phi\dot{E})+\delta_a^j (u_j+\phi \dot{W}_j)\,,\\  \label{gauge-field-pert-2}
\delta A^a_{~i}&=&\delta^a_i (Q-\frac{\dot{\phi}}{H} C)+\delta^{ak}\partial_{ik}(\tilde{Z}+\phi E)+g\phi\epsilon^{a~k}_{~i}\partial_{k}(Z-\tilde{Z})\nonumber\\
&+&\delta_a^j\partial_i (v_j+\phi W_j)+\phi\epsilon^{a~j}_{~i}w_j+\delta^{aj}\phi(\tilde{h}_{ij}+\frac{1}{2}h_{ij})\,,
\eea
where following our discussions in \ref{gauge-flation-general-setup}, we have identified the gauge indices with the
local Lorentz indices and  the perturbation is done around the background ansatz \eqref{gauge-ansatz-summary}.
In \eqref{gauge-field-pert-1} and \eqref{gauge-field-pert-2},  we have parameterized the gauge field perturbations
by four scalars $Y, Q, Z, \tilde{Z}$, three divergence-free vector modes $u_j, v_j, w_j$ and one symmetric,
traceless divergence-free tensor mode $\tilde{h}_{ij}$, adding up to $4+3\times 2+2=12$. However, as we see explicitly in
\eqref{gauge-field-pert-1} and \eqref{gauge-field-pert-2}, due to the gauge (vector) field nature of $A^a_{~\mu}$,
it turns out to be more
convenient to parameterize the gauge field perturbations in a way which also involves metric perturbations.
To be more precise, in \eqref{gauge-field-pert-2} we recalled \eqref{gauge-ansatz-triad} and hence
\be
A^a_{~i}=\psi e^a_{~i}\quad\rightarrow\quad \delta A^a_{~i}=\psi \delta e^a_{~i}+\delta\!_{_{gf}}A^a_{~i},
\ee
where $\delta e^a_{~i}$ is the perturbation of the spatial triad and induces the metric perturbations on the
gauge field, while $\delta\!_{_{gf}}A^a_{~i}$ represents the spatial fluctuations of the gauge field itself.
Thus, in the $\delta A^a_{~i}$, all the genuine gauge field fluctuations are coming from the $\delta\!_{_{gf}}A^a_{~i}$
and the metric fluctuations are from the $ \delta e^a_{~i}$ term.
Similarly, for the $\delta A^a_{~0}$ in \eqref{gauge-field-pert-1}, $Y$ and $u_j$
are chosen in such a way that they are invariant under any infinitesimal space-time transformations, that is
\be
A^a_{~0}=0\quad\rightarrow\quad \delta A^a_{~0}=\phi\delta^{j}_a(\partial_j\dot{E}+\dot{W}_j)+\delta\!_{_{gf}}A^a_{~0},
\ee
where $\delta\!_{_{gf}}A^a_{~0}$ is invariant under any infinitesimal space-time transformations.

Because of the gauge symmetries of the problem, not all 25 metric and gauge field perturbations are physical.
The gauge transformations in our case, in addition to the standard ones discussed in section
\ref{Inflation-review-section} which are generated by infinitesimal \textit{space-time} transformations \eqref{coor}
we also have the infinitesimal \textit{internal gauge field} transformations
\be \label{gua}
\delta A^a_{~\mu}\rightarrow \delta A^a_{~\mu}-\frac{1}{g}\partial_\mu\lambda^a-\epsilon^a_{~bc}\lambda^b A^c_{~\mu}\,,
\ee
where the gauge parameters $\lambda^a$ can be decomposed into a scalar and a divergence-free vector
\be
\lambda^a=\delta^{ai}\partial_i\lambda+\delta^a_i\lambda^{~i}_{V}\,.
\ee
In what follows to distinguish the above two gauge freedoms
we call the space-time gauge transformations  \eqref{coor} ``$x^\mu$-gauge"
and the gauge field internal gauge transformations  \eqref{gua} ``$A^a$-gauge" where
\begin{itemize}
\item{\textit{$x^\mu$-gauge} act on the perturbed metric $\delta g_{\mu\nu}$, hence $A, B, C ,E, W_i$ and $V_i$ are
transformed under the space-time gauge transformations, while they are invariant under the internal gauge
transformations \eqref{gua}.}
\item{\textit{$A^a$-gauge} act on the the genuine gauge field perturbations, $\delta\!_{_{gf}}A^a_{~\mu}$, thus
$Q, Y, \tilde Z, Z, u_i, v_i$ and $w_i$ are transformed under internal gauge transformations, but they are invariant
under the infinitesimal space-time gauge transformations.}
\item{The two tensor modes $h_{ij}$ and $\tilde h_{ij}$ are invariant under both type of the gauge transformations,
and hence are physical quantities.}
\end{itemize}

\paragraph{Field equations.} After discussing the decomposition of metric and gauge field perturbations we study their
dynamics governed by perturbed field equations. As in the general multi-scalar field case reviewed in section
\ref{Inflation-review-section} these equations are nothing but the perturbed Einstein equations and perturbed gauge field
equations
\be\label{perturbed-Einstein+gauge-field}
\delta G_{\mu\nu}=\delta T_{\mu\nu}\,,\quad \delta\big(\frac{\partial(\sqrt{-g}\mathcal{L})}{\partial A^a_{~\mu}}\big)=0\, ,
\ee
where by $\delta$ in the above we mean first order in field perturbations.
Nonetheless, as in the scalar-driven case the integrability condition of perturbed Einstein equations
$\delta(\nabla^\mu G_{\mu\nu})=0$ implies \emph{some} of the gauge field equations of motion, but of course not all.

To see how many gauge field equations are needed to be added to perturbed Einstein equations to obtain the full set of equations,
let us recall the counting of these equations. Among the ten perturbed equations, there are \emph{four} scalars, \emph{two vectors}
and \emph{one} tensor (\emph{cf.} discussions of section \ref{Inflation-review-section}). So, as far as scalar and vector modes
are concerned, we need one more scalar and one more vector equation. These extra equations may be provided by
the $\mu=0$ component of perturbed gauge field equations. The equation of motion for $A^a_{~0}$ component is a constraint
enforcing the gauge invariance of the action.\footnote{Dealing with a gauge invariant action, $\dot A^a_{~0}$
does not appear in the Lagrangian density $\mathcal{L}$, the momentum conjugate to $A^a_{~0}$,
given as $\frac{\partial\mathcal{L}}{\partial \dot A^a_{~0}}$, is identically zero.} Therefore, this extra constraint is
independent of the Einstein equations. The full set of equations needed to deal with scalar and vector perturbations are hence
provided through\footnote{Note that for a single scalar case
the perturbed Einstein equations are enough and they contain the equation of motion for the perturbations of inflaton field \cite{Inflation-Books}, whereas
for the multi-scalar cases with $n$ number of inflatons we need to consider $n-1$ perturbed equations for inflatons.
Similarly, for the models with one vector gauge field, again the equation of motion for gauge field perturbation is included in
perturbed Einstein equations (this may be seen from our discussions of section \ref{prim-magnet-NG-section}), while if we have
more than one gauge field we need to consider gauge field equations.}
\bea
\label{perturbed-Einstein}
\delta G_{\mu\nu}=\delta T_{\mu\nu}\,,\quad \delta\big(\frac{\partial(\sqrt{-g}\mathcal{L})}{\partial A^a_{~0}}\big)=0\, .
\eea
We stress that to deal with the tensor perturbations, we need to consider another equation among
perturbed gauge field equations
$\delta\big(\frac{\partial(\sqrt{-g}\mathcal{L})}{\partial A^a_{~i}}\big)=0$.

Instead of working with perturbed equations of motion at first order, one may alternatively consider action expanded
up to second order in field perturbations; the equations of motion for second order action should obviously reproduce \eqref{perturbed-Einstein+gauge-field}.
Working out second order action is also needed for the process of quantization of perturbations, to read the canonical momentum conjugate to
the dynamical fields. Therefore, in what follows we also present the part of (but not full)
second order action which is require for our analysis.  In terms of the second order action, the gauge field equation
in \eqref{perturbed-Einstein} may be written as
\bea
\label{constraint-2nd-order-action}
\delta^{k}_a\partial_k\big(\frac{\partial\delta\!_{_{2}}(\sqrt{-g}\mathcal{L})}{\partial Y }\big)=0\, , \quad
\delta^a_i\big(\frac{\partial\delta\!_{_{2}}(\sqrt{-g}\mathcal{L})}{\partial u_i }\big)=0\,,
\eea
where $\delta_2$ stands for second order in perturbations. The equation of motion for the tensor mode $\tilde{h}_{ij}$ will also be
obtained from the corresponding part of the second order action.

To work with perturbed Einstein equations we use the same decomposition developed in section \ref{Inflation-review-section} and
decompose perturbed energy-momentum as in \eqref{pert-Tmunu}:%
\begin{align}
\delta T_{ij}=&\bar P\delta g_{ij}+a^2\left(\delta_{ij}(\delta P-\frac13\nabla^2\pi^S)+\partial_{ij}\pi^S
+\partial_i\pi^V_j+\partial_j\pi^V_i+\pi^T_{ij}\right)\,,\cr
\delta T_{i0}=&\bar P\delta g_{i0}-(\partial_i\delta q+\delta q_i^V)\,,\cr
\delta T_{00}=&-\bar\rho\delta g_{00}+\delta \rho\,,\nn
\end{align}
where $\bar\rho$ and $\bar P$ are the background quantities and $\pi^S$,
$\pi^V_i$ ,~ $\pi^T_{ij}$ represent the \textit{anisotropic inertia} and
characterize departures from the perfect fluid form of the
energy-momentum tensor; $\pi^V_{~i}$ and $\delta q^V_{~i}$ are divergence-free 3-vectors and a tensor mode,
$\pi^T_{~ij}$ is a symmetric, divergence-free traceless tensor.

 We summarize the above discussion in the following table. In the left hand side of the table we have the fields d.o.f, while in the right hand side we summarized the number of independent equations governing the dynamics of each part of the system:
\begin{center}
\begin{tabular}{|cp{1cm}|p{1.4cm}|p{1.4cm}|p{1.4cm}|p{1.5cm}|p{1.7cm}||p{1.5cm}|p{1.5cm}|p{1cm}|}
\hline
&  & $\ \  \ \delta g_{\mu\nu}$ &\begin{footnotesize}
$\ \ \delta\!_{gf}A^a_{\mu}$
\end{footnotesize}  & \begin{footnotesize}
$x^\mu$-gauge
\end{footnotesize} & \begin{footnotesize}
$A^a$-gauge
\end{footnotesize} &\begin{footnotesize}
Gauge-invariant
\end{footnotesize} &\begin{footnotesize}
Einstein Eqs
\end{footnotesize}  &\begin{footnotesize}
Gauge Field Eqs
\end{footnotesize}  & \begin{scriptsize}
Total Eqs
\end{scriptsize}\\ [2.5ex]
\hline
 &\begin{footnotesize}
Scalar
\end{footnotesize}  & 4 & 4 & $-2$ & $-1$ & 5 & 4 & 1 & 5    \\ [1.1ex] \hline
& \begin{footnotesize}
Vector
\end{footnotesize} & 2 & 3 & $-1$ & $-1$ & 3 & 2 & 1 & 3    \\[1.1ex]  \hline
& \begin{footnotesize}
Tensor
\end{footnotesize} & 1 & 1 & ~~0  & ~~0 & 2 & 1  & 1 & 2     \\ [1.1ex] \hline
&\begin{scriptsize}
Total d.o.f
\end{scriptsize} & 10 & 12 & $-4$ & $-3$ & 15 \\[1.1ex]  \cline{1-7}
\end{tabular}
\vskip 0.05 cm
\textbf{Table I: Gauge-flation perturbation modes}
\end{center}

In the table $\delta\!_{gf}A_{\mu\nu}$ represents the genuine gauge field fluctuations, ``$x^\mu$-gauge"
denotes the space-time gauge transformations and the ``$A^a$-gauge" represents the internal gauge field transformations.
Note also that, in the gauge field Eqs column we have only counted the equations which are independent and not included in
Einstein or other gauge field equations.

\subsubsection{Treating the equations of motion, classical solutions and quantization}

After determining the field equations,  we are now ready to eliminate the gauge degrees of freedom and
solve the equations. In the following, first by constructing the gauge-invariant combinations we remove the gauge
d.o.f, then we solve the equations for scalar, vector and tensor perturbations.

\subsubsection*{$\bullet$ Scalar modes}

In the scalar sector of the perturbations, $A$, $B$, $C$, $E$ are coming from the perturbed metric and,
$Q$, $Y$, $Z$ and $\tilde{Z}$ from the perturbations of the gauge field.
Under the action of the transformation \eqref{coor}, the scalar fluctuations of the metric transform as
\be
\begin{split}
&A \rightarrow A-\dot{\delta t}\,,\qquad \qquad \quad
C\rightarrow C+H\delta t\,,\\
&B \rightarrow B+\frac{\delta t}{a}-a\dot{\delta x}\,,\qquad
E\rightarrow E-\delta x\,.
\end{split}
\ee
while the genuine gauge scalars ($Q$, $Y$, $Z$ and $\tilde{Z}$) are invariant under the coordinate transformations.
On the other hand, under the action of the internal gauge field transformation of the form \eqref{gua}, the gauge field perturbations transform as
\be
\begin{split}
Q \rightarrow Q\,& ,\qquad
Y \rightarrow Y-\frac{1}{g}\dot{\lambda}\,,\\
Z\rightarrow Z\,&,\qquad
\tilde{Z}\rightarrow \tilde{Z}-\frac{1}{g}\lambda\,.
\end{split}
\ee
We note that $Q$ and $Z$ are invariant under both internal and space-time gauge-transformations.
Then, we can construct five independent gauge-invariant combinations. One such choice is as follows.
The standard Bardeen potentials \eqref{Bardeen-potentials}
\bea
\Psi&=&C+a^2H(\dot{E}-\frac{B}{a})\,,\\
\Phi&=&A-\frac{d}{dt}\left(a^2(\dot{E}-\frac{B}{a})\right)\,,
\eea
 from the metric perturbations, in addition to
\bse
\begin{align}
Q=&Q,\\
 M=&\frac{g^2\phi^3}{a^2}Z,\\
\tilde{M}=&\dot\phi(\dot{\tilde{Z}}-Y)\,,
\end{align}
\ese
which are the three gauge invariant combinations coming from the genuine gauge field fluctuations.

The scalar part of the first-order perturbations of the gauge field strength is given as%
\bea \label{Fmunu-scalar}
\delta F^a_{~0i}&=&\delta^a_i(Q-\frac{\dot\phi}{H} C\dot{)}+\delta^{aj}\partial_{ij}(\dot{\tilde{Z}}-Y+\dot{\phi}E)+g\epsilon^{a~j}_{~i}\partial_j((\phi Z\dot{)}-\phi(\dot{\tilde{Z}}-Y)-\dot{\phi}\tilde{Z}+\phi^2\dot{E})\,,\nonumber\\
\delta F^a_{~ij}&=&2\delta^a_{[j}\partial_{i]}(Q-\frac{\dot\phi}{H} C-g^2\phi^2(Z-\tilde{Z}))+2g\phi\epsilon^{ak}_{~~[i}\partial_{j]k}(Z+\phi E)-2g\phi\epsilon^a_{~ij}(Q-\frac{\dot\phi}{H} C)\,,\nonumber
\eea
 which is made of gauge-invariant combinations ($Q$, $M$ and $\tilde{M}$) as well as $\tilde{Z}$ which is not gauge-invariant.
 However, note that $\delta F^a_{\mu\nu}$ are not gauge-invariant and under an arbitrary gauge transformation, they transform as
\be
 F^a_{\mu\nu}\to F^a_{\mu\nu}-\epsilon^a_{\ bc} \lambda^b F^c_{\mu\nu}.
 \ee
It is straightforward to show that in the gauge field strengths $\delta F^a_{\mu\nu}$ the gauge freedom part can be used to
remove $\tilde{Z}$ terms and therefore, in the equations govern the dynamics of the system
$\tilde{Z}$ terms cancel out.

Since we are working with gauge-invariant combinations, we should write the Einstein equations in a gauge-invariant form. In order this we note that $\delta T_{\mu\nu}$ has four gauge-invariant scalar parts
$\dre_g$, $\dpe_g$, $\dqe_g$ \eqref{diff-inv-Tmunu}
\bea
\dre_g&=&\dre-\dot{\bar\rho}a^2(\dot{E}-\frac{B}{a})\,,\quad\dpe_g=\dpe-\dot{\bar P}a^2(\dot{E}-\frac{B}{a})\,,\\
\dqe_g&=&\dqe+(\bar\rho+\bar P)a^2(\dot{E}-\frac{B}{a})\,,
\eea
and $a^2\pi^S$ \cite{Inflation-Books}.
Perturbing the energy-momentum tensor to the first order, we obtain
\bea\label{PiS}
a^2\pi^S&=&2(M-\tM),\\
\dqe_g&=&-2(\dot M+3H\delta\times M-HM-\frac{g^2\phi^3}{\dot{\phi}a^2}\tM+\frac{\dot{\phi}}{a}(\frac{Q}{a}-\frac{\dot\phi}{aH}\Psi)),\\
\dre_g&=&3(1+\kk)\frac{\dot \phi}{a^2}(\dot Q-\frac{\dot\phi}{H}\dot\Psi)+6(1+\dk)\frac{g^2\phi^3}{a^3}\frac{Q}{a}-3(1+\kk)\frac{\dot\phi^2}{a^2}\Phi\nonumber\\
&+&3\epsilon\YM2\Psi-(1+\kk)\frac{k^2}{a^2}\tM-2(1+\dk)\frac{k^2}{a^2}M,\\
\dpe_g&=&(1-3\kk)\frac{\dot \phi}{a^2}(\dot Q-\frac{\dot\phi}{H}\dot\Psi)+2(1-3\dk)\frac{g^2\phi^3}{a^3}\frac{Q}{a}-(1-3\kk)\frac{\dot\phi^2}{a^2}\Phi\nonumber\\
&-&(4\frac{\dot\phi^2}{a^2}+3\frac{g^2\phi^4}{a^4})\epsilon\Psi-(\frac13-\kk)\frac{k^2}{a^2}\tM-2(\frac13-\dk)\frac{k^2}{a^2}M.
\eea
The perturbed Einstein equations then take the form
\bea%
\label{pi^s}
&~&a^2\partial_{ij}\pi^s=\partial_{ij}(\Psi-\Phi)\,,\\
\label{dq} &~&\partial_{i}(\dqe_g+2(\dot{\Psi}+H\Phi))=0\,,\\
\label{drho} &~&\dre_g-3H\dqe_g+2\frac{k^2}{a^2}\Psi=0\,,\\
\label{dP}
&~&\dpe_g+\dot{\dqe}_g+3H\dqe_g+2\epsilon H^2\Phi-
\frac23\frac{k^2}{a^2}(\Psi-\Phi)=0\,. %
\eea%
Although it is not independent of the Einstein equations, here for later convenience we also write the equation of
energy conservation
\be\label{energy-conservation}
\delta\dot\rho_g-3H\delta\dot q_g+3\epsilon H^2\delta q_g-6H\frac{k^2}{a^2}\Psi+H\frac{k^2}{a^2}(\Psi-\Phi)+2\frac{k^2}{a^2}(\dot\Psi+H\Phi)=0.
\ee
As mentioned before, the four scalar modes of the Einstein equations do not suffice to deal with five
gauge-invariant scalar degrees of freedom and we need one more equation.
This last equation is provided  with the scalar part of \eqref{constraint-2nd-order-action}, or by the equation of motion for
$\tilde{M}$ field coming from the second order action for the scalar perturbations
\bea
&~&~~~~~~\delta_{_{2}}S_{_{tot}}=\int ~ a^3 d^4x~ \big[\frac32\left(1+\kk\right)\frac{\dot{Q}^2}{a^2}-\bigg((1+\kk)\frac{k^2}{a^2}\frac{\tM}{\dot{\phi}/a}+3(1+\kk)\frac{\phi}{a}\dot\Psi\nonumber\\
&+&(1+\kk)\frac{\dot\phi}{a}(6\Psi+3\Phi)+4\frac{\kappa \dot\phi^2}{a^2}\frac{k^2}{a^2}\frac{M}{\dot{\phi}/a}\bigg)\frac{\dot Q}{a}+\bigg(3\frac{g^2\phi^2}{a^2}+12\frac{\dot\phi^2}{\phi^2}+3\dk\frac{g^2\phi^2}{a^2}-\frac{k^2}{a^2}\bigg)\frac{Q^2}{a^2}\nonumber\\
&+&\bigg(-12\kk\frac{\dot\phi}{a}\dot\Psi+2(3-2\dk)\frac{k^2}{a^2}\frac{M}{\phi/a}+(12\frac{g^2\phi^2}{a^2}(1-\dk)+2\frac{k^2}{a^2})\frac{\phi}{a}\Psi-6\frac{g^2\phi^2}{a^2}(1+\dk)\frac{\phi}{a}\Phi\nonumber\\&-&2\frac{\kappa g^2\phi^4}{a^4}
\frac{k^2}{a^2}\frac{\tM}{\phi/a}-2\frac{\kappa\dot\phi^2}{a^2}\frac{k^2}{a^2}\frac{\dot{M}}{\dot{\phi}/a}\bigg)\frac{Q}{a}-\frac32\frac{g^2\phi^4}{a^4H^2}\dot{\Psi}^2-3\frac{g^2\phi^4}{a^4H}\dot\Psi\Phi-\frac32\frac{g^2\phi^4}{a^4}\Psi^2-\frac32\frac{g^2\phi^4}{a^4}\Phi^2\nonumber\\
&+&\frac{k^2}{a^2}\Psi^2-\frac{2k^2}{a^2}\Psi\Phi+\frac{k^2}{a^2}\frac{\dot{M}^2}{g^2\phi^4/a^4}+2\dk\frac{k^2}{a^2}\frac{\phi}{\dot{\phi}}\Psi\dot{M}+\left(H^2(2-\epsilon)+\frac{k^2}{a^2}(-1+\frac23\dk)\right)\frac{k^2}{a^2}\frac{M^2}{g^2\phi^4/a^4}
\nonumber\\&+&\left(\frac{g^2\phi^2}{\dot{\phi}^2}+\frac{1}{2\dot{\phi}^2/a^2}(1+\frac13\kk)\frac{k^2}{a^2}\right)\frac{k^2}{a^2}\tM^2+\bigg((1+\kk)(\frac{\phi}{\dot\phi}\dot{\Psi}+\Phi)+(2-\epsilon\YM2)\Psi-2\frac{\dot M}{\phi\dot\phi}\nonumber\\
&+&2\frac{M}{\phi^2/a^2}+\frac23\dk\frac{k^2}{a^2}\frac{M}{\dot\phi^2/a^2}\bigg)\frac{k^2}{a^2}\tM+\left(4\dk\frac{\phi}{\dot\phi}\dot\Psi+4(-1+\dk)\Psi+2(1+\dk)\Phi\right)\frac{k^2}{a^2}M\big].\nonumber
\eea
This equation after using \eqref{drho} takes the following form\footnote{The constraint \eqref{A0eq} is equal to the gauge field constraint equation $D_\mu\big(\frac{\partial\cal L}{\partial F^a_{~0\mu}}\big)=0$.}
\be\label{A0eq}
\epsilon H^2\frac{Q}{\phi}+\delta\times H(\dot\Psi+H\Phi)-\frac{1}{2}\YM2\epsilon\Psi+\frac16\frac{k^2}{a^2}(\Psi+\Phi)=0.
\ee
With the above constraint we have enough number of equations for the gauge-invariant scalar perturbations. Note that the rest of field equations does not lead to any new equation. In particular, as expected, the field equation of $\Phi$ is equal to \eqref{drho}, while the field equations of $M$ and $\tilde{M}$ reduce to \eqref{A0eq}. On the other hand, using  the constraint \eqref{A0eq}, the field equation of $Q$ is equal to the energy conservation equation \eqref{energy-conservation} while field equation of $\Psi$ is identical to \eqref{dP} plus the field equation of Q.

From the combination of \eqref{dq}-\eqref{dP}
and \eqref{PiS}, we obtain
\bea\label{Psi-eq}
\ddot\Psi+H\dot\Phi-2\frac{g^2\phi^4}{a^4}\Phi+\frac{k^2}{a^2}(2M-\Psi)-2\frac{\dot \phi}{a}(\frac{\dot Q}{a}-\frac{\dot\phi}{Ha}\dot\Psi)-4\frac{g^2\phi^3}{a^3}\frac{Q}{a}+2\frac{\dot\phi^2}{a^2}\epsilon\Psi=0.
\eea
Furthermore, one can write \eqref{drho} as
\bea\label{drho-eq}
&~&(6H^2-3\YM2)\frac{\dot Q}{\dot\phi}+(12H^2-6\frac{\dot\phi^2}{a^2})\frac{Q}{\phi}-\frac12\frac{k^2}{a^2}(1+\kk)(\Psi+\Phi)+\frac{k^2}{a^2}(3+\kk)\Psi\nonumber\\
&~&-\frac{k^2}{a^2}(3+\kk+2\dk)M+3\YM2\Phi+3\YM2\frac{\Dot\Psi}{H}+3\epsilon\YM2\Psi=0.
\eea
Also using \eqref{PiS}, we can omit $\tM$ in \eqref{dq} and obtain
\be\label{dq-eq}
\dot\Psi+H(1+\frac\gamma2)\Phi-\dot{M}+H(1+\gamma)M-(\frac{\gamma}{2}-\frac{\phi^2}{a^2})H\Psi-\frac{\dot\phi}{a}\frac{Q}{a}\simeq0.
\ee
Equations \eqref{A0eq}, \eqref{Psi-eq}, \eqref{drho-eq} and \eqref{dq-eq}  provide enough number of equations for
$Q, M, \Psi$ and $\Phi$. Note that the first three equations are exact in slow-roll parameters, while the last one
has been written in first order in slow-roll. Using \eqref{PiS}, one can then determine $\tM$ in terms of the rest of variables.

In order to solve the equations and determine the dynamics, we first write equations in the two asymptotic
limits of subhorizon ($\frac{k}{a}\gg H$) and superhorizon scales ($\frac{k}{a}\ll H$) and then combining them together,
we derive the closed form differential equations governing the dynamics of our dynamical variable $Q$.
For convenience, we will rewrite the equations in conformal time $\tau$ ($\tau=\int\frac1a dt$).

\paragraph{\textit{  Asymptotic past limit:}}
In the asymptotic past limit $k\tau\gg1$, the equations \eqref{A0eq} and \eqref{dq-eq} take the following forms respectively
\be\label{eq-in-A-P}
k^2(\Psi+\Phi)=0\quad \textmd{and}\quad M'-\Psi'+\frac{\dot\phi}{a}Q=0,
\ee
where prime represents a derivative respect to conformal time. Using the above constraints in \eqref{energy-conservation} and \eqref{Psi-eq}, we can omit $\Psi$ and $M$ in terms of $Q$ and $\Phi$ which leads to the following set of coupled equations for $\Phi$ and $Q$ respectively
\bea\label{Q-Psi-1}
&~&\frac{\dot\phi}{a}Q''+k^2\big(\frac{\gamma+2}{3\gamma}\frac{\dot\phi}{a}Q+\frac{2}{3\gamma}\Phi'\big)=0,\\\label{Q-Psi-2}
&~&(2\frac{\dot\phi}{a}Q+\Phi')''+k^2\big(2\frac{\dot\phi}{a}Q+\Phi'\big)=0.
\eea
The above equations, then imply that $Q\propto k\Phi$ in the $k\tau\rightarrow-\infty$ limit. Moreover, \eqref{eq-in-A-P}
indicates that we have a non-zero scalar anisotropic inertia $a^2\pi^S$:
\be\label{piS-A-P}
a^2\pi^S=-2\Phi.
\ee
Note that regardless of the details, for all the scalar inflationary models in the context of GR the anisotropic stress
$a^2\pi^S$ is identically zero.

In the set of equations \eqref{Q-Psi-1}-\eqref{Q-Psi-2}, while the first one has a complicated form, the second one is simply a wave equation for $~2\frac{\dot\phi}{a}Q+\Phi'~$
with a sound speed equal to one.
Then, multiplying the former by a factor of $\gamma+1$ and subtracting the result from the latter we obtain
\be
(\gamma-1)\frac{\dot\phi}{a}Q''-(\Phi')''+(\frac{\gamma-2}{3\gamma})k^2\big((\gamma-1)\frac{\dot\phi}{a}Q-\Phi'\big)=0,
\ee
which is a wave equation for the variable $~(\gamma-1)\frac{\dot\phi}{a}Q-\Phi'~$ with a sound speed square equal to $(\frac{\gamma-2}{3\gamma})$. In fact, in terms of
\be\label{the-two-scalar-modes}
X_1=2\frac{\dot\phi}{a}Q+\Phi'\quad \textmd{and} \quad X_2=(\gamma-1)\frac{\dot\phi}{a}Q-\Phi',
\ee
 the set of equations \eqref{Q-Psi-1}-\eqref{Q-Psi-2} is diagonalized into two wave equations for $X_1$ and $X_2$ with sound speeds $c_1^2=1$ and $c_2^2=(\frac{\gamma-2}{3\gamma})$ respectively.

Correspondingly, in the asymptotic past limit:
\begin{itemize}
\item{One can decompose $Q$ as $Q=Q_1+Q_2$ where $Q_{1,2}$ satisfy
\bea
\label{Q+-}
Q_{1}''+k^2Q_{1}= 0\,,\quad  \quad Q_{2}''+\frac{\gamma-2}{3\gamma}k^2Q_{2}= 0.
\eea
}
\item{Then, we can decompose $\Phi$ as $\Phi=\Phi_1+\Phi_2$ such that
\bea
\label{Psi+-}
\Phi_{1}''+k^2\Phi_{1}= 0\,,\quad  \quad \Phi_{2}''+\frac{\gamma-2}{3\gamma}k^2\Phi_{2}= 0.
\eea
Besides that, we also have the following two constraints
\be
\Phi'_1=(\gamma-1)H\psi Q_1\,,\quad \quad  \Phi'_2=-2H\psi Q_2,
\ee
which couple $\Phi$ and $Q$ fields in $k\tau\rightarrow-\infty$ limit. }
 \end{itemize}
We note that, although the sound speed of $Q_2$ is negative for $\gamma\in(0,2)$, after fitting gauge-flation with the cosmic data, we will see that this region is not physically interesting and the data requires $\gamma>2$ values.

Up to this point, we worked out the equations governing the dynamics of system in the asymptotic past limit and the only quantity which is left to be determined at this limit is the the canonical normalized field.
In order to read the canonical normalized field, here we determine the form of the 2nd order action at the asymptotic past limit which after using constraints and in terms of $Q_{1}$ and $Q_{2}$ can be written as
 \bea\label{2ndS-asymtotic-past}%
 \delta_2S_{_{tot}}&\simeq&\int
d\tau d^3x
 \Biggl[(1+\gamma)(Q_1'^2-k^2Q_1^2)+3\frac{(\gamma+1)}{(\gamma-2)}(Q_2'^2-\frac{(\gamma-2)}{3\gamma}k^2Q_2^2)\Biggl].
 \eea%
 Thus, our canonically normalized field is given as
 \be\label{Canonically-norm}
 Q_{_{norm}}=\sqrt{2(1+\gamma)}Q_1.
 \ee

\paragraph{\textit{The superhorizon limit:}}
In the superhorizon limit $k\tau\ll1$, \eqref{A0eq} takes the following form
\be
\label{A0eq-S-H}
\frac{Q}{\phi}+\frac{\delta}{\epsilon}\times(\Phi+\frac{\dot\Psi}{H})-\frac12\frac{g^2\phi^4}{a^4H^2}\Psi=0.
\ee
On the other hand, up to the leading terms in slow-roll parameters,  \eqref{drho-eq} leads to
\be
\label{drho-eq-S-H}
\frac{Q}{\phi}+\frac16(\epsilon-\eta)(\Phi+\frac{\dot\Psi}{H})\simeq0.
\ee
Recalling the slow-roll background relation \eqref{eta-x} ($\delta\simeq\frac16(\epsilon-\eta)\epsilon$),
and comparing \eqref{A0eq-S-H} and \eqref{drho-eq-S-H}, we obtain
\be
\label{super-cons}
\frac{Q}{\phi}\simeq-\frac16(\epsilon-\eta)\Phi\quad\textmd{and}\quad \frac{Q}{\phi}\sim\Psi.
\ee
Use of the above result in \eqref{Psi-eq} leads to the following equation for $\Phi$
\be
\label{super-Phi-eq}
\Phi''-2(\epsilon-\eta)\mH^2\Phi\simeq0.
\ee
Moreover, form \eqref{energy-conservation}, we  have the following equation for $Q$
\bea
\label{super-Q-eq}
Q''-\mH^2(2+8\epsilon+6(\epsilon-\eta))Q\simeq 0,
\eea
which indicates that the superhorizon behavior of $Q$ is similar to the Sasaki-Mukhanov variable
$v$ \eqref{SM-variable-def}. As we see, in this limit, we have only one equation for both of $Q_1$ and $Q_2$,
similarly both of $\Phi_1$ and $\Phi_2$ are described by the same equation.
Moreover, the Bardeen potentials
$\Phi$ and $\Psi$ are both constant on super-Hubble scales $(k\tau\ll1)$, similar to all the other adiabatic
perturbations.

Eq. \eqref{super-cons} reveals that at the superhorizon limit, the scalar anisotropic stress
$a^2\pi^S$ is non-vanishing and is given by
\be
a^2\pi^S\simeq-\Phi.
\ee
Before this, \eqref{piS-A-P} showed that gauge-flation has a non-zero $a^2\pi^S$ at the asymptotic past limit.
The above relation indicates that this quantity has a non-zero value also at the superhorizon, thus is an observable
quantity which we will come back to it later. This is a unique and specific feature of the non-Abelain gauge field
inflation, not shared by any scalar-driven
inflationary model.

Working out the field equations of $Q$ and $\Phi$ in asymptotic past and the superhorizon limits, in the following
we combine them and read the closed form differential equations corresponding to each field and study the system.
As we see in \eqref{Canonically-norm}, $Q$ is our quantum field, while $\Psi$ is the classical field.

\subsubsection*{$\blacktriangleright$ Quantization of the scalar perturbations}

Combining \eqref{Q+-} and \eqref{super-Q-eq},  field equations of $Q_{1,2}$ take the form
\bea
\label{Q+}
Q_{1}''+\big(k^2-\frac{z''}{z}\big)Q_{1}&\simeq &0,\\
\label{Q-}
Q_{2}''+\big(\frac{\gamma-2}{3\gamma}k^2-\frac{z''}{z}\big)Q_{2}&\simeq &0,
\eea
with the following algebraic constraint at superhorizon scales \eqref{super-cons}
\be\label{algebraic-Q-eq}
Q_{1}+Q_{2}=\mathcal{O}(\epsilon)\phi\Phi\quad \textmd{at}\quad  k\tau\ll1\,.
\ee
The effective mass term is given as $$\frac{z''}{z}\simeq (2+8\epsilon+6(\epsilon-\eta))\mH^2.$$
On the other hand, up to the leading orders in slow-roll, we have $\mH\simeq-(1+\epsilon)/\tau$,
which makes it possible to write $\frac{z''}{z}$ as
\be\label{nu-R}
\frac{z''}{z}=\frac{\nu_Q^2-\frac14}{\tau^2}\quad \textmd{where}\quad \nu_Q\simeq\frac32+2(3\epsilon-\eta).
\ee
The general solution to the equation \eqref{Q+} is a linear combination of Hankel functions
\be
Q_{1}(k,\tau)\simeq \frac{\sqrt{\pi\vert\tau\vert}}{2}e^{i(1+2\nu_Q)\pi/4}\big(q_1 H^{(1)}_{\nu_Q}(k\vert\tau\vert)+\tilde{q}_1 H^{(2)}_{\nu_Q}(k\vert\tau\vert)\big).
\ee
On the other hand, the general solution of \eqref{Q-} is
\bea\label{Q2}
Q_{2}(k,\tau)\simeq \left\{\begin{array}{cc}\frac{\sqrt{\vert\tau\vert}}{\sqrt{\pi}}\big(q_2 K_{\nu_Q}(\sqrt{\frac{\vert 2-\gamma\vert}{3\gamma}}k\vert\tau\vert)+\tilde{q}_2 I_{\nu_Q}(\sqrt{\frac{\vert 2-\gamma\vert}{3\gamma}}k\vert\tau\vert)\big), &\quad \gamma-2<0\\
\frac{\sqrt{\pi\vert\tau\vert}}{2}\big(iq_2 H^{(1)}_{\nu_Q}(\sqrt{\frac{\vert\gamma-2\vert}{3\gamma}}k\vert\tau\vert)+\tilde{q}_2 H^{(2)}_{\nu_Q}(\sqrt{\frac{\vert\gamma-2\vert}{3\gamma}}k\vert\tau\vert)\big),&\quad \gamma-2>0,
\end{array}\right.
\eea
which as we see in case that $\gamma-2<0$, it is expressed as a linear combination of modified Bessel functions,
otherwise it is expressed in terms of Hankel functions. Note that in \eqref{Q2}, the coefficients are chosen such
that in both cases, $Q_{2}$ has the same superhorizon value. Later fitting the gauge-flation with cosmic data, we will see that $\gamma>2$. Hence the region $\gamma\in(0,2)$ in which the square sound speed of $Q_2$ becomes negative, is physically uninteresting and is excluded by the cosmic data.

The undetermined coefficients $q_i$ is fixed by the initial conditions.
Note that, despite of dealing with two second order differential equations \eqref{Q+} and \eqref{Q-},
which in general needs four initial conditions to be fully determined, the algebraic constraint \eqref{algebraic-Q-eq}
relates $Q_1$ and $Q_2$ such that we need only two initial conditions, as in the single scalar field models.
Imposing the usual Minkowski (Bunch-Davis) vacuum state in the
asymptotic past limit ($k\tau\rightarrow-\infty$) of the canonical normalized field \eqref{Canonically-norm}, we obtain
\be
Q_{_{norm}}=\sqrt{2(\gamma+1)}Q_1\rightarrow\frac{e^{-ik\tau}}{\sqrt{2k}},
\ee
which corresponds to
\be\label{q1}
q_1=\frac{1}{\sqrt{2(\gamma+1)}},\quad \textmd{and} \quad \tilde{q}_1=\tilde{q}_2=0.
\ee
From \eqref{algebraic-Q-eq} and after using the asymptotic forms of $H^{(1)}_\nu(z)$ and $K_\nu(z)$ in the $z\ll 1$ limit\footnote{The $H^{(1)}_\nu(z)$ and $K_\nu(z)$ functions have the following asymptotic forms in the limit of $z\ll 1$:$$H^{(1)}_\nu(z)\simeq-\frac{i}{\pi}\Gamma(\nu)\big(\frac{z}{2}\big)^{-\nu},\quad K_\nu(z)\simeq\frac{1}{2}\Gamma(\nu)\big(\frac{z}{2}\big)^{-\nu}.$$}, one can read $q_2$ as
\be\label{q2}
q_2\simeq i\bigg(\frac{\vert\gamma-2\vert}{3\gamma}\bigg)^\frac{3}{4}q_1=\frac{i}{\sqrt{2(1+\gamma)}}\bigg(\frac{\vert\gamma-2\vert}{3\gamma}\bigg)^\frac{3}{4}.
\ee
After obtaining $q_{1,2}$ and $\tilde{q}_{1,2}$, the quantum field $Q$ is fully determined and
we turn to find the classical field $\Phi$ sourced by $Q$.

\subsubsection*{$\blacktriangleright$ Classical scalar perturbations}

To determine $\Phi=\Phi_1+\Phi_2$, we note that $\Phi_{1,2}$  are governed by
\bea
\label{Phi+}
\Phi_{1}''+\big(k^2-\frac{\theta''}{\theta}\big)\Phi_{1}&\simeq &0,\\
\label{Phi-}
\Phi_{2}''+\big(\frac{\gamma-2}{3\gamma}k^2-\frac{\theta''}{\theta}\big)\Phi_{2}&\simeq &0,
\eea
dynamical equations the solutions of which are subject to the constraint equation
\be\label{constraint-Phi-eq}
\Phi'_1\simeq(\gamma-1)\psi HQ_1,\quad \textmd{at}\quad  \Phi'_2\simeq-2\psi HQ_2\,,
\ee
in the asymptotic past limit. Here $\frac{\theta''}{\theta}=2\mH^2(\epsilon-\eta)$, which can be written as
\be\label{nu-Phi}
\frac{\theta''}{\theta}=\frac{\nu^2_{R}-\frac14}{\tau^2},\quad \textmd{where} \quad \nu_{R}\simeq\frac12+2(\epsilon-\eta).
\ee
Similar to $Q_{1,2}$, the general solution of the above equations are given as
\be
\Phi_{1}(k,\tau)\simeq \frac{\sqrt{\pi\vert\tau\vert}}{2k}\big(b_1 H^{(1)}_{\nu_R}(k\vert\tau\vert)+\tilde{b}_1 H^{(2)}_{\nu_R}(k\vert\tau\vert)\big),
\ee
and
\bea\label{Phi2}
\Phi_{2}(k,\tau)\simeq \left\{\begin{array}{cc}\frac{\sqrt{\vert\tau\vert}}{\sqrt{\pi}k}\big(b_2 K_{\nu_R}(\sqrt{\frac{\vert 2-\gamma\vert}{3\gamma}}k\vert\tau\vert)+\tilde{b}_2 I_{\nu_R}(\sqrt{\frac{\vert 2-\gamma\vert}{3\gamma}}k\vert\tau\vert)\big), &\quad \gamma-2<0\\
\frac{\sqrt{\pi\vert\tau\vert}}{2k}\big(ib_2 H^{(1)}_{\nu_R}(\sqrt{\frac{\vert \gamma-2\vert}{3\gamma}}k\vert\tau\vert)+\tilde{b}_2 H^{(2)}_{\nu_R}(\sqrt{\frac{\vert\gamma-2\vert}{3\gamma}}k\vert\tau\vert)\big),&\quad \gamma-2>0.
\end{array}\right.
\eea
In \eqref{Phi2}, the coefficients are chosen in such a way that for both cases $\Phi_2$ satisfies \eqref{constraint-Phi-eq} with the same value.
Putting \eqref{q1} and \eqref{q2} into \eqref{constraint-Phi-eq} and after using the asymptotic form of Bessel functions\footnote{The $H^{(1)}_\nu(z)$ and $K_\nu(z)$ functions have the following asymptotic forms in the limit of $z\gg 1$:$$H^{(1)}_\nu(z)\simeq\sqrt{\frac{2}{z\pi}}e^{-\frac{i\pi}{4}(2\nu+1)} e^{iz},\quad K_\nu(z)\simeq\sqrt{\frac{2}{z\pi}}e^{-z}.$$} in the $k\tau\rightarrow-\infty$, we have $\tilde{b}_1=\tilde{b}_2=0$, while
\be
b_1\simeq-\frac{(\gamma-1)}{\sqrt{2(\gamma+1)}}H\psi,\quad \textmd{and}\quad b_2\simeq \frac{2i}{\sqrt{2(1+\gamma)}}\bigg(\frac{\vert\gamma-2\vert}{3\gamma}\bigg)^{\frac14}H\psi.
\ee
Having the coefficients above and using the background slow-roll relation $\epsilon\simeq(1+\gamma)\psi^2$,  we obtain
\footnote{In the asymptotic past limit $\Phi_2$
is a mode which can have negative $c_s^2$ for $\gamma<2$. Nonetheless, our analysis above shows explicitly that this
does not render our perturbation theory analysis unstable, because what is physical is the total $\Phi$ \emph{after} imposing
the constraint equations on the \emph{superhorizon}
scales. In other words, $\Phi_2$ mode in the asymptotic past is fixed by the constraints on the dynamical equations
and not an independent mode. Note also that what matters in the CMB data, is the superhorizon field values.\label{footnote-gamma2}}
\be\label{after-horizon-Phi}
\Phi=\Phi_1+\Phi_2\simeq\frac{i\sqrt{\epsilon}}{2k^{3/2}}H\big(\frac{k\vert\tau\vert}{2}\big)^{\frac12-\nu_R},\quad k\vert\tau\vert\ll1.
\ee

We are now ready to compute the power spectrum of $\Phi$ and curvature perturbations.
The power spectrum for the Bardeen potential $\Phi$ is given by %
\be
\Delta^2_\Phi=\frac{4\pi k^3}{(2\pi)^3}|\Phi|^2\,,
\ee
which after using \eqref{after-horizon-Phi} and \eqref{nu-Phi}, has the following form on the large scales ($k\ll aH$) %
\be
\Delta^2_\Phi\simeq\frac{\epsilon}{8}\left(\frac{H}{\pi}\right)^2
\left(\frac{k|\tau|}{2}\right)^{3-2\nu_{R}}\,.
\ee
Recalling \eqref{R} and using \eqref{super-cons}, we learn that the comoving curvature perturbation is given by
${\cal R}\simeq\frac{\Phi}{\epsilon}$ in the superhorizon scales.
The power spectrum of the comoving curvature perturbation ${\cal R}$ is then given as%
\be\label{PR}
\Delta_s^2\simeq\frac{1}{8\epsilon}\left(\frac{H}{\pi}\right)^2\bigg{|}_{k=aH}\,,
\ee%
which becomes constant on super-Hubble scales. Note that the scalar power spectrum in our model is exactly equal to the power spectrum of the
comoving curvature perturbation in the standard single scalar field model \eqref{Power-single-scalar}. The spectral index of the curvature perturbations, $n_{s}-1=3-2\nu_{R}$, to the
leading order in the slow-roll parameters is%
\be\label{nr}
n_{s}-1\simeq -2(\epsilon-\eta)\,.
\ee
We note that the spectral tilt \eqref{nr} is always negative in our model.

In addition to the power spectrum of the scalar and its spectral tilt, as one of the specific feature of the
non-Abelian gauge field inflation, we have a non-zero scalar anisotropic stress value with the following power spectrum
\be\label{PpiS}
\Delta_{a\!^2\!\pi^S}^2\simeq\frac{\epsilon}{8}\left(\frac{H}{\pi}\right)^2\!\!\bigg{|}_{k=aH}\,,
\ee%
which becomes constant on super-Hubble scales and is hence a physical observable.
This is in contrast with all kinds of scalar inflationary models in the general relativity,
in which $a^2\pi^S$ is identically zero.

A non-zero $a^2\!\!\pi_S$ is a feature present in the inflationary/cosmological models with modified gravity
\cite{non-zero-pis} and causes difference between the two Bardeen potentials $\Phi$ and $\Psi$, this provides a way to trace its observable effects by affecting the structure formation analysis, for a more detailed discussion on this point see \cite{non-zero-pis} and papers referring to it.

\subsubsection*{$\bullet$ Vector modes}

In the vector sector, we have $V_i$, $W_i$, $u_i$, $v_i$ and $w_i$ which
under the action of an infinitesimal ``vector'' coordinate transformation \eqref{veccoor}
$$
x^i\rightarrow \tilde x^i=x^i+\delta x_V^i\,, \quad(\partial_i\delta x^V_i=0)\,,
$$
transform as
\be
\begin{split}
V_i\rightarrow V_i-a\delta\dot{x}_V^i\,&,\qquad W_i \rightarrow W_i-\delta x_V^i\,,\\
\end{split}
\ee
while $u_i$ and $v_i$ as our genuine gauge field fluctuations remain invariant.
On the other hand, under the vector part of infinitesimal gauge transformation \eqref{gua},
\be
u_i \rightarrow u_i-\frac1g\dot{\lambda}_V^i\,,\quad
v_i \rightarrow v_i-\frac1g\lambda_V^i\,,\quad
w_i \rightarrow w_i+\lambda_V^i\,,
\ee
while  $V_i$ and $W_i$ obviously remain invariant.

We can construct three gauge invariant divergence-free vector perturbations, one from the metric fluctuation
\be
\mathcal{Z}_i=a\dot{W}_i-V_i\,,
\ee
and two from our genuine gauge field perturbations
\bea
\mathcal{U}_i=\frac1g\dot{w}_i+u_i\,,\quad \textmd{and} \quad \mathcal{V}_i=\frac1g w_i+v_i\,.
\eea

The vector sector of the first-order gauge field strength perturbations are
\be \label{Fmunu-vector}
\begin{split}
\delta F^a_{~0i}=&\delta_a^j\partial_i(\dot{\mathcal{V}}_j-\mathcal{U}_j+\dot\phi W_j)
+\epsilon^{a~j}_{~i}(g\phi \mathcal{U}_j+g\phi^2 \dot{W}_j+\dot{\phi}w_j)\,,\\ 
\delta F^a_{~ij}=&2\epsilon^{a~~k}_{~[j}\partial_{i]}(g\phi \mathcal{V}_k+g\phi^2 W_k)+2g\phi^2\delta^a_{~\;[i}w_{j]}\,.
\end{split}\ee
As mentioned before, $\delta F^a_{\mu\nu}$ are not gauge-invariant,
but transform as $F^a_{\mu\nu}\to F^a_{\mu\nu}-\epsilon^a_{\ bc} \lambda^b F^c_{\mu\nu}$.
Thus,  $w_i$ terms in the field strength tensor above are pure gauge terms such that
all the physical quantities are independent of $w_i$.

In order to investigate the dynamics of this sector we first work out vector parts of
the perturbed energy-momentum tensor, $\delta q_i^V$ and
$\pi^V_i$. At the linear order in perturbation theory only the vector perturbations contribute to the vector part of energy-momentum tensor perturbations, thus we have%
\bea\label{deltaq}
\delta q^V_i&=&-2\frac{g^2\phi^3}{a^2}\left(\mathcal{U}_i+\frac{\phi}{a}\mathcal{Z}_i\right)+\frac{g\phi^2}{a^2}
\left(\nabla\times(\dot{\vec{\mathcal{V}}}-\vec{\mathcal{U}})\right)_i
-\frac{g\phi\dot{\phi}}{a^2}\left(\nabla\times\vec{\mathcal{V}}\right)_i\,,\\
\label{piv}a\pi_i^V&=&\frac{g^2\phi^3}{a^3}\mathcal{V}_i
+\frac{\dot{\phi}}{a}(\mathcal{U}_i-\dot{\mathcal{V}}_i)\,. \eea
As mentioned before, both of them are gauge-invariant, hence can be written in
terms of gauge invariant variables.

The perturbed Einstein equations involves two vector equations, one constraint and one dynamical equation,
given as \eqref{firstV-2}%
\bea \label{firstV}
&~&\partial_i\left(2a^2\pi^V_j-\frac{1}{a}(a^2\mathcal{Z}_j\dot{)}\right)=0\,,\\\label{delta-q-V}
&~&2a\delta q_i^V+\nabla^2\mathcal{Z}_i=0\,.
\eea
On the other hand, dealing with three unknowns,
the two Einstein equations are not enough to fully determine the system and we need one more equation, which as discussed
in section \ref{section-5.6.1} is provided by the vector part of the second order action is the field equation of $\delta A^a_{~0}$.
Explicitly, once we write down the second-order action for the gauge field perturbations, the momentum
conjugate to $u_i$ is vanishing (the vector part of \eqref{constraint-2nd-order-action}), yielding
\be\label{2nd-vector}
-2\frac{g^2\phi^3}{a^2}(\mathcal{U}_i+\frac{\phi}{a}\mathcal{Z}_i)
+\frac{g\phi^2}{a^2}\big(\vec\nabla\times(\dot{\vec{\mathcal{V}}}
+\frac{\phi}{a}\vec{\mathcal{Z}})\big)_i-\frac{g\phi\dot\phi}{a^2}(\vec\nabla\times\vec{\mathcal{V}})_i
-\frac{\phi}{a^2}\nabla^2(\mathcal{U}_i-\dot{\mathcal{V}}_i)=0.
\ee
Using \eqref{delta-q-V}, the above equations leads to the following equation
\be\label{vec-const}
\frac{g\phi^2}{a^2}\big(\vec\nabla\times(\vec{\mathcal{U}}+\frac{\phi}{a}\vec{\mathcal{Z}})\big)_i-
\frac{\phi}{a^2}\nabla^2(\mathcal{U}_i-\dot{\mathcal{V}}_i)+\frac{1}{2a}\nabla^2{\cal Z}_i=0 \,.
\ee
This completes the set of equations we need for solving vector perturbations.
Then, the combination of \eqref{firstV}-\eqref{delta-q-V} and \eqref{vec-const} indicates that $\cal Z$ is damping exponentially during the inflation.
From the combination of  \eqref{deltaq} and \eqref{firstV}, we then learn that  $\mathcal{Z}_i$ vanishes
after horizon crossing.

 The above is the usual result of the scalar-driven inflationary models that the vector modes are
diluted away by the (exponential) accelerated expansion of the Universe during inflation. In our model, despite
of having vector gauge fields as inflaton, the power spectrum of the vector modes are unimportant in inflationary cosmology.

\subsubsection*{$\bullet$ Tensor modes}

The tensor perturbations $h_{ij}$ and $\tilde{h}_{ij}$, as symmetric, traceless and divergence-free tensors, are both gauge invariant with two degrees of freedom.
 The contribution of these modes to the linear order perturbed gauge field strength is
\be\label{Fmunu-tensor}%
\begin{split}
\delta F^a_{~0i}=&\delta^{aj}\bigg(\phi(\tilde{h}_{ij}+\frac12 h_{ij})\dot{\bigg)}\,,\\
\delta F^a_{~ij}=&2\phi\delta^{ak}(\partial_{[i}\tilde{h}_{j]k}+\frac12\partial_{[i}h_{j]k})-2g\phi^2\epsilon^{ak}_{~~~[j}(\tilde{h}_{i]k}-\frac12h_{i]k})\,.
\end{split}
\ee

The perturbed Einstein equations involve one equation for $h_{ij}$, sourced by the contribution of $\tilde{h}_{ij}$ to the energy-momentum tensor, reads as \eqref{T}%
\be \label{T-gf}
\ddot{h}_{ij}+3H\dot{h}_{ij}+\frac{k^2}{a^2}h_{ij}=2\pi^T_{ij}\,.
\ee
{}Computing the linear order energy-momentum tensor, we obtain $\pi^T_{ij}$%
\be \label{piT}
\pi^T_{~ij}=\bigg(2(\frac{g^2\phi^4}{a^4}-\frac{\dot{\phi}^2}{a^2})\tilde{h}_{ij}-\frac{\dot{\phi}\phi}{a^2}(2\dot{\tilde{h}}_{ij}
+\dot{h}_{ij})+\frac{g\phi^3}{a^3}\partial_k\big(2\epsilon^{kl}_{~~(i}\tilde{h}_{j)l}+\epsilon^{kl}_{~~(i}h_{j)l}\big)\bigg)\,. \ee
This equation and the equation for the tensor perturbation of the gauge field $\tilde h_{ih}$
is provided by the second order action of the tensor modes which after using
the slow-roll approximation $(\dot{\phi}\simeq H\phi)$, is given as\footnote{A similar result and feature is also obtained in the chromo-natural model \cite{Adshead-Wyman-Martinec-2013}.}
\bea
\label{2ndts}
&~&\hspace*{-7mm}\delta S^{(2)}_T\simeq\frac{1}{2}\int d^3x dt a^3\biggl(\frac{1}{4}(1+\frac{\phi^2}{a^2})(\dot{h}_{ij}^2-(\frac{1}{a^2}{\partial_k h_{ij}} )^2)+\big(-6\frac{\dot{\phi}^2}{a^2}\tilde{h}_{ij}-\frac{\dot{\phi}\phi}{a^2}(2\dot{\tilde{h}}_{ij}
+\dot{h}_{ij})\big)h_{ij}-\frac32\frac{\dot\phi^2}{a^2}h^2_{ij}\nonumber\\
&+&\frac{\phi^2}{a^2}(\dot{h}_{ij}+\dot{\tilde{h}}_{ij})\dot{\tilde{h}}_{ij}-2(2\frac{\dot\phi^2}{a^2}
+\frac{g^2\phi^4}{a^4})\tilde{h}_{ij}^2-\frac{\phi^2}{a^4}{\partial_k}(h_{ij}+\tilde{h}_{ij}){\partial_k}
\tilde{h}_{ij}+\frac14((\frac{\kappa g\phi^2\dot\phi}{a^3}\dot{)}+2\frac{g\phi}{a})\frac{\phi^2}{a^3}\epsilon^{ijk}h_{kl}{\partial_i }h_{jl}\nn\\
&+&((\frac{\kappa g\phi^2\dot\phi}{a^3}\dot{)}-2\frac{g\phi}{a})\frac{\phi^2}{a^3}\epsilon^{ijk}\tilde h_{kl}{\partial_i }\tilde h_{jl}+\frac12(\frac{\kappa g\phi^2\dot\phi}{a^3}\dot{)}\frac{\phi^2}{a^2}\epsilon^{ijk}(\tilde h_{kl}\partial_i \tilde h_{jl}+h_{kl}\partial_i \tilde h_{jl})
\biggr)\,.
\eea

Having two degrees of freedom, each of $h_{ij}$ and $\tilde{h}_{ij}$ can be expressed in terms of the
standard plus and cross polarizations which are the eigenvalues of $\nabla^2$ ($\nabla^2 e^{+,\times}_{~ij}=-k^2e^{+,\times}_{~ij}$).
However, they are not eigenvalues of parity-violating operator ( $e^{ijk}\partial_k$) and in order  to decouple two degrees of freedoms in each of $h_{ij}$ and $\tilde{h}_{ij}$, we need to write them in terms of the right and left circular polarizations.  We can use the following parametrization for $h_{ij}$
\bea
h_{ij}=\frac{1}{2a}
  \begin{pmatrix}
   h_{_{R}}+h_{_{L}}& ~-i( h_{_{R}}-h_{_{L}})&~0 \\
   \\
  -i( h_{_{R}}- h_{_{L}})  & ~ -( h_{_{R}}+h_{_{L}})&~0\\
   \\
 ~ 0& ~0&~0
  \end{pmatrix}
\eea
where working with Fourier modes, we chose  $k^i=(0,0,k)$ and imposed the transversality condition.
In a similar way, one can parameterize $\tilde h_{ij}$ and $\pi^T_{ij}$ in terms of right and left circular polarizations $\tilde{h}_{_{R,L}}$ and $\pi^T_{_{R,L}}$.

In terms of $h_{_{R,L}}$ and $\tilde{h}_{_{R,L}}$,  \eqref{T-gf} reads
\bea\label{h}
h_{_{R,L}}''+\big(k^2-(2-\epsilon)\mH^2\big)h_{_{R,L}}\simeq 2a^2\pi^T_{_{R,L}},
\eea
where
\be
a^2\pi^T_{_{R,L}}\simeq\psi^2(-2\mH \tilde h_{_{R,L}}'+2\gamma \mH^2 \tilde{h}_{_{R,L}}
\mp2k\mH\sqrt{\gamma}(\tilde h_{_{R,L}}+\frac12 h_{_{R,L}})-\mH h_{_{R,L}}'+\mH^2h_{_{R,L}}).
\ee
Here $\tau$ is the conformal time $dt=a d\tau$, $\psi=\phi/a$ is the background effective inflaton field, $\mH=\dot{a}$, prime denotes
derivative with respect to the conformal time, and $\gamma\equiv\frac{g^2\psi^2}{H^2}$.
Equation \eqref{h} implies that in the superhorizon limit, we have $h_{_{R,L}}\propto a$.

On the other hand, the field equations of $\tilde{h}_{_{R,L}}$ are obtained from the second order action \eqref{2ndts} which after using \eqref{phi-eom} leads to
\bea\label{eq-tildeh}
\tilde{h}_{_{R,L}}''+\left(k^2+\big(2(1+\gamma)+\epsilon\big)\mH^2\mp2k\mH\frac{(1+2\gamma)}{\sqrt{\gamma}}\right)\tilde{h}_{_{R,L}}\simeq
 \mH(h_{_{R,L}}'-\mH h_{_{R,L}}) \pm k\mH\frac{(\gamma+1)}{\sqrt{\gamma}}h_{_{R,L}}.~~~~~
\eea
Note that using the superhorizon scale behavior of $h_{_{R,L}}$ ($h_{_{R,L}}\propto a$),
from \eqref{eq-tildeh}, we learn that while $\tilde{h}_{_{R,L}}$ behaves like a plane-wave at subhorizon scales,
it is exponentially damped like $\tilde{h}_{_{R,L}}\propto a^{-(1+\gamma)}$ at superhorizon scales. However, the parity violating term plays an important role just before the horizon crossing, leading to tachyonic growth of $\tilde{h}_{R}$ around the horizon crossing.

\begin{figure}[t]
\begin{center}
\includegraphics[angle=0, width=80mm, height=70mm]{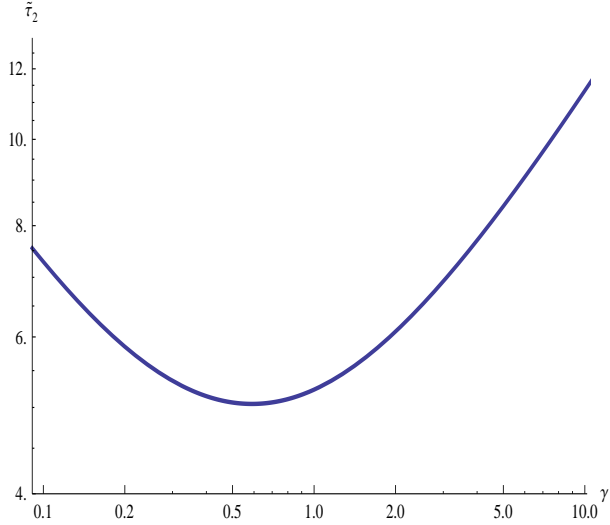}
\caption{$\tilde{h}_{R}$ undergoes a tachyonic growth phase in $\tilde{\tau}_2(\gamma)\geq -k\tau \geq 1$ (\emph{cf}.  \eqref{tilde-h-simple} and \eqref{tilde-h-omega}). In this figure, we have depicted $\tilde{\tau}_2$ vs. $\gamma$. The minimum is $\tilde\tau=5$ which is at $\gamma\simeq0.6$.}\label{omega-root-fig}
\end{center}
\end{figure}

\begin{figure}
\includegraphics[angle=0, width=85mm, height=85mm]{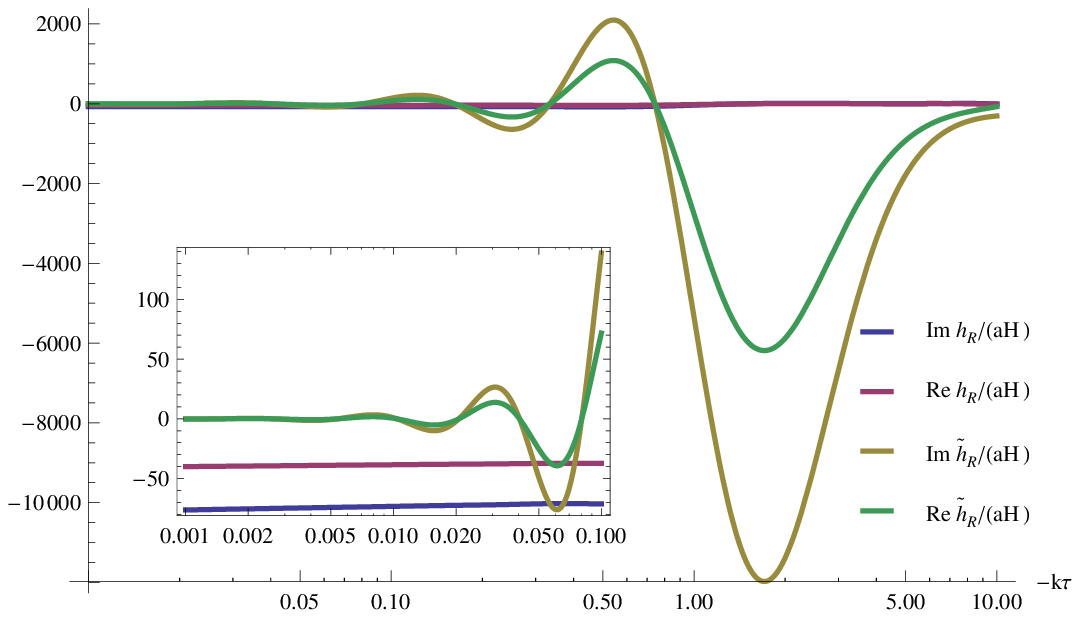}
\includegraphics[angle=0, width=85mm, height=80mm]{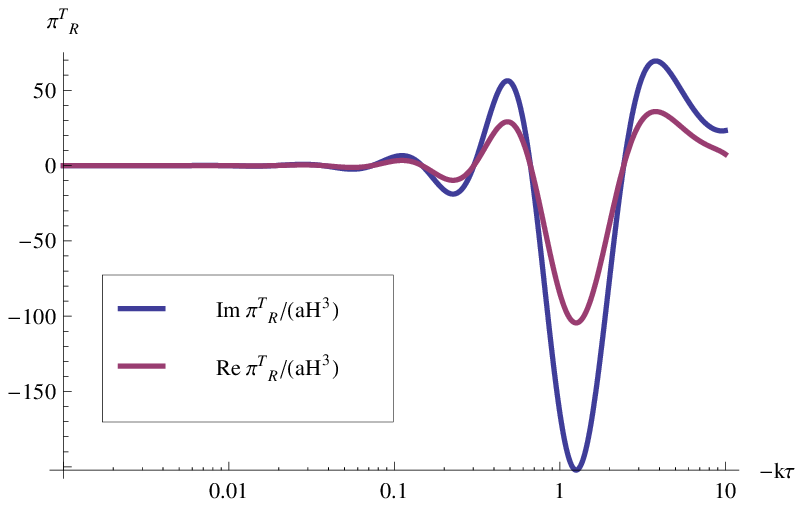}\\
\includegraphics[angle=0,width=85mm, height=80mm]{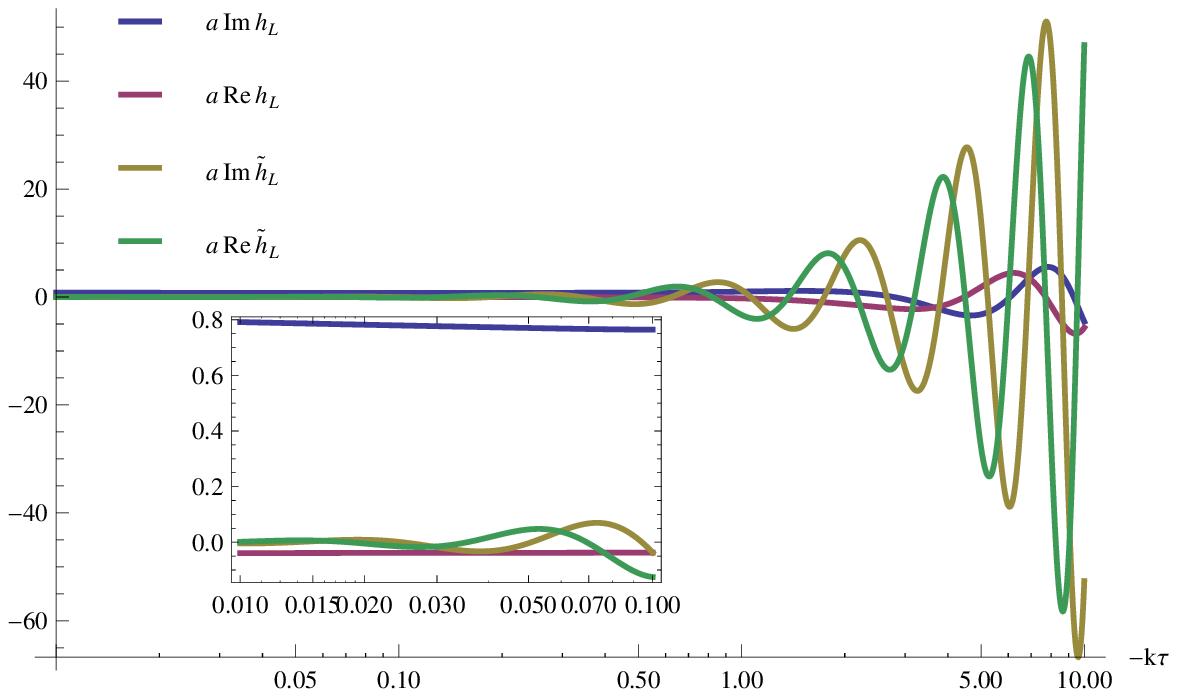}
\includegraphics[angle=0,width=80mm, height=80mm]{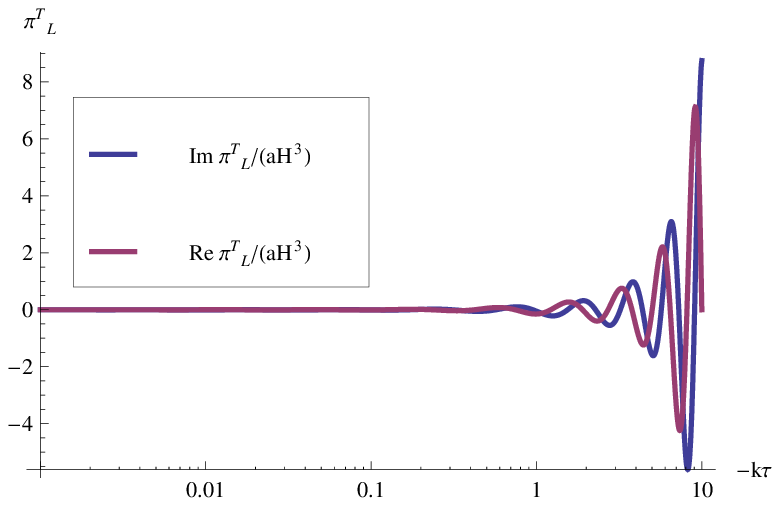}
\caption{This figure presents the tensor modes solution for $\psi=5\times 10^{-2}$, $\gamma=10$ and $H_0=10^{-6}$. In the top-left panel, we have the tensor field values $\frac{h_{_{R}}}{aH}$ and $\frac{\tilde h_{_{R}}}{aH}$ versus $-k\tau$, where \textit{Re} and \textit{Im} denote read and imaginary parts of the corresponding quantity. The small box presented the superhorizon behavior of the fields. The top-right panel shows $\frac{\pi^T_{_{R}}}{aH^3}$. In the bottom panels we presented the left-handed polarizations.}\label{Tensor-Mode-figures}
\end{figure}

Before the superhorizon scales, we can neglect $h_{R,L}$ terms in the RHS of \eqref{eq-tildeh}, which leads to the follwoing wave equation for $\tilde{h}_{R,L}$
\be\label{tilde-h-simple}
\partial^2_{\tilde{\tau}}\tilde{h}_{R,L}+\omega^2_{R,L}(\tilde{\tau},\gamma)\tilde{h}_{R,L}\simeq0,
\ee
where $\tilde{\tau}=-k\tau$ and $\omega^2_{R,L}(\tilde{\tau},\gamma)$ is given as
\be\label{tilde-h-omega}
\omega^2_{R,L}(\tilde{\tau},\gamma)=\big(1+\frac{2(1+\gamma)}{\tilde{\tau}^2}\mp2\frac{(1+2\gamma)}{\sqrt{\gamma}\tilde{\tau}}\big).
\ee
While $\omega^2_{L}$ is always positive, there is an interval $\tilde{\tau}\in(\tilde\tau_1,\tilde\tau_2)$  in which $\omega^2_{R}$ becomes negative. This short interval which has a negative $\omega^2_{R}$, leads to the tachyonic growth of $\tilde{h}_{R}$ (Fig. \ref{Tensor-Mode-figures}).  In Fig. \ref{omega-root-fig}, we presented  $\tilde\tau_2$ vs. $\gamma$ and $\tilde\tau_1$ is almost one.

The source term $\pi^T_{_{R,L}}$ (the RHS of \eqref{h}) vanishes at superhorizon scales $k\tau\ll1$. Nonetheless,  due to the tachyonic growth of $\tilde{h}_R$ at the vicinity of the horizon-crossing, $\pi^T_{_{R}}$  has the behavior of an impulse function in that region (see Fig. \ref{Tensor-Mode-figures}), inducing the growth in $h_{_{R}}$ and enhancing its superhorizon value. On the other hand, $\pi^T_{_{L}}$ is small at the horizon crossing and has negligible effect on the superhorizon value of $h_{_{L}}$.

\begin{figure}[t]
\includegraphics[angle=0, width=80mm, height=70mm]{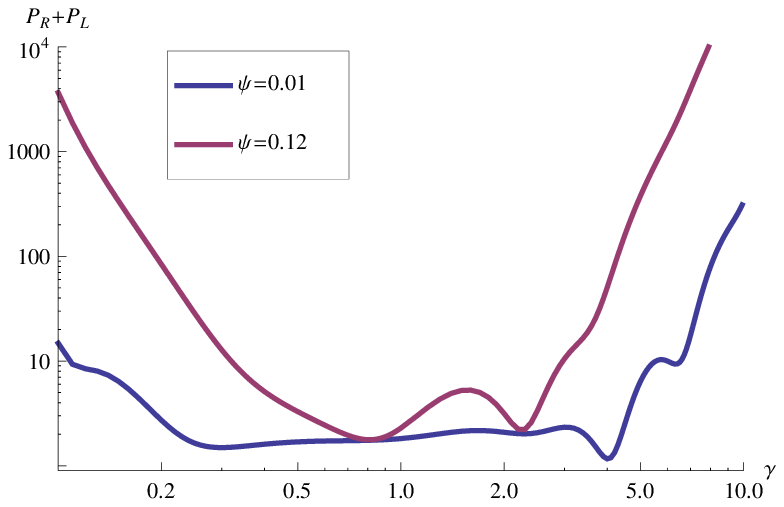}
\includegraphics[angle=0, width=80mm, height=70mm]{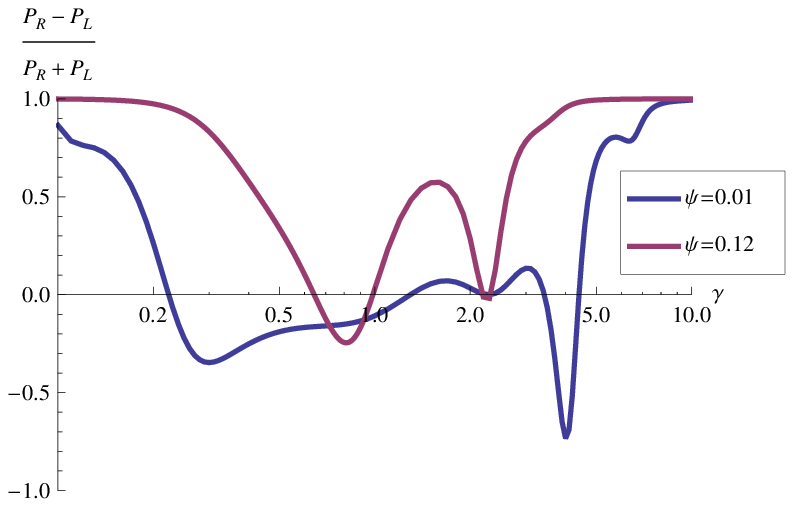}\\
\caption{In the left panel we have depicted the sum of Left and Right gravity wave power spectra ${P_{_{R}}}+{P_{_{L}}}$. In the standard scalar-driven inflationary models ${P_{_{R}}}={P_{_{L}}}=1$. The right panel the parity violating factor $\frac{P_{_{R}}-P_{_{L}}}{P_{_{R}}+P_{_{L}}}$ versus $\gamma$ for $\psi=10^{-2}$ and $\psi=0.12$ is shown. The power spectra have been calculated at $k\tau=-0.01$, long enough after modes have crossed the horizon and behave quite classically. As we see in the right panel,  for very small and very large $\gamma$ values $P_{_{R}}\gg P_{_{L}}$.}\label{Tensor-Mode-power-figures}
\end{figure}

In the leading  slow-roll approximation  the standard Minkowski (Bunch-Davis)
vacuum normalization for the canonically normalized fields (\emph{cf}. \eqref{2ndts})
takes the form
\be
h_{_{R,L}}\rightarrow\frac{e^{-ik\tau}}{\sqrt{k}}\quad \textmd{and} \quad\tilde{h}_{_{R,L}}\rightarrow\frac{e^{-ik\tau}}{2\psi\sqrt{k}}
\,,\qquad k\tau \to -\infty\,.
\ee
The power spectra for the Left and Right moving gravitational wave modes is then obtained to be
\bea\label{PT}%
\Delta_{T_{R}}^2&\simeq & P_{_R} \left(\frac{H}{\pi}\right)^2\!\!\bigg\vert_{k=aH},\\%
\Delta_{T_{L}}^2&\simeq & P_{_L} \left(\frac{H}{\pi}\right)^2\!\!\bigg\vert_{k=aH}\,,%
\eea%
where $P_{_R},\ P_{_L}$ are  functions of the parameters of the slow-roll background, in particular $\gamma, \psi$. Numerical analysis reveals that $P_{_L}$ is a function very close to one (ranging from $1.0$ at low $\gamma$ to $1.25$ at $\gamma=10$) while $P_{_R}$ varies quite considerably in its range. Since their explicit analytic expression is not illuminating we have only presented their ratio as function of $\gamma$ has been depicted in Fig. \ref{Tensor-Mode-power-figures}. In our model, we hence expect to see birefringent gravity waves.

Moreover, the spectral index of tensor perturbations, $n_T$ is given by %
\be %
n_T\simeq -2\epsilon\,, %
\ee%
which as expected are equal to their corresponding quantities in the standard scalar inflationary models,
given in \eqref{power-tensor-single-scalar} and \eqref{tilt-tensor-single-scalar}.
Note that due its exponential suppression on the superhorizon scales, the $\tilde{h}$ mode does not contribute to
the tensor power spectrum.
For our model tensor-to-scalar ratio $r$ is%
\be\label{ratio}%
r=8(P_{_R}+{P_{_L}})\epsilon\,,%
\ee
which can be written as $r=-4(P_{_R}+{P_{_L}})n_T$. That is, our model respects a modified version of the Lyth consistency relation \cite{Lyth:1996im}. This feature is also shared by the chromo-natural model \cite{Adshead-Wyman-Martinec-2013} to be reviewed in the
next section.

\subsubsection{Summary of gauge-flation cosmic perturbation theory }\label{gauge-flation-perturbation-summary}
Let us summarize the main observational predictions of gauge-flation.
Although it can be described by an effective scalar field at the level of the background,
at the perturbation level the gauge-flation model deals with many fields and has specific and interesting features:
 \begin{itemize}
 \item{Scalar modes: we have 5 gauge-invariant scalar modes, two of which are the standard Bardeen potentials and three of them are coming from the gauge field perturbations. As has been explicitly shown here and in \cite{arXiv:1102.1932}, due to the different constraint structure that we are dealing with, our scalar perturbations are adiabatic. In other words, similar to the scalar-driven inflationary models, all of the 5 physical scalar fields gets constant at the superhorizon scales. On the other hand, being a non-Abelian gauge field inflationary model, gauge-flation predicts a non-zero scalar anisotropic stress $a^2\pi^S\neq0$, in contrast with the scalar-driven models.}
\item{Vector modes: there are 3 gauge-invariant vector modes which are exponentially damping, similar to the scalar-driven inflationary models.}
 \item{Tensor modes: we have two tensor modes, one of which is the usual metric tensorial part, while the other is coming from the contribution of the gauge field. The former gets constant at the superhorizon scales, however, the later is diluted away at the superhorizon limit, thus unimportant in inflationary cosmology. Due to the existence of parity violating terms in the tensor sector,the right-handed and left-handed circular polarizations are not equal, but the right-handed mode is enhanced by the large value of its corresponding tensor anisotropic inertia $\pi^T$ at the horizon crossing.}

\item It is instructive to state the above in a different wording: In our gauge-flation action we have metric and $SU(2)$ gauge fields, which in the standard flat background terminology these yield one divergence-free traceless two-tensor (the graviton) and three transverse gauge fields $W_\mu^a$'s, since each gauge field has two polarizations, altogether we have eight propagating degrees of freedom. On the FLRW inflationary background and in our decomposition, where we have \emph{identified} the gauge (internal) and spacial indices, the graviton appears as before, while the three gauge fields will further decompose into a propagating transverse vector, two propagating scalars and one propagating tensor. Explicitly, let us focus on $W_\mu^a$. The temporal components may just be set to zero, and hence we deal with $W_i^a$, such that $\nabla^i W_i^a=0$. The propagating vector is  the part of $W_i^a$, ${\tilde W}_i$, where $W_i^a=\delta^{ja}\partial_j{\tilde W}_i$. The remaining ${\hat W}_i^a$ is divergence-free on both of $i$ and $a$ indices and may hence be decomposed as
    $$ {\hat W}_i^a=\delta_i^a W_1+\epsilon_i^{\ aj}\partial_j W_2+{\tilde h}_i^a\,,$$
    where $W_1, W_2$ are the two propagating scalars and ${\tilde h}_i^a$ is the divergence-free traceless tensor. One can directly identify these propagating degrees of freedom in terms of our earlier decompositions: The two scalars $W_1, W_2$ correspond to the two modes $Q_1, Q_2$ introduced in \eqref{the-two-scalar-modes}. The propagating vector is a linear combination of $\mathcal{U}_i$ and $\mathcal{V}_i$ which is governed by \eqref{firstV}-\eqref{delta-q-V} and \eqref{vec-const}. (Note that these equations only involve first time derivatives of $\mathcal{U}_i$ and $\mathcal{V}_i$, nonetheless they are mixing the two and once we eliminate one of the variables we end up with a single mode with second-order time derivative equation.) Finally, the tensor mode is exactly the mode governed by \eqref{eq-tildeh}.
\end{itemize}

 In the following table we have recollected all the results of this subsection and compared them with their
corresponding values in the standard single scalar model which has been reviewed in section \ref{Inflation-review-section}.

\begin{center}
\begin{tabular}{|cp{2cm}|c p{3cm}|c p{3.2cm} |}
\hline
&  & & Gauge-fation $~~~~~$model &  & Single-scalar $~~~~$model \\ [2.5ex]\hline
& $\Delta_s^2$ & &$\frac{1}{8\pi^2\epsilon}\left(\frac{H}{\mpl}\right)^2$ & &$\frac{1}{8\pi^2\epsilon}\left(\frac{H}{\mpl}\right)^2$ \\ [2.5ex]
\hline
& $n_s-1$ & & $-2(\epsilon-\eta)$ &  &$-4\epsilon+2\eta$ \\ [2.5ex]
\hline
& $r$ & & $8(P_{_R}+{P_{_L}})\epsilon$ & &$16\epsilon$ \\ [2.5ex]
\hline
& $n_T$ & & $-2\epsilon$ & &$ -2\epsilon$ \\ [2.5ex]
\hline
& $\Delta_{a\!^2\!\pi^S}^2$ & ~ & $\frac{\epsilon}{8\pi^2}\left(\frac{H}{\mpl}\right)^2$ & & identically zero \\ [2.5ex]
\hline
\end{tabular}\label{gauge-flation-summary-table}
\vskip 0.5 cm
\textbf{Table II: Gauge-flation summary results}
\end{center}
As we see, these two modes predict the same scalar power spectra, but different scalar
spectral tilts and tensor power spectra footnote{ Note that $n_s$ is written in terms of the Hubble slow-roll parameter $\eta$
\eqref{epsilon-eta-def}.}.
Moreover, as unique and specific features of the non-Abelian gauge field inflation, not shared by any scalar-driven
inflationary model, gauge-flation predicts \textit{non-zero}  scalar and tensor anisotropic inertias, $a^2\pi^S$ and $a^2\pi^T$. Regardless of the details, the anisotropic inertia $a^2\pi^S$ is \textit{ identically zero} in all   scalar inflationary
models in the context of general relativity \cite{Inflation-Books} and a non-zero anisotropic
inertia power spectrum $\Delta_{a\!^2\!\pi^S}^2$ in our mode is a  consequence of having gauge fields in the
system. As we see for our model
\be
\frac{\Delta_{a\!^2\!\pi^S}^2}{\Delta_s^2}=\epsilon^2\,.
\ee
As we see, the anisotropic inertia power spectrum, unlike what is usually perceived, is not necessarily
attributed to modified gravity and in particular Galileon models \cite{Galileon-cosmology} and these models can have different predictions for
$\Delta_{a\!^2\!\pi^S}^2/\Delta_s^2$, than the gauge-flation. It is desirable to study the anisotropic stress and its cosmological
observable consequences more thoroughly.

The existence of a non-zero tensor anisotropic inertia $\pi^T_{_{R,L}}$ with different values for $\pi^T_{_{R}}$ and $\pi^T_{_{L}}$ at the horizon crossing leads to the interesting effect of parity violating gravity wave power spectra, i.e.  the right-handed polarization of the tensor modes gets enhanced  while the left-handed mode is almost equal to the tensor modes in the standard scalar driven inflationary models. The observational consequences of such parity violation on CMB has been discussed in \cite{Lue:1998mq,Alexander:2004wk}.

\subsection{Confronting with CMB observations}\label{testing-gauge-flation-section}

Having worked out the basic observables of the gauge-flation model, i.e.  power spectrum of curvature perturbations, spectral tilt, power spectrum of gravity waves and its tilt, we are now ready to confront the model with the data and restrict the parameter space of the model using the WMAP7 results \cite{Komatsu:2010fb}, reviewed and summarized in section \ref{Inflation-review-section}. First, we note that in order for inflation to solve the flatness and horizon problems it should have lasted for a minimum number of e-folds $N_e$. This amount of course depends on the scale of inflation and somewhat to the details of physics after inflation ends \cite{Inflation-Books}. However, for a large inflationary scale, like $H\sim 10^{-4}-10^{-5}\mpl$, it is usually demanded that $N_e\simeq 60$.
As for the CMB data, one may use the best-fit values given in \eqref{PR-WMAP7}-\eqref{r-bound-WMAP7}. As is seen from Fig. \ref{ns-r}, however the range for $n_s$ in the best-fit values, is associated with $r=0$ case. However, to perform a more precise analysis, we consider the allowed parameter space in $n_s-r$ plane depicted in Fig. \ref{ns-r}, together with the COBE normalization for $\Delta^2_s$.

\begin{figure}[h]
\begin{center}
\includegraphics[angle=0, width=90mm, height=65mm]{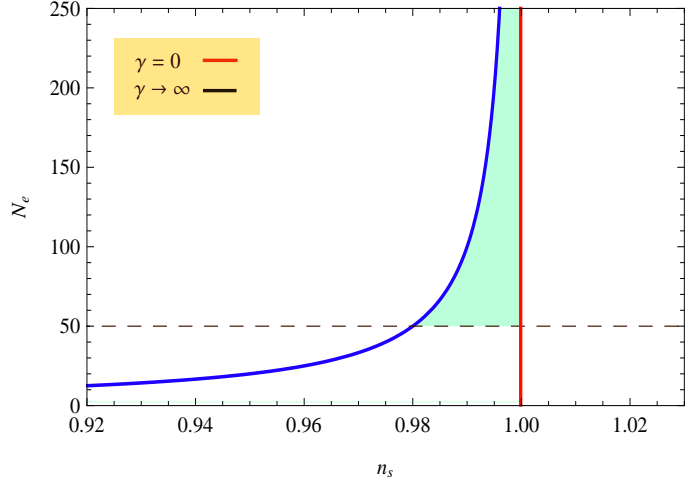}
\caption{The shaded region exhibits the region which leads to $N_e>50$. As we see this may happen for any value of $\gamma$ parameter. This also shows that our spectral tilt is always in $0.98\leq n_s< 1$ range. Moreover, our model allows for arbitrary large $N_e$.}\label{fig-ns-vs-Ne}
\end{center}
\end{figure}

To perform the analysis, we use the results obtained in the slow-roll regime. We then note that the gauge-flation data are summarized in Table II are functions of $\epsilon, \eta$ and $H$. Number of e-folds $N_e$ \eqref{Ne-gauge-flation} may also be written in terms of these three recalling that $\gamma\simeq (\epsilon-\eta)/\eta$. Note also that $H$ appears only through the COBE normalization and its value may be fixed once we obtain the bound on $\epsilon$ and $\eta$ using the other data. So for the moment we focus on $n_s,\ N_e,\ r$ and choose to work with $\epsilon$ and $\gamma$ as variables to perform analysis. Recalling our previous results we have
\begin{align}
n_s&=1-\frac{f(\gamma)}{N_e}\,, \qquad f(\gamma)\equiv \gamma \ln(\frac{\gamma+1}{\gamma})\leq 1\,\\
n_s&=1-\frac{r}{4(P_{_{R}}+P_{_{L}})} \frac{\gamma}{\gamma+1}\,.
\end{align}
Since $0\leq f(\gamma)\leq 1$, $n_s\geq 1-\frac{1}{N_e}$. That is, in gauge-flation spectrum cannot be very red. For example, as depicted in \ref{fig-ns-vs-Ne}, for $N_e\geq 50$,
\be
n_s^{(50)}\geq 0.98\,,
\ee
and of course we have a less red spectrum (i.e. lowest value for $n_s$) if we demand larger $N_e$.

\begin{figure}[h]
\includegraphics[angle=0, width=90mm, height=70mm]{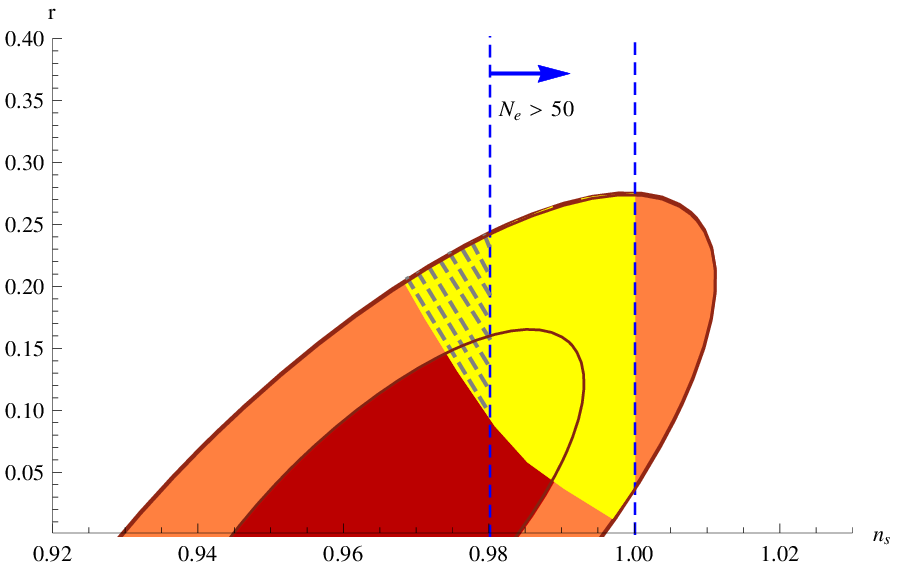}
\includegraphics[angle=0, width=80mm, height=80mm]{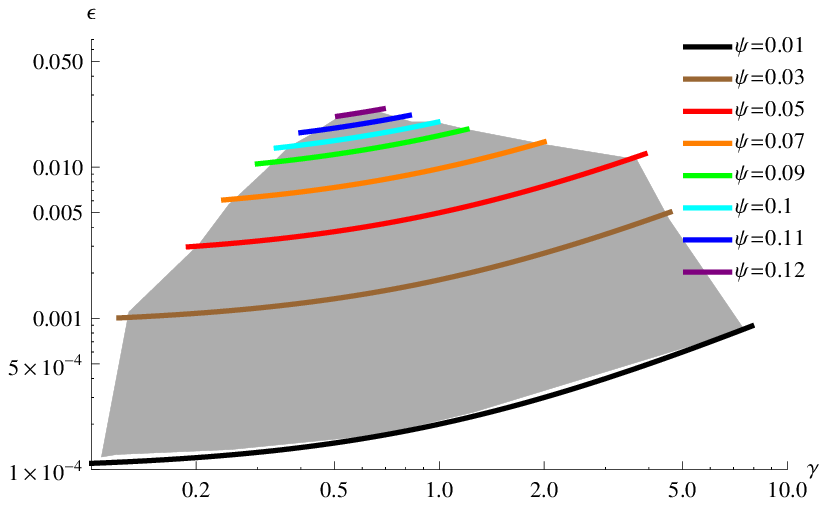}
\caption{The left panel shows  $1\sigma$ and $2\sigma$ contour bounds of 7-year WMAP+BAO+H0. The yellow area (region with lighter color) represents the gauge-flation predictions for $\psi\in(0.01,0.12)$ range. As depicted in Fig. \ref{fig-ns-vs-Ne}, the region with enough number of e-folds restricts us to $n_s>0.98$ region, that is on the right-side of the $N_e=50$ line. Therefore, the allowed region is the highlighted region between $N_e=50$ and $n_s=1$ lines. The shaded region in right panel shows the allowed values for $\epsilon$ and $\psi$, given in \eqref{psi-range} and \eqref{bound-epsilon-gamma}. }\label{ns-r-gf}
\end{figure}

Similarly, as depicted in the left panel of Fig. \ref{ns-r-gf}, we see that  our model predicts a minimum value for $r$
\be\label{r-bound-gf}
0.02\leq r\leq 0.28\,.
\ee
This is a very specific prediction of our model and gauge-flation may be falsified by the upcoming Planck satellite results.
In the allowed region
\be\label{psi-range}
\psi\simeq (0.01-0.1)\mpl,
\ee
where the \textit{max}   and \textit{min} possible values of $r$  respectively correspond to $(\psi=0.01, \gamma=5, P_{_{R}}+P_{_{L}}=6.3)$ and $(\psi=0.01, \gamma=8, P_{_{R}}+P_{_{L}}=77)$. Moreover , in the
right panel of Fig. \ref{ns-r-gf}, we present the allowed region in terms of $\epsilon$ and $\gamma$ which indicates that
\be\label{bound-epsilon-gamma}
\epsilon=(10^{-4} - 2\times 10^{-2}),\quad \gamma=(0.1 - 8).
\ee
Using the above and the COBE normalization, we can read the value of $H$:
\be
\left(\frac{H}{\mpl}\right)^2=\frac{r~\pi^2}{P_{_{R}}+P_{_{L}}} \Delta_s^2=2\times 10^{-7}\epsilon,\nonumber
\ee
and hence
\be\label{H-value-gf}
H=(0.45 - 6.3)\times 10^{-5}\ \mpl\,.
\ee

To determine the two parameters of our model,  $\kappa$ and $g$, we can read them in terms of the parameter $\gamma$
\bse\label{psi-kappa-g}
\begin{align}
\gamma=\frac{g^2\psi^2}{H^2}\quad \Rightarrow\quad  \frac{g^2}{4\pi}=& 2\pi\Delta_s^2\gamma(\gamma+1)=1.5\times 10^{-8} \gamma(\gamma+1)\,,\\
\kappa g^2\psi^6 \simeq 2 \quad \Rightarrow \quad \kappa\simeq&\frac{2(1+\gamma)^2}{\epsilon^2H^2\gamma}=10^{7} \times \frac{(\gamma+1)^2}{\gamma\epsilon^3}.
\end{align}
\ese
As we see from the left panel of figures \ref{ns-r-gf}, the value of $\gamma$ is restricted as \eqref{bound-epsilon-gamma}, which gives
\be
g\simeq(0.15 - 3.7)\times 10^{-3},\qquad \Lambda\sim (10^{-5}-10^{-4})\mpl\,,\quad \kappa\equiv \Lambda^{-4}\,.
\ee
 As we see from (\ref{psi-kappa-g}a), smallness of the gauge coupling is directly related to the smallness of the power spectrum of CMB curvature fluctuations. This is a notable and interesting feature of our model.

Nonetheless, if we restrict ourselves to 1$\sigma$ contour in the left panel of Fig. \ref{ns-r-gf}, we find more stringent bounds on $r$, $n_s$ and hence $H$:
\be\label{2sigma-values}
0.98 \leq n_s\leq 0.99,\qquad 0.05<r<0.15,\qquad H\simeq (3.4-5.4)\times 10^{-5}\mpl,
\ee
which leads to the following bounds for $\gamma$ and $\psi$
$$0.04\leq\psi\leq0.1\quad \textmd{and} \quad 0.6\times 10^{-2}\leq\epsilon\leq1.5\times 10^{-2}.$$

As we see the field $\psi$ is sub-Planckian and of order $10^{17}$ GeV. Moreover, recalling the connection to massive axion model (\emph{cf}. discussions of section \ref{gauge-flation-theoretical-motivation}) and in particular \eqref{kappa-axion}, one can see that $\mu\sim 5\times 10^{14}$GeV and $\lambda\sim 2.5\times 10^4$. We will discuss in the next section, these values are all in a very natural range for energy scales and coupling of axions in beyond standard models.

\section{Chromo-natural inflation}\label{Chromo-natural-section}

One of the main theoretical issues concerning most of inflationary models discussed in the literature is the ``naturalness'' issue. For example, in the simplest $m^2\phi^2$ or $\lambda\phi^4$ chaotic models, the field values are super-Planckian, $\phi\sim 10-20\ \mpl$, while the $m\sim 10^{-6}\mpl$ or $\lambda\sim 10^{-14}$. (In these models the Hubble during inflation $H$ is saturating the current observation bound, $H\sim 10^{-5}\mpl$.) The naturalness issue, besides its theoretical unappealing feature, brings puzzles and problems in embedding these models within other high energy physics models, like GUT's or string theory. One of the issues regarding the naturalness of inflationary potentials $V$ is coming from the slow-roll conditions, that $V\sim (10^{-3}-10^{-2}\ \mpl)^4$, while its derivatives in Hubble units should be small, of percent level or lower. Without embedding in a UV completed theory one cannot guarantee protection of the inflaton potential against quantum loop corrections.
Therefore, there has been many efforts in building inflationary models which are natural, specifically within a given consistent high energy theory setting, e.g. see \cite{inflection-point,Baumann:2009ds}.

One of the main theoretical obstacles in building ``natural'' inflationary models is to find a setting which avoids fine-tuning in the potential (coming by demanding the slow-roll conditions) by relating it to an approximate symmetry in the potential. The simplest such approximate symmetry which does the job is the translation in the field space. Motivated by this idea the ``natural inflation'' model  was proposed \cite{Natural-Inflation}. This model proposes using non-Abelian gauge theory axions as inflaton field. The axion field $\chi$ classically couples to $F\wedge F$ term of non-Abelian gauge theory and is hence invariant under $\chi\to \chi +\chi_0$ for arbitrary $\chi_0$ shifts. This shift symmetry is, however, broken to a discrete subgroup considering quantum effects (instanton contributions), which induce a cosine-type potential $V(\chi)\sim \mu^4 (1+\cos\chi/f)$, where $\mu$ and $f$ are two scales \cite{Weinberg-QFT-II}. The axion potential is pertubatively exact.

The natural inflation model of \cite{Natural-Inflation}, although ameliorates the naturalness problem by seeking shift symmetry and that the axion potential does not receive perturbative quantum corrections,  does not fully resolve it. As successful slow-roll inflation model is obtained for super-Planckian $f$ parameter (while $\mu^4\sim H^2\mpl^2$) which is not a natural scale within particle physics models. It is, however, noted that considering non-Abelian gauge fields and their coupling to axions can resolve this problem \cite{chromo-natural-short}. In this section we review the chromo-natural inflation model of \cite{chromo-natural-short,chromo-natural-long} and study its connection with the gauge-flation model of previous section.

\subsection{The basic setup}

Axions, although gauge singlets, couple to the topological $F\wedge F$ term of the non-Abelian gauge theories and have a nonperturbatively induced potential. This system describes the chromo-natural inflation model with the Lagrangian
\be\label{chromo-narutal-action}
{\cal L}_{c.n.} =-\bigg(-\frac{R}{2}+\frac{1}{4}F^a_{~\mu\nu}F_a^{~\mu\nu}+\frac12(\partial_\mu\tchi)^2  +\mu^4(1+\cos\frac{\tchi}{f})
-\frac{\lambda}{8f}\tchi\ (\epsilon^{\alpha\beta\mu\nu}F^a_{\alpha\beta}F^a_{\mu\nu})\biggr)\,
\ee
where we have set $8\pi G\equiv \mpl^{-2}=1$, $\mu$ and $f$ are parameters of dimension of energy, $\lambda$ is an order one dimensionless coupling. In our conventions the axion field $\tchi$ take values in $[0,\pi f]$ range.

As discussed in \cite{chromo-natural-short}, slow-roll inflationary dynamics for the above action can happen for $\mu, f \ll \mpl$, only if we also turn on a ``rotationally invariant gauge field'' of the form \cite{arXiv:1102.1513,Galtsov-inflation}
\be\label{A-ansatz-background}
A^a_{~\mu}=\left\{
\begin{array}{ll} a(t)\psi(t)\delta^a_i\, ,\qquad  &\mu=i
\\
0\,, \qquad &\mu=0\,,
\end{array}\right.
\ee%
in the temporal gauge. This is of course of the same form as discussed in the previous section in the gauge-flation setup. Our discussions of section 5.1 regarding the preservation of isotropy and homogeneity with the above gauge field configuration  almost trivially extends to the chromo-natural model too. In a similar manner one can show that reduction to the ``isotropic and homogeneous sector'' described by $\psi$ and $\tchi$ fields is a consistent reduction and the reduced Lagrangian may be simply obtained by plugging the gauge field ansatz \eqref{A-ansatz-background} into the action \eqref{chromo-narutal-action}, to obtain
\be\label{chromo-natural-reduced-action}
\cL_{reduced}=\cL_{gr}+ a^3\left[\frac32(\dot\psi+H\psi)^2-\frac32 g^2\psi^4+\frac12\dot\tchi^2-\mu^4(1+\cos\frac{\tchi}{f})-\frac{3\lambda g}{f} \tchi\psi^2(\dot\psi+H\psi)\right]\,,
\ee
where dot denotes the derivative w.r.t. the comoving time $t$. One may then work out the equations of motion for the above system:
\bse\label{c-n.e.o.m}
\begin{align}
&\ddot\tchi+3H\dot\tchi-\frac{\mu^4}{f}\sin\frac{\tchi}{f}=-3\frac{\lambda g}{f} \psi^2(\dot\psi+H\psi)\,,\\
&\ddot\psi+3H\dot\psi+(H^2+\dot H)\psi+2g^2\psi^3=\frac{\lambda g}{f}\psi^2\dot\tchi\,,\\
&H^2=\frac12\left[(\dot\psi+H\psi)^2+g^2\psi^4\right]+\frac16\dot\tchi^2+\frac{\mu^4}{3}(1+\cos\frac{\tchi}{f})\,,\\
&\dot H=-\left[(\dot\psi+H\psi)^2+g^2\psi^4\right]-\frac12\dot\tchi^2\,.
\end{align}
\ese
We note that the axion-gauge field coupling term, the $\lambda$-term, is metric independent and hence does not contribute to the energy momentum tensor of the the system. The effects of this term appears in the dynamics of the system through the terms in the right hand side of (\ref{c-n.e.o.m}a), (\ref{c-n.e.o.m}b). The $\lambda$-term plays the important role  of taming and flattening  the steepness of the axion potential, without requiring the parameter $f$ to be super-Planckian. In other words, considering the axion-gauge field coupling and turning on the gauge fields makes the model technically natural, justifying the name ``chromo-natural'' inflation model. This setup is particularly interesting because the axion potential and coupling terms are perturbatively and non-perturbatively exact.

We will show in this section that equations \eqref{c-n.e.o.m} indeed allow for  standard slow-roll trajectories for which $\epsilon=-\dot H/H^2$ and $\dot\tchi/(H\mpl),\ \dot\psi/H\psi $ remain small during the inflationary period. Moreover, interestingly this happens for a large region of parameter space and initial values which are  technically natural. In our analysis we will first assume possibility of slow-roll dynamics and then check the validity of assumption, both analytically and numerically.  We can now look for slowly rolling inflationary solutions of this system of equations assuming $\ddot\tchi$, $\ddot \psi$, and $\dot H \simeq 0$
and studying the resulting equations. In this limit (\ref{c-n.e.o.m}a) reduces to a simple equation
\be\label{chi-psi-trajectory}
\sin\frac{\chi}{f}\simeq -\frac{\lambda}{8\mu^4}\epsilon^{\alpha\beta\mu\nu}F^a_{\alpha\beta}F^a_{\mu\nu}\simeq +3{\lambda g}\frac{H\psi^3}{\mu^4}\,,
\ee
where, as before, $\simeq$ stands for equality up to the first order in slow-roll parameters $\epsilon, \eta$, and for axion field values which satisfy \eqref{chi-psi-trajectory} we will drop the tilde and use $\chi$, and
to obtain the second equality we have dropped $\dot\psi$ against $H\psi$.

Alternatively, one may recall the potential for the axion field $\tchi$:
\be\label{axion-potential}
U(\tchi,\Xi)=\mu^4\left(1+\cos\frac{\tchi}{f}+\Xi\frac{\tchi}{f}\right)\,,
\ee
where $\Xi= -\frac{\lambda}{8\mu^4 } \epsilon^{\alpha\beta\mu\nu}F^a_{\alpha\beta}F^a_{\mu\nu}$.
This potential has an extremum at $\tchi=\chi$, which to first order in slow-roll is given by \eqref{chi-psi-trajectory}. The potential has extrema if $\Xi\leq 1$. This latter puts an upper bound on $\psi$ (for given $g$ and $H$).\footnote{We note that in the discussions of \cite{chromo-natural-short} \eqref{chi-psi-trajectory} has been viewed as the condition minimizing effective potential for the $\psi$ field.} Eq.\eqref{chi-psi-trajectory} has two solutions, $\chi/f$ and $\pi-\chi/f$ and  our slow-roll trajectories are such that the axion field rolls down along the minimum of the potential and inflation ends when $\chi$ is close to $\pi f$. It is easily seen that the solution with $\chi>\pi f/2$ corresponds to the minimum of potential \eqref{axion-potential} and the one with $\chi<\pi f/2$ to maximum.
This suggests two ``natural'' initial conditions for axion field $\chi_0$ to start slow-roll: a ``small axion model'' where we start close to the top of the potential, and ``large axion model'' one with $\chi_0/f$ close to $\pi$. The former is the model discussed in \cite{chromo-natural-short} while the latter is nothing but the gauge-flation model discussed in the previous section (\emph{cf}. discussions of section \ref{gauge-flation-theoretical-motivation}). However, as has been shown in \cite{Maleknejad-Zarei} and we will review in section \ref{chromo-natural-slow-roll-section},  it is possible to get a successful slow-roll inflation for any arbitrary value of $\chi_0$. In addition, although both $\psi$ and $\chi$ fields are slowly rolling, their rolling is ``adiabatic'' in the sense that during slow-roll inflation $\chi$ and $\psi$ fields vary such that the minimizing condition \eqref{chi-psi-trajectory} holds.

As may be seen from (\ref{c-n.e.o.m}a) or from \eqref{chi-psi-trajectory} during the slow-roll  the energy is generically taken from the axion sector and  injected into the Yang-Mills gauge field sector. In this process, which is the new ingredient of the chromo-natural model compared to the natural inflation model of \cite{Natural-Inflation}, the speed of axion rolling is reduced through the coupling to a specific combination of gauge field condensate. This condensate is proportional to the instanton number density $F\wedge F$ (though in the Minkowski signature) and it the cosmological setups (for $K=1$ FLRW cosmologies) this has been dubbed as ``cosmological sphaleron'' \cite{cosmoligical-sphaleron,Galtsov-inflation}. Nonvanishing cosmological sphaleron configurations in Abelian and non-Abelian gauge theories can be a basis for lepto/baryo genesis model building. The former has been discussed in \cite{Alexander:2011hz}  and the latter in \cite{Noorbala:2012fh}.

\subsection{Chromo-natural vs gauge-flation}\label{chromo-natural-vs-gauge-flation-section}

If the value of $\tchi$ at minimum, $\chi$, is very close to $\pi f$ one may expand the axion potential \eqref{axion-potential} and to second order in $\chi$ we obtain the potential for a massive axion discussed in
\eqref{massive-axion-YM} and hence the chormo-natural model will reduce to the gauge-flation, if the initial value of the $\chi$ field $\chi_0$ is close to $\pi f$. That is, for the ``large axion'' region, the chromo-natural model is equivalent to gauge-flation. This can be made in more robust technical setting  by integrating out axion field.
To this end, one may expand the axion theory around the minimum of its potential at $\tchi=\chi$ (for $\chi>\pi/2$) given in \eqref{chi-psi-trajectory} and, integrate out the fluctuations \cite{Peskin}. This may be done by writing $\tchi=\chi+\delta\chi$, expanding the action up to the second order in $\delta\chi$ and computing the one loop effective potential \cite{gauge-flationVs.chormo-natural} \footnote{We would like to comment that here we use the Wilsonian effective action language in which field renormalizations are reabsorbed into the (re)definition of fields. Field renormalizations should be considered if we used ``renormalized perturbation theory" language \cite{Peskin}.}
\be\label{one-loop-potential}
U_{eff}(\chi)=U(\chi)+ \frac{i}{2}\ln \det(\Box+\frac{\mu^4}{f^2}\cos\frac{\chi}{f})\,,
\ee
where $U(\chi)$ is the potential \eqref{axion-potential} computed at the minimum $\tchi=\chi$ and  $\Box$ is the Laplacian computed on the (almost-de Sitter) inflationary  background.

To compute the $\ln\det$ term explicitly, as in standard field theory treatment, one should integrate over the spectrum of the  $\Box+\frac{\mu^4}{f^2}\cos\frac{\chi}{f}$ operator. However, since we are doing this loop analysis on a non-flat space, only \emph{subhorizon modes}, modes with physical momenta larger than $H$, which are quantum modes will contribute. For  these subhorizon modes one may approximate the $\Box$ term with its flat space value.
We also note that in the region of parameter space we will be interested in the gauge-flation (\emph{cf.} discussions of section \ref{c-n-perturbation-theory}) $\mu^4\gg H^2 f^2$ and $H\ll\mu\lesssim f \ll \mpl$.  The integral over the momenta can be cut off at a UV cutoff $\Lambda$ which may be taken equal to $f$. Using equations in \cite{Peskin}, and in the $\mathrm{\overline{MS}}$ scheme, we find
$$
U_{eff}(\chi)=U(\chi)+\delta U
$$
where $\sin\frac{\chi}{f}=\Xi= -\frac{\lambda}{8\mu^4 } \epsilon^{\alpha\beta\mu\nu}F^a_{\alpha\beta}F^a_{\mu\nu}$ and
\bea\label{U-Xi}
U(\chi)&=&-\mu^4\left[1-\sqrt{1-\Xi^2}-\Xi\arcsin\Xi\right]\,,\\
\label{deltaU}
\delta U&=&\frac14\frac{1}{(4\pi^2)}\left(\frac{\mu^4}{f^2}\cos\frac{\chi}{f}\right)^2\left(\ln(\frac{\mu^4}{f^4}
\cos\frac{\chi}{f})-\frac32\right)\,.
\eea
Note that to obtain the expression for $U$ and $\delta U$, we have evaluated for the $\chi>\pi/2$ (corresponding to the minimum of potential \eqref{axion-potential}) solution of \eqref{chi-psi-trajectory}, for which
$$
\cos\frac{\chi}{f}=-\sqrt{1-\Xi^2}\,.
$$
$U$ in \eqref{U-Xi} is the minimum value of the axion potential \eqref{axion-potential}, while $\delta U$ is the one-loop quantum corrections to this potential. The validity of one-loop analysis and integrating out  axion around the minimum of its potential is quantified by the $\delta U/U$ ratio; one can see that $\delta U/U\sim few\times \mu^4/f^4$. Therefore, if $\mu/f \lesssim 1$ one can safely approximate the potential by its classical value $U(\chi)$.
To summarize, chromo-natural model, upon integrating out the axion field during the slow-roll evolution \emph{occurring around the minimum of the potential}, reduces to a gauge theory whose action consists of the $F^2$ Yang-Mills term and $U(\Xi)$. This theory is not exactly our gauge-flation model, as the potential $U(\Xi)$ is not just the $\Xi^2$ term. Nonetheless, for $|\Xi|\lesssim 1$, one may expand $U$ to obtain gauge-flation $\kappa$-term (\emph{cf.} discussions of section \ref{gauge-flation-theoretical-motivation}).
Finally we comment that, our analysis of next subsection conforms that standard ``slow-roll intuition'' holds for our case and  for slow-roll trajectories  inflation is mainly driven by the $U(\Xi)$, i.e. $3H^2\mpl^2\simeq U(\Xi)$.

In the above we focused on the field configurations and regions of parameter space for which the axion field is massive while the gauge field (at least some of its components, specifically the effective inflaton field $\psi$) is light.
One may consider another regime in parameter space for which the opposite situation happens. That is, gauge fields become massive (with effective mass of $\psi$ field to be around Hubble $H$) while the axions are light. This happens when $\gamma=g^2\psi^2/H^2\gg 1$.
In this regime it is more appropriate to integrate out the gauge fields and obtain a theory which only involves axions. This has been carried out in \cite{gauge-field-integrate-out} and leads to a natural inflation model \cite{Natural-Inflation} with $(\partial\chi)^4$-type additions.
Explicitly, for $\gamma \gg 1$, chromo-natural model below the cutoff scale $\Lambda\sim f\sqrt{g}/\lambda$ reduces to a single scalar K-inflation model \cite{ArmendarizPicon:1999rj} with the action \cite{gauge-field-integrate-out}
\be\label{chormo-natural-K-inflation-eff-action}
S_{\chi}=\int d^4x\sqrt{-g}\big(\frac{R}{2}-\frac12(\partial\chi)^2+\frac{1}{4\Lambda^4}(\partial\chi)^4-
\mu^4(1+\cos(\frac{\chi}{f}))\big)+\cdots\,,
\ee
here $\cdots$ denotes corrections suppressed by $\gamma^{-1}$ and $\Lambda$.

\subsection{Slow-roll trajectories}\label{chromo-natural-slow-roll-section}
The slow-roll parameters \eqref{epsilon-eta-def} should remain small, $\epsilon, \eta\ll 1$, during the slow-roll.
This together with \eqref{c-n.e.o.m}  gives the following approximation for $H^2$ during inflation
\be\label{H2-sl}
H^2\simeq\frac13\mu^4(1+\cos\frac{\tchi}{f}),
\ee
where $\simeq$ means equality up to the first order in $\epsilon$.
Dividing (\ref{c-n.e.o.m}c) and (\ref{c-n.e.o.m}d) gives $\epsilon$
\be
\epsilon=\left((1-\delta)^2+\gamma\right)\psi^2+\frac12\frac{\dot\tchi^2}{H^2},
\ee
where $\delta\equiv-\frac{\dot\psi}{H\psi}$ and $\gamma\equiv\frac{g^2\psi^2}{H^2}$ are  defined as in \eqref{delta-def} and \eqref{x-def}. It turns out useful to separate the gauge field and axion field contributions to $\epsilon$, \textit{i.e},
\be\label{epsilon-dec}
\epsilon=\epsilon_\psi+\epsilon_\chi\,,
\ee
where
\be
\epsilon_\psi=\left((1-\delta)^2+\gamma\right)\psi^2,\quad \epsilon_\chi=\frac12\frac{\dot\tchi^2}{H^2}\,.
\ee
Since both of $\epsilon_\psi$ and $\epsilon_\chi$ are positive, slow-roll condition demands that each of them should be very small during inflation. Also combing \eqref{epsilon-eta-def} and \eqref{epsilon-dec}, one can determine $\eta$
\be
\eta=\eta_\chi+\eta_\psi\,,
\ee
where again $\eta_\chi$, $\eta_\psi$ are respectively $\chi$ and $\psi$ contributions to $\eta$,
\bea
\eta_\chi&=&-\frac{\ddot\tchi}{H\dot\tchi}\frac{\epsilon_\chi}{\epsilon },\\
\label{eta-psi}
\eta_\psi&=&\left(\frac{2\delta}{\epsilon}+\frac{(1-\delta)^2}{((1-\delta)^2+\gamma)}\bigl(\frac{\frac{\dot\delta}{\epsilon H}}{1-\delta}+1-\frac{\delta}{\epsilon}\bigr)\right)\epsilon_\psi.~~~
\eea
A successful slow-roll inflation hence requires  $\epsilon_\psi,\ \epsilon_\chi,\ \eta_\chi\ll 1$ and $\delta\lesssim\epsilon$. We will show below that depending on the initial value of $\tchi$ and $\chi_0$,
$\delta$ should be a quantity of the order $\epsilon^2$ to $\epsilon$.

We now consider the $\tchi$ and $\psi$ equations of motion (\ref{c-n.e.o.m}a,b), which in the leading slow-roll approximation leads to \eqref{chi-psi-trajectory}. One may then use \eqref{H2-sl} to replace $H$ for $\chi$ to find $\psi$ during the inflation
\be\label{psi-sl}
\psi^3\simeq\left(\frac{\Upsilon}{\lambda }\right)^{3/2}\ \sin\frac{\chi}{2f}\,,
\ee
where
\be
\Upsilon^3\equiv \frac{2\lambda\mu^4}{3g^2}\,.
\ee
The above equation implies that $\delta\propto\cos\frac{\chi}{2f}$. On the other hand, \eqref{eta-psi} indicates that to have a slow-roll inflation we need $\delta\lesssim\epsilon$. Thus, in cases with small $\frac{\chi}{f}$, that is the ``small axion model'' discussed in \cite{chromo-natural-short,chromo-natural-long}, $\delta$ should be of the order $\epsilon$, while for case with $\frac{\chi}{f}=\pi+\mathcal{O}(\epsilon)$, the case the model effectively reduces to the gauge-flation model \cite{gauge-flationVs.chormo-natural}, we have $\delta\sim\epsilon^2$ ( Fig. \ref{delta-chi}).

\begin{figure}[h]
\begin{center}
\includegraphics[angle=0, width=90mm, height=80mm]{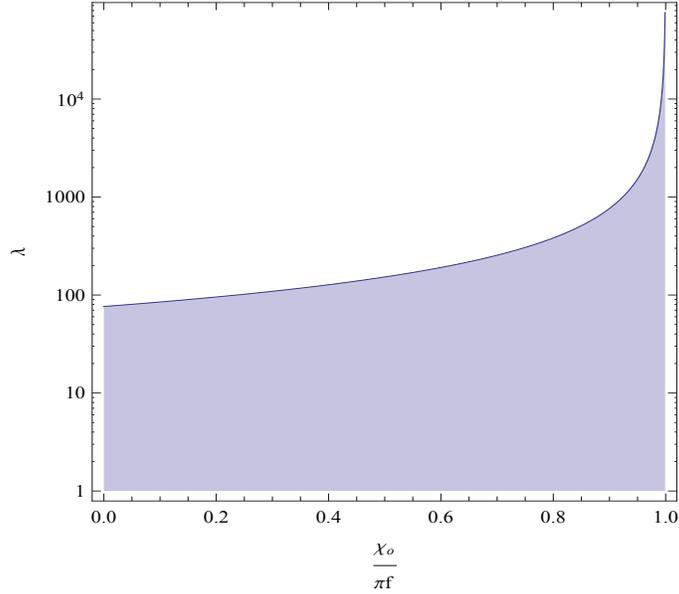}
\caption{In the above we have set $N_e=60$ in the inequality \eqref{ubn} and plotted $\frac{\chi_0}{\pi f}$ versus $\lambda$. The shaded area indicates parts of the $\lambda$-${\chi_0}$ plane which is excluded for successful slow-roll inflation. }\label{shaded}
\end{center}
\end{figure}

Within the slow-roll approximation the $\psi$ equation of motion (\ref{c-n.e.o.m}b) yields
\be\label{epsilon-psi}
\epsilon_\psi\simeq\frac12\frac{g\lambda}{fH}\psi^3\frac{\dot\chi}{H}\,.
\ee
On the other hand, during the slow-roll inflation (\ref{c-n.e.o.m}a)
$\frac{\dot\chi}{H}\ll\frac{g\lambda}{fH}\psi^3$, which leads to
\be\label{epsilon-chi}
\epsilon_\chi\ll\epsilon_\psi\,,
\ee
and consequently
\be\label{eps}
\epsilon\simeq\epsilon_\psi\,.
\ee
In other words, in the course of slow-roll inflation the energy budget of the Universe which is dominated by the axion potential \eqref{H2-sl}, is slowly, but predominantly  converted into Yang-Mills gauge field energy rather than the kinetic energy of axion itself.\footnote{There are, however,  regions in the parameter space where $\dot\chi$ can have a sudden change during inflation leading to interesting features in the dynamics of the system \cite{Maleknejad-Zarei}.}

We can now compute number of e-folds using (\ref{c-n.e.o.m}b) and \eqref{eps}:
\be\label{N-chi-psi}
N_e\simeq\frac{\lambda\Upsilon}{\sqrt[3]{4}}\int^\pi _{\frac{\chi_0}{f}}\frac{(1+\cos x)^{2/3}(\sin x)^{1/3}}{\Upsilon^2(1+\cos x)^{4/3}+(2\sin x)^{2/3}}dx,
\ee
where  ``$0$''  subscript denotes  an  initial  value  and $x\equiv\frac{\chi}{f}$. Before giving the complete analytic result for the above integral it is useful to note that, upon calling
$$
X\equiv \Upsilon(1+\cos\frac{\chi}{f})^{2/3}\,, \qquad Y\equiv (2\sin\frac{\chi}{f})^{1/3},
$$
the integrand takes the form $\frac{XY}{X^2+Y^2}$, which is always less than $1/2$ and hence
\be\label{ubn}
N_e\leq\frac{\lambda}{4}(\pi-\frac{\chi_0}{f}).
\ee
We note that the above relation depends only on $\lambda$ and not $\Upsilon$. Thus, setting $N_e \sim60$ in the inequality \eqref{ubn}, for the hilltop-type small $\chi_0$ case \cite{chromo-natural-short} in which $\pi-\frac{\chi_0}{f}\simeq\pi$, we should have $\lambda\sim100$. For the gauge-flation case \cite{arXiv:1102.1513} in which $\pi-\frac{\chi_0}{f}=\mathcal{O}(\epsilon)$, we need larger $\lambda$ values, $\lambda\sim10^4$ \cite{gauge-flationVs.chormo-natural}. Having the very simple relation \eqref{ubn}, is very useful in restricting the parameter space of the model. Regardless of the values of other parameters, the shaded area in Fig.\ref{shaded} does not lead to a successful slow-roll inflation with enough number of e-folds.

\subsubsection{Exploring the parameter space}

Up to this point, we demonstrated that the chromo-natural model can lead to slow-roll inflation in the entire region of $\chi_0\in(0,\pi)$. Furthermore, having a preliminary relation between the value of $\chi_0$, $\lambda$ and $N_e$, we could restrict the parameter space.
Here, first working out the explicit form of the number of e-folds $N_e$, we determine the parameter space in which successful slow-roll inflation ($N_e>60$) is possible. Next, studying the system numerically, we show that the slow-roll analysis presented before is in excellent agreement with the dynamics of the system during inflation.

The integral of number of e-folds \eqref{N-chi-psi} may be computed analytically. To this end we introduce
$y\equiv\sin^{2/3}(\frac{\chi}{2f})$, as
\be\label{N-e-exact-chromo-natural}
N_e\simeq \frac32\Upsilon\lambda\int^{1} _{y_0}\frac{y}{\Upsilon^2(1-y^3)+y}dy\,,\qquad y_0={\sin(\frac{\chi_0}{2f})}\,.
\ee
The denominator always has a real and positive root, $\bar{y}$
\be\label{ybar}
\Upsilon^{-2}=\bar{y}^2-\frac{1}{\bar{y}},
\ee
since $\Upsilon^2>0$, then $\bar{y}>1$.
Furthermore, the other roots $y_{\pm}$ are given as below
\be
y_{\pm}=-\frac{\bar{y}}{2}(1\mp\sqrt{1-\frac{4}{\bar{y}^3}}),
\ee
which depending on the value of $\Upsilon$,  can be either  both negative or both complex numbers:
\begin{itemize}
\item{$\Upsilon\leq \sqrt[6]{\frac{4}{27}}\simeq 0.73$,\ $(\bar{y}\geq4^{\frac13})$, $y_{\pm}$ are two real valued negative roots. Thus, the value of  the integral is
\be\label{N1}
N_e\simeq\frac{3\lambda}{2\Upsilon}\left[\left(\frac{y_+\ln(y-y_+)}{(\bar{y}-y_+)}-\frac{y_-\ln(y-y_-)}{(\bar{y}-y_-)}\right)\frac{1}{(y_+-y_-)}
+\frac{\bar{y}\ln(\bar{y}-y)}{(\bar{y}-y_+)(\bar{y}-y_-)}\right]^{\sin\frac{\chi_0}{2f}}_{1}.
\ee
}
\item{$\Upsilon> \sqrt[6]{\frac{4}{27}}\simeq 0.73$, $(\bar{y}\leq4^{\frac13})$, $y_{\pm}$ are  complex valued, $y_{\pm}=-y_R\pm iy_I$, with $y_R=\frac{\bar y}{2}$ and $y_I=\frac{\bar y}{2}\sqrt{\frac{4}{\bar y^3}-1}$. We then have
\be\label{N2}
N_e\simeq\frac{3\lambda}{2\Upsilon}\frac{4\bar y^2}{5\bar y^3+16}\left[\frac{1}{2}\ln((y+y_R)^2+y_I^2)+\ln(\bar{y}-y)-(\frac{3\bar y}{4y_I}+\frac{y_I}{\bar{y}})\arctan(\frac{y+y_R}{y_I})\right]^{\sin\frac{\chi_0}{2f}}_{1}\,.
\ee
}
\end{itemize}
The above expressions determine $N_e$ as a function of $\Upsilon$, $\lambda$ and $\chi_0/f$ which make it possible to explore the inflationary parameter space analytically.
Note that in deriving the integral \eqref{N-chi-psi}, we used the slow-roll attractor solution of $\psi$  \eqref{chi-psi-trajectory}, which determines it as a function of $\chi$. Thus, in the expression of $N_e$, only $\chi_0$ appreases explicitly.

\begin{figure*}[h] 
\includegraphics[angle=0, width=80mm, height=80mm]{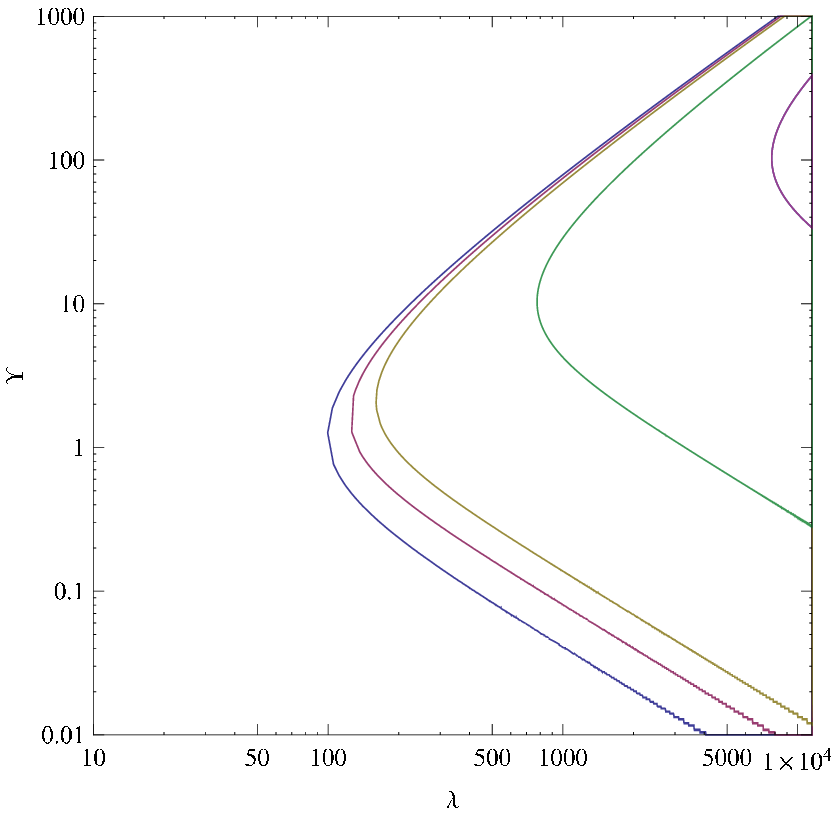}  \includegraphics[angle=0, width=80mm, height=80mm ]{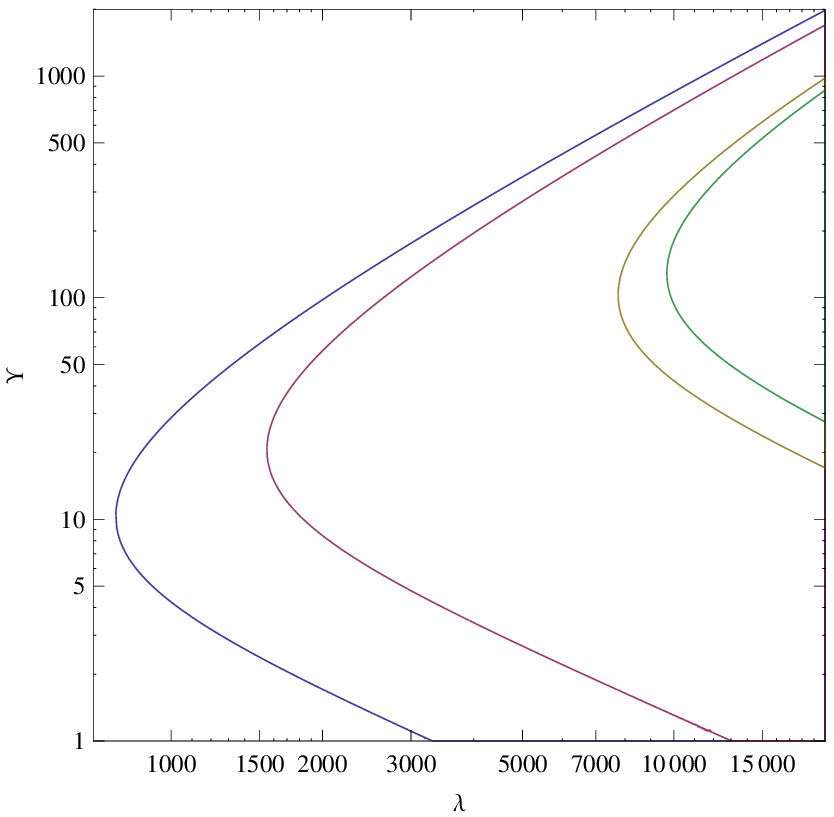}
\caption{
In the above plots we have plotted the parameter space, $\lambda$ vs. $\Upsilon$ for $N_e=60$. The curves on the Left figure correspond to various initial values of axion field $\chi_0$; from left to right corresponding to $\frac{\chi_0}{f}=10^{-2}\pi,\ \pi/3, \ \pi/2, \ 0.9\pi$ and $0.99\pi$. The Right plot shows the region on the parameter space which chromo-natural model is effectively equal to the gauge-flation, and the curves from left to right corresponds to $\frac{\chi_0}{f}=0.9\pi, \ 0.95\pi\ , 0.99\pi$ and $0.992\pi$. The ``outer region of the curves correspond to $N_e<60$ and are hence excluded. Note that $\lambda$ is typically (e.g. the minimum value of $\lambda$) of order $10^2$ for the ``small axion values'' (in the left figure), while it is of order $10^4$ for the  large axion values, in the right figure.
The plots imply that $\log(\frac{\lambda}{\lambda_{_{min}}})\geq(\log(\frac{\Upsilon_{_{min}}}{\Upsilon}))^2$ while we have $\lambda_{_{min}}\sim 10^2\Upsilon_{_{min}}$.}
   \label{parameterSpace}
\end{figure*}

In order for inflation to solve the flatness and horizon problems it should have lasted for a minimum number of e-folds $N_e$ which depends on the scale of inflation and somewhat to the details of physics after inflation ends \cite{Inflation-Books}. However, it is usually demanded that $N_e\simeq 60$. Using the explicit form of $N_e$ in \eqref{N1} and \eqref{N2}, and after considering $N_e>60$, one can determine the accessible parameter space corresponding to each initial value of axion field, $\chi_0/f\in(0,\pi)$.
In Fig.\ref{parameterSpace} setting $N_e=60$, we presented the parameter space, $\lambda$ versus $\Upsilon$ for several $\chi_0/f$ values between $10^{-2}$, which is a ``hilltop-type model" \cite{chromo-natural-short}, to $0.99\pi$ that is effectively reduced to ``gauge-flation" \cite{gauge-flationVs.chormo-natural}.  The outer region of the curves correspond to $N_e<60$ and are hence excluded. As discussed before, $\lambda$ is typically of order $10^2$ for the small axion values which is the minimum value of $\lambda$, while it is of order $10^4$ for the large axion values.
Fig. \ref{parameterSpace} indicates that as the initial value of the axion field increases, the value of $\lambda$ and $\Upsilon$ increase such that
$\log(\frac{\lambda(\frac{\chi_0}{f})}{\lambda_{_{min}}(\frac{\chi_0}{f})})\geq(\log(\frac{\Upsilon_{_{min}}(\frac{\chi_0}{f})}{\Upsilon(\frac{\chi_0}{f})}))^2$ while we have $\lambda_{_{min}}(\frac{\chi_0}{f})\sim 10^2\Upsilon_{_{min}}(\frac{\chi_0}{f})$.

\begin{figure*}[h] 
  \includegraphics[angle=0, width=80mm, height=50mm ]{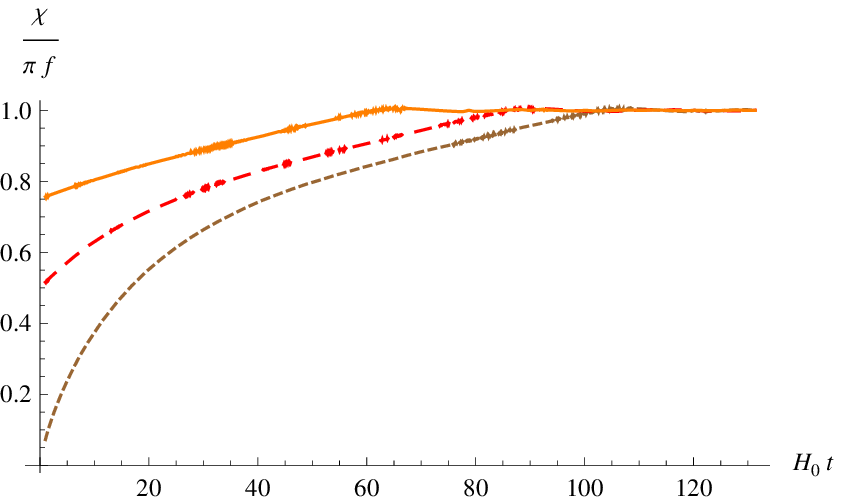} \includegraphics[angle=0, width=80mm, height=50mm]{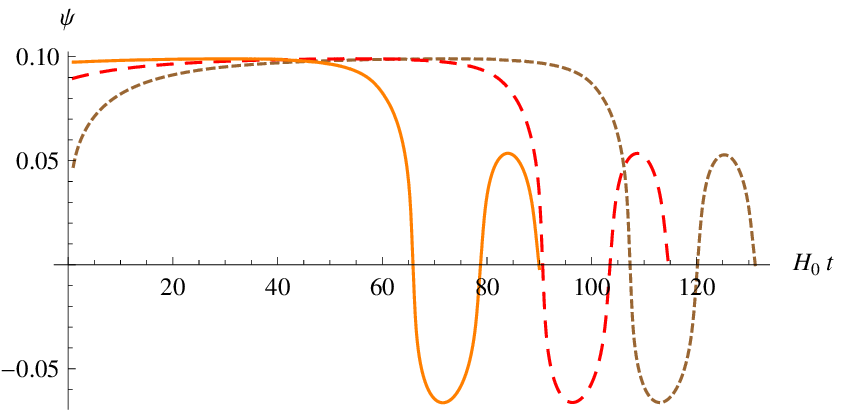}
\caption{In this figure we presented the classical trajectories with g=$10^{-6}$, $\lambda$=400, $\mu$=7$\times10^{-4}$, $f=0.009$, started from different axion initial values, $\frac{\chi_0}{f}$. In both panels, the solid (orange) lines, the dashed (red) lines and the dotted (brown) lines correspond to $\frac{\chi_0}{f}$ values equal to $3\pi/4$, $\pi/2$ and $10^{-2} $, respectively. In the left panel, we see that axion field slowly increases during the inflation, then it gets equal to $f\pi$ and inflation ends. As we see in the right panel, $\psi$  gradually increases during the inflation and after the inflation ends, it starts oscillating just like the gauge-flation model in Figs.  \ref{x=6.35-slow-roll-figures} and \ref{N200-slow-roll-figures}. Note that as the initial value of the axion field increases, $\delta$ gets smaller, in agreement with our slow-roll analysis. We also note that, as our analytical calculations show as we decrease initial value of $\chi$ for the given $(g,\lambda, \mu, f)$ values we get a larger number of e-folds. In particular, as we see, the set we have chosen gives 60 number of e-folds for $\chi_0/f=3\pi/4$, while it gives about $N_e=100$ for $\chi_0/f=10^{-2}$. }
   \label{delta-chi}
\end{figure*}

After working out the parameter space which leads to sufficient inflation, at this point, we present the numerical result of studying the equations \eqref{c-n.e.o.m} for a set of parameters with different $\chi_0/f$ values: $10^{-2}, \pi/2, 3\pi/4$ (Fig.\ref{delta-chi}).
As we see in the left panel of Fig. \ref{delta-chi}, starting from $\chi_0$, the axion field increases slowly during the slow-roll inflation and inflation ends when $\chi/f$ gets equal to $\pi$. Furthermore, the right panel of Fig. \ref{delta-chi}, confirms that during the slow-roll $\psi$ is given by \eqref{psi-sl} and $\delta$ which is associated with  each initial value of axion decreases as $\chi_0/f$ gets closer to $\pi$. Remarkably, regardless of the value of parameters and the initial value of fields, after slow-roll inflation ends all trajectories lead to an oscillatory $\psi$, as in  gauge-flation case.

\subsection{A quick treatment of chromo-natural cosmic perturbation theory}\label{c-n-perturbation-theory}

So far, we showed that regardless of the initial axion field value,  chromo-natural model can lead to a successful slow-roll inflation with enough number of e-folds in a vast part of its natural parameter space.  On the other hand, almost all of the precision data of cosmology is related to perturbations, and we hence need to study the cosmic perturbation theory.
Full analysis requires considering all the physical degrees of freedoms, \ie the gauge field and metric perturbations as was discuss in section \ref{gauge-flation-cosmic-perturbation-theory-section}, and  perturbation of the axion field $\delta\chi$.
A through and complete analysis which takes into account all the details of interactions, as we saw in section \ref{gauge-flation-cosmic-perturbation-theory-section},  has not yet appeared in the literature, see however \cite{Adshead-Wyman-Martinec-2013,Dimastrogiovanni:2012ew}. We hope to discuss this in a later publication \cite{progress}. Here, we first  show that vector and tensor perturbations in chromo-natural and gauge-flation models are exactly the same. Then, we review the cosmic perturbation analysis presented in \cite{chromo-natural-short} and \cite{chromo-natural-long}.

\subsubsection{Vector and tensor perturbations in chromo-natural model}

Here we explicitly show that in vector and tensor sectors, the chromo-natural model is exactly the same as gauge-flation model which has been studied in the previous section.

The matter Lagrangian density of gauge-flation and chromo-natural models (have been introduced in \eqref{The-model} and \eqref{chromo-narutal-action} respectively) are as below
\bea
\label{model-gf}
^{^{Gf}}\!\cL_{m}&=&\cL_{_{YM}}+\cL_{_{Gf}},\\
\label{model-cn}
^{^{Cn}}\!\cL_{m}&=&\cL_{_{YM}}+\cL_{_{Cn}},
\eea
where $\cL_{_{YM}}$ is the Yang-Mills term, $\cL_{_{Gf}}=\frac{\kappa}{384}(F\wedge F)^2$  and $\cL_{_{Cn}}=\frac{\lambda\chi}{8f}F\wedge F$.

Since the axion-gauge field interaction term in the chromo-natural model is metric independent, it it does not contribute to the energy-momentum tensor, and  in this model we have
\be\label{}
^{^{C\! n}}\! T^{\mu}_{~\nu}=-\frac{2g^{\mu\lambda}}{\sqrt{-g}}\frac{\delta (\sqrt{-g}\cL_{_{Y\! M}})}{\delta g^{\lambda\nu}}.
\ee
On the other hand, since the metric dependence of $\cL_{_{Gf}}$ is only through $
 det{g}$, $T^{\mu}_{~\nu}$ of the gauge-flation model is given by
\bea
^{^{Gf}}\! T^{\mu}_{~\nu}=-\frac{2g^{\mu\lambda}}{\sqrt{-g}}\frac{\delta (\sqrt{-g}\cL_{_{Y\! M}})}{\delta g^{\lambda\nu}}-\delta^\mu_{\nu}\cL_{_{G\! f}},
\eea
which indicates that $\cL_{_{G\! f}}$ term can only contribute to the scalar sector of the perturbations and the vector and tensor parts are coming from the Yang-Mills term. Hence,  the first order tensor and vector parts of the Einstein equations are exactly the same in the gauge-flation and chromo-natural models.
On the other hand, to have a complete set of equations, we need the field equation of the tensor mode perturbation of the gauge field  $\tilde h_{ij}$ which is provided by  the tensor second order action. Similarly, enough number of  equations for vector modes is provided by the field equation of $A^a_{~0}$ as well as the vector parts of the Einstein equations. The former is obtained from the second order action of the vector perturbations.

Considering the tensor perturbations, up to the linear order we have $\delta(\sqrt{-g}F\wedge F)=0$. Thus the contribution of $\mathcal{L}_{_{Cn}}$ and $\mathcal{L}_{_{Gf}}$ to the tensor second order action are both given by $\delta^2(\sqrt{-g}F\wedge F)$, where $\delta^2$ denotes second order of perturbations. After an integration by part and removing the total derivative terms, contributions of these terms to the second order tensor action is obtained as
\bea
^{\!(2)}\!\delta(\!\sqrt{-g}\cL_{_{Cn}}\!)_{T} &=&-\frac{\lambda\dot\chi}{2f}\bigg(\frac{g\phi^3}{a^3}X_{ij}^2+\frac{\phi^2}{a^2}\epsilon^{ijk}X_{kl}\partial_i X_{jl}\bigg),\nn\\
^{\!(2)}\!\delta(\!\sqrt{-g}\cL_{_{Gf}}\!)_{T} &=&(\frac{\kappa g\phi^2\dot\phi}{2a^3}\dot{)}\bigg(\frac{g\phi^3}{a^3}X_{ij}^2+\frac{\phi^2}{a^2}\epsilon^{ijk}X_{kl}\partial_i X_{jl}\bigg),\nn
\eea
where $X_{ij}\equiv\tilde h_{ij}+\frac12 h_{ij}$. Using (6.4b) and (5.33c), we find that they are identically the same. Thus, the second order  action for tensor perturbations of the chromo-natural and gauge-flations model are exactly the same. The tensor mode perturbations of chromo-natural model has also appeared in \cite{Adshead-Wyman-Martinec-2013}.

It is straightforward to prove, in a similar manner, that $\mathcal{L}_{_{Cn}}$ and $\mathcal{L}_{_{Gf}}$ has no contribution to the second order vector action, which indicates that chromo-natural and gauge-flation models has the same vector second order action and hence similar vector mode power spectra.

\subsubsection{Scalar modes}

Following \cite{chromo-natural-short,gauge-field-integrate-out}, we assume that the non-Abelian gauge field has a mass of order $H$ during inflation and being very massive has a negligible effect on the quantum fluctuations. Therefore,  the perturbations are predominantly coming from the axion field $\chi$.
One may then use the $\delta N$-formulation (see Appendix \ref{deltaN-appendix} for a review) and after neglecting the contribution of the gauge field perturbations, from \eqref{delta-N-phi} we have
\be
\delta N=\frac{\partial N(\chi_0)}{\partial \chi_0}\delta\chi+\cdots\,,
\ee
where $N$ is given in \eqref{N-e-exact-chromo-natural}.
Then, this leads us to the following rough estimation of the scalar power spectrum \cite{chromo-natural-short}
\be
\Delta^2_{s}(k)\sim\frac{1}{4\pi^2}\frac{H^2}{\chi'^2},
\ee
and scalar spectral tilt
\be
n_s\sim 1-2\epsilon+2\frac{\chi''}{\chi'}\,.
\ee
In the regime where gauge fields are very massive one may also use the effective action obtained from integrating out the gauge fields \eqref{chormo-natural-K-inflation-eff-action} and use K-inflation perturbation theory results \cite{ArmendarizPicon:1999rj}.

Here for completeness we finish by summarizing two sets of possible values of parameters and initial fields  of chromo-natural model which leads to successful model.
\paragraph{Small axion model.} This is the range of parameters discussed in \cite{chromo-natural-short,chromo-natural-long}
\be\label{chromo-natural-parameter-range}
\begin{split}
&\mu^2=10^{-7}\,,\quad f=10^{-2}\,,\quad H\simeq \sqrt{\frac23}\mu^2=8\times 10^{-8}\\
&\lambda=200\,,\quad g=2\times 10^{-6}\,,\quad \epsilon\simeq 1.8\times 10^{-3}\,,\\
&\eta\simeq 1.4\times 10^{-3}\,,\quad \dot\psi\sim -1\times 10^{-6}H,\quad \psi\simeq 3.3\times 10^{-2}\,.
\end{split}
\ee
(For the above values of parameters to get enough number of e-folds, inflation should start when $\chi_0/f< 0.4\pi$.
\paragraph{Large axion region} which is essentially the gauge-flation region covered by parameters \cite{gauge-flationVs.chormo-natural}
\begin{equation}
\label{gauge-flation}
\begin{aligned}
&\chi_0=\pi f- 5\times 10^{-4}, & &f=10^{-2}, & &\lambda=2\times 10^4,\\
&H\simeq 3.3\times 10^{-5}, & &\mu^2=1.6\times 10^{-3}, & &g=10^{-3}, \\
&\psi\simeq 4\times 10^{-2}, & &\epsilon\simeq 4\times 10^{-3}, & &\eta\simeq 1.6\times 10^{-3}.
\end{aligned}
\end{equation}

As we discussed in the previous subsection, however, it is possible to get slow-roll inflation with enough number of e-folds for all values of $\chi_0$. Note that all dimensionful quantities are measured in units of $\mpl$.

\section{Summary and outlook }\label{discussion-section}

This article is devoted to reviewing various analysis and discussions appeared so far in the literature about the role gauge fields may have played in the dynamics of the early Universe, specifically in inflationary cosmology. Here we classified and discussed several such effects. We reviewed some models in which gauge fields are the main driving force for the inflationary background. Here we did not discuss the ``vector inflation models'' \cite{vector-inflation}, as these models involve vector fields and not gauge fields. It is an established common knowledge in quantum field theory that in the absence of gauge symmetry one should, in principle, worry about the consistency vector field theories and in particular absence of propagating ghosts. Presence of ghosts will spoil unitarity and/or positivity of the energy (Hamiltonian) of the model, and hence rendering the model inconsistent \cite{vector-inflation-ghost}.

To avoid  inconsistencies with the ghosts in dealing with vector fields, presence of gauge symmetry is essential. Gauge symmetry, among other things, forbids addition of arbitrary ``potential'' terms for the gauge fields. Moreover, one should in principle worry about the rotation asymmetry caused by turning vector gauge fields in the background.
As we discussed these issues can be dealt with, if we consider non-Abelian gauge fields. This was done through the gauge-flation model reviewed in section \ref{Gauge-flation-section}. Here we discussed one specific inflationary model within the non-Abelian gauge inflation setup. The gauge invariant actions are not of course limited to the action we discussed; the other terms may be added to the action because of the loop effects and/or coupling to other fields.
One such case, which is closely related to the gauge-flation model, was reviewed in section \ref{Chromo-natural-section}. There are many more possibilities which may be explored.

One of the appealing features of the models with non-Abelian gauge fields turned on at the background level is the stability of the \emph{isotropic} inflationary trajectories with respect to the choice of initial conditions as well as in the parameter space, despite the fact that we have vector gauge fields in the background. Providing a setup for a stable inflationary background, we discussed specific feature of the cosmic perturbation theory within the gauge-flation and chromo-natural model. It is desirable to analyze in further detail the perturbation theory of these models, e.g. studying non-Gaussianity and possible observable statistical anisotropic effects of gauge-flation or its extension.

One of the motivations to study non-Abelian gauge field inflation was its presence in all high energy particle physics models and/or string theory. In fact, there is a general argument that all the symmetries in a theory of quantum gravity, like string theory, are gauged and these theories provide a wealth of gauge theories of various matter fields. The next natural task is hence try to embed our models in the particle physics or string theory setups. First steps in this direction has been taken in \cite{Martinec-1,Martinec-2} and a more detailed discussion is still needed.

Two other lines of research in gauge fields and inflation has been followed where gauge fields do not have the dominant contribution to the energy budget of the Universe during inflation, nonetheless, they have very interesting observational consequences. These two lines were discussed in sections \ref{prim-magnet-NG-section} and \ref{Anisotropic-inflation-section}. As discussed quantum fluctuations of gauge fields can, in principle, provide the seed for primordial magnetic fields, although no successful mechanism is known yet.
Remarkably, gauge fields can leave observable effects on the power spectrum and bi-spectrum (and hence on non-Gaussianity) on the CMB anisotropies. Depending on various couplings that gauge fields can have with other matter fields, and in particular the inflaton field(s), these observable effects will have their own specific features. More interestingly, gauge fields with axion coupling can produce
primordial gravitational waves even for the low scale inflation.

The gauge fields, Abelian or non-Abelian, if they have certain kinetic coupling to inflaton fields can lead to anisotropic Bianchi-type quasi-de Sitter cosmology. This is interesting in two different ways: theoretical and observational. Theoretically, this provides examples of inflationary models with cosmic hair, evading the cosmic no hair conjecture \cite{no-hair-conjecture} and Wald's cosmic no hair theorem \cite{Wald:1983ky}. This latter, of course, happens because these models, although successful (quasi-de Sitter) inflationary models, do not strictly obey the assumptions of Wald's theorem. In fact, motivated by these examples, we proved an extension of Wald's theorem which all models of inflation are subject to it. This theorem is reviewed in Appendix \ref{No-hair-extension-appendix}.
According to this theorem inflationary models in principle can have a cosmic hair, but the amount of anisotropy has an upper bound. This upper bound is stronger for slow-roll inflationary models and in fact models discussed in section \ref{Anisotropic-inflation-section} saturate the bound.

From the CMB observations viewpoint the anisotropic inflationary models have their own significance. There are already bounds on the amount of statistical anisotropy of the CMB temperature fluctuations and these bounds will be improved by the upcoming CMB data. Hence, the anisotropy of the expansion have to be sufficiently small. However, the anisotropic coupling induced by vector-hair can generate the statistical anisotropy, the cross correlation between curvature perturbations and primordial gravitational waves, and the observable non-Gaussianity. As discussed in some detail in section \ref{Anisotropic-inflation-section}, the consistency relation among observables including non-Gaussianity can then provide a tool to discriminate between anisotropic inflationary models. Or conversely, if the consistency relations are confirmed, it will be a strong case for the models discussed in section \ref{Anisotropic-inflation-section}, motivating further studies of these models or building more inflationary models with anisotropic hair.

Gauge fields in inflation is a subject which has received a lot of attention especially in recent years and as we discussed, it has many interesting  theoretical, model building and observational aspects and we expect it to attract more attention in the coming years. This review contains a summary of the current results  and
we believe can set the stage for the researchers who want to work in this field.

\section*{Acknowledgement}
We would like to thank A. Ashoorioon for comments on the draft. 
MMSHJ would like to thank M. Noorbala for discussions. JS would like to thank S.Kanno, K. Murata, K. Yamamoto, S.Yokoyama, and M.Watanabe for fruitful collaboration on this subject. We would like to thank Nicola Bartolo, Sabino Matarrese, Marco Peloso, Angelo Ricciardone for explaining their work, and M. Zarei for contributions to the results quoted in section 6.
Work of AM is supported in part by the grants from \emph{Boniad Melli Nokhbegan of Iran}.   JS is supported by  the
Grant-in-Aid for  Scientific Research Fund of the Ministry of Education, Science and Culture of Japan (C) No.22540274, the Grant-in-Aid for  Scientific Research on Innovative Area No.21111006. 

\appendix

\section{Conventions and notations}\label{Convention-appendix}

Here we fix the conventions used in this article.
\begin{itemize}
\item \emph{Metric signature:} We use mostly plus convention for metric, $(-+++)$ signature.
\item \emph{Volume form}: In our conventions $\epsilon_{\mu\nu\alpha\beta}$ is the Levi-Civita totally anti-symmetric tensor, where $\epsilon_{0123}=-\sqrt{-\det g}$ and hence $\epsilon^{0123}=+\frac{1}{\sqrt{-\det g}}$.
\item \emph{Natural units:} Unless explicitly mentioned, we use units in which speed of light $c$ and Planck constant $\hbar$ are set to one. In these units reduced Planck mass $\mpl$, $8\pi G_N=\mpl^{-2}$, is $$\mpl=2.43\times 10^{18}\ \mathrm{GeV}.$$
It will be convenient to choose $\mpl$ our energy/mass unit and set it to 1. That is, all the dimensionful quantities are measured in Planck units.
\item In our conventions Einstein equation is the equation of motion for the four dimensional Einstein-Hilbert action:
\be\label{EH-action}
S_{EH}=\int d^4x \sqrt{\det g} (\frac12 R)\,,
\ee
where $R$ is the Ricci scalar.
\item For the flat FLRW metric \eqref{FLRW} non-zero components of Ricci tensor are,
\be\label{Ricci-FLRW}
R^0_{\ 0}=\frac{3\ddot a}{a}=3(H^2+\dot H)\,, \quad R^i_{\ j}= \left(\frac{\ddot a}{a}+2(\frac{\dot a}{a})^2+2\frac{K}{a^2}\right)\delta^i_{\ j}=(3H^2+\dot H+2\frac{K}{a^2})\delta^i_{\ j}
\ee
and hence $R=6(2H^2+\dot H+\frac{K}{a^2})$.

\end{itemize}

\section{A quick review on $\delta N$ formulation}\label{deltaN-appendix}

In this appendix, we review the $\delta N$ formalism which is an efficient method for calculating power spectra and non-Gaussianity. This can be regarded as the leading order truncation of the long wavelength approximation which is valid on superhorizon scales. Here we will be very brief and for further readings we refer the reader to \cite{delta-N}. Let us start with FLRW metric in conformal time coordinate $\tau$ with scalar perturbations. Recalling \eqref{metric-pert}, that is
\begin{eqnarray}
   ds^2 = a^2 (\tau ) \left[ - (1+ 2A) d\tau^2  + 2\partial_i B dx^i d\tau + \left\{ (1-2C ) \delta_{ij} + 2\partial_{ij} E \right\} dx^i dx^j  \right] \ ,
\end{eqnarray}
The spatial curvature can be calculated as
\begin{eqnarray}
   \delta_s R = \frac{4}{a^2} \nabla^2 C \ .
\end{eqnarray}
Therefore, we can interpret $C$ as the curvature perturbations.
Using the unit  normal vector $n_\mu$ with components $n_0 =-a (1+A) \ , n_i = 0$, we can compute the shear tensor as
\begin{eqnarray}
   \sigma_{ij} = a \partial_{ij} \sigma_{g}
- \frac{1}{3} a \delta_{ij}\nabla^2 \sigma_g \ ,
\end{eqnarray}
where we defined the shear perturbations $\sigma_g = E' - B$.
The expansion $\theta \equiv n^\mu{}_{;\mu}$ can be calculated as
\begin{eqnarray}
   \theta = 3 \frac{a'}{a^2} \left[  1-A - \frac{a}{a'} C' \right]
+\frac{1}{a} \nabla^2 \sigma_g
\end{eqnarray}
On superhorizon scales, the shear perturbations $\sigma_g$ can be negligible compared to other terms.
Thus, using the proper time $T$ defined by
\begin{eqnarray}
 dT = a (1+A) d\tau \ ,
\end{eqnarray}
we can compute the e-fold number ${\cal N}$
\begin{eqnarray}
 {\cal N} &=& \frac{1}{3}   \int \theta  d T
          = \int \frac{\cal H}{a} \left[ 1-A - \frac{a}{a'} C' \right] a(1+A) d\tau
          = \int \left[ {\cal H} - C'  \right] d\tau \nonumber\\
          &=&  N - \int C' d\tau
\end{eqnarray}
The e-fold number perturbation $\delta N$ may then be defined as
\begin{eqnarray}
   \delta N = {\cal N} - N(t_f) = - C (\tau_f ) + C (\tau_i ) \ ,
\label{delta}
\end{eqnarray}
where $N(t_f)= \int_i^f {\cal H} d\tau = \int_i^f Hdt$, is the background e-fold number at the final time $t_f$.

So far, we have not specified any time slicing.
Now, let us take the flat slicing gauge $ C =0$ on the \emph{initial} hypersurface and fix the comoving gauge $\delta \phi =0$ on the \emph{final} hypersurface.
Then, (\ref{delta}) reads
\begin{eqnarray}
\delta  N = - C (\tau_f ) \ .
\end{eqnarray}
Recalling the definition of the comoving curvature perturbation
\begin{eqnarray}
  {\cal R}_c = C + \frac{\cal H}{\phi'} \delta \phi  \ ,
\end{eqnarray}
${\cal R}_c (\tau_f ) = C (\tau_f ) $ on the final hypersurface, and hence
\begin{eqnarray}
  \delta N = - {\cal R}_c (\tau_f )  \ .
\end{eqnarray}

Under the assumption made in the above derivation, we can read off the functional dependence
\begin{eqnarray}
 {\cal N} = {\cal N} \left( \phi (\tau_i , x^i ) , \tau_f \right)  \ .
\end{eqnarray}
In principle, we also need to take into account the $\phi'$-dependence of ${\cal N}$.
However, we can neglect this dependence  for the slow-roll models.
Thus, we finally obtain the useful relation
\begin{eqnarray}\label{delta-N-phi}
  - {\cal R}_c (\tau_f ) = \delta  N
                         = \frac{\partial   N}{\partial \phi}  \delta \phi (\tau_i , x^i ) + \frac{1}{2}\frac{\partial^2  N}{\partial \phi^2 } \delta \phi (\tau_i , x^i )^2
+ \cdots  \ ,
\end{eqnarray}
where we have used the fact that fluctuations of the $e$-fold number ${\cal N}$ comes from
 fluctuations of the inflaton.

Given this formalism, we can calculate the power spectrum as
\begin{eqnarray}
  \left< {\cal R}_c ({\bf k}) {\cal R}_c ({\bf k}') \right>
= \left( \frac{\partial  N}{\partial \phi}\right)^2
     \left< \delta \phi ({\bf k})\delta \phi ({\bf k}')\right> \ .
\end{eqnarray}
For example, for the quadratic potential  $V=m^2 \phi^2 /2$,
the $e$-fold number can be calculated as
\begin{eqnarray}
    N (\phi)  = \frac{\phi^2}{4\mpl^2} + const. \ .
\end{eqnarray}
Hence, we obtain
\begin{eqnarray}
   \left( \frac{\partial  N}{\partial \phi}\right)^2
   = \frac{\phi^2}{4M_p^4}  = \frac{1}{2\epsilon_V \mpl^2} \ ,
\end{eqnarray}
where we used the slow roll parameter
\begin{eqnarray}
   \epsilon_V = 2 \frac{\mpl^2}{\phi^2} \ .
\end{eqnarray}
Since we know the power spectrum of the canonically normalized scalar field at the horizon crossing
\begin{eqnarray}
    \left< \delta \phi ({\bf k})\delta \phi ({\bf k}')\right>
  = (2\pi)^3 \delta ({\bf k} + {\bf k}' ) \frac{2\pi^2}{k^3}
             \left( \frac{H}{2\pi} \right)^2 \ ,
\end{eqnarray}
we finally obtain
\begin{eqnarray}
  \left< {\cal R}_c ({\bf k}) {\cal R}_c ({\bf k}') \right>
 = (2\pi)^3 \delta ({\bf k} + {\bf k}' ) \frac{2\pi^2}{k^3}
             \frac{1}{8\pi^2 \epsilon_V}\left( \frac{H}{\mpl} \right)^2 \ .
\end{eqnarray}
Similarly, we can calculate the bispectrum. In the case of the quadratic potential, we see that the parameter $f_{\rm NL}$ can be estimated as
\begin{eqnarray}
  f_{\rm NL} \sim
\frac{\frac{\partial^2 N}{\partial \phi^2 } }{   \left( \frac{\partial  N}{\partial \phi}\right)^2 } = \epsilon_V \ ,
\end{eqnarray}
which is the famous result obtained in \cite{Maldacena:2002vr}.

It is straightforward to extend this formalism to multi-field cases. Even in the presence of other fields, as long as the expansion is isotropic, the $\delta N$  formalism would be useful.

\section{Bianchi models, a quick review}\label{Bianchi-appendix}

As it is standard textbook material e.g. see \cite{Ellis:1968vb, Hervik-GR, mathematical-cosmology}, FLRW metrics are the only isotropic and homogeneous spacetimes. Anisotropic, but spatially homogeneous geometries   have also been classified and known as Bianchi models.\footnote{The Kantowski-Sachs model is also a spatially homogeneous model which has a four dimensional symmetry group \cite{Hervik-GR} However, since this solution is unstable we do not consider it here.}
The Bianchi family are geometries with spatially homogeneous (constant $t$) surfaces which are invariant under the action of a  three dimensional Lie group, the symmetry group and  can hence be foliated into the spatial homogeneous hypersurfaces $\Sigma_t$,
$$M=\mathbb{R}\times\Sigma_t,$$
here $\mathbb{R}$ is the time variable. In each of these spatially homogeneous surfaces there exist a set of basis vectors $e_i$ (Killing vectors) that spans a Lie algebra as
\be\label{basis}
[e_i,e_j]=C^k_{~ij}e_k.
\ee
A Bianchi metric can always be written as (e.g. see \cite{MacCallum-Stephani})
\be\label{metr}
ds^2=-dt^2+h_{ij}(t)e^i\otimes e^j,
\ee
where $i,j=1,2,3$ label the coordinates in homogeneous space-like hypersurfaces, $h_{ij}$ is the homogeneous spatial metric and $e^i$ are one-forms dual to basis  $e_i$. Recalling \eqref{basis}, we find the following property for $e^i$s
\be\label{cartan}
\textmd{d}e^i=-\frac12C^i_{~jk}e^j\wedge e^k.
\ee
Moreover, the spatial scalar curvature ${}^{^{(3)\!\!}}R$ is given as
\be\label{R3}
{}^{^{(3)\!\!}}R=-C^i_{~ij}C^{k~j}_{~k}+\frac12C^i_{~jk}C^{k~j}_{~i}-\frac14C_{ijk}C^{ijk},
\ee
which indicates that each three dimensional Lie algebra corresponds to a spatially homogeneous cosmological model.

Bianchi classification categorized the family of three dimensional Lie algebras into 9 different classes in which each algebra is labeled by a number \textit{I}-\textit{IX}. For instance, Bianchi type \textit{I}, which generalized the flat FLRW metric with the symmetry group described by $C^k_{~ij}=0$, corresponds to flat hypersurfaces. Meanwhile, Bianch type \textit{IX} corresponds to the $so(3)$ Lie algebra with a positive ${}^{^{(3)\!\!}}R$. In fact all Bianchi models, expect type \textit{IX}, have a non-positive spatial scalar curvature \cite{Wald:1983ky, MacCallum-Stephani}
\be
{}^{^{(3)\!\!}}R\leq0.
\ee

It is common to write the structure constants $C^k_{~ij}$ in terms of a trace part and a trace-free part, the so called Behr decomposition \cite{Ellis-MacCallum},
\be
C^k_{~ij}=A_{[i}\delta^k_{j]}+\epsilon_{ijk}M^{kl},
\ee
where $M^{jk}$ is a symmetric tensor. The models with $C^j_{~ij}=0$ are called class A models, whiles modes with $C^j_{~ij}\neq0$ are called class B models.

\section{Cosmic no-hair theorem and its extensions}\label{No-hair-extension-appendix}

By the current data the Universe as cosmological scales looks homogeneous and isotropic up to one in $10^5$ part. On the other hand, one of the theoretically attractive feature of inflationary cosmology has been detaching, to a good extent, the outcome of inflation from the initial conditions. These initial conditions, unless severely fine-tuned,   are not homogeneous and isotropic. Therefore, it is very appealing if one can seek a ``dynamical'' explanation for this homogeneity and isotropy. That is, to show that isotropic and homogeneous Universe is an attractor of the cosmic evolution. Within the standard model of cosmology  cosmic evolution is governed by the Einstein gravity coupled to the cosmic fluid, which is not necessarily a perfect fluid and hence if this latter idea is true, homogeneity and isotropy should be an outcome of  Einstein equations. The first such attempt, which was dubbed as \emph{cosmic no-hair conjecture} was made in \cite{no-hair-conjecture} arguing that the late-time behavior of any  accelerating Universe  is an isotropic Universe. The first attempt to prove this conjecture was presented in Wald's seminal paper \cite{Wald:1983ky}, which is called (Wald's) \emph{cosmic no-hair theorem}.

\paragraph{Wald's cosmic no-hair theorem \cite{Wald:1983ky}}states that Bianchi-type models (except Bianchi IX) with the total  energy momentum tensor of the form
$$
T_{\mu\nu}=-\Lambda_0 g_{\mu\nu}+ \tilde T_{\mu\nu},
$$
and with a constant positive $\Lambda_0$ and a $\tilde T_{\mu\nu}$ satisfying \emph{Strong and Dominant Energy Conditions}, respectively SEC and DEC, approach de Sitter space exponentially fast, within a few Hubble times $H^{-1}=\sqrt{3/\Lambda_0}$.  The fate of Bianchi type \textit{IX} is similar, if $\Lambda_0$ is large enough compared with spatial curvature terms. As a result, in these systems\textit{ inflation never ends} and Universe will appear to be an empty de Sitter space. For a more thorough historical review and other related works on cosmic no-hair conjecture/theorem see \cite{mathematical-cosmology} and references therein.

As stated and formulated, Wald's cosmic no-hair theorem is not primarily meant for inflationary models and in fact inflationary models better not obey the fate of Universe predicted by Wald's theorem, because to match the observations, inflation must end and we should enter a phase which is not (quasi) de Sitter at the end of inflation.
This is of course no surprise, because inflationary models, by construct, do not satisfy the DEC and SEC conditions of Wald's theorem (to have a built-in mechanism for terminating inflation). So, Wald's theorem, \emph{per se}, is not  applicable to inflationary models. In \cite{Wald-extended-theorem}, and motivated by the anisotropic inflationary models discussed in detail in section \ref{Anisotropic-inflation-section}, Wald's cosmic no-hair theorem was extended to include inflationary models too. This was done by modifying SEC and DEC to milder conditions respected by requirement of having accelerated expansion.

\paragraph{Inflationary extended cosmic no-hair theorem \cite{Wald-extended-theorem}}states that for general inflationary systems of all Bianchi type with energy momentum tensor of the form
$$
T_{\mu\nu}=-\Lambda(t)g_{\mu\nu}+\mathcal{T}_{\mu\nu},
$$
where $\Lambda(t)$ is a cosmological term which \emph{decreases} by time and $\mathcal{T}_{\mu\nu}$ satisfies SEC and WEC, in contrast to the cosmic no-hair conjecture (and also Wald's cosmic no-hair theorem), anisotropy may grow
nonetheless there is an upper bound on the growth of anisotropy. For slow-roll inflationary models this upper bound is of the order of slow-roll parameters.

In what follows, we first present the general setup in which  Wald's cosmic no-hair theorem and its inflationary extension can be formulated and then present the two theorems.

\subsection{Cosmic no-hair theorems, a general setup}

Consider a general Bianchi model and let $n^\mu$ be the unit tangent vector field of the congruence of
time-like geodesics orthogonal to the homogeneous space-like
hypersurfaces $\Sigma_t$. Then, we obtain the following covariant
form for the spatial metric $h_{\mu\nu}$ \eqref{metr}
\be
h_{\mu\nu}=g_{\mu\nu}+n_\mu n_\nu,
\ee
where $g_{\mu\nu}$ is the
metric of the spacetime. Moreover, $h_{ij}=h_{ij}(t)$ can be decomposed as
\be\label{hdcom}
h_{ij}=e^{2\alpha}e^{2\beta_{ij}},
\ee
where $e^{\alpha}$ is the isotropic scale factor and $\beta_{ij}$ is a traceless matrix which describes the anisotropy.

Extrinsic curvature of $\Sigma_t$ is
defined as
\be
K_{\mu\nu}\equiv \frac12\mathfrak{L}_n
h_{\mu\nu}=\frac12\dot{h}_{\mu\nu}\,,
\ee
where the dot represents derivative with respect to the time coordinate $t$. One can decompose
$K_{\mu\nu}$ into  trace, and  traceless parts
\be\label{decomk}
K_{\mu\nu}=\frac13 Kh_{\mu\nu}+\sigma_{\mu\nu}\,,
\ee
where
\be
\label{hubble}
K(t)=K_{\mu\nu}h^{\mu\nu}=3H(t)\,.
\ee
Here $H(t)=\dot\alpha$ is the Hubble parameter corresponding to the
homogeneous scale factor $e^{\alpha}$ in \eqref{hdcom}. The shear of the time-like
geodesic congruences $\sigma_{\mu\nu}$ is related to the time
derivative of $\beta_{ij}$ and is a symmetric, traceless and purely
spatial tensor:
\be\label{prosigma}
h_{\mu\nu}\sigma^{\mu\nu}=n_\mu\sigma^{\mu\nu}=0\,.
\ee

We now analyze Einstein's equation for Bianchi models
which can be decomposed into  four
constraint equations
\bea \label{initialvalue}
T_{\mu\nu}n^\mu n^\nu&=&\frac12{}^{^{(3)\!\!}}R-\frac12\sigma_{\mu\nu}\sigma^{\mu\nu}+\frac13K^2\,,\\
\label{constraint}
T_{\sigma\lambda}h^\sigma_{~i}n^\lambda&=&K^\sigma_{~j}C^j_{~\sigma
i}+K^\sigma_{~i}C^j_{~\sigma j}\,,
\eea
and six dynamical equations
\bea \label{dyneinst}
(T_{\sigma\lambda}-\frac12h_{\sigma\lambda}T)h^\sigma_{~i}h^\lambda_{~j}=\mathfrak{L}_n\sigma_{ij}+\frac13(\dot Kh_{ij}+K^2h_{ij}+K\sigma_{ij})-2\sigma_{i\lambda}\sigma^\lambda_{~j}+{}^{^{(3)\!\!}}R_{ij}\,.~~~~~~~
\eea
Here ${}^{^{(3)\!\!}}R_{ij}$ is the spatial Ricci tensor and
can be written in terms of the structure-constant tensor $C^i_{jk}$
\eqref{cartan} as
\bea\label{Ricci3}
{}^{^{(3)\!\!}}R_{ij}=\frac14C_{ikl}C_j^{~kl}-C^k_{~kl}C_{(ij)}^{~~~l}-C_{kli}C^{(kl)}_{~~j}\,,
\eea
where $i,j$ indices are raised and lowered with metric $h_{ij}$, and
${}^{^{(3)\!\!}}R={}^{^{(3)\!\!}}R_{ij}h^{ij}$ is the spatial
curvature of $\Sigma_t$  \eqref{R3}.
As mentioned in the Appendix \ref{Bianchi-appendix}, all but one (Bianchi IX) of Bianchi models have  negative spatial
curvature.

Combining trace of \eqref{dyneinst} with \eqref{initialvalue}, we
obtain Raychaudhuri equation
\be\label{ray}
(T_{\mu\nu}-\frac12g_{\mu\nu}T)n^\mu n^\nu=-\dot
K-\frac13K^2-\sigma_{\mu\nu}\sigma^{\mu\nu},
\ee
where $\dot K\equiv\mathfrak{L}_n K$, and contracting \eqref{dyneinst} with
$h^{ij}$ and removing the trace part, we obtain the equation
for the shear tensor:
\be\label{anisotropy}
\dot{\sigma}^i_{~j}+K\sigma^i_{~j}+{}^{^{(3)\!\!}}S^{i}_{~j}= T_{kl}h^{ki}h^l_{~j}-\frac13T_{kl}h^{kl}h^i_{~j}\,,
\ee
where ${}^{^{(3)\!\!}}S^{i}_{~j}$ is the anisotropic part of the spatial 3-curvature
 \be
 {}^{^{(3)\!\!}}S^{i}_{~j}={}^{^{(3)\!\!}}R^{i}_{~j}-\frac13\ {}^{^{(3)\!\!}}Rh^i_{~j}\,.
 \ee
Integrating equation \eqref{anisotropy}, we obtain the following
integral equation for $\sigma^i_{~j}(t)$
\be\label{anisotInt}
\sigma^i_{~j}(t)=e^{-3\alpha(t)}\int_{t_0} ^t
\frac{d(e^{3\alpha(t')})}{K(t')}\big(T_{kl}h^{ki}h^l_{~j}
-\frac13T_{kl}h^{kl}h^i_{~j}-{}^{^{(3)\!\!}}S^{i}_{~j}\big)+c_{~j}^ie^{-3\alpha(t)}\,,
\ee
 with integration constants $c^i_{~j}$. In order to determine
$\sigma^i_{~j}(t)$ we need more information about the energy
momentum tensor $T_{\mu\nu}$.
Before that, let us remind three common energy conditions which are defined as:
\begin{itemize}
\item\textbf{Strong energy condition (SEC)} is satisfied  when for energy momentum tensor $\mathcal{T}_{\mu\nu}$ and for all time-like $t^\mu$
\be\label{SEC}
(\mathcal{T}_{\mu\nu}-\frac12g_{\mu\nu}\mathcal{T})t^\mu t^\nu\geq0.
\ee
\item\textbf{Dominant energy condition (DEC)}  stipulates that for all future-directed causal vectors $t^\mu, t'^\nu$
\be\label{DEC}
\mathcal{T}_{\mu\nu}t^\mu t'^\nu\geq0.
\ee
\item \textbf{Weak energy condition (WEC)} is defined through the dominant energy condition for $t'^\mu=t^\mu$. That is,
\be\label{WEC}
\mathcal{T}_{\mu\nu}t^\mu t^\nu\geq0.
\ee
\end{itemize}

At this point, we consider two different forms for $T_{\mu\nu}$, one leads to the Wald's cosmic no-hair theorem and the other one is the most general form of the energy momentum tensor for an inflationary setting and investigate the dynamical evaluation of anisotropy for each case.

\subsection{Wald's cosmic no-hair theorem}

Following \cite{Wald:1983ky}, here we consider initially expanding general Bianchi models with the following energy momentum tensor
\be\label{WaldT}
T_{\mu\nu}=-\Lambda_0 g_{\mu\nu}+\mathcal{T}_{\mu\nu},
\ee
in which $\Lambda_0$ is a positive cosmological constant and $\mathcal{T}_{\mu\nu}$ satisfies SEC and DEC.
Putting \eqref{WaldT} into \eqref{initialvalue} and \eqref{ray}, we have
\bea\label{k2}
K^2&=&3\Lambda_0+\frac32\sigma^{\mu\nu}\sigma_{\mu\nu}-\frac32\ {}^{^{(3)\!\!}}R+3\mathcal{T}_{\mu\nu}n^\mu n^\nu,\\
\label{dotk}
\dot K&=&\Lambda_0-\frac13K^2-\sigma^{\mu\nu}\sigma_{\mu\nu}-(\mathcal{T}_{\mu\nu}-\frac12g_{\mu\nu}\mathcal{T})n^\mu n^\nu.
\eea
After using \eqref{SEC}, \eqref{WEC} and \eqref{Ricci3}, from the combination of \eqref{k2} and \eqref{dotk} we obtain the following inequalities for all Bianchi types except \textit{IX}
\be
K^2\geq 3\Lambda_0 \quad\textmd{and}\quad \dot K\leq\Lambda-\frac13K^2\leq0.
\ee
Then integrating the inequality as well as using the lower bound on $K$, gives
\be
(3\Lambda_0)^{\frac12}\leq K \leq\frac{(3\Lambda_0)^{\frac12}}{\tanh((\frac{\Lambda_0}{3})^{\frac12}t)}.
\ee
As we see, $K$ is sandwiched between its lower bound $(3\Lambda_0)^{\frac12}$ and an upper bound which exponentially approaches $(3\Lambda_0)^{\frac12}$ with a time scale $(\frac{3}{\Lambda_0})^{\frac12}$.
Applying this result on \eqref{k2} again, we obtain
\be
\sigma^{\mu\nu}\sigma_{\mu\nu}\leq\frac{2\Lambda_0}{\sinh^2((\frac{\Lambda_0}{3})^{\frac12}t)}\quad \textmd{and}\quad  T_{\mu\nu}n^\mu n^\nu\leq\frac{\Lambda_0}{\sinh^2((\frac{\Lambda_0}{3})^{\frac12}t)}.
\ee
Thus, the shear of the homogeneous hypersurfaces as well as the energy density of $\mathcal{T}_{\mu\nu}$ rapidly approach zero. Since $\mathcal{T}_{\mu\nu}$ satisfies DEC, $\mathcal{T}_{\mu\nu}n^\mu n^\nu$ is larger or equal
than the rest of orthonormal frame elements of $\mathcal{T}_{\mu\nu}$. Hence, the homogeneous hypersurfaces and all components of $\mathcal{T}_{\mu\nu}$ exponentially approach zero, as such the geometry, regardless of the initial conditions, evolve toward the de Sitter solution.

\subsection{Inflationary extended cosmic no-hair theorem}
Following \cite{Wald-extended-theorem}, here we consider inflationary Bianchi models with the following energy momentum tensor
\be\label{ourT}
T_{\mu\nu}=-\Lambda(t) g_{\mu\nu}+\mathcal{T}_{\mu\nu},
\ee
in which $\Lambda(t)$ is a \emph{positive but decreasing in time} cosmological term and $\mathcal{T}_{\mu\nu}$ satisfies SEC and WEC. Energy momentum tensors of inflationary models fall into the class we consider here.

Without loss of generality $\mathcal{T}_{\mu\nu}$ in \eqref{ourT} may be decomposed
as
\be\label{tildeT}
\mathcal{T}_{\mu\nu}(t)=\tilde\rho(t) n_\mu
n_\nu+\tilde P(t) h_{\mu\nu}+\pi_{\mu\nu}(t),
\ee
 where $\pi_{\mu\nu}$ is the anisotropic stress tensor which is symmetric,
traceless, and purely spatial:
\be
\pi_{\mu\nu}h^{\mu\nu}=0\,\,\,\,\&\,\,\,\, \pi_{\mu\nu}n^\mu=0\,.
\ee
We also define $\Lambda_0\equiv 3H^2_0$, where subscript $0$
denotes the initial value (at the beginning of inflation) and
\be\label{deLambda}
 \Lambda(t)=\Lambda_0+\delta\Lambda(t)\,.
 \ee
Moreover, the spatial curvature
${}^{^{(3)\!\!}}R$ has a time dependence proportional to
$e^{-2\alpha}$ and this term is damped quickly in inflationary
systems. Therefore, using \eqref{ourT} and \eqref{tildeT} in \eqref{initialvalue} and \eqref{ray}, after a few e-folds ${}^{^{(3)\!\!}}R$ is negligible and we approximately have
\bea\label{1}
-\frac{\delta\Lambda(t)}{3H^2(t)}-\frac{\tilde \rho(t)-\tilde P(t)}{6H^2(t)}&\simeq&\frac{\epsilon(t)}{3}+\frac{H_0^2}{H^2(t)}-1,\\
\label{2} \frac{\tilde\rho(t)+\tilde
P(t)}{2H^2(t)}+\frac{\sigma_{\mu\nu}\sigma^{\mu\nu}}{2H^2(t)}&\simeq&\epsilon(t). \eea

On the other hand, $\mathcal{T}_{\mu\nu}$ satisfies SEC which
choosing $t_\mu=n_\mu+s_\mu$, where $s_\mu$ is a normalized
arbitrary space-like 4-vector orthogonal to $n_\mu$, implies
\be\label{DEC-SEC}
\begin{split}
\tilde\rho(t)+\tilde P(t)+\pi_{ij}(t)s^is^j\geq0,\quad i,j=1,2,3\,,
\end{split}
\ee where $s^i=s_\mu h^{i\mu}$. Recalling the fact that $\pi_{ij}$
is traceless and the above inequality should  hold for all $s^i$,
one obtains\footnote{Although we have not used here, we remind that
WEC on $\mathcal T_{\mu\nu}$  imply $\tilde \rho\geq 0$.}
\be\label{upPi}
\frac{|\pi^i_{~j}(t)|}{H^2(t)}\leq -4\frac{\dot
H(t)}{H^2(t)}=4\epsilon(t)\,\quad \forall i,j. \ee
{}From the
combination of \eqref{1} and \eqref{2} and noting that
\eqref{DEC-SEC} implies $\tilde\rho(t)+\tilde P(t)\geq 0$ we obtain
\be\label{sigma2-bound}
\begin{split}
\sigma_{\mu\nu}\sigma^{\mu\nu}&\leq 2H^2(t)\epsilon(t)\,,\\
\delta\Lambda(t) &\leq -(\tilde\rho
(t)+\frac{\sigma_{\mu\nu}\sigma^{\mu\nu}}{2})\leq 0\,.
\end{split}
\ee The above already implies a weak upper bound on anisotropy
\be\label{sigma-weak-bound} |\sigma^i_{~j}|\leq
H(t)\sqrt{2\epsilon}\,. \ee As we will see later,
assuming slow-roll leads to a stronger bound.

During inflation (on the average) $\epsilon$ is an increasing
quantity and this result among other things shows the possibility of
growth for the anisotropic part of the stress tensor. This is in
contrast with behavior of systems described by the Wald's theorem,
in which all elements of $\mathcal{T}_{\mu\nu}$, including
$\pi^i_{~j}(t)$'s, are damped exponentially with a time scale
$H_0^{-1}$ \cite{Wald:1983ky}.

Inserting \eqref{tildeT} into \eqref{anisotInt}, we have the
following form for the Hubble-normalized shear tensor
 \be\label{diagFinal}
\frac{\sigma^i_{~j}(t)}{H(t)}\simeq\frac{1}{3H(t)e^{3\alpha(t)}}\int^t_{\tilde
t_0}{\frac{d(H(t')e^{3\alpha(t')})}{(1+\frac{\dot
H(t')}{3H^2(t')})}\ \frac{\pi^i_{~j}(t')}{H^2(t')}}\,,
 \ee
  where $\tilde t_0$ is a few e-folds after the beginning of inflation in which the spatial curvature terms has become  negligible. As
we see $\frac{\pi^i_{~j}(t)}{H^2(t)}$ acts as a source term for anisotropy $\sigma^i_{\  j}$ and its
five degrees of freedom may be determined  for specific  models of
inflation.

For models with perfect fluid  $T_{\mu\nu}$, in which
$\pi^i_{~j}(t)$ is identically zero, shear tensor $\sigma^i_{~j}(t)$
is  damped exponentially fast. Thus, in this kind of systems
inflation washes away any initial anisotropy and isotropizes the
system. These models hence comply with the cosmic no-hair conjuncture. However, this is
\emph{not} necessarily the case in a general inflationary model with
non-zero anisotropic stress tensor  $\pi^i_{~j}(t)$. In fact, in presence of anisotropic stress tensor $\pi^i_{~j}$
elements of shear tensor $\sigma^i_{~j}(t)$ \emph{are allowed} to
grow by inflationary dynamics, which is against the cosmic no-hair
conjuncture. Although $\sigma^i_{~j}(t)$ can increase, as one
expects from \eqref{upPi}, inflation enforces an upper bound
on how large they can grow.

\subsection*{Quasi-de Sitter expansion}

In the following we focus on the slow-roll models. We  investigate the dynamics of system  assuming that ${\pi^i_{~j}(t)}/{H^2(t)}$ always saturates its maximum value,
and determine the possible upper bound value of anisotropies during slow-roll inflation.

From \eqref{sigma2-bound} and \eqref{upPi}, we find that $\mathcal T_{\mu\nu}$, $\sigma_{\mu\nu}\sigma^{\mu\nu}$ and ${}^{^{(3)\!\!}}R$ are at most of the order $\epsilon H^2$.
 Hence recalling \eqref{deLambda} and using \eqref{initialvalue}, we obtain
\bea
H(t)&\simeq&H_0\big(1-\epsilon_0H_0t\big),
\eea
where $\simeq$ means to the first order in slow-roll parameters, subscript 0 denotes the initial value and $H_0=\sqrt{\Lambda_0/3}$.  Inserting the above relation in \eqref{1} and \eqref{2}, we obtain \footnote{Note that we are considering simple slow-roll models where $\epsilon(t)$ is an always increasing function during inflation, $\dot\epsilon>0$. From the definition of $\epsilon$ \eqref{epsilon-eta-def} we learn that $\frac{\dot\epsilon}{H\epsilon}=2(\epsilon-\eta)$.}
\bea\label{sloweq}
\frac{\delta\Lambda}{H^2}+\frac{\tilde\rho(t)-\tilde P(t)}{2H^2(t)}&\simeq&-\epsilon_0(1+6 H_0 t),\\
\label{freeeq}
\frac{\tilde\rho(t)+\tilde P(t)}{2H^2(t)}+\frac{\sigma_{\mu\nu}\sigma^{\mu\nu}}{2H^2(t)}&\simeq&\epsilon_0\big(1+2(\epsilon_0-\eta_0)H_0t\big).
\eea
As we see in \eqref{freeeq}, $\sigma^{\mu\nu}\sigma_{\mu\nu}$ is not directly related to the slow-roll dynamics, but the combination of $\frac{\tilde\rho(t)+\tilde P(t)}{2H^2(t)}+\frac{\sigma_{\mu\nu}\sigma^{\mu\nu}}{2H^2(t)}$ should be slow varying during inflation. As a result,  $\sigma_{\mu\nu}$ is not fully determined by the slow-roll dynamics, while \eqref{sloweq} implies that the isotropic part of the system is governed by the slow-roll. Thus, assuming slow-roll inflation, anisotropies can still evolve quickly in time.

Applying slow-roll approximation in \eqref{diagFinal}, we obtain the following form for the late time behavior of the Hubble-normalized shear tensor $\frac{\sigma^i_{~j}(t)}{H(t)}$
 \be\label{sigmasl}
\frac{\sigma^i_{~j}(t)}{H(t)}\simeq H_0(1+\epsilon_0H_0t)e^{-3H_0t}\int^t_{t_0}{dt'e^{3H_0t'}
(1-2\epsilon_0H_0t')\frac{\pi^i_{~j}(t')}{H^2(t')}}.
\ee
In order to determine $\frac{\sigma^i_{~j}(t)}{H(t)}$, we need $\pi^i_{~j}(t)$'s five degrees of freedom which should be provided by  independent equations. However, regardless of the particular inflationary model which we consider, recalling \eqref{upPi}, slow-roll enforces an upper bound on the anisotropy growth. Rewriting \eqref{upPi} in terms of slow-roll parameters, we obtain
\be\label{SLupPi}
 \frac{|\pi^i_{~j}(t)|}{H^2(t)} \leq 4\epsilon(t)\simeq4\epsilon_0(1+2(\epsilon_0-\eta_0)H_0t))\,.
 \ee
As mentioned before, from the combination of \eqref{2} and \eqref{diagFinal}, we realize that during the slow-roll inflation, ${\pi^i_{~j}(t)}/{H^2(t)}$ can saturate its upper bound and grow in time.
Then, assuming that ${\pi^i_{~j}}(t)/{H^2(t)}$ always saturates its upper bound during the slow-roll, we can find an upper bound value on the value of ${\sigma^i_{~j}}(t)/{H(t)}$ enforced by the slow-roll dynamics.

Inserting \eqref{SLupPi}
into \eqref{sigmasl}, we obtain the upper bound value on the  anisotropies at the end of slow-roll inflation
\be\label{upperSigma}
\frac{|\sigma^i_{~j}|}{H}\bigg|_{t_{sl}}\leq\frac83(\epsilon_0-\eta_0).
\ee
where $t_{sl}$ is the end of slow-roll inflation and approximately is $H_0\epsilon_0t_{sl}\simeq1$.
Thus, in a general model of slow-roll inflation we end up with an almost isotropic Universe, with a shear of the order $\epsilon_0$. Note that slow-roll dynamics has improved the bound \eqref{sigma-weak-bound} to \eqref{upperSigma}.

To summarize, we have proved an inflationary extended  no-hair theorem, according which despite the fact that anisotropies can in principle grow during inflation (when the anisotropic part of energy momentum tensor of the inflaton sector is non-zero\footnote{For the scalar-driven inflation models there is no anisotropic stress and hence \eqref{diagFinal} implies exponential damping for anisotropy $\sigma_{\mu\nu}$. Recalling the equivalence of $f(R)$ gravity models with general relativity plus a scalar field matter, the same conclusion also applies for the scalar-driven inflationary models within $f(R)$ modified gravity models.}), having quasi-de Sitter expansion puts an upper bound on anisotropy: Hubble-normalized anisotropic stress ${\pi^i_{~j}(t)}/{H^2(t)}$,
proportional to the slow-roll parameter $\epsilon(t)$ and consequently an upper on the dimensionless anisotropies
${\sigma^i_j(t)}/{H(t)}\leq 2\sqrt{\epsilon}$. Slow-roll assumption improves this bound to $\frac{|\sigma^i_{~j}|}{H}\lesssim \epsilon_0$.

Finally, we end this Appendix with some comments:\\ 1){Note that $\epsilon$ is generically a growing function as inflation progresses, and hence in principle, anisotropy can grow, signaling the possibility of violation of cosmic no-hair conjecture. In fact, as reviewed in section \ref{Anisotropic-inflation-section}, explicit models of inflation have been constructed exhibiting this \cite{Watanabe:2009ct}. All the models discussed in section \ref{Anisotropic-inflation-section} nicely comply with our inflationary extended no-hair theorem.}\\
2) Even assuming slow-roll inflation, the dynamics of anisotropy does not necessarily follow the slow-roll dynamics and they can generally  evolve quickly, and\\ 3) our analysis is general and our upper bounds is expected to hold for all Bianchi models, including type IX.

\end{document}